\pdfoutput=1

\documentclass[11pt,a4paper]{article}
\usepackage{mciteplus}
\usepackage{amssymb}
\usepackage{amsthm}
\usepackage{amstext}
\usepackage{amsmath}
\usepackage{amsfonts}
\usepackage{verbatim}
\usepackage{geometry}
\usepackage{latexsym}
\usepackage{t1enc}
\usepackage{graphicx}
\usepackage[all]{xy}
\usepackage{makeidx}
\usepackage{slashed}
\usepackage{multicol}
\UseRawInputEncoding
\usepackage[numbers, square, comma, sort&compress]{natbib}

\numberwithin{equation}{section}

\begin{document}

\begin{titlepage}
\hbox{Saclay IPhT--T12/027} \hbox{UUITP-12/12}

\vskip 20mm

\begin{center}
\Large{\textbf{Global Poles of the Two-Loop Six-Point \\[0.5mm] $\mathcal{N}=4$ SYM integrand}}
\end{center}

\vskip 6mm

\begin{center}
Kasper J. Larsen

\vskip 8mm

\textit{Department of Physics and Astronomy,} \\[2mm]
\textit{Uppsala University, SE-75108 Uppsala, Sweden} \\[5mm]
\textit{Institut de Physique Théorique, CEA-Saclay,} \\[2mm]
\textit{F-91191 Gif-sur-Yvette cedex, France}\\[5mm]
\textit{School of Natural Sciences}\\[2mm]
\textit{Institute for Advanced Study, Princeton, NJ 08540, USA}\\[8mm]

{\texttt{Kasper.Larsen@cea.fr}}\\[11mm]

{\Large Abstract}

\end{center}

\vskip -1mm

\noindent Recently, a recursion relation has been developed, generating the
four-dimensional integrand of the amplitudes of $\mathcal{N}=4$ supersymmetric Yang-Mills
theory for any number of loops and legs. In this paper, I provide a comparison
of the prediction for the two-loop six-point maximally helicity-violating (MHV) integrand
against the result obtained by use of the leading singularity method. The
comparison is performed numerically for a large number of randomly selected
momenta and in all cases finds agreement between the two results to high numerical
accuracy.

\end{titlepage}

\section{Introduction}

Scattering amplitudes in gauge theories are fascinating quantities
partly because they provide a direct link between theory and
experiment and partly because their investigation continues to
uncover rich structures in quantum field theory. Our understanding
of computing amplitudes has undergone a revolution over the past
decade and a half, owing in large part to the development of
on-shell recursion relations for tree-level amplitudes
\cite{Britto:2004ap,Britto:2005fq} and a purely on-shell formalism
for loop-level amplitudes, the modern unitarity method
\cite{Bern:1994zx,Bern:1994cg,Bern:1996je,Bern:1995db,Bern:1997sc,Bern:1996ja,
Britto:2004nc,Britto:2004nj,Bidder:2004tx,Bidder:2005ri,Bidder:2005in,Bern:2005hh,
BjerrumBohr:2007vu,Bern:2005cq,Britto:2005ha,Britto:2006sj,Mastrolia:2006ki,
Brandhuber:2005jw,Ossola:2006us,Bern:2007dw,Forde:2007mi,Badger:2008cm,
Anastasiou:2006jv,Anastasiou:2006gt,Giele:2008ve,Britto:2006fc,Britto:2007tt,
Britto:2008vq,Britto:2008sw,Berger:2009zb,Bern:2010qa,Kosower:2011ty}.
These powerful modern methods have to a large extent made the more
traditional approach of Feynman diagrams obsolete for tree-level
and one-loop amplitudes. Thus, the current frontier is the
development of systematic approaches for computing two-loop
amplitudes.

The maximally supersymmetric gauge theory in four dimensions, the
$\mathcal{N}=4$ super Yang-Mills theory, has attracted a great
deal of effort over the past several years, serving as a
laboratory for testing new ideas. This has lead to the discovery
of a surprising new symmetry in the planar limit, the so-called
dual conformal symmetry, which, although not inherited from the
theory's Lagrangian in any obvious way, is nonetheless a property
of its amplitudes. This symmetry combines with the standard superconformal
symmetry, dictated by the Lagrangian, to form an
infinite-dimensional symmetry algebra, the Yangian of the
superconformal group. This integrable structure has been
intensively studied
\cite{Drummond:2008vq,Berkovits:2008ic,Beisert:2008iq,Drummond:2009fd,
Bargheer:2009qu,Brandhuber:2009xz,Korchemsky:2009hm,Sever:2009aa,
Alday:2009dv,Beisert:2010gn,Alday:2010vh,Drummond:2010uq,
Korchemsky:2010ut,Alday:2010ku,Henn:2011xk,Bullimore:2011kg} and
has recently been exploited successfully to determine the theory's
scattering amplitudes recursively in the number of loops
\cite{CaronHuot:2011kk}. Another spectacular recent advance,
following earlier insights based on a reformulation of the S
matrix of $\mathcal{N}=4$ SYM as a contour integral on a
Grassmannian manifold
\cite{ArkaniHamed:2009dn,Mason:2009qx,ArkaniHamed:2009vw,Bullimore:2009cb,
Kaplan:2009mh,ArkaniHamed:2009dg,Bourjaily:2010kw,Bullimore:2010pa}
--- but this time extendable to any planar supersymmetric gauge
theory --- has been the development of recursion relations à la
BCFW for the (strictly four-dimensional) loop integrand
\cite{ArkaniHamed:2010kv,Boels:2010nw,ArkaniHamed:2010gg,ArkaniHamed:2010gh}.
\\
\\
Ensuring the soundness of new ideas requires performing careful
tests of the results they produce. In this spirit, the question we
wish to address in this paper is whether the result in refs.
\cite{ArkaniHamed:2010kv,ArkaniHamed:2010gh} for the two-loop
six-point MHV integrand of $\mathcal{N}=4$ SYM theory can be
reproduced by more traditional methods. In the
generalized-unitarity based approach, an $L$-loop amplitude is
expressed as a linear combination of known basis integrals, plus
terms that are rational functions of the external momenta (and which will
not be discussed in this paper),
\begin{equation}
{\rm Amplitude} = \sum_{j\in {\rm Basis}}
  {\rm coefficient}_j {\rm Integral}_j +
{\rm Rational} \: . \label{BasicEquation}
\end{equation}
The coefficients of the integrals (evaluated in dimensional regularization) are obtained by changing the
integration range from $(\mathbb{R}^{4-2\epsilon})^{\otimes L}$ into specific
contours $\sigma$ of real dimension $4L$, embedded in
$\mathbb{C}^{4L}$ and encircling the points where the maximum
number of denominators of the integrand become zero. Unlike the
path followed in the leading singularity method
\cite{Buchbinder:2005wp,Cachazo:2008dx,Cachazo:2008vp,Cachazo:2008hp,Spradlin:2008uu}
in which one allows any choice of contour $\sigma$ (by virtue of
first having expanded the amplitude in an artfully chosen,
typically overcomplete, basis), in generalized unitarity
the contours $\sigma$ are subject to the constraint that any
function which integrates to zero on $(\mathbb{R}^{4-2\epsilon})^{\otimes L}$
must also integrate to zero on $\sigma$ \cite{Kosower:2011ty}. As
argued in this paper, multidimensional contours $\sigma$
satisfying this consistency condition are guaranteed to produce
correct results for scattering amplitudes in any gauge theory, not
only $\mathcal{N}=4$ SYM theory. The change of integration contour
has the effect of transforming the integrals in eq.
(\ref{BasicEquation}) into contour integrals (which are easily
evaluated by taking residues). By making the various allowed
choices of contours $\sigma$ one produces a set of linear
equations satisfied by the integral coefficients which can then be
solved to determine them uniquely.

The set of linear relations between two-loop integrals includes
the set of all integration-by-parts (IBP) identities
\cite{Gluza:2010ws,Tkachov:1981,Chetyrkin:1981,Laporta:2000dc,Laporta:2001dd,
Gehrmann:1999as,Lee:2008tj,Anastasiou:2004vj,Smirnov:2008iw,Studerus:2009ye,
Smirnov:2010hn,Brown:1952,Brown:1961,Petersson:1965,Kallen:1965,Melrose:1965,
Passarino:1979,Neerven:1984,Oldenborgh:1990} between the various
tensor integrals arising from the Feynman rules of gauge theory;
however, at present a complete knowledge of such relations
involving six-point two-loop integrals is not available
\cite{David:privcomm}. For this reason, in this paper we will
carry out the analysis following the leading singularity method,
allowing any contour $\sigma$. This approach was shown to
reproduce the result of a unitarity-based calculation
\cite{Bern:2008ap} for the parity-even part of the two-loop
six-point MHV amplitude in ref. \cite{Cachazo:2008hp}. By expressing
the full two-loop amplitude (i.e., including both parity-even and
odd parts) in terms of the same basis as used in ref.
\cite{Cachazo:2008hp}, we therefore expect the leading singularity
method to also produce the correct result for the parity-odd part
of this amplitude.

The results of refs. \cite{ArkaniHamed:2010kv,ArkaniHamed:2010gh}
are expressed in terms of a different basis from that of ref.
\cite{Cachazo:2008hp}, and thus it is not meaningful to check
agreement between individual integral coefficients in either
representation. A quantity that can be meaningfully compared is
the two-loop integrand: in general, in the planar limit of any
field theory, the loop integrand is a well-defined rational
function of the external momenta (which for example can be thought
of as being produced by the Feynman rules). We have evaluated the
integrand for a large number of randomly selected momenta and in
all cases find agreement with the recent literature to high
numerical accuracy \cite{Simon:privcomm}.

An interesting spinoff of the calculation in this paper are the
potential applications of the intermediate results (in particular,
the enumeration of the global poles of the integrand and the
expressions for the cut integrals). Once all the necessary IBP
relations do become available, we expect these results to
greatly facilitate the task of determining the maximal-cut
contours that allow the extraction of integral coefficients in any
two-loop six-point gauge theory amplitude.

A complementary approach is that of
refs.~\cite{Mastrolia:2011pr,Badger:2012dp,Mastrolia:2012an,Kleiss:2012yv,Badger:2012dv},
in which the heptacut integrand is reconstructed by
polynomial matching in similarity with the OPP approach \cite{Ossola:2006us}.
A recent paper by Zhang~\cite{Zhang:2012ce} adds tools required in
such an approach for reducing integrands to a basis of monomials.
\\
\\
This paper is organized as follows. In Section
\ref{sec:notation_and_conventions}, we explain conventions and
introduce notation used throughout the paper. In Section
\ref{sec:two-loop_heptacuts}, we compute the heptacuts of the
general double box integral and of the factorized double box with
six massless legs. In Section \ref{sec:global_poles_of_integrand},
we apply these heptacuts to the ansatz for the two-loop six-point
MHV amplitude of $\mathcal{N}=4$ SYM and employ the leading
singularity method to obtain linear equations satisfied by the
integral coefficients. We discuss how these equations can be used
to directly obtain the parity-even part of the integrand. We then
explain how to evaluate numerically the full $\mathcal{N}=4$ SYM
integrand (i.e., including both parity-even and odd parts) as
obtained by the leading singularity method and report agreement
with the result in refs.
\cite{ArkaniHamed:2010kv,ArkaniHamed:2010gh}. Finally, in Section
\ref{sec:conclusions} we provide conclusions and suggest
directions for future investigation. In Appendix
\ref{sec:appendix}, we provide full details of all the heptacuts
of the two-loop six-point MHV amplitude.

\section{Notation and conventions}\label{sec:notation_and_conventions}

In this section we explain conventions and introduce notation used
throughout the paper.

All external momenta in an amplitude are outgoing and will be
denoted by $k_i$. We will make use of the spinor helicity
formalism
\cite{Berends:1981,DeCausmaecker:1982,Xu:1984,Kleiss:1985,
Gunion:1985,Xu:1987,Mangano:1991,Dixon:1996wi} in which a given
massless four-dimensional momentum is written as a tensor product
of two massless Weyl spinors,
\begin{equation}
k_i^\mu \hspace{1mm}=\hspace{1mm} \overline{u_+ (k_i)}
\hspace{0.3mm} \sigma^\mu \hspace{0.2mm} u_+ (k_i)
\hspace{1mm}=\hspace{1mm} \overline{u_- (k_i)} \hspace{0.3mm}
\sigma^\mu \hspace{0.2mm} u_- (k_i) \: .
\end{equation}
We define the spinors
\begin{equation}
\lambda_i = u_+ (k_i) \: , \hspace{10mm} \widetilde{\lambda}_i =
u_- (k_i)
\end{equation}
and the Lorentz invariant inner products formed out of the
spinors,
\begin{equation}
\langle i j \rangle = \langle i^- | j^+ \rangle = \overline{u_-
(k_i)} \hspace{0.3mm} u_+ (k_j) \: , \hspace{10mm} [i j] = \langle
i^+ | j^- \rangle = \overline{u_+ (k_i)} \hspace{0.3mm} u_- (k_j)
\end{equation}
which satisfy
\begin{equation}
\langle ij \rangle [ji] = 2k_i \cdot k_j \: .
\end{equation}
Spinor strings are defined as follows
\begin{align}
\langle K_i^- | \slashed{P} + \slashed{Q} | K_j^- \rangle &=
\langle K_i^- | \slashed{P} | K_j^- \rangle + \langle K_i^- | \slashed{Q} | K_j^- \rangle \\
\langle K_i^- | \slashed{P} | K_j^- \rangle &= \langle K_i P
\rangle [P K_j] \hspace{6mm} \mbox{if \hspace{2mm} $P^2 = 0$} \: .
\end{align}
We will use the following notation for sums and invariant masses
of external momenta,
\begin{eqnarray}
k_{i_1 \cdots i_n} &\equiv& k_{i_1} + \cdots + k_{i_n} \\
s_{i_1 \cdots i_n} &\equiv& (k_{i_1} + \cdots + k_{i_n})^2 \\
S_i &\equiv& K_i^2 \: .
\end{eqnarray}
Throughout we will make use of the ``flattened'' momenta
introduced in ref.~\cite{Ossola:2006us,Forde:2007mi}: for a
pair of momenta $K_1, K_2$, define the quantity
\begin{equation}
\gamma_{1 \pm} = (K_1 \cdot K_2) \pm \sqrt{\Delta_1} \: ,
\hspace{10mm} \Delta_1 = (K_1 \cdot K_2)^2 - K_1^2 K_2^2
\label{eq:def_of_gamma_1_pm}
\end{equation}
which can take two different values if both momenta are massive
(i.e., if $S_1 S_2 \neq 0$). For a given value of $\gamma_1$ one
defines a pair of massless ``flattened'' momenta as follows
\begin{equation}
K_{1 \pm}^\flat = \frac{K_1 - (S_1/\gamma_{1 \pm}) K_2}{1 - S_1
S_2/\gamma_{1 \pm}^2} \: , \hspace{10mm} K_{2 \pm}^\flat =
\frac{K_2 - (S_2/\gamma_{1 \pm}) K_1}{1 - S_1 S_2/\gamma_{1
\pm}^2} \: . \label{eq:def_of_Kflat_1_and_Kflat_2_pm}
\end{equation}
If one of the momenta $K_1$ or $K_2$ is massless, $\gamma_{1 \pm}$
can only take one value, and we will use the following abbreviated
notation:
\begin{equation}
S_1 S_2 \hspace{0.5mm}=\hspace{0.5mm} 0 \hspace{6mm}
\Longrightarrow \hspace{6mm} \left\{ \begin{aligned} \gamma_1
\hspace{0.5mm}&=\hspace{0.5mm} 2 K_1 \cdot K_2
\\ K_1^\flat
\hspace{0.5mm}&=\hspace{0.5mm} K_1 - (S_1/\gamma_1) K_2 \\
K_2^\flat \hspace{0.5mm}&=\hspace{0.5mm} K_2 - (S_2/\gamma_1) K_1
\: .
\end{aligned} \right.  \label{eq:def_of_gamma_1}
\end{equation}
\noindent Similarly, we will use the notation
\begin{equation}
\gamma_{2 \pm} = (K_4 \cdot K_5) \pm \sqrt{\Delta_2} \: ,
\hspace{10mm} \Delta_2 = (K_4 \cdot K_5)^2 - K_4^2 K_5^2
\label{eq:def_of_gamma_2_pm}
\end{equation}
\begin{equation}
K_{4 \pm}^\flat = \frac{K_4 - (S_4/\gamma_{2 \pm}) K_5}{1 - S_4
S_5/\gamma_{2 \pm}^2} \: , \hspace{10mm} K_{5 \pm}^\flat =
\frac{K_5 - (S_5/\gamma_{2 \pm}) K_4}{1 - S_4 S_5/\gamma_{2
\pm}^2} \: . \label{eq:def_of_Kflat_4_and_Kflat_5_pm}
\end{equation}
\begin{equation}
S_4 S_5 \hspace{0.5mm}=\hspace{0.5mm} 0 \hspace{6mm}
\Longrightarrow \hspace{6mm} \left\{
\begin{aligned} \gamma_2 \hspace{0.5mm}&=\hspace{0.5mm} 2 K_4 \cdot K_5 \\ K_4^\flat \hspace{0.5mm}&=\hspace{0.5mm}
K_4 - (S_4/\gamma_2) K_5 \\ K_5^\flat
\hspace{0.5mm}&=\hspace{0.5mm} K_5 - (S_5/\gamma_2) K_4 \: .
\end{aligned} \right. \label{eq:def_of_gamma_2}
\end{equation}
Finally, we will denote the elements of the dihedral group $D_6$
as follows
\begin{equation}
\begin{array}{rrr}
\sigma_1 = (1,2,3,4,5,6) \hspace{0.8cm} & \sigma_2 = (2,3,4,5,6,1)
\hspace{0.8cm} & \sigma_3 = (3,4,5,6,1,2) \phantom{\: .} \\
\sigma_4 = (4,5,6,1,2,3) \hspace{0.8cm} & \sigma_5 = (5,6,1,2,3,4)
\hspace{0.8cm} & \sigma_6 = (6,1,2,3,4,5) \phantom{\: .} \\
\sigma_7 = (6,5,4,3,2,1) \hspace{0.8cm} & \sigma_8 = (5,4,3,2,1,6)
\hspace{0.8cm} & \sigma_9 = (4,3,2,1,6,5) \phantom{\: .} \\
\sigma_{10} = (3,2,1,6,5,4) \hspace{0.8cm} & \sigma_{11} =
(2,1,6,5,4,3) \hspace{0.8cm} & \sigma_{12} = (1,6,5,4,3,2) \: .
\end{array}
\end{equation}

\section{Heptacut two-loop integrals}\label{sec:two-loop_heptacuts}

As preparation for computing the generalized-unitarity cuts of the two-loop
six-point amplitude, in this section we compute the heptacuts of the general double box
integral and the factorized double box integral, illustrated below
in Figs. \ref{fig:General_double_box} and
\ref{fig:factorized_double_box}. These integrals are part of the
linear bases in which the two-loop six-point $\mathcal{N}=4$ SYM
helicity amplitudes were expanded in ref.~\cite{Bern:2008ap,Cachazo:2008hp,Kosower:2010yk}. In this paper we
shall use the (overcomplete) basis in ref.~\cite{Cachazo:2008hp}
(illustrated for convenience in Fig.
\ref{fig:6-point_integral_basis} in Section
\ref{sec:global_poles_of_integrand}). As explained further in Section
\ref{sec:heptacut_1}, the heptacuts of the remaining integrals in
this basis are easily obtained from the heptacuts of the double
box integrals by multiplying additional factors (involving
propagators and possibly numerator insertions) onto the integrand
of the latter. From the knowledge of the generalized cuts of all basis
integrals, the cuts of the amplitude are easily obtained, as will be explained
further in Section \ref{sec:global_poles_of_integrand}.

We emphasize that the generalized cuts considered throughout this
paper are strictly four-dimensional (as opposed to
$(4-2\epsilon)$-dimensional). Moreover, we will suppress the
Feynman $i\varepsilon$-prescription when writing propagators.
Finally, the internal lines in all diagrams are taken to be
massless.

\subsection{The maximal cut of the general double box}\label{sec:general_double_box}

In this section we will compute the maximal cut of
the double box integral

\begin{figure}[!h]
\begin{center}
\includegraphics[angle=0, width=0.45\textwidth]{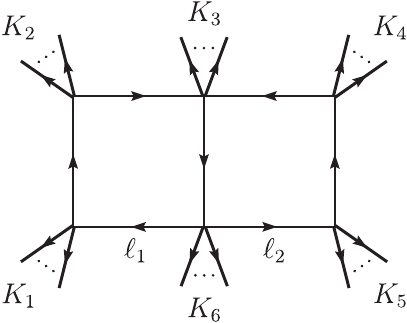}
\caption{The general double-box integral.}\label{fig:General_double_box}
\end{center}
\end{figure}

\noindent where the vertex momenta $K_i$ shown in Fig. \ref{fig:General_double_box}
are typically sums of several lightlike momenta, $K_i = k_{i_1} + \cdots + k_{i_n}$ with
$k_{i_j}^2 = 0$. Setting $D=4-2\epsilon$, this integral is defined by
\begin{equation}
\int \frac{d^D \ell_1}{(2 \pi)^D} \frac{d^D \ell_2}{(2 \pi)^D}
\left( \frac{1}{\ell_1^2} \frac{1}{(\ell_1-K_1)^2}
\frac{1}{(\ell_1-K_1-K_2)^2} \frac{1}{(\ell_1+\ell_2+K_6)^2}
\frac{1}{\ell_2^2} \frac{1}{(\ell_2-K_5)^2}
\frac{1}{(\ell_2-K_4-K_5)^2} \right)
\label{eq:scalar-doublebox}
\end{equation}
where the integration is over real Minkowski space for both loop
momenta. By considering the heptacut of the general double box integral
we can easily obtain the heptacuts of all the double boxes appearing
in the basis in Fig. \ref{fig:6-point_integral_basis} (in Section
\ref{sec:global_poles_of_integrand}) by taking
appropriate vertex momenta to be massless. As will be explained in Section
\ref{sec:assembling_heptacut_from_its_contrbituons}, in the case
of six massless external momenta, there are three qualitatively distinct
ways of distributing the momenta at the vertices of the double
box. In order to streamline the presentation we will first
compute the heptacut of the completely general double box integral, making
no assumptions about masslessness of any vertex momenta. Subsequently,
we will describe each of the three cases in turn.
\\
\\
Formally speaking, the four-dimensional heptacut of
(\ref{eq:scalar-doublebox}) is obtained by replacing each of the
seven propagators by a $\delta$-function whose argument is the
denominator of the propagator in question, and replacing the $D$-dimensional
integration measure by the corresponding four-dimensional measure (up to factors
of $2\pi$, depending on conventions). Thus, the heptacut
replaces the double box integral in (\ref{eq:scalar-doublebox}) by
the integral
\begin{eqnarray}
J_\mathrm{formal} \hspace{-1mm} &=& \hspace{-1mm} \int d^4 \ell_1
d^4 \ell_2 \hspace{0.9mm} \delta(\ell_1^2) \hspace{0.7mm}
\delta\big((\ell_1-K_1)^2\big) \hspace{0.7mm}
\delta\big((\ell_1-K_1 - K_2)^2\big)
\hspace{0.7mm} \delta\big((\ell_1+\ell_2+K_6)^2\big) \nonumber \\
&\phantom{=}& \hspace{22mm} \times \hspace{0.7mm} \delta(\ell_2^2)
\hspace{0.7mm} \delta\big((\ell_2-K_5)^2\big) \hspace{0.7mm}
\delta\big((\ell_2-K_4 - K_5)^2\big) \: .
\label{eq:formal_general_heptacut}
\end{eqnarray}
However, this integral only
receives contributions from regions of integration space where the
loop momenta solve the joint on-shell constraints
\begin{eqnarray}
                 \ell_1^2 &=& 0 \label{eq:on-shell_constraint_1}\\
         (\ell_1 - K_1)^2 &=& 0 \label{eq:on-shell_constraint_2}\\
   (\ell_1 - K_1 - K_2)^2 &=& 0 \label{eq:on-shell_constraint_3}\\
                 \ell_2^2 &=& 0 \label{eq:on-shell_constraint_4}\\
           (\ell_2-K_5)^2 &=& 0 \label{eq:on-shell_constraint_5}\\
   (\ell_2 - K_4 - K_5)^2 &=& 0 \label{eq:on-shell_constraint_6}\\
(\ell_1 + \ell_2 + K_6)^2 &=& 0\label{eq:on-shell_constraint_7}\:,
\end{eqnarray}
which in general only have solutions for complex loop momenta
$(\ell_1, \ell_2) \in \mathbb{C}^4 \times \mathbb{C}^4$.

The natural definition of $\delta$-functions with complex
arguments involves contour integrals -- integrating out a variable
$q$ in an integrand involving $\delta$-functions will fix $q$ to
some value $q_0$; in the language of contour integrals, this
corresponds to integrating in the complex $q$-plane along a small
circle centered at $q_0$. Indeed, as observed in ref.~\cite{Roiban:2004yf,Vergu:2006np}, Cauchy's residue theorem
implies that the localization property
\begin{equation}
\int dq \hspace{0.8mm} \delta(q-q_0) f(q)
\hspace{1mm}=\hspace{1mm} f(q_0)
\end{equation}
remains to hold if we define $\delta(q-q_0) \equiv \frac{1}{2\pi
i} \frac{1}{q - q_0}$ and take the integral to be a contour
integral along a small circle in the complex $q$-plane centered at
$q_0$.

By analogy, taking the four-dimensional heptacut of the double box integral
(\ref{eq:scalar-doublebox}) should really be understood as a
change of integration range from $\mathbb{R}^D \times
\mathbb{R}^D$ to a surface (of real dimension 8) embedded in
$\mathbb{C}^4 \times \mathbb{C}^4$ while leaving the integrand in
eq. (\ref{eq:scalar-doublebox}) unchanged. The maximal-cut
integral is thus a multidimensional contour integral whose contour
is in general a linear combination of tori encircling the
so-called \emph{global poles} of the integrand. These are points
$(\ell_1, \ell_2) \in \mathbb{C}^4 \times \mathbb{C}^4$ where all
seven propagators in (\ref{eq:scalar-doublebox}) become on-shell.\footnote{To
be more accurate, we will use the terminology ``global poles'' to refer
to points where, in addition to the seven on-shell constraints
(\ref{eq:on-shell_constraint_1})-(\ref{eq:on-shell_constraint_7}) being solved,
there is an additional singularity, for example coming from
the Jacobians arising from linearizing these cut constraints.
A more precise definition will be given in Section \ref{sec:extraction_of_integral_coeffs}.}
The change of contour away from real Minkowski space renders the
double box integral IR and UV finite, and one can therefore
disregard the dimensional regulator part of the measure
$d^{-2\epsilon} \ell_1 \hspace{0.4mm} d^{-2\epsilon} \ell_2$ and the
$(-2\epsilon)$-dimensional components of the loop momenta.

In the following we will continue to write multidimensional
contours symbolically in terms of $\delta$-functions, as in
eq. (\ref{eq:formal_general_heptacut}), as we find the latter notation
more suggestive. As it turns out, in all cases considered in this paper,
the only respect in which the
multidimensional contour integrals do not behave like integrals
of $\delta$-functions is the transformation formula for changing variables:
Given a holomorphic function $f = (f_1, \ldots, f_n) : \mathbb{C}^n \to
\mathbb{C}^n$ with an isolated zero\footnote{A function $f = (f_1, \ldots, f_n) : \mathbb{C}^n \to
\mathbb{C}^n$ is said to have
an isolated zero at $a \in \mathbb{C}^n$ iff by choosing a small
enough neighborhood $U$ of $a$ one can achieve $f^{-1}(0) \cap U = \{ a\}$.}
at $a \in \mathbb{C}^n$, the residue
at $a$ is computed by the integral over the contour
$\Gamma_\epsilon (a) = \{ z \in \mathbb{C}^n : |f_i (z)| = \epsilon_i,
\hspace{1mm} i = 1, \ldots, n \}$. This contour integral satisfies the transformation formula
\begin{equation}
\frac{1}{(2\pi i)^n} \int_{\Gamma_\epsilon (a)} \frac{h(z) \hspace{0.4mm} dz_1 \wedge \cdots
\wedge dz_n}{f_1 (z) \cdots f_n (z)} \hspace{1mm}=\hspace{1mm} \frac{h(a)}
{\det_{i,j} \frac{\partial f_i}{\partial z_j}} \label{eq:contour_integration_transf_formula}
\end{equation}
which, crucially, does not involve taking the absolute value of the
inverse Jacobian. This ensures
that this factor is analytic in any variables on which it depends, so that further contour
integrations can be carried out.
\\
\\
In order to visualize the multidimensional tori in question,
it turns out to be convenient to use the following parametrization
of the loop momenta:
\begin{eqnarray}
\ell_1^\mu \hspace{-1.5mm}&=&\hspace{-1.5mm} \alpha_1 K_1^{\flat
\mu} + \alpha_2 K_2^{\flat \mu} + \alpha_3 \langle K_1^{\flat -} |
\gamma^\mu | K_2^{\flat -} \rangle + \alpha_4 \langle K_2^{\flat
-} | \gamma^\mu | K_1^{\flat -} \rangle \label{eq:l1_parametrized} \\
\ell_2^\mu \hspace{-1.5mm}&=&\hspace{-1.5mm} \beta_1 K_4^{\flat
\mu} + \beta_2 K_5^{\flat \mu} + \beta_3 \langle K_4^{\flat -} |
\gamma^\mu | K_5^{\flat -} \rangle + \beta_4 \langle K_5^{\flat -}
| \gamma^\mu | K_4^{\flat -} \rangle \:
.\label{eq:l2_parametrized}
\end{eqnarray}
The virtue of this parametrization is that it linearizes as many
of the cut constraints as possible, and in turn it becomes easy to
locate the positions of the global poles of the integrand in the
coordinates $\alpha_1, \ldots, \alpha_4, \beta_1, \ldots,
\beta_4$. The multidimensional tori discussed above are then
easily obtained as products of small circles each encircling one
of the entries of a given global pole.

After changing variables from the components of the loop momenta
$\ell_1^\mu$ and $\ell_2^\nu$ to the parameters $\alpha_i$ and
$\beta_j$, the heptacut of the double box integral becomes
\begin{eqnarray}
J \hspace{-2mm} &=& \hspace{-2mm} \frac{1}{\gamma_1^3
\gamma_2^3} \int \prod_{i=1}^4 d\alpha_i d\beta_i \left(
\det_{\mu, i} \frac{\partial \ell_1^\mu}{\partial \alpha_i}
\right) \left( \det_{\nu, j} \frac{\partial \ell_2^\nu}{\partial
\beta_j} \right) \delta(\alpha_1 \alpha_2 - 4 \alpha_3 \alpha_4)
\hspace{0.8mm} \delta\hspace{-1mm} \left( (\alpha_1 -1)
\left(\alpha_2- \textstyle{\frac{S_1}{\gamma_1}} \right)-
4\alpha_3 \alpha_4\right)\nonumber \\
&\phantom{=}& \hspace{1.5cm} \times \hspace{0.5mm}
\delta\hspace{-1mm} \left( \left( \alpha_1 -
\textstyle{\frac{S_2}{\gamma_1}} - 1 \right)\left(\alpha_2 -
\textstyle{\frac{S_1}{\gamma_1}} - 1\right) - 4 \alpha_3 \alpha_4
\right) \delta(\beta_1 \beta_2 - 4\beta_3 \beta_4) \nonumber \\
&\phantom{=}& \hspace{1.5cm} \times \hspace{0.5mm} \delta
\hspace{-1mm} \left( \left( \beta_1 -
\textstyle{\frac{S_5}{\gamma_2}} \right) (\beta_2 -1)-4\beta_3
\beta_4\right) \delta \hspace{-1mm} \left( \left( \beta_1 -
\textstyle{\frac{S_5}{\gamma_2}} -1 \right) \left(\beta_2
- \textstyle{\frac{S_4}{\gamma_2}} -1 \right) -4\beta_3 \beta_4\right) \nonumber \\
&\phantom{=}& \hspace{1.5cm} \times \hspace{0.5mm}
\delta\big(\left. (\ell_1 + \ell_2 + K_6)^2 \right|_\mathrm{param}
\big) \label{eq:heptacut-double-box-8-param-integrand}
\end{eqnarray}
where the subscript ``param'' on the argument of the
$\delta$-function in the last line indicates that it is to be
evaluated in the parametrization
(\ref{eq:l1_parametrized})-(\ref{eq:l2_parametrized}) and where
the Jacobians associated with the change of variables are,
respectively\footnote{These identities can be shown by writing the
determinants as square roots of Gram determinants (since both
of these equal the volume of the spanned parallelotope, up to a complex
phase) and using the
special properties of the vectors $\{K_1^{\flat \mu}, K_2^{\flat
\mu}, \langle K_1^{\flat -} | \gamma^\mu | K_2^{\flat -} \rangle,
\langle K_2^{\flat -} | \gamma^\mu | K_1^{\flat -} \rangle\}$ and
$\{ K_4^{\flat \mu}, K_5^{\flat \mu}, \langle K_4^{\flat -} |
\gamma^\mu | K_5^{\flat -} \rangle, \langle K_5^{\flat -} |
\gamma^\mu | K_4^{\flat -} \rangle \}$. This will fix the
determinants up to an overall factor of $i^k$ which can be found
numerically.}
\begin{equation}
\det_{\mu, i} \frac{\partial \ell_1^\mu}{\partial \alpha_i}
\hspace{0.5mm}=\hspace{0.5mm} i\gamma_1^2 \: , \hspace{10mm}
\det_{\nu, j} \frac{\partial \ell_2^\nu}{\partial \beta_j}
\hspace{0.5mm}=\hspace{0.5mm} i\gamma_2^2 \: .
\label{eq:Jacobians_from_comps_of_l1l2_to_alpha_beta}
\end{equation}
In the parametrization
(\ref{eq:l1_parametrized})-(\ref{eq:l2_parametrized}), the six
on-shell constraints
(\ref{eq:on-shell_constraint_1})-(\ref{eq:on-shell_constraint_6})
which only involve a single loop momentum are solved by setting
\begin{eqnarray}
\alpha_1 \hspace{-2mm} &=& \hspace{-2mm} \frac{\gamma_1(S_2 +
\gamma_1)}{\gamma_1^2 - S_1 S_2} \: , \hspace{5mm} \alpha_2 =
\frac{S_1 S_2 (S_1 + \gamma_1)}{\gamma_1(S_1 S_2 - \gamma_1^2)} \:
, \hspace{5mm} \alpha_3 \alpha_4 = -\frac{S_1 S_2 (S_1 +
\gamma_1)(S_2 + \gamma_1)}{4(\gamma_1^2 - S_1 S_2)^2}
\phantom{aaa} \label{eq:onshell-values-alpha-param1} \\
\beta_1 \hspace{-2mm} &=& \hspace{-2mm} \frac{S_4 S_5 (S_5 +
\gamma_2)}{\gamma_2(S_4 S_5 - \gamma_2^2)} \: , \hspace{5mm}
\beta_2 = \frac{\gamma_2 (S_4 + \gamma_2)}{\gamma_2^2 - S_4 S_5}
\: , \hspace{5mm} \beta_3 \beta_4 = -\frac{S_4 S_5 (S_4 +
\gamma_2)(S_5 + \gamma_2)}{4(\gamma_2^2 - S_4 S_5)^2} \: .
\phantom{aaa} \label{eq:onshell-values-beta-param1}
\end{eqnarray}
We observe that the integrations of the
$\delta$-functions in
(\ref{eq:heptacut-double-box-8-param-integrand}) unambiguously fix the values of $\alpha_1, \alpha_2,
\beta_1, \beta_2$. After imposing the last on-shell constraint
(\ref{eq:on-shell_constraint_7}), the four variables $\alpha_3,
\alpha_4, \beta_3, \beta_4$ are subject to three relations, and
one is free to choose either one of them to be an unconstrained
free complex parameter $z$. Each of these four choices give rise
to different contributions to the heptacut of the double box.
As will be explained further in Section \ref{sec:assembling_heptacut_from_its_contrbituons},
these contributions correspond to the various existing classes of solutions
to the on-shell constraints
(\ref{eq:on-shell_constraint_1})-(\ref{eq:on-shell_constraint_7}),
and the heptacut (\ref{eq:heptacut-double-box-8-param-integrand})
is an appropriately weighted sum of these contributions.

In Section \ref{sec:leaving_alpha_3_as_the_free_parameter} below
we compute in detail the contribution to the heptacut obtained by
letting $z=\alpha_3$ be the unconstrained parameter. The results
for the remaining contributions are quoted in Sections
\ref{sec:leaving_alpha_4_as_the_free_parameter}-\ref{sec:leaving_beta_4_as_the_free_parameter}
and are obtained in an entirely analogous way. Finally, in Section
\ref{sec:assembling_heptacut_from_its_contrbituons} we explain how
to assemble the heptacut of the double box integral from its
various contributions.

\subsubsection{Leaving $z=\alpha_3$ as the free parameter}\label{sec:leaving_alpha_3_as_the_free_parameter}

In this example, we aim to leave $z=\alpha_3$ as the unconstrained
parameter, and we will therefore integrate out $\alpha_1,
\alpha_2, \alpha_4$ and $\beta_1, \beta_2, \beta_3, \beta_4$. This
will proceed in three stages: first we integrate out the three
$\delta$-functions involving only the $\alpha$-variables; then we
integrate out the three $\delta$-functions involving only the
$\beta$-variables; finally, we integrate out the remaining
$\beta$-variable.

Thus, we start by considering the following integral whose
integrand consists of all the $\delta$-functions in
(\ref{eq:heptacut-double-box-8-param-integrand}) that only involve
the $\alpha$-variables,
\begin{eqnarray}
J_\alpha \hspace{-2mm} &=& \hspace{-2mm} \int d\alpha_1 d\alpha_2
d\alpha_4 \hspace{0.7mm} \delta(\alpha_1 \alpha_2 - 4 \alpha_3
\alpha_4) \hspace{0.7mm} \delta \hspace{-0.8mm} \left( (\alpha_1
-1) \left(\alpha_2- \textstyle{\frac{S_1}{\gamma_1}} \right)-
4\alpha_3 \alpha_4\right) \nonumber \\
&\phantom{=}& \hspace{2.5cm} \times \hspace{0.8mm} \delta
\hspace{-0.8mm} \left( \left(\alpha_1 -
\textstyle{\frac{S_2}{\gamma_1}} - 1 \right) \left(\alpha_2 -
\textstyle{\frac{S_1}{\gamma_1}} - 1 \right) - 4 \alpha_3 \alpha_4
\right) \: . \label{eq:J_alpha_unintegrated}
\end{eqnarray}
This integral is the inverse Jacobian associated with the change
of integration variables from the $\alpha$-parameters to the arguments of the
$\delta$-functions. It is straightforwardly evaluated to yield
\begin{equation}
J_\alpha \hspace{1mm}=\hspace{1mm} \frac{1}{4 \left(1 - \frac{S_1
S_2}{\gamma_1^2} \right) \alpha_3} \label{eq:Ialpha-param1}
\end{equation}
where we recall that the $\delta$-functions in eqs.
(\ref{eq:heptacut-double-box-8-param-integrand}) and (\ref{eq:J_alpha_unintegrated})
are a short-hand notation for multi\-dimensional contour integrations. As these are subject to the
transformation formula in (\ref{eq:contour_integration_transf_formula}),
which involves the Jacobian of the transformation rather than its absolute value,
there is no absolute value in eq. (\ref{eq:Ialpha-param1}).

Second, we consider the following integral whose integrand
consists of all the $\delta$-functions in
(\ref{eq:heptacut-double-box-8-param-integrand}) that only involve
the $\beta$-variables,
\begin{eqnarray}
J_\beta \hspace{-2mm} &=& \hspace{-2mm} \int d\beta_1 d\beta_2
d\beta_4 \hspace{0.7mm} \delta(\beta_1 \beta_2 - 4\beta_3 \beta_4)
\hspace{0.9mm} \delta \hspace{-0.6mm} \left( \left( \beta_1 -
\textstyle{\frac{S_5}{\gamma_2}} \right)
(\beta_2-1) -4\beta_3 \beta_4\right) \nonumber \\
&\phantom{=}& \hspace{2cm} \times \hspace{0.5mm}
\delta\hspace{-1mm}\left( \left(\beta_1 -
\textstyle{\frac{S_5}{\gamma_2}} -1 \right) \left(\beta_2
-\textstyle{\frac{S_4}{\gamma_2}} -1 \right) -4\beta_3
\beta_4\right) \: .
\end{eqnarray}
In addition to integrating out $\beta_1, \beta_2$, we have here
chosen to integrate out $\beta_4$. Alternatively, we could have
imagined integrating out $\beta_3$; but this would ultimately lead
to the same final result. Once again, the integral $J_\beta$ is a
Jacobian and is straightforwardly evaluated to yield
\begin{equation}
J_\beta \hspace{1mm}=\hspace{1mm} -\frac{1}{4 \left(1 - \frac{S_4
S_5}{\gamma_2^2} \right) \beta_3} \: . \label{eq:Ibeta-param1}
\end{equation}
Putting together the partial results in eqs.
(\ref{eq:Jacobians_from_comps_of_l1l2_to_alpha_beta}),
(\ref{eq:Ialpha-param1}), (\ref{eq:Ibeta-param1}) and applying the
loop momentum parametri\-zation to the argument of the remaining
factor $\delta\big(\left. (\ell_1 + \ell_2 + K_6)^2
\right|_\mathrm{param} \big)$ in eq.
(\ref{eq:heptacut-double-box-8-param-integrand}) one finds the
expression
\begin{equation}
\frac{\gamma_1 \gamma_2}{32 (\gamma_1^2 - S_1 S_2) (\gamma_2^2 -
S_4 S_5)} \int \frac{d\alpha_3 d\beta_3}{\alpha_3 \beta_3}
\hspace{0.8mm} \delta\big( B_1 \beta_3 + B_0 +
B_{-1}\beta_3^{-1}\big)
\end{equation}
where
\begin{eqnarray}
B_1 \hspace{-2mm} &=& \hspace{-2mm} \langle K_4^{\flat-} |
\gamma_\mu | K_5^{\flat -} \rangle \Big( \alpha_1 K_1^{\flat \mu}
+ \alpha_2 K_2^{\flat \mu} + \alpha_3 \langle K_1^{\flat -} |
\gamma^\mu | K_2^{\flat -} \rangle + \alpha_4 \langle K_2^{\flat
-} | \gamma^\mu |
K_1^{\flat -} \rangle + K_6^\mu \Big) \phantom{aaaaa} \label{eq:coeff_B1} \\
B_0 \hspace{-2mm} &=& \hspace{-2mm} \Big( \beta_1 K_{4\mu}^\flat +
\beta_2 K_{5\mu}^\flat + K_{6\mu} \Big) \nonumber \\
&\phantom{=}& \hspace{0.4cm} \times \hspace{0.5mm} \Big( \alpha_1
K_1^{\flat \mu} + \alpha_2 K_2^{\flat \mu} + \alpha_3 \langle
K_1^{\flat -} | \gamma^\mu | K_2^{\flat -} \rangle + \alpha_4
\langle K_2^{\flat -} | \gamma^\mu | K_1^{\flat
-} \rangle + K_6^\mu \Big) - \textstyle{\frac{1}{2}} S_6 \label{eq:coeff_B0} \\
B_{-1} \hspace{-2mm} &=& \hspace{-2mm} -\frac{S_4 S_5(S_4 +
\gamma_2)(S_5 + \gamma_2) \langle K_5^{\flat -} |\gamma_\mu |
K_4^{\flat -} \rangle}{4(\gamma_2^2 - S_4 S_5)^2} \nonumber \\
&\phantom{=}& \hspace{0.9cm} \times \hspace{0.5mm} \Big( \alpha_1
K_1^{\flat \mu} + \alpha_2 K_2^{\flat \mu} + \alpha_3 \langle
K_1^{\flat -} | \gamma^\mu | K_2^{\flat -} \rangle + \alpha_4
\langle K_2^{\flat -} | \gamma^\mu | K_1^{\flat -} \rangle +
K_6^\mu \Big) \: . \label{eq:coeff_B-1}
\end{eqnarray}
We can integrate out $\beta_3$ using
\begin{equation}
\int \frac{d\beta_3}{\beta_3} \hspace{0.4mm} \delta\big( B_1
\beta_3 + B_0 + B_{-1} \beta_{3}^{-1} \big)
\hspace{1mm}=\hspace{1mm} \big( B_0^2 - 4B_1 B_{-1} \big)^{-1/2}
\: .
\end{equation}
In conclusion, leaving $z=\alpha_3$ unconstrained and integrating
out the remaining loop momentum parameters from
(\ref{eq:heptacut-double-box-8-param-integrand}) produces the
following contribution to the heptacut of the double box,
\begin{equation}
J\big|_{z=\alpha_3} \hspace{1mm}=\hspace{1mm} \frac{\gamma_1
\gamma_2}{32 (\gamma_1^2 - S_1 S_2) (\gamma_2^2 - S_4 S_5)} \oint
\frac{dz}{z} \Big( B_0(z)^2 - 4B_1 (z) B_{-1}(z) \Big)^{-1/2}
\label{eq:heptacut_double_box_alpha_3}
\end{equation}
where we have relabeled $\alpha_3 \equiv z$ and made the
dependence of $B_1, B_0, B_{-1}$ on $z$ explicit. Although
the factor $\big( \cdots \big)^ {-1/2}$ appears to have a branch
cut, the radicand turns out to be a perfect square in all cases
considered in this paper, and the integrand in eq. (\ref{eq:heptacut_double_box_alpha_3})
always takes the form $\frac{1}{z (z - P)}$. As we will
discuss further in Section \ref{sec:global_poles_of_integrand}, we will allow the integration
contour in eq. (\ref{eq:heptacut_double_box_alpha_3}) to encircle any individual singularity of the integrand.

Note that the form of the final result
(\ref{eq:heptacut_double_box_alpha_3}) does not depend on the
order of integration: if one instead chooses to integrate out
$\alpha_1, \alpha_2, \alpha_4$ and $\beta_1, \beta_2, \beta_3$
from (\ref{eq:heptacut-double-box-8-param-integrand}) and
subsequently integrates out $\beta_4$, one finds
(\ref{eq:heptacut_double_box_alpha_3}) as may easily be checked.

\subsubsection{Leaving $z=\alpha_4$ as the free parameter}\label{sec:leaving_alpha_4_as_the_free_parameter}

Integrating out $\alpha_1, \alpha_2, \alpha_3$ and $\beta_1,
\beta_2, \beta_3, \beta_4$ from
(\ref{eq:heptacut-double-box-8-param-integrand}) produces
\begin{equation}
J\big|_{z=\alpha_4} \hspace{1mm}=\hspace{1mm} \frac{\gamma_1
\gamma_2}{32 (\gamma_1^2 - S_1 S_2) (\gamma_2^2 - S_4 S_5)} \oint
\frac{dz}{z} \Big( B^\bullet_0(z)^2 - 4B^\bullet_1 (z)
B^\bullet_{-1}(z) \Big)^{-1/2}
\label{eq:heptacut_double_box_alpha_4}
\end{equation}
where
\begin{eqnarray}
B^\bullet_1(z) \hspace{-2mm} &=& \hspace{-2mm} \langle
K_4^{\flat-} | \gamma_\mu | K_5^{\flat -} \rangle \Big( \alpha_1
K_1^{\flat \mu} + \alpha_2 K_2^{\flat \mu} + \alpha_3 \langle
K_1^{\flat -} | \gamma^\mu | K_2^{\flat -} \rangle + z \langle
K_2^{\flat -} |
\gamma^\mu | K_1^{\flat -} \rangle + K_6^\mu \Big) \phantom{aaaaa} \label{eq:coeff_tildeB1} \\
B^\bullet_0(z) \hspace{-2mm} &=& \hspace{-2mm} \Big( \beta_1
K_{4\mu}^\flat + \beta_2 K_{5\mu}^\flat + K_{6\mu} \Big) \nonumber
\\ &\phantom{=}& \hspace{0.4cm} \times \hspace{0.5mm} \Big(
\alpha_1 K_1^{\flat \mu} + \alpha_2 K_2^{\flat \mu} + \alpha_3
\langle K_1^{\flat -} | \gamma^\mu | K_2^{\flat -} \rangle + z
\langle K_2^{\flat -} | \gamma^\mu |
K_1^{\flat -} \rangle + K_6^\mu \Big) - \textstyle{\frac{1}{2}} S_6 \label{eq:coeff_tildeB0} \\
B^\bullet_{-1}(z) \hspace{-2mm} &=& \hspace{-2mm} -\frac{S_4
S_5(S_4 + \gamma_2)(S_5 + \gamma_2) \langle K_5^{\flat -}
|\gamma_\mu | K_4^{\flat -} \rangle}{4(\gamma_2^2 - S_4 S_5)^2} \nonumber \\
&\phantom{=}& \hspace{0.9cm} \times \hspace{0.5mm} \Big( \alpha_1
K_1^{\flat \mu} + \alpha_2 K_2^{\flat \mu} + \alpha_3 \langle
K_1^{\flat -} | \gamma^\mu | K_2^{\flat -} \rangle + z \langle
K_2^{\flat -} | \gamma^\mu | K_1^{\flat -} \rangle + K_6^\mu \Big)
\: . \label{eq:coeff_tildeB-1}
\end{eqnarray}
Again this result is independent of the order of the integrations.

\subsubsection{Leaving $z=\beta_3$ as the free parameter}\label{sec:leaving_beta_3_as_the_free_parameter}

Integrating out $\alpha_1, \alpha_2, \alpha_3, \alpha_4$ and
$\beta_1, \beta_2, \beta_4$ from
(\ref{eq:heptacut-double-box-8-param-integrand}) produces
\begin{equation}
J\big|_{z=\beta_3} \hspace{1mm}=\hspace{1mm} \frac{\gamma_1
\gamma_2}{32 (\gamma_1^2 - S_1 S_2) (\gamma_2^2 - S_4 S_5)} \oint
\frac{dz}{z} \Big( A_0(z)^2 - 4A_1 (z) A_{-1}(z) \Big)^{-1/2}
\label{eq:heptacut_double_box_beta_3}
\end{equation}
where
\begin{eqnarray}
A_1(z) \hspace{-2mm} &=& \hspace{-2mm} \langle K_1^{\flat -} |
\gamma^\mu | K_2^{\flat -} \rangle \Big( \beta_1 K_{4\mu}^\flat +
\beta_2 K_{5\mu}^\flat + z \langle K_4^{\flat -} | \gamma_\mu |
K_5^{\flat -} \rangle + \beta_4 \langle K_5^{\flat -} | \gamma_\mu
| K_4^{\flat
-} \rangle + K_{6 \mu} \Big) \phantom{aaaaa} \label{eq:coeff_A1} \\
A_0(z) \hspace{-2mm} &=& \hspace{-2mm} \Big( \alpha_1 K_1^{\flat
\mu} + \alpha_2 K_2^{\flat \mu} + K_6^\mu
\Big) \nonumber \\
&\phantom{=}& \hspace{0.4cm} \times \hspace{0.5mm} \Big( \beta_1
K_{4 \mu}^\flat + \beta_2 K_{5 \mu}^\flat + z \langle K_4^{\flat
-} | \gamma_\mu | K_5^{\flat -} \rangle + \beta_4 \langle
K_5^{\flat -} | \gamma_\mu | K_4^{\flat -} \rangle + K_{6 \mu} \Big) - \textstyle{\frac{1}{2}} S_6 \label{eq:coeff_A0} \\
A_{-1}(z) \hspace{-2mm} &=& \hspace{-2mm} -\frac{S_1 S_2(S_1 +
\gamma_1)(S_2 + \gamma_1) \langle K_2^{\flat -} |\gamma^\mu |
K_1^{\flat -} \rangle}{4(\gamma_1^2 - S_1 S_2)^2} \nonumber \\
&\phantom{=}& \hspace{0.9cm} \times \hspace{0.5mm} \Big( \beta_1
K_{4 \mu}^\flat + \beta_2 K_{5 \mu}^\flat + z \langle K_4^{\flat
-} | \gamma_\mu | K_5^{\flat -} \rangle + \beta_4 \langle
K_5^{\flat -} | \gamma_\mu | K_4^{\flat -} \rangle + K_{6 \mu}
\Big) \: . \label{eq:coeff_A-1}
\end{eqnarray}
Again this result is independent of the order of the integrations.

\subsubsection{Leaving $z=\beta_4$ as the free parameter}\label{sec:leaving_beta_4_as_the_free_parameter}

Integrating out $\alpha_1, \alpha_2, \alpha_3, \alpha_4$ and
$\beta_1, \beta_2, \beta_3$ from
(\ref{eq:heptacut-double-box-8-param-integrand}) produces
\begin{equation}
J\big|_{z=\beta_4} \hspace{1mm}=\hspace{1mm} \frac{\gamma_1
\gamma_2}{32 (\gamma_1^2 - S_1 S_2) (\gamma_2^2 - S_4 S_5)} \oint
\frac{dz}{z} \Big( A^\bullet_0(z)^2 - 4A^\bullet_1 (z)
A^\bullet_{-1}(z) \Big)^{-1/2}
\label{eq:heptacut_double_box_beta_4}
\end{equation}
where
\begin{eqnarray}
A^\bullet_1(z) \hspace{-2mm} &=& \hspace{-2mm} \langle K_1^{\flat
-} | \gamma^\mu | K_2^{\flat -} \rangle \Big( \beta_1
K_{4\mu}^\flat + \beta_2 K_{5\mu}^\flat + \beta_3 \langle
K_4^{\flat -} | \gamma_\mu | K_5^{\flat -} \rangle + z \langle
K_5^{\flat -} | \gamma_\mu |
K_4^{\flat -} \rangle + K_{6 \mu} \Big) \phantom{aaaaa} \label{eq:coeff_tildeA1} \\
A^\bullet_0(z) \hspace{-2mm} &=& \hspace{-2mm} \Big( \alpha_1
K_1^{\flat \mu} + \alpha_2 K_2^{\flat
\mu} + K_6^\mu \Big) \nonumber \\
&\phantom{=}& \hspace{0.4cm} \times \hspace{0.5mm} \Big( \beta_1
K_{4 \mu}^\flat + \beta_2 K_{5 \mu}^\flat + \beta_3 \langle
K_4^{\flat -} | \gamma_\mu | K_5^{\flat -} \rangle + z \langle
K_5^{\flat -} | \gamma_\mu |
K_4^{\flat -} \rangle + K_{6 \mu} \Big) - \textstyle{\frac{1}{2}} S_6 \label{eq:coeff_tildeA0} \\
A^\bullet_{-1}(z) \hspace{-2mm} &=& \hspace{-2mm} -\frac{S_1
S_2(S_1 + \gamma_1)(S_2 + \gamma_1) \langle K_2^{\flat -}
|\gamma^\mu | K_1^{\flat -} \rangle}{4(\gamma_1^2 - S_1 S_2)^2} \nonumber \\
&\phantom{=}& \hspace{0.9cm} \times \hspace{0.5mm} \Big( \beta_1
K_{4 \mu}^\flat + \beta_2 K_{5 \mu}^\flat + \beta_3 \langle
K_4^{\flat -} | \gamma_\mu | K_5^{\flat -} \rangle + z \langle
K_5^{\flat -} | \gamma_\mu | K_4^{\flat -} \rangle + K_{6 \mu}
\Big) \: . \label{eq:coeff_tildeA-1}
\end{eqnarray}
Again this result is independent of the order of the integrations.
\\
\\
In the section below we will explain how to
assemble the heptacut double box integral from the individual
contributions in eqs. (\ref{eq:heptacut_double_box_alpha_3}),
(\ref{eq:heptacut_double_box_alpha_4}),
(\ref{eq:heptacut_double_box_beta_3}),
(\ref{eq:heptacut_double_box_beta_4}).

\subsubsection{Assembling the heptacut double box from its
contributions}\label{sec:assembling_heptacut_from_its_contrbituons}

As alluded to in the beginning of this section, there are three
qualitatively distinct ways of distributing six massless external
momenta at the vertices of the double box. In this section we
shall describe the classification of these cases, and how to
assemble the heptacut double box integral from the individual
contributions in eqs. (\ref{eq:heptacut_double_box_alpha_3}),
(\ref{eq:heptacut_double_box_alpha_4}),
(\ref{eq:heptacut_double_box_beta_3}),
(\ref{eq:heptacut_double_box_beta_4}) in each of these cases.
Before proceeding to describe the classification of the three
cases, it is useful to appreciate that the individual
contributions to the heptacut arise from the different existing classes of
solutions to the joint on-shell constraints
(\ref{eq:on-shell_constraint_1})-(\ref{eq:on-shell_constraint_7}).
Indeed, these seven constraints  leave one unfixed degree of
freedom $z$ in the two loop momenta $\ell_1, \ell_2$, and in
general there are several classes of solutions. Each class is
parametrized by a free complex variable $z$ and has the remaining
seven loop momentum parameters $\alpha_i, \beta_j$ fixed to
specific values. Two classes $\mathcal{S}$ and $\mathcal{S}'$ are
identical if and only if there exists an invertible holomorphic
function $\varphi(z)$ that maps one class into the other; that is,
$\varphi \left( \alpha_i \big|_{\mathcal{S}} \right) = \alpha_i
\big|_{\mathcal{S}'}$ and $\varphi \left( \beta_i
\big|_{\mathcal{S}} \right) = \beta_i \big|_{\mathcal{S}'}$ for $i
= 1, \ldots, 4$.

\begin{figure}[!h]
\begin{minipage}[b]{0.45\linewidth}
\begin{center} \includegraphics[scale=0.83]{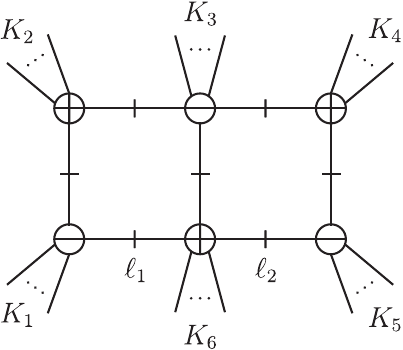} \end{center}
{\center \vspace{-3mm} Solution $\mathcal{S}_5$, obtained by
setting \vspace{-2mm}
\begin{equation*} \begin{array}{ll} \alpha_3 = 0 &
\hspace{7mm} \beta_3 = \beta_3(z) \\
\alpha_4 = z & \hspace{7mm} \beta_4 = 0
\end{array} \end{equation*}}
\end{minipage}
\hspace{8mm}
\begin{minipage}[b]{0.45\linewidth}
\begin{center} \includegraphics[scale=0.83]{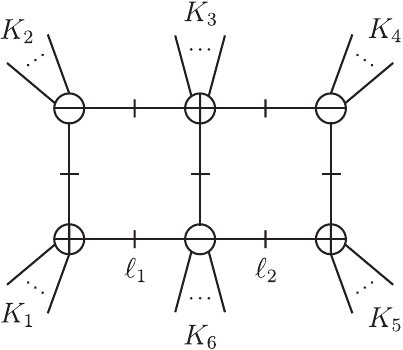} \end{center}
{\center \vspace{-3mm} Solution $\mathcal{S}_6$, obtained by
setting \vspace{-2mm}
\begin{equation*} \begin{array}{ll} \alpha_3 = z & \hspace{7mm}
\beta_3 = 0 \\
\alpha_4 = 0 & \hspace{7mm} \beta_4 = \beta_4(z)
\end{array} \end{equation*}}
\end{minipage} \vspace{-5mm} \caption{The six kinematical
solutions to the heptacut constraints for the double box topology
in case I. For all solutions, the loop momentum parameters
$(\alpha_1, \alpha_2, \beta_1, \beta_2)$ are set equal to the values
given in eqs. (\ref{eq:onshell-values-alpha-param1})-(\ref{eq:onshell-values-beta-param1}).
Any blob connecting more than three legs does
not have a well-defined chirality and its sign should be ignored.
For $\mathcal{S}_5$ and $\mathcal{S}_6$,
the parameters $\beta_3$ and $\beta_4$ are determined by solving
the on-shell constraint $(\ell_1 + \ell_2 + K_6)^2 = 0$ for the
respective
parameter.
The on-shell values $P_1, Q_1$ etc. are functions of the external
momenta; examples may be found in Sections \ref{sec:heptacut_1},
\ref{sec:heptacut_2}-\ref{sec:heptacut_5},
\ref{sec:heptacut_8}.}\label{fig:kinematical_solutions_case_I}\vspace{-45mm}
\end{figure}

\begin{figure}[!h]
\begin{minipage}[b]{0.45\linewidth}
\begin{center} \includegraphics[scale=0.83]{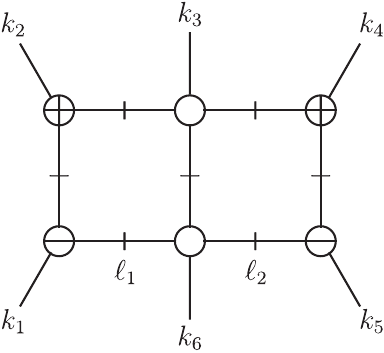} \end{center}
{\center \vspace{-3mm} Solution $\mathcal{S}_3$, obtained by
setting \vspace{-2mm}
\begin{equation*} \begin{array}{ll} \alpha_3 = 0 &
\hspace{4mm} \beta_3 = \frac{Q_2^\bullet (P_2 - P_6) (z - P_1)}
{(P_2 - P_1) (z - P_6)} \\
\alpha_4 = z & \hspace{4mm} \beta_4 = 0
\end{array} \end{equation*}}
\end{minipage}
\hspace{8mm}
\begin{minipage}[b]{0.45\linewidth}
\begin{center} \includegraphics[scale=0.83]{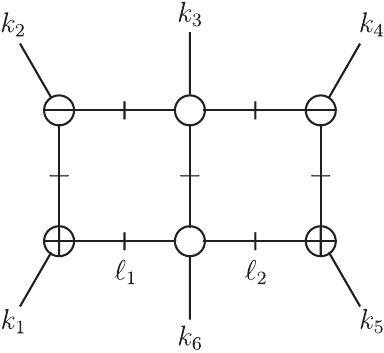} \end{center}
{\center \vspace{-3mm} Solution $\mathcal{S}_4$, obtained by
setting \vspace{-2mm}
\begin{equation*} \begin{array}{ll} \alpha_3 = z &
\hspace{4mm} \beta_3 = 0 \\
\alpha_4 = 0 & \hspace{4mm} \beta_4 = \frac{Q_2 (P_2^\bullet -
P_6^\bullet) (z - P_1^\bullet)} {(P_2^\bullet - P_1^\bullet) (z -
P_6^\bullet)}
\end{array} \end{equation*}}
\end{minipage} \caption{The four kinematical
solutions to the heptacut constraints for the double box topology
in case II. For all solutions, the loop momentum parameters
$(\alpha_1, \alpha_2, \beta_1, \beta_2)$ are set equal to the values
given in eqs. (\ref{eq:onshell-values-alpha-param1})-(\ref{eq:onshell-values-beta-param1}).
The on-shell values $P_1, Q_1$ etc. are functions of the external
momenta and may be found in Section \ref{sec:heptacut_6}.}
\label{fig:kinematical_solutions_case_II}
\end{figure}

\begin{figure}[!h]
\begin{minipage}[b]{0.45\linewidth}
\begin{center} \includegraphics[scale=0.83]{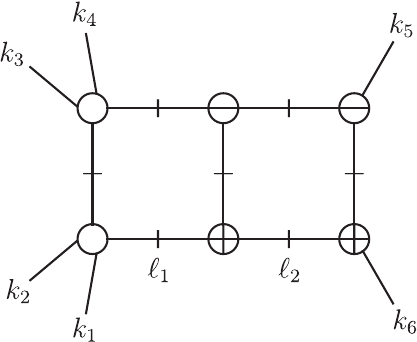} \end{center}
{\center \vspace{-3mm} Solution $\mathcal{S}_3$, obtained by
setting \vspace{-2mm}
\begin{equation*} \begin{array}{ll} \alpha_3^\pm = \frac{\xi^\pm}{P_1^\pm} &
\hspace{7mm} \beta_3 = 0 \\[3mm]
\alpha_4^\pm = P_1^\pm & \hspace{7mm} \beta_4 = z
\end{array} \end{equation*}}
\end{minipage}
\hspace{8mm}
\begin{minipage}[b]{0.45\linewidth}
\begin{center} \includegraphics[scale=0.83]{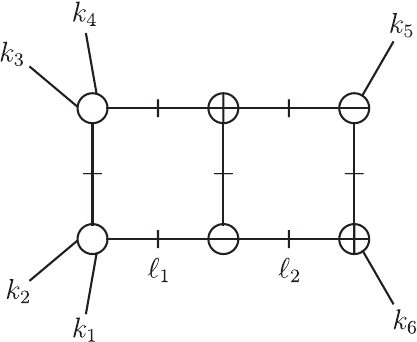} \end{center}
{\center \vspace{-3mm} Solution $\mathcal{S}_4$, obtained by
setting \vspace{-2mm}
\begin{equation*} \begin{array}{ll} \alpha_3^\pm = \frac{\xi^\pm}{\alpha_4^\pm} &
\hspace{3mm} \beta_3 = 0 \\[3mm]
\alpha_4^\pm = \frac{\xi^\pm}{P_1^{\pm \bullet}} \frac{Q_2^\pm
}{Q_2^\mp} \frac{z - Q_2^\mp}{z - Q_2^\pm} & \hspace{3mm} \beta_4
= z
\end{array} \end{equation*}}
\end{minipage} \caption{The four kinematical
solutions to the heptacut constraints for the double box topology
in case III. In this case, because both $K_1$ and $K_2$ are
massive, there are two solutions for $\gamma_1^\pm$ and therefore
two pairs of flattened momenta $(K_{1\pm}^\flat, K_{2\pm}^\flat)$.
For all solutions, the loop momentum parameters
$(\alpha_1, \alpha_2, \beta_1, \beta_2)$ are set equal to the values
given in eqs. (\ref{eq:onshell-values-alpha-param1})-(\ref{eq:onshell-values-beta-param1}).
The on-shell values $P_1^\pm, Q_2^\pm$ etc. are functions of the external
momenta and may be found in Section \ref{sec:heptacut_7}.
The quantity $\xi^\pm$ is defined as
}\label{fig:kinematical_solutions_case_III}
\end{figure}

The classification into the three cases is given as stated below.
\begin{itemize}
\item Case I: all three vertical propagators in Fig. \ref{fig:General_double_box}
are part of some three-particle vertex. There are six classes of kinematical solutions to the
on-shell constraints (\ref{eq:on-shell_constraint_1})-(\ref{eq:on-shell_constraint_7}),
illustrated in Fig. \ref{fig:kinematical_solutions_case_I}. The heptacut double box
integral is $J = \sum_{i=1}^6 J_i$ where
\begin{equation}
\begin{array}{rlll}
J_1 \hspace{-1mm}&=&\hspace{-1mm}  J\big|_{z = \beta_3}&
\hspace{6mm} \mbox{(given in eq. (\ref{eq:heptacut_double_box_beta_3}))} \\[0.7mm]
J_2 = J_6 \hspace{-1mm}&=&\hspace{-1mm}  J\big|_{z = \alpha_3}&
\hspace{6mm} \mbox{(given in eq. (\ref{eq:heptacut_double_box_alpha_3}))} \\[0.7mm]
J_3 \hspace{-1mm}&=&\hspace{-1mm}  J\big|_{z = \beta_4}&
\hspace{6mm} \mbox{(given in eq. (\ref{eq:heptacut_double_box_beta_4}))} \\[0.7mm]
J_4 = J_5 \hspace{-1mm}&=&\hspace{-1mm} J\big|_{z = \alpha_4}&
\hspace{6mm} \mbox{(given in eq. (\ref{eq:heptacut_double_box_alpha_4}))} \: .
\end{array} \label{eq:case_I_Jacobians}
\end{equation}
This case encompasses heptacuts \#1-5 and \#8, discussed in detail
below in Sections \ref{sec:heptacut_1},
\ref{sec:heptacut_2}-\ref{sec:heptacut_5} and \ref{sec:heptacut_8}
where one can also find explicit results for the on-shell values
$P_1, Q_1$ etc. quoted in Fig.
\ref{fig:kinematical_solutions_case_I}. To determine, for example,
the function $\beta_3 (z)$ in solution $\mathcal{S}_5$, one
expresses the loop momenta appearing in the on-shell constraint
$(\ell_1 + \ell_2 + K_6)^2 = 0$ in their parametrized form
(\ref{eq:l1_parametrized})-(\ref{eq:l2_parametrized}), sets the
values of the parameters $\alpha_i, \beta_j$ equal to those quoted
in eqs.
(\ref{eq:onshell-values-alpha-param1})-(\ref{eq:onshell-values-beta-param1})
and below solution $\mathcal{S}_5$ in Fig.
\ref{fig:kinematical_solutions_case_I} and then solves the
on-shell constraint for $\beta_3$.

\item Case II: the left- and rightmost vertical propagators in Fig. \ref{fig:General_double_box}
are part of some three-particle vertex, but the middle one is not. There are four classes of kinematical solutions to the
on-shell constraints (\ref{eq:on-shell_constraint_1})-(\ref{eq:on-shell_constraint_7}),
illustrated in Fig. \ref{fig:kinematical_solutions_case_II}. The heptacut double box
integral is $J = \sum_{i=1}^4 J_i$ where
\begin{equation}
\begin{array}{rlll}
J_1 \hspace{-1mm}&=&\hspace{-1mm}  J\big|_{z = \beta_3} &
\hspace{6mm} \mbox{(given in eq. (\ref{eq:heptacut_double_box_beta_3}))} \\[0.7mm]
J_2 \hspace{-1mm}&=&\hspace{-1mm}  J\big|_{z = \beta_4} &
\hspace{6mm} \mbox{(given in eq. (\ref{eq:heptacut_double_box_beta_4}))} \\[0.7mm]
J_3 \hspace{-1mm}&=&\hspace{-1mm}  J\big|_{z = \alpha_4} &
\hspace{6mm} \mbox{(given in eq. (\ref{eq:heptacut_double_box_alpha_4}))} \\[0.7mm]
J_4 \hspace{-1mm}&=&\hspace{-1mm}  J\big|_{z = \alpha_3} &
\hspace{6mm} \mbox{(given in eq. (\ref{eq:heptacut_double_box_alpha_3}))} \: .
\end{array}\label{eq:case_II_Jacobians}
\end{equation}
This case encompasses heptacut \#6, discussed in detail below in Section
\ref{sec:heptacut_6}
where one can also find explicit results for the on-shell values $P_1, Q_1$ etc. quoted in Fig.
\ref{fig:kinematical_solutions_case_II}.

\item Case III: the two rightmost vertical propagators in Fig. \ref{fig:General_double_box}
are part of some three-particle vertex, but the leftmost is not.  There are four classes of kinematical solutions to the
on-shell constraints (\ref{eq:on-shell_constraint_1})-(\ref{eq:on-shell_constraint_7}),
illustrated in Fig. \ref{fig:kinematical_solutions_case_III}. The heptacut double box
integral is $J = \sum_{i=1}^4 J_i$ where
\begin{equation}
\begin{array}{rlllll}
J_1 \hspace{-1mm}&=&\hspace{-1mm} J_2 \hspace{-1mm}&=&\hspace{-1mm}  J\big|_{z = \beta_3} &
\hspace{6mm} \mbox{(given in eq. (\ref{eq:heptacut_double_box_beta_3}))} \\
J_3 \hspace{-1mm}&=&\hspace{-1mm} J_4 \hspace{-1mm}&=&\hspace{-1mm}  J\big|_{z = \beta_4} &
\hspace{6mm} \mbox{(given in eq. (\ref{eq:heptacut_double_box_beta_4}))} \: .
\end{array}\label{eq:case_III_Jacobians}
\end{equation}
This case encompasses heptacut \#7, discussed in detail below in
Section \ref{sec:heptacut_7} where one can also find explicit
results for the on-shell values $P_1^\pm, Q_2^\pm$ etc. quoted in Fig.
\ref{fig:kinematical_solutions_case_III}. In this case, because
both $K_1$ and $K_2$ are massive, there are two solutions for
$\gamma_1^\pm$ (given in eq. (\ref{eq:def_of_gamma_1_pm})) and
therefore two pairs of flattened momenta $(K_{1\pm}^\flat,
K_{2\pm}^\flat)$ (see eq.
(\ref{eq:def_of_Kflat_1_and_Kflat_2_pm})). The on-shell values of
the loop momenta are independent of which sign is chosen,
\begin{eqnarray}
\ell_1 (\alpha_1^+, \alpha_2^+, \alpha_3^+, \alpha_4^+)
\hspace{-1mm}&=&\hspace{-1mm} \ell_1
(\alpha_1^-, \alpha_2^-, \alpha_3^-, \alpha_4^-) \\
\ell_2 (\beta_1^+, \beta_2^+, \beta_3^+, \beta_4^+)
\hspace{-1mm}&=&\hspace{-1mm} \ell_2 (\beta_1^-, \beta_2^-,
\beta_3^-, \beta_4^-) \: .
\end{eqnarray}
\end{itemize}

\clearpage
\begin{figure}[!h]
\begin{minipage}[b]{0.45\linewidth}
\begin{center} \includegraphics[scale=0.83]{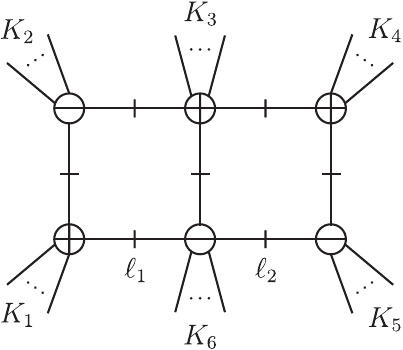} \end{center}
{\center \vspace{-3mm} Solution $\mathcal{S}_1$, obtained by
setting \vspace{-2mm}
\begin{equation*}
\begin{array}{ll} \alpha_3 = P_1^\bullet &
\hspace{7mm} \beta_3 = z \\
\alpha_4 = 0 & \hspace{7mm} \beta_4 = 0
\end{array} \end{equation*}}
\end{minipage}
\hspace{8mm}
\begin{minipage}[b]{0.45\linewidth}
\begin{center} \includegraphics[scale=0.83]{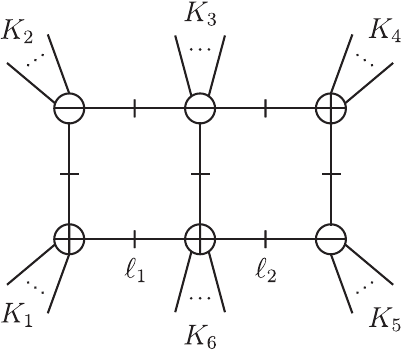} \end{center}
{\center \vspace{-3mm} Solution $\mathcal{S}_2$, obtained by
setting \vspace{-2mm}
\begin{equation*} \begin{array}{ll} \alpha_3 = z &
\hspace{7mm} \beta_3 = Q_1^\bullet \\
\alpha_4 = 0 & \hspace{7mm} \beta_4 = 0
\end{array} \end{equation*}}
\end{minipage}
\end{figure}
\vspace{-10mm}
\begin{figure}[!h]
\begin{minipage}[b]{0.45\linewidth}
\begin{center} \includegraphics[scale=0.83]{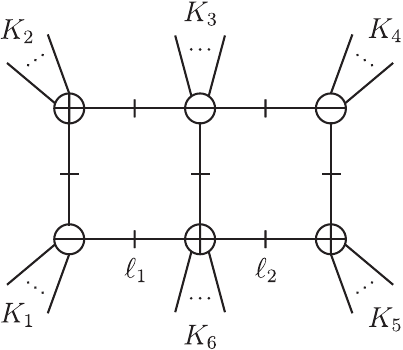} \end{center}
{\center \vspace{-3mm} Solution $\mathcal{S}_3$, obtained by
setting \vspace{-2mm}
\begin{equation*} \begin{array}{ll} \alpha_3 = 0 &
\hspace{7mm} \beta_3 = 0 \\
\alpha_4 = P_1 & \hspace{7mm} \beta_4 = z
\end{array} \end{equation*}}
\end{minipage}
\hspace{8mm}
\begin{minipage}[b]{0.45\linewidth}
\begin{center} \includegraphics[scale=0.83]{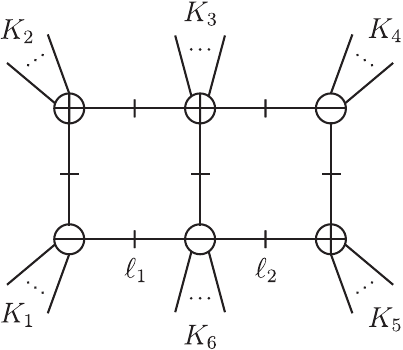} \end{center}
{\center \vspace{-3mm} Solution $\mathcal{S}_4$, obtained by
setting \vspace{-2mm}
\begin{equation*} \begin{array}{ll} \alpha_3 = 0 &
\hspace{7mm} \beta_3 = 0 \\
\alpha_4 = z & \hspace{7mm} \beta_4 = Q_1
\end{array} \end{equation*}}
\end{minipage}
\end{figure}
\vspace{-10mm}
\clearpage

\begin{figure}[!h]
\begin{minipage}[b]{0.45\linewidth}
\begin{center} \includegraphics[scale=0.83]{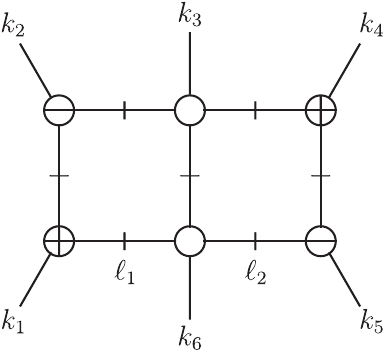} \end{center}
{\center \vspace{-3mm} Solution $\mathcal{S}_1$, obtained by
setting \vspace{-2mm}
\begin{equation*}
\begin{array}{ll} \alpha_3 = -\frac{\langle k_4^- |
\slashed{k}_{61} | k_5^- \rangle (z - Q_1^\bullet)}{2 \langle k_1
k_4 \rangle [k_5 k_2] (z - Q_5^\bullet)} & \hspace{2mm} \beta_3 = z \\
\alpha_4 = 0 & \hspace{2mm} \beta_4 = 0
\end{array} \end{equation*}}
\end{minipage}
\hspace{8mm}
\begin{minipage}[b]{0.45\linewidth}
\begin{center} \includegraphics[scale=0.83]{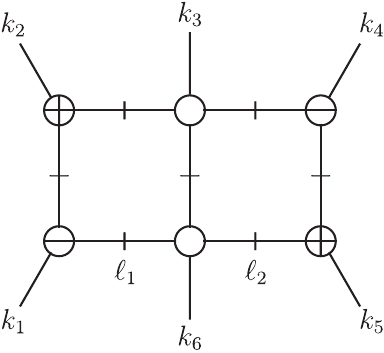} \end{center}
{\center \vspace{-3mm} Solution $\mathcal{S}_2$, obtained by
setting \vspace{-2mm}
\begin{equation*} \begin{array}{ll} \alpha_3 = 0 &
\hspace{2mm} \beta_3 = 0 \\
\alpha_4 = -\frac{\langle k_5^- | \slashed{k}_{61} | k_4^- \rangle
(z - Q_1)}{2[k_1 k_4] \langle k_5 k_2 \rangle (z - Q_5)} &
\hspace{2mm} \beta_4 = z
\end{array} \end{equation*}}
\end{minipage}
\end{figure}
\vspace{-5mm}

\clearpage

\begin{figure}[!h]
\begin{minipage}[b]{0.45\linewidth}
\begin{center} \includegraphics[scale=0.83]{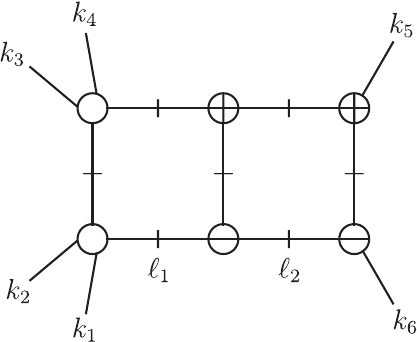} \end{center}
{\center \vspace{-3mm} Solution $\mathcal{S}_1$, obtained by
setting \vspace{-2mm}
\begin{equation*}
\begin{array}{ll} \alpha_3^\pm = P_1^{\pm \bullet} &
\hspace{7mm} \beta_3 = z \\[3mm]
\alpha_4^\pm = \frac{\xi^\pm}{P_1^{\pm \bullet}} & \hspace{7mm}
\beta_4 = 0
\end{array} \end{equation*}}
\end{minipage}
\hspace{8mm}
\begin{minipage}[b]{0.45\linewidth}
\begin{center} \includegraphics[scale=0.83]{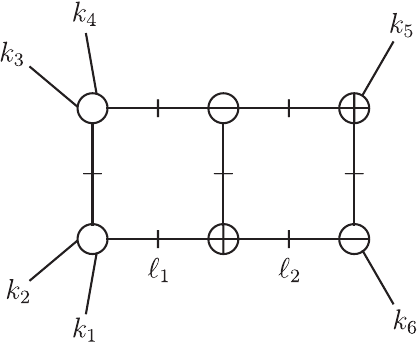} \end{center}
{\center \vspace{-3mm} Solution $\mathcal{S}_2$, obtained by
setting \vspace{-2mm}
\begin{equation*} \begin{array}{ll} \alpha_3^\pm = \frac{\xi^\pm}{P_1^\pm}
\frac{Q_2^{\pm \bullet}}{Q_2^{\mp \bullet}} \frac{z - Q_2^{\mp
\bullet}}{z - Q_2^{\pm \bullet}} & \hspace{3mm} \beta_3 = z \\[3mm]
\alpha_4^\pm = \frac{\xi^\pm}{\alpha_3^\pm} & \hspace{3mm} \beta_4
= 0
\end{array} \end{equation*}}
\end{minipage}
\end{figure}
\vspace{-5mm}
{\vskip -5mm}
\begin{equation}
\xi^\pm \hspace{1mm}=\hspace{1mm} - \frac{S_1 S_2 (S_1 +
\gamma_1^\pm) (S_2 + \gamma_1^\pm)}{4(\gamma_1^{\pm 2} - S_1
S_2)^2} \: . \label{eq:def_of_xi}
\end{equation}

\clearpage

\subsection{Heptacut of the factorized double box}\label{sec:factorized_double_box}

For the double box integral considered in Section
\ref{sec:general_double_box} there are various ways of
distributing six external momenta at the vertices, and we
therefore computed the heptacut with an arbitrary
number of external legs. In contrast, the factorized double box
integral
\begin{eqnarray}
&\phantom{=}& \hspace{-7mm} \left( \int \frac{d^D
\ell_1}{(2\pi)^D} \frac{1}{\ell_1^2} \frac{1}{(\ell_1 - k_1)^2}
\frac{1}{(\ell_1 - k_{12})^2} \frac{1}{(\ell_1 - k_{123})^2}
\right) \nonumber \\
&\phantom{=}& \hspace{7mm} \times \left( \int \frac{d^D
\ell_2}{(2\pi)^D} \frac{1}{\ell_2^2} \frac{1}{(\ell_2 - k_6)^2}
\frac{1}{(\ell_2 - k_{56})^2} \frac{1}{(\ell_2 - k_{456})^2}
\right)
\end{eqnarray}
with which we shall be concerned in this section, admits a unique
way of distributing six (cyclically ordered) external momenta at
its vertices, and we therefore restrict ourselves to this case.
The factorized double box with six massless legs is illustrated in
Fig. \ref{fig:factorized_double_box} below.

\begin{figure}[!h]
\begin{center}
\includegraphics[angle=0,
width=0.55\textwidth]{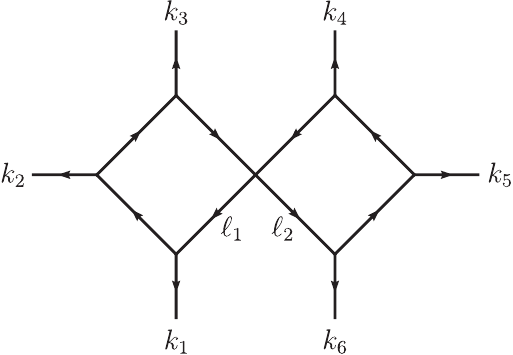}
\caption{The factorized double box integral with six massless
external momenta.}\label{fig:factorized_double_box}
\end{center}
\end{figure}

\noindent We consider the heptacut defined by imposing the
following joint on-shell constraints
\begin{eqnarray}
            \ell_1^2 &=& 0 \label{eq:on-shell_constraint_factorized_1}\\
    (\ell_1 - k_1)^2 &=& 0 \label{eq:on-shell_constraint_factorized_2}\\
 (\ell_1 - k_{12})^2 &=& 0 \label{eq:on-shell_constraint_factorized_3}\\
            \ell_2^2 &=& 0 \label{eq:on-shell_constraint_factorized_4}\\
      (\ell_2-k_6)^2 &=& 0 \label{eq:on-shell_constraint_factorized_5}\\
 (\ell_2 - k_{56})^2 &=& 0 \label{eq:on-shell_constraint_factorized_6}\\
(\ell_2 - k_{456})^2 &=& 0 \: .
\label{eq:on-shell_constraint_factorized_7}
\end{eqnarray}
This is not a maximal cut: we deliberately leave one degree of
freedom $z$ in the loop momenta unfrozen to make manifest the
singularities in the Jacobians arising from changes of variables.
This in turn makes the global poles of the integrand easy to
identify.

We will use the following parametrization of the loop momenta
\begin{eqnarray}
\ell_1^\mu \hspace{-1mm} &=& \hspace{-1mm} \alpha_1 k_1^\mu +
\alpha_2 k_2^\mu + \alpha_3 \langle k_1^- | \gamma^\mu | k_2^-
\rangle + \alpha_4 \langle k_2^-
| \gamma^\mu | k_1^- \rangle \label{eq:loop_param_factDB1}\\
\ell_2^\mu \hspace{-1mm} &=& \hspace{-1mm} \beta_1 k_5^\mu +
\beta_2 k_6^\mu + \beta_3 \langle k_5^- | \gamma^\mu | k_6^-
\rangle + \beta_4 \langle k_6^- | \gamma^\mu | k_5^- \rangle \: .
\label{eq:loop_param_factDB2}
\end{eqnarray}
The constraints
(\ref{eq:on-shell_constraint_factorized_1})-(\ref{eq:on-shell_constraint_factorized_6})
form a special case of eqs.
(\ref{eq:on-shell_constraint_1})-(\ref{eq:on-shell_constraint_6})
after appropriately relabeling the external momenta. From eqs.
(\ref{eq:onshell-values-alpha-param1})-(\ref{eq:onshell-values-beta-param1})
we then find that the constraints are satisfied by setting
\begin{equation}
\begin{array}{rrr} \alpha_1 = 1 \: , & \hspace{8mm} \alpha_2 = 0 \: , & \hspace{8mm}
\alpha_3 \alpha_4 = 0 \: \phantom{.} \\ \beta_1 = 0 \: , &
\hspace{8mm} \beta_2 = 1 \: , & \hspace{8mm} \beta_3 \beta_4 = 0
\: .
\end{array} \label{eq:factorized_double_box_on-shell_values_of_alpha12_and_beta12}
\end{equation}
The final on-shell constraint
(\ref{eq:on-shell_constraint_factorized_7}) combined with eq.
(\ref{eq:factorized_double_box_on-shell_values_of_alpha12_and_beta12})
turns out to have four classes of solutions which can be
conveniently expressed by defining the spinor ratios
\begin{equation}
Q_1 = \frac{\langle 45 \rangle}{2\langle 46 \rangle} \: ,
\hspace{10mm} Q_1^\bullet = \frac{[45]}{2[46]} \: .
\label{eq:factorized_DB_def_of_Q1_and_Q1_conj}
\end{equation}
Thus, the four classes of solutions to the joint on-shell
constraints
(\ref{eq:on-shell_constraint_factorized_1})-(\ref{eq:on-shell_constraint_factorized_7})
are
\begin{equation}
\begin{array}{l} \mbox{$\mathcal{S}_1$:} \hspace{3mm} \left\{ \hspace{-0.5mm} \begin{array}{ll} \alpha_3 = z \: , & \hspace{4mm} \beta_3 = Q_1^\bullet \\
\alpha_4 = 0 \: , & \hspace{4mm} \beta_4 = 0 \end{array} ; \right.
\hspace{12mm} \mbox{$\mathcal{S}_2$:} \hspace{3mm} \left\{ \hspace{-0.5mm} \begin{array}{ll} \alpha_3 = z \: , & \hspace{4mm} \beta_3 = 0 \\
\alpha_4 = 0 \: , & \hspace{4mm} \beta_4 = Q_1 \end{array} \right.
\\[6mm]
\mbox{$\mathcal{S}_3$:} \hspace{3mm} \left\{ \hspace{-0.5mm} \begin{array}{ll} \alpha_3 = 0 \: , & \hspace{4mm} \beta_3 = Q_1^\bullet \\
\alpha_4 = z \: , & \hspace{4mm} \beta_4 = 0 \end{array} ; \right.
\hspace{12mm} \mbox{$\mathcal{S}_4$:} \hspace{3mm} \left\{ \hspace{-0.5mm} \begin{array}{ll} \alpha_3 = 0 \: , & \hspace{4mm} \beta_3 = 0 \\
\alpha_4 = z \: , & \hspace{4mm} \beta_4 = Q_1 \end{array} \right.
\end{array}\phantom{lll} \label{eq:factorized_double_box_kinematical_solutions}
\end{equation}
where the parameters $\alpha_1, \alpha_2, \beta_1, \beta_2$ are
put equal to the values quoted in eq.
(\ref{eq:factorized_double_box_on-shell_values_of_alpha12_and_beta12}).
The solutions are illustrated below in Fig.
\ref{fig:solutions_to_heptacut_constaints_for_factorized_DB}.

Since we are not cutting all propagators, the relevant quantity to
compute is not the heptacut of the factorized double box integral
itself, but rather the Jacobian
\begin{eqnarray}
J \hspace{-1mm} &=& \hspace{-1mm} \left( \int d^4 \ell_1
\hspace{0.5mm} \delta(\ell_1^2) \hspace{0.5mm} \delta\big( (\ell_1
- k_1)^2\big) \hspace{0.5mm} \delta\big( (\ell_1 - k_{12})^2\big) \right) \nonumber \\
&\phantom{=}& \hspace{7mm} \times \hspace{0.5mm} \left( \int d^4
\ell_2 \hspace{0.5mm} \delta(\ell_2^2) \hspace{0.5mm} \delta\big(
(\ell_2 - k_6)^2\big) \hspace{0.5mm} \delta\big( (\ell_2 -
k_{56})^2\big) \hspace{0.5mm} \delta\big( (\ell_2 -
k_{456})^2\big) \right) \phantom{aaaa}
\label{eq:factorized_DB_Jacobian}
\end{eqnarray}
since from this one can easily obtain the heptacut of any given
object, as we will see below. The $\delta$-functions in eq.
(\ref{eq:factorized_DB_Jacobian}) may be integrated out along the
lines described in Section
\ref{sec:leaving_alpha_3_as_the_free_parameter}, and one finds
that for all four kinematical solutions $\mathcal{S}_i$ the
contribution to the Jacobian $J$ is given by
\begin{equation}
J\big|_{\mathcal{S}_i} \hspace{1.5mm}=\hspace{1.5mm} -\frac{1}{16
s_{12} s_{45} s_{56}} \oint_{\Gamma_i} \frac{dz}{z} \hspace{6mm}
\mathrm{for} \hspace{3mm} i = 1, \ldots, 4 \: .
\end{equation}
Then the heptacut factorized double box integral is
\begin{equation}
- \frac{1}{16 s_{12} s_{45} s_{56}} \sum_{i=1}^4 \oint_{\Gamma_i}
\frac{dz}{z} \left. \frac{1}{(\ell_1 - k_{123})^2}
\right|_{\mathcal{S}_i}
\end{equation}
where the subscript $\big(\cdots\big)\big|_{\mathcal{S}_i}$
indicates that the propagator is to be evaluated in the
parametrization
(\ref{eq:loop_param_factDB1})-(\ref{eq:loop_param_factDB2}) with
the parameters set equal to the values in eqs.
(\ref{eq:factorized_double_box_on-shell_values_of_alpha12_and_beta12})
and (\ref{eq:factorized_double_box_kinematical_solutions}).
Likewise, the heptacut
(\ref{eq:on-shell_constraint_factorized_1})-(\ref{eq:on-shell_constraint_factorized_7})
of the two-loop amplitude is
\begin{equation}
- \frac{1}{16 s_{12} s_{45} s_{56}} \sum_{i=1}^4 \oint_{\Gamma_i}
\frac{dz}{z} \left. \prod_{j=1}^6 A_j^\mathrm{tree} (z)
\right|_{\mathcal{S}_i} \: .
\end{equation}

\clearpage

\begin{figure}[!h]
\begin{minipage}[b]{0.45\linewidth}
\begin{center} \includegraphics[scale=0.78]{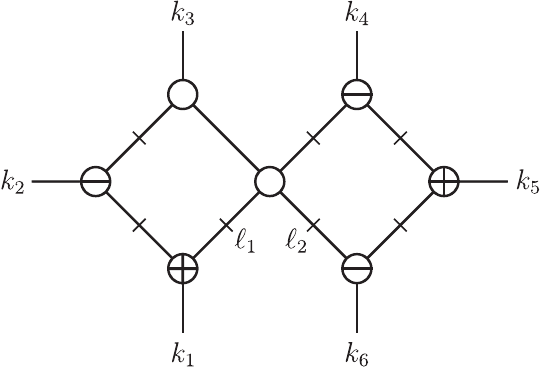} \end{center}
{\center Solution $\mathcal{S}_1$, obtained by setting
\vspace{-2mm}
\begin{equation*} \begin{array}{ll} \alpha_3 = z \: , & \hspace{4mm} \beta_3 = Q_1^\bullet \\
\alpha_4 = 0 \: , & \hspace{4mm} \beta_4 = 0
\end{array} \end{equation*}}
\end{minipage}
\hspace{8mm}
\begin{minipage}[b]{0.45\linewidth}
\begin{center} \includegraphics[scale=0.78]{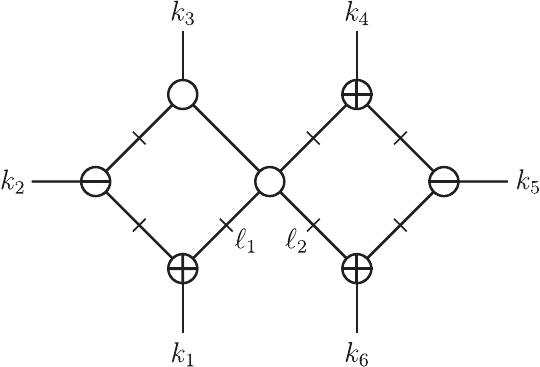} \end{center}
{\center Solution $\mathcal{S}_2$, obtained by setting
\vspace{-2mm}
\begin{equation*} \begin{array}{ll} \alpha_3 = z \: , & \hspace{4mm} \beta_3 = 0 \\
\alpha_4 = 0 \: , & \hspace{4mm} \beta_4 = Q_1
\end{array} \end{equation*}}
\end{minipage}
\end{figure}
\vspace{-3mm}
\begin{figure}[!h]
\begin{minipage}[b]{0.45\linewidth}
\begin{center} \includegraphics[scale=0.78]{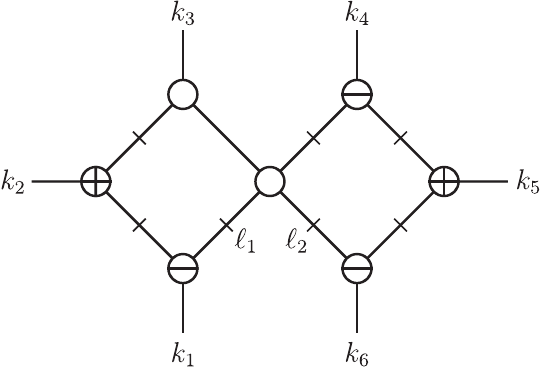} \end{center}
{\center Solution $\mathcal{S}_3$, obtained by setting
\vspace{-2mm}
\begin{equation*} \begin{array}{ll} \alpha_3 = 0 \: , & \hspace{4mm} \beta_3 = Q_1^\bullet \\
\alpha_4 = z \: , & \hspace{4mm} \beta_4 = 0
\end{array} \end{equation*}}
\end{minipage}
\hspace{8mm}
\begin{minipage}[b]{0.45\linewidth}
\begin{center} \includegraphics[scale=0.78]{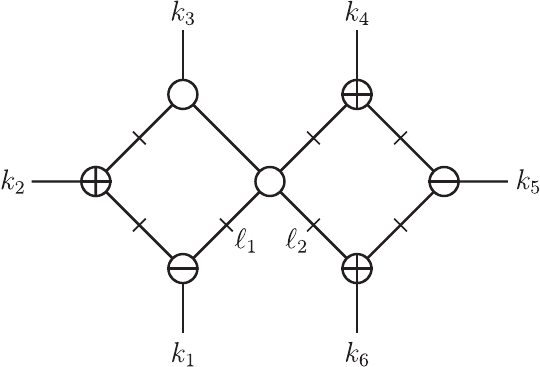} \end{center}
{\center Solution $\mathcal{S}_4$, obtained by setting
\vspace{-2mm}
\begin{equation*} \begin{array}{ll} \alpha_3 = 0 \: , & \hspace{4mm} \beta_3 = 0 \\
\alpha_4 = z \: , & \hspace{4mm} \beta_4 = Q_1
\end{array} \end{equation*}}
\end{minipage} \caption{The four kinematical solutions to the heptacut constraints
given in eqs.
(\ref{eq:on-shell_constraint_factorized_1})-(\ref{eq:on-shell_constraint_factorized_7})
for the factorized double box with six massless external momenta.
The on-shell values $Q_1$ and $Q_1^\bullet$ are defined in eq.
(\ref{eq:factorized_DB_def_of_Q1_and_Q1_conj}).}\label{fig:solutions_to_heptacut_constaints_for_factorized_DB}
\end{figure}

\clearpage

\section{Global poles of the two-loop six-point integrand}\label{sec:global_poles_of_integrand}

In this section we apply the heptacuts discussed in Section
\ref{sec:two-loop_heptacuts} to compute the two-loop six-point MHV
integrand in $\mathcal{N}=4$ SYM theory. The parity-even part of
this integrand was first calculated in ref. \cite{Bern:2008ap} and
was later reexamined using the leading singularity method in ref.
\cite{Cachazo:2008hp}. In Section \ref{sec:heptacut_1} we give a
pedagogical review of the use of the leading singularity method to
determine the two-loop six-point integrand. In particular, we
discuss how one can set up linear equations to determine the
parity-even part of the integrand directly. The approach here is
similar to that of ref. \cite{Cachazo:2008hp}, but differs in the
use of the loop-momentum parametrization
(\ref{eq:l1_parametrized})-(\ref{eq:l2_parametrized}) and
(\ref{eq:loop_param_factDB1})-(\ref{eq:loop_param_factDB2}) which
has the virtue of making the multidimensional contours associated
with the heptacuts completely explicit. It is also worth remarking on the similarity between the approach
followed here and that of Forde in ref.~\cite{Forde:2007mi}: the
heptacut at two loops, in analogy with the triple cut at one
loop, leaves one a priori undetermined contour integration.
The associated contour may be chosen to encircle the various
poles coming from the measure or from additional propagators,
allowing the determination of the integral coefficients by
use of the residue theorem.

The main result of this
paper is contained in Section \ref{sec:loop_level_recursion_test}
in which we report on a numerical check that the result produced
by the leading singularity method for the full (i.e., parity-even
and odd) two-loop six-point MHV integrand is in agreement with
recent predictions in the literature \cite{ArkaniHamed:2010kv,ArkaniHamed:2010gh}.
\\
\\
In general one can apply integral reductions to any two-loop
six-point amplitude to express it as a linear combination of
integrals in some sufficiently large basis. Due to the special
symmetries of $\mathcal{N}=4$ SYM theory amplitudes, it is natural
to include integrals in the basis that reflect these symmetries
(in particular, dual pseudoconformal integrals), and in this paper
we shall use the basis in ref. \cite{Cachazo:2008hp}, for
convenience illustrated in Fig. \ref{fig:6-point_integral_basis}
below. Note that in order to express the parity-odd part of the
amplitude, this basis contains additional (non-dual conformal
invariant) integrals beyond those included in the bases of refs.
\cite{Bern:2008ap,Kosower:2010yk} and as such is
over-complete. We shall discuss the consequences of the
over-completeness in detail in Section \ref{sec:heptacut_1}.

Thus, we will use the following ansatz for the planar two-loop six-point
MHV amplitude of $\mathcal{N}=4$ SYM theory
\begin{equation}
A_{6, \hspace{0.4mm} \mathrm{MHV}}^{(2)} \hspace{1mm}=\hspace{1mm}
\frac{1}{4} \sum_{\substack{i=1,\hspace{0.2mm}\ldots, \hspace{0.2mm} 24 \\
j=1, \hspace{0.2mm} \ldots, \hspace{0.2mm} 12}} r_i c_{i,\sigma_j}
I_{i,\sigma_j} \: , \label{eq:6gluonampansatz}
\end{equation}
in which it is expressed as a linear combination of the basis
integrals illustrated in Fig. \ref{fig:6-point_integral_basis}
whose symmetry factors $r_i$ are given as follows,
\begin{equation}
\begin{array}{rll}
(r_1, r_2, r_3, r_4, r_5, r_6) \hspace{-1mm}&=&\hspace{-1mm}
\left( \textstyle{\frac{1}{4}}, 1, \textstyle{\frac{1}{2}},
\textstyle{\frac{1}{2}}, 1, 1 \right) \\[1mm]
(r_7, r_8, r_9, r_{10}, r_{11}, r_{12})
\hspace{-1mm}&=&\hspace{-1mm} \left(\textstyle{\frac{1}{4}},
\textstyle{\frac{1}{2}}, 1, 1, 1, \textstyle{\frac{1}{2}} \right) \\[1mm]
(r_{13}, r_{14}, r_{15}, r_{16}, r_{17}, r_{18})
\hspace{-1mm}&=&\hspace{-1mm} \left( \textstyle{\frac{1}{2}},
\textstyle{\frac{1}{4}}, \textstyle{\frac{1}{2}},
\textstyle{\frac{1}{2}}, \textstyle{\frac{1}{2}}, 1 \right) \\[1mm]
(r_{19}, r_{20}, r_{21}, r_{22}, r_{23}, r_{24})
\hspace{-1mm}&=&\hspace{-1mm} \left( \textstyle{\frac{1}{2}}, 1,
1, 1, \textstyle{\frac{1}{4}}, 1 \right) \: .
\end{array} \label{eq:symmetry_factors}
\end{equation}

Thus, the task of computing the two-loop amplitude reduces to
determining the coefficients $c_{i,\sigma_j}$ as functions of the
external momenta. This is achieved by applying generalized cuts to
both sides of eq. (\ref{eq:6gluonampansatz}) which has the effect
of turning either side into a contour integral in the complex
plane. More specifically, the left hand side of eq.
(\ref{eq:6gluonampansatz}) is turned into a product of tree
amplitudes whose external states coincide with either the external
states of the two-loop amplitude or with the states traveling
along the propagators that are becoming on-shell. On the right
hand side of eq. (\ref{eq:6gluonampansatz}), the generalized cut
has the effect of removing all the integrals that do not contain
the propagators being cut, and turning the remaining integrals
into contour integrals.
\begin{figure}[!h]
\begin{center}
\includegraphics[angle=0, width=0.93\textwidth]{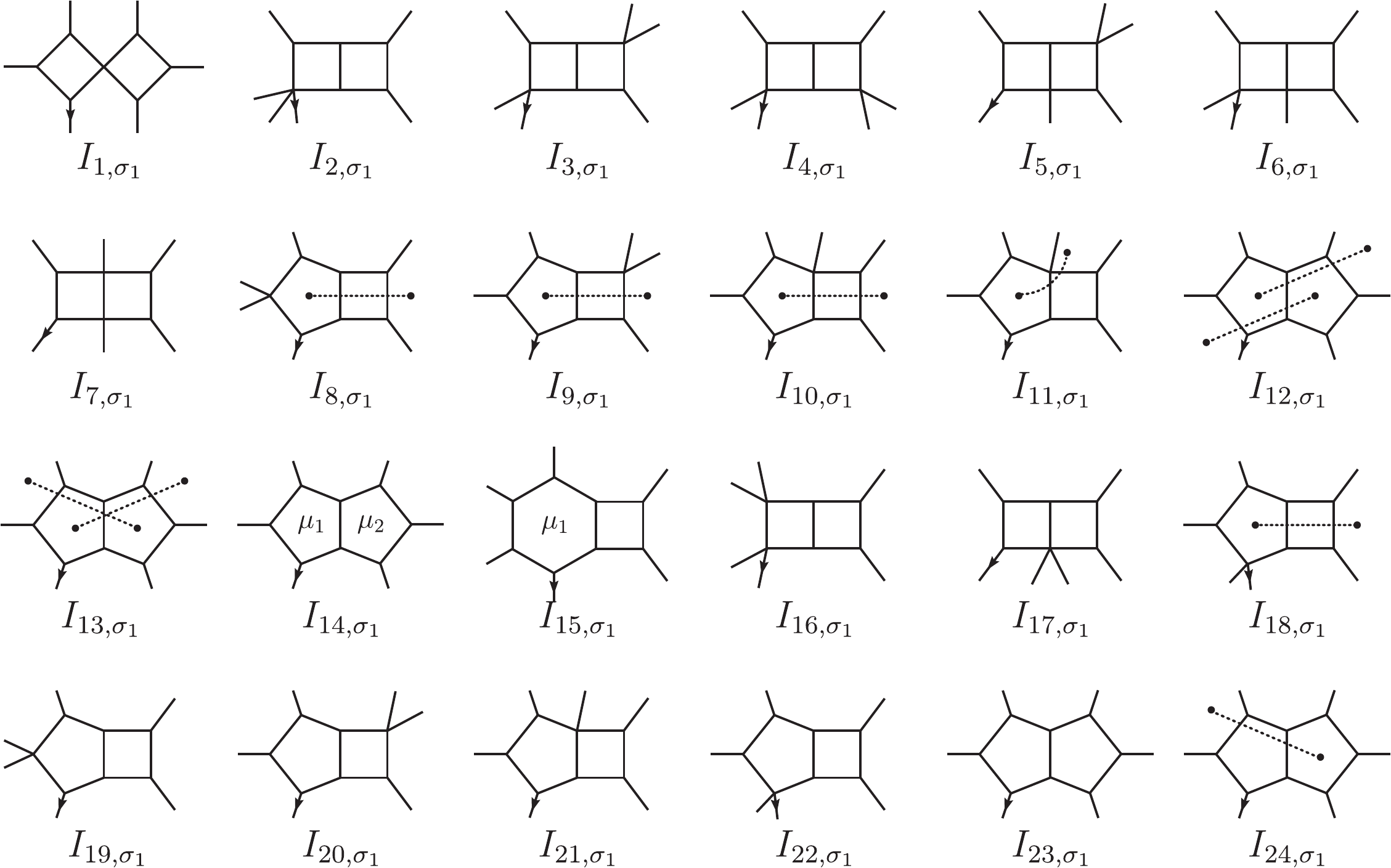}
\caption{The integral basis in terms of which the two-loop
six-point amplitude is expressed. The integrals shown here are
recorded in the $\sigma_1$ permutation where the external momenta
are labeled clockwise starting with $k_1$ at the position of the
arrow. The basis is overcomplete, giving rise to non-uniqueness of
the integral coefficients in eq.
(\ref{eq:6gluonampansatz}).}\label{fig:6-point_integral_basis}
\end{center}
\end{figure}

To limit the number of terms remaining on the right hand of eq.
(\ref{eq:6gluonampansatz}) as much as possible, it would therefore
be natural to start by imposing as many simultaneous
four-dimensional cut constraints as possible, which at two loops
would lead us to consider octacuts. Thus having obtained the
coefficients of integrals that admit an octacut, we could then
proceed to relax one cut constraint to allow the contributions of
the double box integrals in Fig. \ref{fig:6-point_integral_basis}
and in turn determine their coefficients. This, however, is not
the route we will be following: instead, we will focus our
attention on the heptacuts of eq. (\ref{eq:6gluonampansatz}).
Indeed, imposing seven on-shell constraints on two
four-dimensional loop momenta leaves an unconstrained parameter
$z$, and this makes it easy to survey the global poles of the
integrand (to be defined properly in Section
\ref{sec:extraction_of_integral_coeffs}) on the right hand side of eq.
(\ref{eq:6gluonampansatz}). This in turn will make it
straightforward to determine the tori encircling the global poles,
which we shall refer to as leading singularity contours.

It is important to distinguish between these leading singularity
contours and the maximal-cut contours. Indeed, as explained in
ref. \cite{Kosower:2011ty}, the maximal-cut contours are
particular linear combinations $\sigma$ of these tori whose
coefficients are determined by the requirement that any function
that integrates to zero on $\mathbb{R}^D \times \mathbb{R}^D$
should also integrate to zero on $\sigma$. This constraint ensures
that two Feynman integrals which are equal, possibly through some
non-trivial relations, will also have equal maximal cuts. As
argued in ref. \cite{Kosower:2011ty}, multidimensional contours
$\sigma$ satisfying this consistency condition are guaranteed to
produce correct results for scattering amplitudes in any gauge
theory, not only $\mathcal{N}=4$ SYM theory. It would therefore be
very interesting to determine these maximal-cut contours.

Such an analysis is, however, not currently possible: the set of
linear relations between two-loop integrals includes the set of
all integration-by-parts (IBP) identities between the various
tensor integrals\footnote{By ``tensor integral'' is meant an
integral whose integrand's numerator contains powers of the loop
momenta contracted into external vectors; the corresponding
standard Feynman integral with a 1 in the numerator is referred to
as a ``scalar integral'' in this terminology.} arising from the
Feynman rules of gauge theory, and at present a complete knowledge
of these relations is not available. Nevertheless, all the details
needed to carry out this analysis (in particular, the enumeration
of the global poles of the integrand and the expressions for the
cut integrals) are provided in the intermediate results in Section
\ref{sec:heptacut_1} and in Appendix \ref{sec:appendix}, and the
correct heptacut contours can therefore be determined immediately
once the IBP relations do become available.

With the cyclic ordering of the external momenta shown in Fig.
\ref{fig:6-point_integral_basis}, the two-loop six-point heli\-city
amplitudes in $\mathcal{N}=4$ SYM theory admit nine different
heptacuts, dictated by the eight double box integrals and the
factorized double box integral in Fig.
\ref{fig:6-point_integral_basis}, respectively $I_{2, \sigma_1},
I_{3, \sigma_1}, I_{4, \sigma_1}, I_{5, \sigma_1},$ $I_{6,
\sigma_1}, I_{7, \sigma_1}, I_{16, \sigma_1}, I_{17, \sigma_1}$
and $I_{1, \sigma_1}$. We label these heptacuts respectively $\#1,
\ldots, \#9$ and study them in turn in Sections
\ref{sec:heptacut_1} and
\ref{sec:heptacut_2}-\ref{sec:heptacut_9}.
\\
\\
In attempting to obtain the integral coefficients $c_{i,\sigma_j}$
in eq. (\ref{eq:6gluonampansatz}) from generalized
four-dimensional cuts, one encounters two technical issues. The
first point is that the basis is overcomplete, and the integral
coefficients are therefore not uniquely defined. This feature will
manifest itself as the appearance of free parameters in the
solutions of the linear equations satisfied by the integral
coefficients. This in turn means that one has to set a subset of
the integral coefficients equal to specific values in order to
obtain unique solutions for the remaining coefficients. The
non-uniqueness of the integral coefficients is accounted for by
the existence of various linear relations between the integrals in
Fig. \ref{fig:6-point_integral_basis}, as was explained carefully
in ref. \cite{Cachazo:2008hp}.

The second point is that the coefficients of the $\mu$-integrals
$I_{14, \sigma_1}$ and $I_{15, \sigma_1}$ (thus called because
their integrands contain factors involving the
$(-2\epsilon)$-dimensional part of the loop momenta) are of
$\mathcal{O}(\epsilon)$ in the dimensional regulator and hence are
not obtainable from four-dimensional cuts \cite{Bern:2000dn}. As
we restrict ourselves to taking four-dimensional generalized cuts
in this paper, we shall therefore not be concerned with these
integral coefficients.

\subsection{Example: evaluation of heptacut \#1}\label{sec:heptacut_1}

This section is intended as a pedagogical example of the use of
the leading singularity method to obtain integral coefficients of
two-loop scattering amplitudes in $\mathcal{N}=4$ SYM theory. In
this example, we evaluate both sides of the two-loop equation
(\ref{eq:6gluonampansatz}) resulting after imposing the simplest
of the nine heptacuts that the six-point amplitude admits, labeled
\#1. This heptacut is defined by the on-shell constraints in eqs.
(\ref{eq:on-shell_constraint_1})-(\ref{eq:on-shell_constraint_7})
with the vertex momenta
\begin{equation}
\begin{array}{lll}
K_1 = k_{123} & \hspace{1.2cm} K_2 = k_4 & \hspace{1.2cm} K_3 = 0 \\
K_4 = k_5 & \hspace{1.2cm} K_5 = k_6 & \hspace{1.2cm} K_6 = 0 \: .
\end{array} \label{eq:def_of_K1_to_K6_heptacut_1}
\end{equation}
After having evaluated this heptacut of eq.
(\ref{eq:6gluonampansatz}), we will set up equations satisfied by
the full integral coefficients (i.e., including both parity-even
and odd parts) and solve these equations explicitly. Along the
way, we also present the 22 tori encircling the global poles of
the integrand associated with this heptacut. Finally, we comment
briefly on how the linear equations obtained from the remaining
heptacuts $\#2, \ldots, \#9$ (details of which are provided in
Appendix \ref{sec:appendix}) can be used to obtain the parity-even
part of the two-loop six-point integral coefficients directly and report
agreement with the results originally found in ref.
\cite{Bern:2008ap}. \clearpage

\subsubsection{Heptacut \#1 of the right hand side of eq. (\ref{eq:6gluonampansatz})}
\label{sec:heptacut_1_of_RHS_of_2-loop_eq}

Applying heptacut \#1 to the right hand side of eq.
(\ref{eq:6gluonampansatz}) will leave the linear combination of
cut integrals shown in Fig. \ref{fig:heptacut_1} below.

\begin{figure}[!h]
\begin{center}
\includegraphics[angle=0,
width=0.9\textwidth]{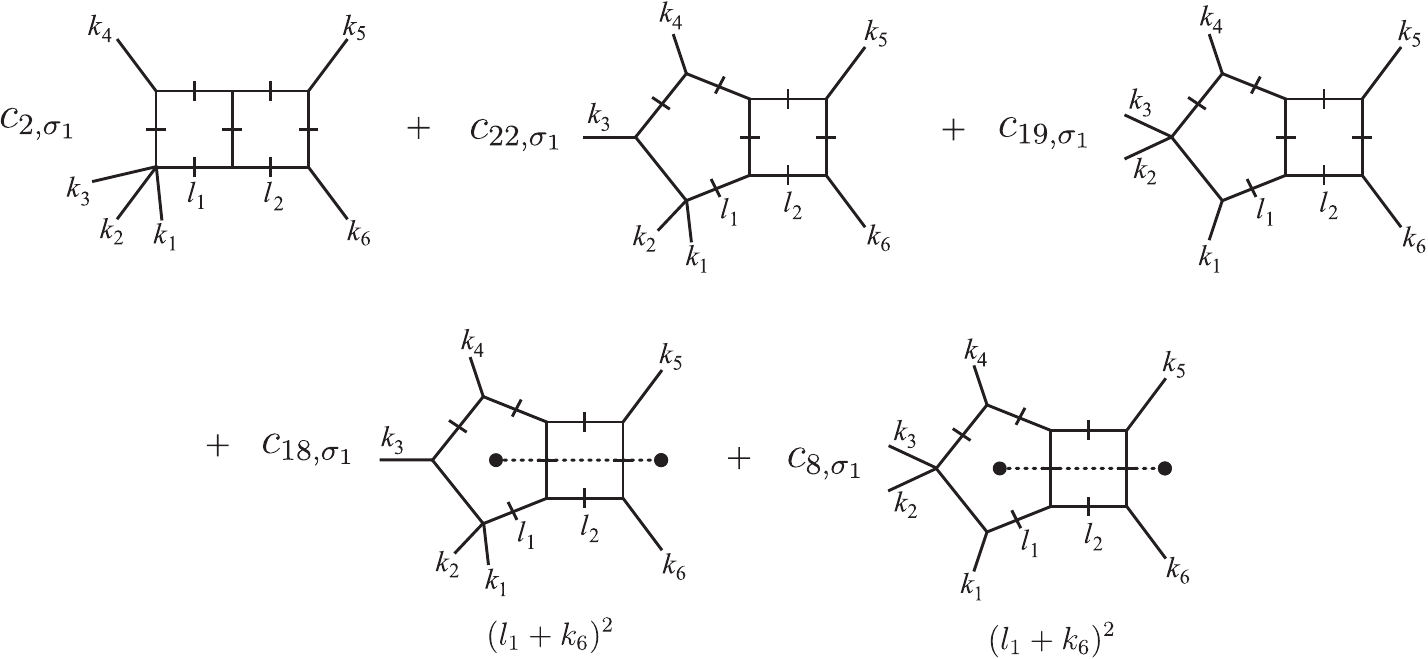}\caption{The
integrals remaining on the right hand side of eq.
(\ref{eq:6gluonampansatz}) after applying the heptacut labeled
\#1. The cut propagators are illustrated by the inclusion of an
additional orthogonal line. This heptacut is defined by the
on-shell constraints in eqs.
(\ref{eq:on-shell_constraint_1})-(\ref{eq:on-shell_constraint_7})
with the vertex momenta given by eq.
(\ref{eq:def_of_K1_to_K6_heptacut_1}). }\label{fig:heptacut_1}
\end{center}
\end{figure}
\vspace{-0.5cm}

We will use the loop momentum parametrizations in eqs.
(\ref{eq:l1_parametrized})-(\ref{eq:l2_parametrized}).
Furthermore, it will be convenient to define the spinor ratios
\begin{equation}
\begin{array}{lll} P_1 = -\frac{\langle K_1^\flat k_6 \rangle}
{2\langle K_2^\flat k_6 \rangle} \: , \hspace{1cm} & P_2 = -
\frac{K_1^\flat \cdot k_{12} - \frac{1}{2} s_{12}}{\langle
K_2^{\flat -} | \slashed{k}_{12} | K_1^{\flat -} \rangle} \: ,
\hspace{1cm} & P_3 = -\frac{\langle K_1^\flat k_1
\rangle} {2\langle K_2^\flat k_1 \rangle} \\[1mm]
P_4 = -\frac{\langle K_1^\flat K_4^\flat \rangle} {2\langle
K_2^\flat K_4^\flat \rangle} \: , & Q_1 = -\frac{[K_1^\flat
K_5^\flat]}{2[K_1^\flat K_4^\flat]} &
\label{eq:heptacut_1_spinor_ratios}
\end{array}
\end{equation}
and their parity conjugates
\begin{equation}
\begin{array}{lll} P_1^\bullet = -\frac{[K_1^\flat k_6]}
{2[K_2^\flat k_6]} \: , \hspace{1cm} & P_2^\bullet = -
\frac{K_1^\flat \cdot k_{12} - \frac{1}{2} s_{12}}{\langle
K_1^{\flat -} | \slashed{k}_{12} | K_2^{\flat -} \rangle} \: ,
\hspace{1cm} & P_3^\bullet = -\frac{[K_1^\flat
k_1]}{2[K_2^\flat k_1]} \\[1mm]
P_4^\bullet = -\frac{[K_1^\flat K_4^\flat]}{2[K_2^\flat
K_4^\flat]} \: , & Q_1^\bullet = -\frac{\langle K_1^\flat
K_5^\flat \rangle}{2\langle K_1^\flat K_4^\flat\rangle} \: . &
\end{array} \label{eq:heptacut_1_conj_spinor_ratios}
\end{equation}
This heptacut belongs to case I treated in Section
\ref{sec:two-loop_heptacuts}: each of the vertical propagators in
the double box integral $I_{2,\sigma_1}$ is part of some
three-particle vertex. There are thus six kinematical solutions
(shown in Fig. \ref{fig:kinematical_solutions_case_I}) to the
on-shell constraints. The heptacut of the double box integral
receives contributions from each of the kinematical solutions.
These contributions were found in Section \ref{sec:two-loop_heptacuts} (see eq.
(\ref{eq:case_I_Jacobians})) and take the form
\begin{equation}
\oint_{\Gamma_i} dz \, J_i(z)
\end{equation}
where, using the notation in eqs. (\ref{eq:def_of_K1_to_K6_heptacut_1})-(\ref{eq:heptacut_1_conj_spinor_ratios}),
the Jacobians are
\begin{equation}
J_i(z) \hspace{2mm} = \hspace{2mm} \frac{1}{32 \gamma_1 \gamma_2}
\times \left\{ \begin{array}{ll} \left( \langle K_1^{\flat -} |
\slashed{K}_5^\flat | K_2^{\flat -}
\rangle \hspace{0.7mm} z (z-P_1^\bullet) \right)^{-1} & \hspace{4mm} \mathrm{for} \hspace{3mm} i=2,6 \\
\left( \langle K_2^{\flat -} | \slashed{K}_5^\flat | K_1^{\flat -}
\rangle \hspace{0.7mm} z (z-P_1) \right)^{-1} & \hspace{4mm} \mathrm{for} \hspace{3mm} i=4, 5\\
\left( \langle K_4^{\flat -} | \slashed{K}_1^\flat | K_5^{\flat -}
\rangle \hspace{0.7mm} z
(z-Q_1^\bullet) \right)^{-1} & \hspace{4mm} \mathrm{for} \hspace{3mm} i=1 \\
\left( \langle K_5^{\flat -} | \slashed{K}_1^\flat | K_4^{\flat -}
\rangle \hspace{0.7mm} z (z-Q_1) \right)^{-1} & \hspace{4mm}
\mathrm{for} \hspace{3mm} i=3 \: .
\end{array} \right.  \label{eq:heptacut_1_Jacobians}
\end{equation}
Accordingly, the heptacut right hand side of eq.
(\ref{eq:6gluonampansatz}), displayed in Fig.
\ref{fig:heptacut_1}, receives contributions from each of the six
kinematical solutions of the form
\begin{equation}
\oint_{\Gamma_i} dz \, J_i (z) \left. \left( c_{2,\sigma_1}
\hspace{0.5mm}+\hspace{0.5mm} \frac{c_{22,\sigma_1}}{(\ell_1 -
k_{12})^2} \hspace{0.5mm}+\hspace{0.5mm}
\frac{c_{19,\sigma_1}}{(\ell_1 - k_1)^2}
\hspace{0.5mm}+\hspace{0.5mm} \frac{c_{18,\sigma_1} (\ell_1 +
k_6)^2}{(\ell_1 - k_{12})^2} \hspace{0.5mm}+\hspace{0.5mm}
\frac{c_{8,\sigma_1} (\ell_1 + k_6)^2}{(\ell_1 - k_1)^2}
\right)\right|_{\mathcal{S}_i}
\end{equation}
where the subscript in $\big(\cdots \big)\big|_{\mathcal{S}_i}$
indicates that the function is to be evaluated in the
parametrization
(\ref{eq:l1_parametrized})-(\ref{eq:l2_parametrized}) with the
parameters set equal to the values
\begin{equation}
\begin{array}{ll}
\alpha_1 = 1 \: , \hspace{8mm} & \beta_1 = 0\\
\alpha_2 = 0 \: , \hspace{8mm} & \beta_2 = 1
\end{array} \label{eq:heptacut_1_on-shell_values_of_alpha_1_to_beta_2}
\end{equation}
and those given in Fig. \ref{fig:kinematical_solutions_case_I} with
the functions $\beta_3 (z)$ and $\beta_4 (z)$ quoted below solutions
$\mathcal{S}_5$ and $\mathcal{S}_6$ being given by
\begin{eqnarray}
\beta_3 (z) \hspace{-1.5mm}&=&\hspace{-1.5mm} - \frac{\langle k_6
k_4 \rangle (z - P_1)}{2 \langle k_5 k_4 \rangle (z - P_4)} \label{eq:heptacut_1_beta_3} \\[1mm]
\beta_4 (z) \hspace{-1.5mm}&=&\hspace{-1.5mm} - \frac{[k_6 k_4] (z
- P_1^\bullet)}{2[k_5 k_4] (z - P_4^\bullet)} \: . \label{eq:heptacut_1_beta_4}
\end{eqnarray}
Instead of displaying all six contributions individually, we can
make use of the fact that the kinematical solutions come in three
parity-conjugate pairs. Namely, the contributions coming from two
parity-conjugate solutions are obtainable from each other by the
replacements
\begin{eqnarray}
\langle ij \rangle &\longleftrightarrow& [ij] \label{eq:parity_conj_rule_1}\\
\langle K_i^- | \slashed{P} | K_j^- \rangle &\longleftrightarrow&
\langle K_j^- | \slashed{P} | K_i^- \rangle \label{eq:parity_conj_rule_2} \\
P_i &\longleftrightarrow& P_i^\bullet \label{eq:parity_conj_rule_3} \\
Q_i &\longleftrightarrow& Q_i^\bullet \label{eq:parity_conj_rule_4} \\
\alpha_3 &\longleftrightarrow& \alpha_4 \label{eq:parity_conj_rule_5} \\
\beta_3 &\longleftrightarrow& \beta_4
\label{eq:parity_conj_rule_6}
\end{eqnarray}
where the replacement rules
(\ref{eq:parity_conj_rule_5})-(\ref{eq:parity_conj_rule_6})
specify that if one solution has, e.g., $z=\alpha_3$ as the free
parameter, then the parity-conjugate solution should be understood
as having $z=\alpha_4$ as the free parameter.
\\
\\
Putting everything together, the result of applying heptacut \#1
to the right hand side of eq. (\ref{eq:6gluonampansatz}) is
\begin{equation}
\frac{1}{4} \sum_{i=1}^6 \oint_{\Gamma_i} dz \, J_i(z) K_i(z)
\end{equation}
where the Jacobians are given in eq. (\ref{eq:heptacut_1_Jacobians})
and the kernels evaluated on the six kinematical solutions
(illustrated in Fig. \ref{fig:kinematical_solutions_case_I}) are
\begin{eqnarray}
K_1 (z) \hspace{-1mm}&=&\hspace{-1mm} c_{2,\sigma_1} - \frac{1}{2}
\frac{c_{22,\sigma_1}}{\langle K_1^{\flat -} | \slashed{k}_{12} |
K_2^{\flat -} \rangle (P_1^\bullet - P_2^\bullet)} - \frac{1}{2}
\frac{c_{19,\sigma_1}}{\langle K_1^{\flat -} | \slashed{k}_1 |
K_2^{\flat -} \rangle (P_1^\bullet - P_3^\bullet)}
\phantom{aaaaaaa}
\label{eq:cut1_6gluons_kinsol5_result} \\[3mm]
K_2 (z) \hspace{-1mm}&=&\hspace{-1mm} c_{2,\sigma_1} - \frac{1}{2}
\frac{c_{22,\sigma_1}}{\langle K_1^{\flat -} | \slashed{k}_{12} |
K_2^{\flat -} \rangle (z - P_2^\bullet)} - \frac{1}{2}
\frac{c_{19,\sigma_1}}{\langle K_1^{\flat -} | \slashed{k}_1 |
K_2^{\flat -} \rangle (z - P_3^\bullet)} \nonumber \\
&\phantom{=}& \hspace{8mm} - \frac{c_{18,\sigma_1} \langle
K_1^{\flat -} | \slashed{k}_6 | K_2^{\flat -} \rangle (z -
P_1^\bullet)}{\langle K_1^{\flat -} | \slashed{k}_{12} |
K_2^{\flat -} \rangle (z - P_2^\bullet)} - \frac{c_{8,\sigma_1}
\langle K_1^{\flat -} | \slashed{k}_6 | K_2^{\flat -} \rangle (z -
P_1^\bullet)}{\langle K_1^{\flat -} | \slashed{k}_1 | K_2^{\flat
-} \rangle (z - P_3^\bullet)} \label{eq:cut1_6gluons_kinsol1_result} \\[3mm]
K_3 (z) \hspace{-1mm}&=&\hspace{-1mm} \mbox{parity conjugate of
$K_1(z)$ \hspace{0.3mm} (obtained by applying eqs.
(\ref{eq:parity_conj_rule_1})-(\ref{eq:parity_conj_rule_6}))} \\[3mm]
K_4 (z) \hspace{-1mm}&=&\hspace{-1mm} \mbox{parity conjugate of
$K_2(z)$ \hspace{0.3mm} (obtained by applying eqs.
(\ref{eq:parity_conj_rule_1})-(\ref{eq:parity_conj_rule_6}))} \\[3mm]
K_5 (z) \hspace{-1mm}&=&\hspace{-1mm} c_{2,\sigma_1} - \frac{1}{2}
\frac{c_{22,\sigma_1}}{\langle K_2^{\flat -} | \slashed{k}_{12} |
K_1^{\flat -} \rangle (z - P_2)} - \frac{1}{2}
\frac{c_{19,\sigma_1}}{\langle K_2^{\flat -} |
\slashed{k}_1 | K_1^{\flat -} \rangle (z - P_3)} \nonumber \\
&\phantom{=}& \hspace{8mm} - \frac{c_{18,\sigma_1} \langle
K_2^{\flat -} | \slashed{k}_6 | K_1^{\flat -} \rangle (z -
P_1)}{\langle K_2^{\flat -} | \slashed{k}_{12} | K_1^{\flat -}
\rangle (z - P_2)} - \frac{c_{8,\sigma_1} \langle K_2^{\flat -} |
\slashed{k}_6 | K_1^{\flat -} \rangle (z - P_1)}{\langle
K_2^{\flat -} | \slashed{k}_1 | K_1^{\flat -} \rangle (z - P_3)}
\label{eq:cut1_6gluons_kinsol3_result} \\[3mm]
K_6 (z) \hspace{-1mm}&=&\hspace{-1mm} \mbox{parity conjugate of
$K_5(z)$ \hspace{0.3mm} (obtained by applying eqs.
(\ref{eq:parity_conj_rule_1})-(\ref{eq:parity_conj_rule_6}))}
\label{eq:cut1_6gluons_kinsol2_result} \: .
\end{eqnarray}

\subsubsection{Heptacut \#1 of the left hand side of eq. (\ref{eq:6gluonampansatz})}
\label{sec:heptacut_1_of_LHS_of_2-loop_eq}

For MHV configurations in $\mathcal{N}=4$ SYM theory, the
ratio of the two-loop amplitude to the corresponding tree-level
amplitude is independent of the distribution of the helicities
of the external states \cite{Grisaru:1977,Nair:1988}. Without loss of generality,
we will therefore assume throughout the paper that the helicities of the external states
are $(1^-, 2^-, 3^+, 4^+, 5^+, 6^+)$. In this case, the result of applying heptacut \#1 to the left
hand side of eq. (\ref{eq:6gluonampansatz}) is
\begin{equation}
i \sum_{i=1}^6 \oint_{\Gamma_i} dz \, J_i(z) \left. \prod_{j=1}^6
A_j^\mathrm{tree}(z) \right|_{\mathcal{S}_i}
\end{equation}
where the cut amplitude evaluated on the six different kinematical
solutions yields
\begin{equation}
\left. \prod_{j=1}^6 A_j^\mathrm{tree}(z) \right|_{\mathcal{S}_i}
\hspace{3mm}=\hspace{3mm} -\frac{i}{16} A^\mathrm{tree}_{--++++}
\times \left\{ \begin{array}{ll} \frac{1}{J_3(z)} \left(
\frac{1}{z} - \frac{1}{z - Q_1} \right) & \hspace{3mm} \mathrm{for} \hspace{3mm} i = 3 \\[2mm]
-\frac{1}{J_4(z)} \left(
\frac{1}{z - P_1} - \frac{1}{z - P_3} \right) & \hspace{3mm} \mathrm{for} \hspace{3mm} i = 4 \\[2mm]
-\frac{1}{J_5(z)} \left(
\frac{1}{z - P_1} - \frac{1}{z - P_3} \right) & \hspace{3mm} \mathrm{for} \hspace{3mm} i = 5 \\[2mm]
0 & \hspace{3mm} \mathrm{for} \hspace{3mm} i = 1,2,6 \: ,
\end{array} \right. \label{eq:heptacut_1_amplitude_cuts}
\end{equation}
with the tree-level amplitude given by
\begin{equation}
A^\mathrm{tree}_{--++++} \hspace{1mm}=\hspace{1mm} \frac{i \langle 12 \rangle^3}
{\langle 23 \rangle \langle 34 \rangle \langle 45 \rangle
\langle 56 \rangle \langle 61 \rangle} \: .
\end{equation}
The expressions in eq. (\ref{eq:heptacut_1_amplitude_cuts}) can be obtained by first exploiting momentum
conservation to simplify the heptacut amplitude $\prod_{j=1}^6 A_j^\mathrm{tree} \big|_{\mathcal{S}_i}$
as much as possible and then substituting the parametrization of the loop momenta
in eqs. (\ref{eq:l1_parametrized})-(\ref{eq:l2_parametrized}) to obtain the heptacut amplitude
as a function of $z$. To further simplify, one can make use of the fact that, for $\mathcal{N}=4$ SYM theory, the
function
\begin{equation}
\varphi_i(z) \hspace{0.5mm}=\hspace{0.5mm} \frac{16 i}{A^\mathrm{tree}_{--++++}} \left. J_i (z)
\prod_{j=1}^6 A_j^\mathrm{tree}(z) \right|_{\mathcal{S}_i}
\end{equation}
only has simple poles in $z$ and has residues $\pm 1$ or $0$ at finite poles, and $0$ at infinity
\cite{Cachazo:2008dx,ArkaniHamed:2009dn}.\footnote{This property holds for
MHV amplitudes (at any loop order) only; however, for $\mbox{N$^k$MHV}$ amplitudes, the residues can be expressed
economically in terms of superconformal $R$-invariants \cite{Drummond:2008vq}.}

Denoting all finite poles of $\varphi_i(z)$ by $X_j$, these facts combined imply that
\begin{equation}
\varphi_i(z) \hspace{1mm}=\hspace{1mm} \sum_j \frac{\mathop{\mathrm{Res}}_{z=X_j} \varphi_i(z)}{z - X_j}.
\end{equation}
It is from this latter form of $\varphi_i(z)$ that the expressions in eq.
(\ref{eq:heptacut_1_amplitude_cuts}) were extracted.

\subsubsection{Extraction of integral coefficients}\label{sec:extraction_of_integral_coeffs}

To summarize the results of Sections
\ref{sec:heptacut_1_of_RHS_of_2-loop_eq} and
\ref{sec:heptacut_1_of_LHS_of_2-loop_eq}, we find that applying
heptacut \#1 to both sides of eq. (\ref{eq:6gluonampansatz})
produces the equation
\begin{equation}
i \sum_{i=1}^6 \oint_{\Gamma_i} dz \, J_i(z) \left. \prod_{j=1}^6
A_j^\mathrm{tree}(z) \right|_{\mathcal{S}_i} \hspace{2mm} =
\hspace{2mm} \frac{1}{4} \sum_{i=1}^6 \oint_{\Gamma_i} dz \,
J_i(z) K_i(z) \label{eq:heptacut_2-loop_equation}
\end{equation}
where the Jacobians $J_i (z)$, the cut amplitude $\prod_{j=1}^6 A_j^\mathrm{tree} (z)$
and the kernels $K_i (z)$ are given in eqs. (\ref{eq:heptacut_1_Jacobians}),
(\ref{eq:heptacut_1_amplitude_cuts}) and (\ref{eq:cut1_6gluons_kinsol5_result})-(\ref{eq:cut1_6gluons_kinsol2_result}),
respectively. The kernels $K_i (z)$ contain the integral coefficients, and
by making various appropriate choices of the contours $\Gamma_i$,
eq. (\ref{eq:heptacut_2-loop_equation}) produces a system of linear equations
which can be solved to obtain the integral coefficients.

Before proceeding to discussing how equation
(\ref{eq:heptacut_2-loop_equation}) may be used to determine the
integral coefficients of an amplitude, let us first remind
ourselves of the relation of this equation to the original
two-loop equation (\ref{eq:6gluonampansatz}). As we found in
Sections \ref{sec:heptacut_1_of_RHS_of_2-loop_eq} and
\ref{sec:heptacut_1_of_LHS_of_2-loop_eq}, the on-shell constraints
in eqs.
(\ref{eq:on-shell_constraint_1})-(\ref{eq:on-shell_constraint_7})
and (\ref{eq:def_of_K1_to_K6_heptacut_1}) for heptacut \#1 are
solved by setting the parameters $\alpha_i, \beta_j$ equal to the
values quoted in eq.
(\ref{eq:heptacut_1_on-shell_values_of_alpha_1_to_beta_2}) and in
Fig. \ref{fig:kinematical_solutions_case_I}, with the spinor
ratios $P_1, Q_1$ etc. and the functions $\beta_3(z), \beta_4(z)$
being given in eqs.
(\ref{eq:heptacut_1_spinor_ratios})-(\ref{eq:heptacut_1_conj_spinor_ratios})
and (\ref{eq:heptacut_1_beta_3})-(\ref{eq:heptacut_1_beta_4}),
respectively. For any of the six solutions to the on-shell
constraints shown in Fig. \ref{fig:kinematical_solutions_case_I},
one of the loop momentum parameters $\alpha_i, \beta_j$ is set
equal to an unconstrained complex parameter $z$.

The leading singularity contour is, by definition, a torus consisting of circle
factors centered around the seven on-shell values of the
parameters left fixed and where, in addition, one makes a choice
of contour for the unconstrained degree of freedom $z$. The
contour in $z$ can be chosen to encircle the poles of the
Jacobians (\ref{eq:heptacut_1_Jacobians}) or the poles of the loop
momentum parametrization
(\ref{eq:l1_parametrized})-(\ref{eq:l2_parametrized}) at which one
of the loop momenta becomes infinite (see eqs.
(\ref{eq:heptacut_1_beta_3})-(\ref{eq:heptacut_1_beta_4})). The point
$(\alpha_1, \ldots, \alpha_4,$ $\beta_1, \ldots, \beta_4) \in \mathbb{C}^4 \times \mathbb{C}^4$ encircled by the torus is referred
to as a \emph{global pole} of the integrand of the right hand side of eq. (\ref{eq:6gluonampansatz}).

\clearpage
Defining
\begin{equation}
\tau \hspace{1mm}=\hspace{1mm} C_{\alpha_1} (1) \times
C_{\alpha_2}(0) \times C_{\beta_1} (0) \times C_{\beta_2} (1) \: ,
\end{equation}
the six sets of on-shell values of the loop momentum parameters
quoted in Fig. \ref{fig:kinematical_solutions_case_I}, combined
with the various possible choices of contours in $z$, give rise to the following 22
tori corresponding to heptacut \#1,
\begin{equation}
\begin{array}{rl}
T_{\mathcal{S}_1, 0} \hspace{-1mm}&= \hspace{2mm} \tau \times
C_{\alpha_3} (P_1^\bullet) \times C_{\alpha_4}(0)
\times C_{\beta_3 = z} (0) \times C_{\beta_4} (0) \\[1mm]
T_{\mathcal{S}_1, P_1^\bullet} \hspace{-1mm}&= \hspace{2mm} \tau
\times C_{\alpha_3} (P_1^\bullet) \times C_{\alpha_4}(0)
\times C_{\beta_3 = z} (P_1^\bullet) \times C_{\beta_4} (0) \\[1mm]
T_{\mathcal{S}_2, 0} \hspace{-1mm}&= \hspace{2mm} \tau \times
C_{\alpha_3 =z} (0) \times C_{\alpha_4}(0)
\times C_{\beta_3} (Q_1^\bullet) \times C_{\beta_4} (0) \\[1mm]
T_{\mathcal{S}_2, P_i^\bullet} \hspace{-1mm}&= \hspace{2mm} \tau
\times C_{\alpha_3 =z} (P_i^\bullet) \times C_{\alpha_4}(0)
\times C_{\beta_3} (Q_1^\bullet) \times C_{\beta_4} (0) \: , \hspace{4mm} i=1,2,3 \\[1mm]
T_{\mathcal{S}_3, 0} \hspace{-1mm}&= \hspace{2mm} \tau \times
C_{\alpha_3} (0) \times C_{\alpha_4}(P_1)
\times C_{\beta_3} (0) \times C_{\beta_4 = z} (0) \\[1mm]
T_{\mathcal{S}_3, P_1} \hspace{-1mm}&= \hspace{2mm} \tau \times
C_{\alpha_3} (0) \times C_{\alpha_4}(P_1)
\times C_{\beta_3}(0) \times C_{\beta_4 = z}(P_1) \\[1mm]
T_{\mathcal{S}_4, 0} \hspace{-1mm}&= \hspace{2mm} \tau \times
C_{\alpha_3} (0) \times C_{\alpha_4 = z}(0)
\times C_{\beta_3} (0) \times C_{\beta_4} (Q_1) \\[1mm]
T_{\mathcal{S}_4, P_i} \hspace{-1mm}&= \hspace{2mm} \tau \times
C_{\alpha_3} (0) \times C_{\alpha_4 = z}(P_i)
\times C_{\beta_3} (0) \times C_{\beta_4} (Q_1) \: , \hspace{4mm} i=1,2,3 \\[1mm]
T_{\mathcal{S}_5, 0} \hspace{-1mm}&= \hspace{2mm} \tau \times
C_{\alpha_3} (0) \times C_{\alpha_4 =z}(0)
\times C_{\beta_3} (\beta_3 (z)) \times C_{\beta_4} (0) \\[1mm]
T_{\mathcal{S}_5, P_i} \hspace{-1mm}&= \hspace{2mm} \tau \times
C_{\alpha_3} (0) \times C_{\alpha_4 =z}(P_i)
\times C_{\beta_3} (\beta_3 (z)) \times C_{\beta_4} (0) \: , \hspace{4mm} i=1,\ldots, 4 \\[1mm]
T_{\mathcal{S}_6, 0} \hspace{-1mm}&= \hspace{2mm} \tau \times
C_{\alpha_3 =z} (0) \times C_{\alpha_4}(0)
\times C_{\beta_3} (0) \times C_{\beta_4} (\beta_4(z)) \\[1mm]
T_{\mathcal{S}_6, P_i^\bullet} \hspace{-1mm}&= \hspace{2mm} \tau
\times C_{\alpha_3 =z} (P_i^\bullet) \times C_{\alpha_4}(0) \times
C_{\beta_3} (0) \times C_{\beta_4} (\beta_4(z)) \: , \hspace{4mm}
i=1,\ldots, 4
\end{array} \label{eq:heptacut_1_leading_singularity_tori}
\end{equation}
where the indices in $T_{\mathcal{S}_i, X}$ refer to the
kinematical solution $\mathcal{S}_i$ associated with the given
torus, and $X$ is the pole around which the $z$-contour is taken.
Furthermore, $C_{\alpha_j}(X)$ (e.g.) denotes a small circle in
the $\alpha_j$-plane centered around $\alpha_j=X$; in addition, we
write $C_{\alpha_j = z}(X)$ whenever the $\alpha_j$-variable in
question is left unfixed by the heptacut constraints. Finally, the
spinor ratios $P_1, Q_1$ etc. and the functions $\beta_3(z),
\beta_4(z)$ are given in eqs.
(\ref{eq:def_of_K1_to_K6_heptacut_1})-(\ref{eq:heptacut_1_conj_spinor_ratios})
and (\ref{eq:heptacut_1_beta_3})-(\ref{eq:heptacut_1_beta_4}),
respectively.

The relation of eq. (\ref{eq:heptacut_2-loop_equation}) to the
original two-loop equation (\ref{eq:6gluonampansatz}) can now
easily be stated: the former equation is obtained from the latter
by changing the integration contour from the real slice
$\mathbb{R}^D \times \mathbb{R}^D$ to an arbitrary linear
combination of the 22 tori given in eq.
(\ref{eq:heptacut_1_leading_singularity_tori}). The contributions
from, e.g., the first two terms in this linear combination $a_{\mathcal{S}_1, 0} T_{\mathcal{S}_1, 0} +
a_{\mathcal{S}_1, P_1^\bullet} T_{\mathcal{S}_1, P_1^\bullet} + \cdots$ are
found by integrating out all parameters except $z = \beta_3$; this
leads to the $\Gamma_1$-integral in eq.
(\ref{eq:heptacut_2-loop_equation}) where $\Gamma_1 =
a_{\mathcal{S}_1, 0} C_{\beta_3 = z} (0) + a_{\mathcal{S}_1,
P_1^\bullet} C_{\beta_3 = z} (P_1^\bullet)$.

Thus having explicitly stated the relation of eq.
(\ref{eq:heptacut_2-loop_equation}) to the original two-loop
equation (\ref{eq:6gluonampansatz}), let us now return to the
question of how the equation (\ref{eq:heptacut_2-loop_equation})
may be used to determine the integral coefficients of an
amplitude. As explained in the beginning of this section, the
class of contours $\sum_{i,j} a_{i,j} T_{\mathcal{S}_i, X_j}$ that
will produce correct results for the integral coefficients in any
gauge theory amplitude are those that annihilate all functions
that integrate to zero on the real slice $\mathbb{R}^D \times
\mathbb{R}^D$. Determining such contours requires the knowledge of
all integration-by-parts identities at six points; however, as a
complete knowledge of all such relations is presently not
available, we will here proceed as in ref. \cite{Cachazo:2008hp}
and assume that \emph{any} contour is valid. Indeed, as already
mentioned above, the purpose of this section is mainly to provide
a pedagogical exposition of the use of the leading singularity method
to obtain integral coefficients of $\mathcal{N}=4$ SYM
amplitudes.

Below we will use the following notation: $C_\epsilon (X_k)$
denotes a circle centered around $z = X_k$ of some appropriately
small radius $\epsilon$ (i.e., small enough to not enclose any
other poles), and $\Gamma_i = \delta_{i,j} C_\epsilon (X_k)$
denotes a contour which is zero on all six sheets except for the
sheet supporting kinematical solution $\mathcal{S}_j$; on this
sheet, the contour is a small circle centered around $X_k$. From
the equation (\ref{eq:heptacut_2-loop_equation}) we then find, for
example, that
\begin{itemize}
\item setting $\Gamma_i = \delta_{i,5} \hspace{0.2mm} C_\epsilon
(P_1)$ produces the equation
\begin{eqnarray}
&\phantom{=}&  \hspace{-15mm} \frac{1}{4} \left( c_{2,\sigma_1} -
\frac{1}{2} \frac{c_{22,\sigma_1}}{\langle K_2^{\flat -} |
\slashed{k}_{12} | K_1^{\flat -} \rangle (P_1 - P_2)} -
\frac{1}{2} \frac{c_{19,\sigma_1}}{\langle K_2^{\flat -} |
\slashed{k}_1 | K_1^{\flat -} \rangle (P_1 - P_3)} \right) \nonumber \\
&=& -2\gamma_1 \gamma_2 P_1 \langle K_2^{\flat-} | \slashed{K}_5 |
K_1^{\flat -} \rangle
A^\mathrm{tree}_{--++++}\label{eq:6gluonMHV_cut1_eq1}
\end{eqnarray}

\item setting $\Gamma_i = \delta_{i,5} \hspace{0.2mm} C_\epsilon
(P_2)$ produces the equation
\begin{equation}
\frac{1}{4} \left( -\frac{1}{2}\frac{c_{22,\sigma_1}}{\langle
K_2^{\flat -}| \slashed{k}_{12} | K_1^{\flat -} \rangle} -
\frac{c_{18,\sigma_1} \langle K_2^{\flat -} | \slashed{k}_6 |
K_1^{\flat -} \rangle (P_2 - P_1)}{\langle K_2^{\flat -} |
\slashed{k}_{12} | K_1^{\flat -} \rangle} \right) \hspace{1mm} =
\hspace{1mm} 0\label{eq:6gluonMHV_cut1_eq2}
\end{equation}
\item setting $\Gamma_i = \delta_{i,5} \hspace{0.2mm} C_\epsilon
(P_3)$ produces the equation
\begin{eqnarray}
&\phantom{=}& \hspace{-2cm} \frac{1}{4} \left( -\frac{1}{2}
\frac{c_{19,\sigma_1}}{\langle K_2^{\flat -} | \slashed{k}_1 |
K_1^{\flat -} \rangle} - \frac{c_{8,\sigma_1} \langle K_2^{\flat
-} | \slashed{k}_6 | K_1^{\flat -} \rangle (P_3 - P_1)}{\langle
K_2^{\flat -} | \slashed{k}_1 | K_1^{\flat -}
\rangle} \right) \nonumber \\
&=& \hspace{-1mm} 2\gamma_1 \gamma_2 P_3 (P_3 - P_1) \langle
K_2^{\flat -} | \slashed{K}_5 | K_1^{\flat -} \rangle
A^\mathrm{tree}_{--++++} \label{eq:6gluonMHV_cut1_eq3}
\end{eqnarray}

\item setting $\Gamma_i = \delta_{i,2} \hspace{0.2mm} C_\epsilon
(P_1^\bullet)$ produces the equation
\begin{equation}
\frac{1}{4} \left( c_{2,\sigma_1} - \frac{1}{2}
\frac{c_{22,\sigma_1}}{\langle K_1^{\flat -} | \slashed{k}_{12}|
K_2^{\flat -} \rangle (P_1^\bullet - P_2^\bullet)} - \frac{1}{2}
\frac{c_{19,\sigma_1}}{\langle K_1^{\flat -} | \slashed{k}_1 |
K_2^{\flat -} \rangle (P_1^\bullet - P_3^\bullet)} \right)
\hspace{1mm} = \hspace{1mm} 0 \label{eq:6gluonMHV_cut1_eq4}
\end{equation}

\item setting $\Gamma_i = \delta_{i,2} \hspace{0.2mm} C_\epsilon
(P_2^\bullet)$ produces the equation
\begin{equation}
\frac{1}{4} \left( -\frac{1}{2}\frac{c_{22,\sigma_1}}{\langle
K_1^{\flat -}| \slashed{k}_{12} | K_2^{\flat -} \rangle} -
\frac{c_{18,\sigma_1} \langle K_1^{\flat -} | \slashed{k}_6 |
K_2^{\flat -} \rangle (P_2^\bullet - P_1^\bullet)}{\langle
K_1^{\flat -} | \slashed{k}_{12} | K_2^{\flat -} \rangle} \right)
\hspace{1mm} = \hspace{1mm} 0 \label{eq:6gluonMHV_cut1_eq5}
\end{equation}

\item setting $\Gamma_i = \delta_{i,2} \hspace{0.2mm} C_\epsilon
(P_3^\bullet)$ produces the equation
\begin{equation}
\frac{1}{4} \left( -\frac{1}{2} \frac{c_{19,\sigma_1}}{\langle
K_1^{\flat -} | \slashed{k}_1 | K_2^{\flat -} \rangle} -
\frac{c_{8,\sigma_1} \langle K_1^{\flat -} | \slashed{k}_6 |
K_2^{\flat -} \rangle (P_3^\bullet - P_1^\bullet)}{\langle
K_1^{\flat -} | \slashed{k}_1 | K_2^{\flat -} \rangle} \right)
\hspace{1mm} = \hspace{1mm} 0 \: . \label{eq:6gluonMHV_cut1_eq6}
\end{equation}
\end{itemize}
We have here deliberately chosen an overcomplete system of
equations as a consistency check on the method. We in fact find
that the six equations
(\ref{eq:6gluonMHV_cut1_eq1})-(\ref{eq:6gluonMHV_cut1_eq6}) are
consistent, and that expressed in terms of the quantity
\begin{equation}
Y \hspace{1mm} = \hspace{1mm} -8 \gamma_1 \gamma_2 P_1 \langle
K_2^{\flat -} | \slashed{K}_5 | K_1^{\flat -} \rangle
A^\mathrm{tree}_{--++++} \left( 1- \frac{\langle K_1^{\flat -} |
\slashed{k}_1 | K_2^{\flat -} \rangle (P_1^\bullet -
P_3^\bullet)}{\langle K_2^{\flat -} | \slashed{k}_1 | K_1^{\flat
-} \rangle (P_1 - P_3)} \right)^{-1} \: ,
\end{equation}
the solution of the equations
(\ref{eq:6gluonMHV_cut1_eq1})-(\ref{eq:6gluonMHV_cut1_eq6}) takes
the following form
\begin{eqnarray}
c_{2,\sigma_1} &=& Y \label{eq:6gluonMHV_cut1_sol1}\\
c_{22,\sigma_1} &=& 0 \label{eq:6gluonMHV_cut1_sol2} \\
c_{19,\sigma_1} &=& 2\langle K_1^{\flat -} | \slashed{k}_1 | K_2^{\flat -} \rangle
(P_1^\bullet - P_3^\bullet) \hspace{0.7mm} Y \label{eq:6gluonMHV_cut1_sol3} \\
c_{18,\sigma_1} &=& 0 \label{eq:6gluonMHV_cut1_sol4} \\
c_{8,\sigma_1} &=& \frac{\langle K_1^{\flat -} | \slashed{k}_1|
K_2^{\flat -} \rangle}{\langle K_1^{\flat -} | \slashed{k}_6|
K_2^{\flat -} \rangle} \hspace{0.7mm} Y
\label{eq:6gluonMHV_cut1_sol5} \: .
\end{eqnarray}
In the following it will be useful to consider the two-loop
amplitude normalized with respect to the tree-level amplitude,
\begin{equation}
M_{6, \hspace{0.4mm} \mathrm{MHV}}^{(2)}
\hspace{0.6mm}=\hspace{0.6mm} \frac{A_{6, \hspace{0.4mm}
\mathrm{MHV}}^{(2)}}{A^\mathrm{tree}_\mathrm{MHV}}
\hspace{1mm}=\hspace{1mm}
\frac{1}{4} \sum_{\substack{i=1,\hspace{0.2mm}\ldots, \hspace{0.2mm} 24 \\
j=1, \hspace{0.2mm} \ldots, \hspace{0.2mm} 12}} r_i
\overline{c}_{i,\sigma_j} I_{i,\sigma_j}
\label{eq:6gluonampansatz_normalized_by_Atree}
\end{equation}
where the normalized coefficients are defined by
\begin{equation}
\overline{c}_{i,\sigma_j} \equiv
\frac{c_{i,\sigma_j}}{A^\mathrm{tree}_\mathrm{MHV}}
\end{equation}
because this object is independent of the distribution of the helicities
of the external states \cite{Grisaru:1977,Nair:1988};
that is, for example, $M_{6, \hspace{0.4mm} --++++}^{(2)} = M_{6, \hspace{0.4mm} -++++-}^{(2)}$. Moreover, the integrand of $M_{6,
\hspace{0.4mm} \mathrm{MHV}}^{(2)}$ can be decomposed into two terms that
are even and odd under parity and which respectively coincide with its
real and imaginary parts.

Thus, the parity-even part of the coefficients found in eqs.
(\ref{eq:6gluonMHV_cut1_sol1})-(\ref{eq:6gluonMHV_cut1_sol5}) is obtained by dividing
by the tree-level amplitude $A^\mathrm{tree}_{--++++}$ and taking
the real part, yielding
\begin{eqnarray}
\mathrm{Re} \hspace{-0.3mm} \left(
\frac{c_{2,\sigma_1}}{A^\mathrm{tree}_{--++++}}\right)
&=& 2 s_{45} s_{56}^2 \label{eq:heptacut_1_parity_even_coeff_1}\\
\mathrm{Re} \hspace{-0.3mm} \left(
\frac{c_{22,\sigma_1}}{A^\mathrm{tree}_{--++++}} \right) &=& 0 \label{eq:heptacut_1_parity_even_coeff_2}\\
\mathrm{Re} \hspace{-0.3mm} \left(
\frac{c_{19,\sigma_1}}{A^\mathrm{tree}_{--++++}}\right) &=& 0  \label{eq:heptacut_1_parity_even_coeff_3}\\
\mathrm{Re} \hspace{-0.3mm} \left(
\frac{c_{18,\sigma_1}}{A^\mathrm{tree}_{--++++}}\right) &=& 0 \label{eq:heptacut_1_parity_even_coeff_4}\\
\mathrm{Re} \hspace{-0.3mm} \left(
\frac{c_{8,\sigma_1}}{A^\mathrm{tree}_{--++++}}\right) &=& 2
s_{56} (s_{123} s_{234} - s_{56} s_{23})
\label{eq:heptacut_1_parity_even_coeff_5}
\end{eqnarray}
where the equalities have been found to hold numerically. The
expressions on the right hand sides of eqs.
(\ref{eq:heptacut_1_parity_even_coeff_1})-(\ref{eq:heptacut_1_parity_even_coeff_5})
are in agreement with the results originally found in ref.
\cite{Bern:2008ap} and reproduced by the leading singularity
method in ref. \cite{Cachazo:2008hp}. Due to a Ward identity
\cite{Grisaru:1977,Nair:1988,Elvang:2009wd} valid for
$\mathcal{N}=4$ supersymmetry, the coefficients of the cyclically
permuted integrals can simply be obtained by cyclic permutation of
the results found here.
\\
\\
For the remaining heptacuts $\#2, \ldots, \#9$ (details of which
are provided in Appendix \ref{sec:appendix}), the leading
singularity method provides similar linear equations satisfied by
the integral coefficients. However, when solving the linear
equations for these more complicated heptacuts, the
overcompleteness of the basis in Fig.
\ref{fig:6-point_integral_basis} entails a problem, carefully
discussed in ref. \cite{Cachazo:2008hp}: due to linear relations
between various integrals in the basis, the integral coefficients
are not unique. This feature will manifest itself as the
appearance of free parameters in the solutions of the linear
equations. Thus, one has to set some of the integral coefficients
equal to specific values in order to obtain unique solutions for
the remaining coefficients.

This ``gauge fixing'' can be easily done for the parity-even part
of the integral coefficients as analytic results for these were
already obtained in ref. \cite{Bern:2008ap}, and one can proceed as
follows. For a given heptacut, encircle a pole and its parity
conjugate to obtain equations of the schematic form
\begin{eqnarray}
\alpha c_1 + \beta c_2 \hspace{-1mm}&=&\hspace{-1mm} \gamma \\
\alpha^* c_1 + \beta^* c_2 \hspace{-1mm}&=&\hspace{-1mm} \delta
\end{eqnarray}
where, for example, $\alpha^*$ denotes the complex conjugate of $\alpha$.
From this pair of equations immediately follows
\begin{eqnarray}
(\alpha + \alpha^*) \hspace{0.6mm} \overline{c}_1^\mathrm{even}
\hspace{0.3mm}+\hspace{0.3mm} (\beta + \beta^*) \hspace{0.6mm}
\overline{c}_2^\mathrm{even} \hspace{-1mm}&=&\hspace{-1mm}
\mathrm{Re} \left(
\frac{\gamma + \delta}{A^\mathrm{tree}_{--++++}} \right) \label{eq:eq_parity_even_part_schematic_1}\\
(\alpha - \alpha^*) \hspace{0.6mm} \overline{c}_1^\mathrm{even}
\hspace{0.3mm}+\hspace{0.3mm} (\beta - \beta^*) \hspace{0.6mm}
\overline{c}_2^\mathrm{even} \hspace{-1mm}&=&\hspace{-1mm} i
\hspace{0.4mm} \mathrm{Im} \left( \frac{\gamma -
\delta}{A^\mathrm{tree}_{--++++}} \right)
\label{eq:eq_parity_even_part_schematic_2}
\end{eqnarray}
where the parity-even part of the coefficients is simply given as
the real part of the (normalized) coefficient,
\begin{equation}
\overline{c}_{i,\sigma_j}^\mathrm{even} \hspace{1mm} \equiv
\hspace{1mm} \mathrm{Re} \hspace{0.8mm} \overline{c}_{i,\sigma_j}
\: .
\end{equation}
The pair of equations
(\ref{eq:eq_parity_even_part_schematic_1})-(\ref{eq:eq_parity_even_part_schematic_2})
thus allows one to solve directly for the parity-even part of the
integral coefficients. In order to obtain unique answers for
these, we make the following ``gauge choices'': we set
\begin{equation}
\left. \begin{array}{rll} \overline{c}_{11,\sigma_j}^\mathrm{even}
\hspace{-1mm}&=&\hspace{-1mm}
s_{\sigma_j(61)} s_{\sigma_j(12)} s_{\sigma_j(123)} \\
\overline{c}_{12,\sigma_j}^\mathrm{even}
\hspace{-1mm}&=&\hspace{-1mm}
s_{\sigma_j(456)} (s_{\sigma_j(345)} s_{\sigma_j(456)} - s_{\sigma_j(12)} s_{\sigma_j(45)}) \\
\overline{c}_{24,\sigma_j}^\mathrm{even}
\hspace{-1mm}&=&\hspace{-1mm} 0
\end{array} \right\} \hspace{5mm} \mathrm{for} \hspace{2mm} j = 1,
\ldots, 12 \label{eq:gauge-fixing_parity_even}
\end{equation}
and also recall that $c_{14, \sigma_j} = c_{15, \sigma_j} = 0$ for
the full coefficients of the $\mu$-integrals. With this choice,
the equations produced by taking leading singularities of both
sides of eq. (\ref{eq:6gluonampansatz}) and then projecting out
the parity-odd part of the coefficients in analogy with eqs.
(\ref{eq:eq_parity_even_part_schematic_1})-(\ref{eq:eq_parity_even_part_schematic_2})
have a unique solution which is in agreement with the
results for the parity-even coefficients originally found in ref.~\cite{Bern:2008ap}; in particular, one finds that
\begin{equation}
\overline{c}_{19, \sigma_j}^\mathrm{even}
\hspace{0.5mm}=\hspace{0.5mm} \cdots \hspace{0.5mm}=\hspace{0.5mm}
\overline{c}_{24, \sigma_j}^\mathrm{even}
\hspace{0.5mm}=\hspace{0.5mm} 0 \: ,
\end{equation}
in agreement with the parity-even part of the amplitude being dual
conformally invariant.

\subsection{Leading singularities vs. loop-level recursion}\label{sec:loop_level_recursion_test}

In this section we report on a comparison between results for the
full (i.e., parity-even and odd) two-loop six-point MHV
$\mathcal{N}=4$ SYM integrand as produced by the leading
singularity method on one hand and recent predictions in the
literature based on a BCFW-like loop-level recursion relation
\cite{ArkaniHamed:2010kv,ArkaniHamed:2010gh} on the other. Because the
results of these papers are expressed in terms of a different
basis from the one used in this paper, it is obviously not
meaningful to check agreement between individual integral
coefficients in the two representations.

A quantity that can be meaningfully compared is the two-loop
integrand: in general, in the planar limit of any field theory,
the loop integrand is a well-defined rational function of the
external momenta (which for example can be thought of as being
produced by the Feynman rules). In our case, the two-loop
integrand is the quantity under the integral sign in eq.
(\ref{eq:6gluonampansatz_normalized_by_Atree}), obtained as the
sum of the integrands of the 24 basis integrals $I_{i,\sigma_j}$,
weighted by the integral coefficients $c_{i,\sigma_j}$ and
symmetry factors $r_i$, where the summation is taken over all
dihedral permutations $\sigma_j$ of the external momentum labels.
Agreement between the two-loop integrand as computed by either
method would imply agreement between the integrated expressions;
that is, the results for the amplitude.\footnote{More accurately,
the object we are comparing is the strictly four-dimensional part
of the integrand. Indeed, as remarked above, the $\mu$-integrals
$I_{14, \sigma_j},I_{15, \sigma_j}$ give rise to contributions to
the integrand which are $\mathcal{O}(\epsilon)$ in the dimensional
regulator and hence not obtainable from evaluating leading
singularities in strictly four dimensions. Thus, our findings of
agreement between the integrands should be interpreted as a
statement concerning the $\mathcal{O}(\epsilon^0)$ part
exclusively.}
\\
\\
We have performed the comparison of the two-loop integrands
numerically, by verifying agreement to high accuracy for a large
number of randomly selected external and internal momenta.
Accordingly, the remainder of this section will be devoted to
discussing how, given a set of randomly generated momenta, one may
evaluate the integrand of eq.
(\ref{eq:6gluonampansatz_normalized_by_Atree}) -- that is, how the
integral coefficients are obtained, and how the integrands of the
basis integrals $I_{i,\sigma_j}$ are added in a meaningful way.

The integral coefficients are determined in analogy with the
procedure explained in Section
\ref{sec:extraction_of_integral_coeffs}. However, in that context
we were concerned with obtaining analytical results for the
coefficients and could find the coefficients of all dihedral
permutations of, for example, $I_{2,\sigma_1}$ by applying the
appropriate permutation to the algebraic expression for
$c_{2,\sigma_1}$. This is obviously not possible when one is
aiming to find numerical results for the coefficients. Instead,
one must act on the external momentum labels implicit in eq.
(\ref{eq:heptacut_2-loop_equation}) with each of the dihedral
permutations $\sigma_j \in D_6$ in turn so as to produce distinct
linear equations for $c_{2,\sigma_j}$. Again, because of linear
relations between the basis integrals, the solutions of the linear
equations will, for the more complicated heptacuts $\#2, \ldots,
\#9$, contain free parameters. Accordingly, one must set some of
the coefficients equal to specific values in order to obtain
unique solutions for the remaining coefficients. In analogy with
eq. (\ref{eq:gauge-fixing_parity_even}), we choose the ``gauge
fixing''
\begin{equation}
\left. \begin{array}{rll} \overline{c}_{11,\sigma_j}
\hspace{-1mm}&=&\hspace{-1mm}
s_{\sigma_j(61)} s_{\sigma_j(12)} s_{\sigma_j(123)} \\
\overline{c}_{12,\sigma_j}  \hspace{-1mm}&=&\hspace{-1mm}
s_{\sigma_j(456)} (s_{\sigma_j(345)} s_{\sigma_j(456)} - s_{\sigma_j(12)} s_{\sigma_j(45)}) \\
\overline{c}_{24,\sigma_j} \hspace{-1mm}&=&\hspace{-1mm} 0
\end{array} \right\} \hspace{5mm} \mathrm{for} \hspace{2mm} j = 1,
\ldots, 12
\end{equation}
whereby $\overline{c}_{11,\sigma_j}^\mathrm{odd} =
\overline{c}_{12,\sigma_j}^\mathrm{odd} =
\overline{c}_{24,\sigma_j}^\mathrm{odd} = 0$. We then find unique
results for the remaining coefficients with the property that
$\mathrm{Re} \hspace{0.8mm} \overline{c}_{i,\sigma_j} =
\overline{c}_{i,\sigma_j}^\mathrm{even}$.

Expressed as functions of internal and external momenta, the
integrands of the basis integrals cannot be added in any
meaningful way as the value of any term would depend on the
labeling of the internal lines of the corresponding graph (i.e.,
which propagators are labeled $\ell_1$ and $\ell_2$). To remedy
this, the integrand must be expressed in terms of dual $x$-space
coordinates, defined by
\begin{equation}
\begin{array}{lrll}
\hspace{5.1mm}x_i - x_{i+1} \hspace{0.9mm}=\hspace{0.9mm}  k_i & \hspace{6mm}
i\hspace{-1.5mm}&=&\hspace{-1.5mm}1, \ldots, 6 \hspace{4mm} \mbox{(mod 6)} \\[1mm]
\left. \begin{array}{rll}
x_{\sigma_j (1)} - x_7  \hspace{-1.2mm}&=&\hspace{-1.2mm}  \ell_1 \\
x_{\sigma_j (1)} - x_8  \hspace{-1.2mm}&=&\hspace{-1.2mm} -\ell_2
\end{array} \right\}& \hspace{6mm} j \hspace{-1.5mm}&=&\hspace{-1.5mm} 1, \ldots, 6 \\[4mm]
\left. \begin{array}{rll}
x_{\sigma_j (6)} - x_7  \hspace{-1.2mm}&=&\hspace{-1.2mm} -\ell_1 \\
x_{\sigma_j (6)} - x_8  \hspace{-1.2mm}&=&\hspace{-1.2mm}  \ell_2
\end{array} \right\}& \hspace{6mm} j \hspace{-1.5mm}&=&\hspace{-1.5mm} 7, \ldots, 12 \\[4mm]
\hspace{15.6mm}x_{ij} \hspace{0.9mm}\equiv\hspace{0.9mm}  x_i -
x_j & \hspace{6mm} i,j\hspace{-1.5mm}&=&\hspace{-1.5mm} 1, \ldots,
8 \: ,
\end{array} \label{eq:def_of_dual_coords}
\end{equation}
with the additional requirement that, for any given graph,
$\ell_1$ and $\ell_2$ be offset by appropriate translations by
external momenta so that all propagators take the form
$\frac{1}{x_{ij}^2}$. Finally, the integrand must be symmetrized
in $x_7$ and $x_8$. Namely, any assignment of these points to a
given graph will fail to be invariant under vertical reflections
of the graph; to ensure that the value of the integrand is not
dependent on how its contributing graphs happen to be drawn, one
must therefore average over the two possible assignments of these
points. The integrands of the basis integrals in Fig.
\ref{fig:6-point_integral_basis} have been presented in Section
\ref{sec:basis_integrands_in_dual_coords} for convenience.

In summary, given a set of random internal and external momenta,
the evaluation of the integrand of eq.
(\ref{eq:6gluonampansatz_normalized_by_Atree}) proceeds in three
steps. First, the integral coefficients are obtained by solving
the linear equations that follow from taking the leading
singularities of eq.
(\ref{eq:6gluonampansatz_normalized_by_Atree}). Second, the
integrands of the basis integrals are computed after converting
the momenta into the dual $x$-space (which is achieved by solving
eq. (\ref{eq:def_of_dual_coords}) and choosing, e.g., the base
point $x_6 = 0$). Finally, the intermediate results are combined,
weighting the contributions by the appropriate symmetry factors
(\ref{eq:symmetry_factors}) of the integrals. This is essentially
the procedure followed by our code \cite{ancillary:file} which is
available online. As a simple consistency check of the code, we
remark that the results produced for the integrand indeed satisfy
crossing symmetry; that is,
\begin{equation}
\mbox{integrand of} \hspace{1mm} M_{6, \hspace{0.4mm}
\mathrm{MHV}}^{(2)} (x_1 , \ldots, x_6) \hspace{2mm}=\hspace{2mm}
\mbox{integrand of} \hspace{1mm} M_{6, \hspace{0.4mm}
\mathrm{MHV}}^{(2)} (x_{\sigma_j(1)} , \ldots, x_{\sigma_j(6)})
\hspace{5mm} \mathrm{for} \hspace{2mm} \sigma_j \in D_6 \: .
\end{equation}
We have evaluated the two-loop six-point MHV integrand for a large
number of randomly selected rational momenta\footnote{To generate
$n$ rational momenta which are lightlike in $(+,-,-,-)$ signature,
the first $n-2$ can be chosen as arbitrary Pythagorean quadruples
(for example, generated by using the parametrization $(m_3^2 +
m_1^2 + m_2^2, m_3^2 - m_1^2 - m_2^2, 2m_1 m_3, 2m_2 m_3)$ with
$m_i \in \mathbb{Z}$) normalized by their $\| \cdot \|_1$-norm. To
ensure that the $n$-th momentum will be lightlike and satisfy
momentum conservation, the $(n-1)$-th momentum is obtained by
generating an additional Pythagorean quadruple $\xi$ of unit $\|
\cdot \|_1$-norm and then rescaling it by the constant $\alpha = -
\frac{\left( \sum_{i=1}^{n-2} k_i\right)^2}{2 \xi^\mu
\sum_{i=1}^{n-2} k_{\mu i}} \in \mathbb{Q}$ whereby $k_n = -
\left( \sum_{i=1}^{n-2} k_i + \alpha \xi \right)$ is lightlike
and rational.} and in all cases find agreement with
\cite{ArkaniHamed:2010kv,ArkaniHamed:2010gh} to high numerical
accuracy \cite{Simon:privcomm}. In Table 1 below we have provided
a few sample points to allow the curious reader to reproduce our
results. Further data points can be generated by the Mathematica
notebook \cite{ancillary:file} available online.

\noindent \begin{center}
\begin{tabular}{|c|c|c|}
\hline & & \\[-3mm] $(x_1,\ldots, x_6)$ & $(x_7, x_8)$ & integrand of $M_{6, \hspace{0.4mm} \mathrm{MHV}}^{(2)}$\\[1.5mm]
\hline & & \\
\hspace{-2mm}$\Big( \hspace{-0.7mm} \left(-\frac{1}{2},
\frac{1}{2}, 0, 0 \right), \left(-\frac{11}{12}, \frac{1}{6},
\frac{1}{4}, 0 \right),$ &\hspace{-2mm} $\Big( \hspace{-0.7mm}
\left(\frac{1}{4}, \frac{1}{4}, 2, 1
\right),$ &  \\
\hspace{-2mm} $\left(-\frac{4}{3}, \frac{5}{12}, \frac{7}{12}, 0
\right), \left(-\frac{7}{4}, \frac{29}{36}, \frac{13}{18},
-\frac{1}{18} \right),$ &\hspace{-2mm} $\phantom{\Big(
\hspace{-0.7mm}} \left(0, 0, 2, 0 \right) \hspace{-0.7mm} \Big) $
& $-\frac{31 \hspace{0.5mm} 230 \hspace{0.5mm} 748 \hspace{0.5mm}
253} {22 \hspace{0.5mm} 094 \hspace{0.5mm} 130 \hspace{0.5mm} 240}
- \frac{994 \hspace{0.5mm} 276 \hspace{0.5mm} 085}
{981 \hspace{0.5mm} 961 \hspace{0.5mm} 344} i$\\
\hspace{-2mm} $\left(-\frac{23}{18}, \frac{35}{54}, \frac{28}{27},
-\frac{10}{27} \right), \left(0, 0, 0, 0 \right)
\hspace{-0.7mm}\Big)$ &  & \\[6mm]
\hspace{-2mm}$\Big( \hspace{-0.7mm} \left(\frac{3}{8},
\frac{7}{24}, \frac{1}{6}, -\frac{1}{6} \right),
\left(\frac{7}{8}, -\frac{5}{24}, \frac{1}{6}, -\frac{1}{6}
\right),$ &\hspace{-2mm} $\Big( \hspace{-0.7mm} \left(1,
\frac{1}{3}, 0, 2 \right),$ &   \\
\hspace{-2mm} $\left(\frac{157}{120}, -\frac{3}{8}, -\frac{7}{30},
-\frac{1}{6} \right), \left(\frac{69}{40}, -\frac{1}{8},
-\frac{17}{30}, -\frac{1}{6} \right),$ &\hspace{-2mm}
$\phantom{\Big( \hspace{-0.7mm}} \left(0,0,1,1 \right)
\hspace{-0.7mm} \Big) $ &  $ \frac{4 \hspace{0.5mm} 777
\hspace{0.5mm} 009 \hspace{0.5mm} 838 \hspace{0.5mm} 357}{201
\hspace{0.5mm} 230 \hspace{0.5mm} 662 \hspace{0.5mm} 913
\hspace{0.5mm} 280}
+ \frac{1 \hspace{0.5mm} 802 \hspace{0.5mm} 603 \hspace{0.5mm} 853 \hspace{0.5mm} 899}
{259 \hspace{0.5mm} 652 \hspace{0.5mm} 468 \hspace{0.5mm} 275 \hspace{0.5mm} 200}i$ \\
\hspace{-2mm} $\left(\frac{73}{90}, \frac{5}{54}, -\frac{53}{135},
\frac{19}{27} \right), \left(0, 0, 0, 0 \right)
\hspace{-0.7mm}\Big)$ &  & \\[6mm]
\hspace{-2mm}$\Big( \hspace{-0.7mm} \left(\frac{5}{12},
\frac{1}{3}, 0, \frac{1}{4} \right), \left(\frac{79}{96},
-\frac{1}{24}, \frac{1}{8}, \frac{5}{32} \right),$ &\hspace{-2mm}
$\Big( \hspace{-0.7mm} \left(\frac{1}{2},\frac{1}{3},\frac{1}{3},0 \right),$ &  \\
\hspace{-2mm} $\left(\frac{59}{48}, -\frac{1}{6}, \frac{7}{32},
-\frac{7}{32} \right), \left(\frac{157}{96}, -\frac{13}{24},
\frac{1}{8}, -\frac{3}{32} \right),$ &\hspace{-2mm}
$\phantom{\Big( \hspace{-0.7mm}} \left(\frac{1}{2},0,0,0 \right)
\hspace{-0.7mm} \Big) $ & $\frac{3 \hspace{0.5mm} 393
\hspace{0.5mm} 545 \hspace{0.5mm} 258 \hspace{0.5mm} 977
\hspace{0.5mm} 272}{16 \hspace{0.5mm} 669 \hspace{0.5mm} 297
\hspace{0.5mm} 265} - \frac{43 \hspace{0.5mm} 045 \hspace{0.5mm}
877 \hspace{0.5mm} 862 \hspace{0.5mm} 533
\hspace{0.5mm} 664}{183 \hspace{0.5mm} 362 \hspace{0.5mm} 269 \hspace{0.5mm} 915}i$ \\
\hspace{-2mm} $\left(\frac{125}{224}, \frac{15}{28}, \frac{1}{8},
-\frac{3}{32} \right), \left(0, 0, 0, 0 \right)
\hspace{-0.7mm}\Big)$ & & \\[4mm]
\hline
\end{tabular}
\end{center}
{\vskip 2mm} Table 1: Values of the two-loop six-point MHV
integrand (normalized with respect to the tree-level amplitude) at
three randomly chosen sets of points $(x_1, \ldots, x_6)$ and
$(x_7, x_8)$ in dual $x$-space, respectively encoding external and
internal momenta as prescribed by eq.
(\ref{eq:def_of_dual_coords}). The parity-even and odd parts of
the integrand of $M_{6, \hspace{0.4mm} \mathrm{MHV}}^{(2)}$
respectively coincide with its real and imaginary parts. Further
data points can be generated by the Mathematica notebook
\cite{ancillary:file} available online.
\\
\\
Finally, let us observe that we only made use of the
assumption that the two-loop amplitude is MHV when evaluating the heptacuts of the
left hand side of eq. (\ref{eq:6gluonampansatz}). The form of
the heptacuts of the right hand side of eq. (\ref{eq:6gluonampansatz}) is independent
of the external helicities, and the results presented in this paper can
therefore straightforwardly be extended to obtain the NMHV integrand as well.
\\
\\
The two-loop six-point integrand as computed in this paper
was expressed in terms of the basis
in Fig.~\ref{fig:6-point_integral_basis} which
does not include integrals containing subloops with less than four
propagators. The exclusion of such integrals from the basis owes
to the observation that cutting two propagators would factor out
a subtriangle or sub-bubble---but the latter integrals are known
not to contribute to one-loop amplitudes in $\mathcal{N}=4$
SYM. This strongly suggests that the uncut two-loop
integral would appear with zero coefficient if included
in the basis expansion.

One possibility which cannot be rigorously ruled out by this
argument, however, occurs when the two-particle cut is shared between
several integrals containing subtriangles or sub-bubbles:
in principle, the coefficients of the respective cut
two-loop integrals could be nonzero, but such that the
contributions cancel.

Ruling out such a scenario completely
would require extending the analysis of this paper to consider
hexacuts, pentacuts etc. Two-loop integrals whose subloops contain
at least three propagators all admit $T^8$-integration contours
(analogous to those in
eq.~(\ref{eq:heptacut_1_leading_singularity_tori})) that
appropriately define such cuts. Moreover, the cut
integrand is a holomorphic function, and the
contour integrations can thus be performed directly by means of the global
residue theorem~\cite{ArkaniHamed:2009dn}. In contrast, two-loop
integrals with bubble-subloops do not admit $T^8$-contours: for
example, the bubble-box integral rather admits a $T^6 \times S^2$
contour. Several approaches are available to deal with
bubble-inherited $S^2$ contours, among the more elegant ones is that
of Mastrolia~\cite{Mastrolia:2009dr}, exploiting Stokes' theorem;
a related approach is that of Arkani-Hamed et al. in ref.~\cite{ArkaniHamed:2008gz}.
However, we leave such extensions for future work.

\section{Conclusions}\label{sec:conclusions}

In this paper we have provided a check that recent results in the
literature \cite{ArkaniHamed:2010kv,ArkaniHamed:2010gh} for the
full (i.e., parity-even and odd) two-loop six-point MHV integrand
of  $\mathcal{N}=4$ SYM theory can be reproduced by the leading
singularity method. Equivalently, assuming the validity of
refs. \cite{ArkaniHamed:2010kv,ArkaniHamed:2010gh}, one can view the
analysis carried out in this paper as a check that the leading
singularities of the $\mathcal{N}=4$ SYM integrand evaluated in
strictly four dimensions (as opposed to in $D = 4-2\epsilon$
dimensions) are sufficient to detect the parity-odd part. This has
already been shown to be the case for the two-loop five-gluon
$\mathcal{N}=4$ SYM amplitude in ref. \cite{Cachazo:2008vp}, but
the six-gluon MHV amplitude provides a much richer testing ground
owing to the larger number of global poles of the integrand and of
integrals in terms of which the amplitude is expressed. As the main part
of the results presented in this paper
are independent of the helicities of the six external states, the NMHV integrand
can be straightforwardly obtained by supplementing the requisite helicity dependent data.
We leave this as an open problem.

For any two-loop integral, the leading singularities are obtained
by changing the integration range from $\mathbb{R}^D \times
\mathbb{R}^D$ into tori of real dimension 8 (embedded in
$\mathbb{C}^4 \times \mathbb{C}^4$) that encircle the global poles
of the integrand. As explained in ref. \cite{Kosower:2011ty}, the
maximal cuts are particular linear combinations $\sigma$ of these
tori whose coefficients are determined by the requirement that any
function that integrates to zero on $\mathbb{R}^D \times
\mathbb{R}^D$ should also integrate to zero on $\sigma$. This
constraint ensures that two Feynman integrals which are equal,
possibly through some non-trivial relations, will also have
identical maximal cuts. As argued in ref. \cite{Kosower:2011ty},
multidimensional contours $\sigma$ satisfying this consistency
condition are guaranteed to produce correct results for scattering
amplitudes in any gauge theory, not only $\mathcal{N}=4$ SYM
theory.

The set of linear relations between two-loop integrals includes
the set of all integration-by-parts (IBP) identities between the
various tensor integrals arising from the Feynman rules of gauge
theory; however, at present a complete knowledge of such relations
is not available. Thus, an interesting open problem to be pursued
once all necessary IBP relations do become available is to
determine the maximal-cut contours that allow the extraction of
integral coefficients in any two-loop six-point gauge theory
amplitude. We expect the intermediate results provided in this
paper to greatly facilitate this task.

\section*{Acknowledgments}

I am indebted to Simon Caron-Huot for assistance in comparing the
$\mathcal{N}=4$ SYM integrand and to Henrik Johansson for
innumerable useful discussions. I would like to thank Gregory Korchemsky and
David Kosower for comments on the manuscript
and for discussions. I would also like to thank Ruth Britto for discussions. I thank the
Niels Bohr Institute for hospitality where part of this work was
carried out. This work is supported by the European Research
Council under Advanced Investigator Grant ERC--AdG--228301.

\clearpage

\appendix

\section{Details of heptacuts $\#2, \ldots, \#9$}\label{sec:appendix}

\subsection{Integrands of basis integrals in dual
coordinates}\label{sec:basis_integrands_in_dual_coords}

The expressions below are the four-dimensional integrands of the
basis integrals in Fig. \ref{fig:6-point_integral_basis},
expressed in dual $x$-space coordinates (related to the internal
and external momenta through eq. (\ref{eq:def_of_dual_coords})).
The results are recorded in the $\sigma_1$ permutation of the
external momentum labels; the integrands for the remaining
dihedral permutations may be obtained by applying eq.
(\ref{eq:dihedral_perm_of_integrand}) below.
\begin{equation}
\begin{array}{rll}
\mathrm{integrand}_{1,\sigma_1} &=& \frac{1}{2} \left( \big(
x_{17}^2 x_{27}^2 x_{37}^2 x_{47}^2
x_{18}^2 x_{68}^2 x_{58}^2 x_{48}^2\big)^{-1} \hspace{0.7mm}+\hspace{0.7mm} (x_7 \longleftrightarrow x_8) \right)\\
\mathrm{integrand}_{2,\sigma_1} &=& \frac{1}{2} \left( \big(
x_{17}^2 x_{47}^2 x_{57}^2 x_{78}^2
x_{18}^2 x_{68}^2 x_{58}^2 \big)^{-1} \hspace{0.7mm}+\hspace{0.7mm} (x_7 \longleftrightarrow x_8) \right)\\
\mathrm{integrand}_{3,\sigma_1} &=& \frac{1}{2} \left( \big(
x_{17}^2 x_{37}^2 x_{47}^2 x_{78}^2
x_{18}^2 x_{68}^2 x_{48}^2 \big)^{-1} \hspace{0.7mm}+\hspace{0.7mm} (x_7 \longleftrightarrow x_8) \right)\\
\mathrm{integrand}_{4,\sigma_1} &=& \frac{1}{2} \left( \big(
x_{17}^2 x_{37}^2 x_{47}^2 x_{78}^2
x_{18}^2 x_{58}^2 x_{48}^2 \big)^{-1} \hspace{0.7mm}+\hspace{0.7mm} (x_7 \longleftrightarrow x_8) \right)\\
\mathrm{integrand}_{5,\sigma_1} &=& \frac{1}{2} \left( \big(
x_{17}^2 x_{27}^2 x_{37}^2 x_{78}^2
x_{68}^2 x_{58}^2 x_{38}^2 \big)^{-1} \hspace{0.7mm}+\hspace{0.7mm} (x_7 \longleftrightarrow x_8) \right)\\
\mathrm{integrand}_{6,\sigma_1} &=& \frac{1}{2} \left( \big(
x_{17}^2 x_{37}^2 x_{47}^2 x_{78}^2
x_{68}^2 x_{58}^2 x_{48}^2 \big)^{-1} \hspace{0.7mm}+\hspace{0.7mm} (x_7 \longleftrightarrow x_8) \right)\\
\mathrm{integrand}_{7,\sigma_1} &=& \frac{1}{2} \left( \big(
x_{17}^2 x_{27}^2 x_{37}^2 x_{78}^2
x_{68}^2 x_{58}^2 x_{48}^2 \big)^{-1} \hspace{0.7mm}+\hspace{0.7mm} (x_7 \longleftrightarrow x_8) \right)\\
\mathrm{integrand}_{8,\sigma_1} &=& \frac{1}{2} \left(
\frac{x_{67}^2}{x_{17}^2 x_{27}^2 x_{47}^2
x_{57}^2 x_{78}^2 x_{18}^2 x_{68}^2 x_{58}^2} \hspace{0.7mm}+\hspace{0.7mm} (x_7 \longleftrightarrow x_8) \right)\\
\mathrm{integrand}_{9,\sigma_1} &=& \frac{1}{2} \left(
\frac{x_{67}^2}{x_{17}^2 x_{27}^2 x_{37}^2
x_{47}^2 x_{78}^2 x_{18}^2 x_{68}^2 x_{48}^2} \hspace{0.7mm}+\hspace{0.7mm} (x_7 \longleftrightarrow x_8) \right)\\
\mathrm{integrand}_{10,\sigma_1} &=& \frac{1}{2} \left(
\frac{x_{67}^2}{x_{17}^2 x_{27}^2 x_{37}^2
x_{47}^2 x_{78}^2 x_{18}^2 x_{68}^2 x_{58}^2} \hspace{0.7mm}+\hspace{0.7mm} (x_7 \longleftrightarrow x_8) \right)\\
\mathrm{integrand}_{11,\sigma_1} &=& \frac{1}{2} \left(
\frac{x_{57}^2}{x_{17}^2 x_{27}^2 x_{37}^2
x_{47}^2 x_{78}^2 x_{18}^2 x_{68}^2 x_{58}^2} \hspace{0.7mm}+\hspace{0.7mm} (x_7 \longleftrightarrow x_8) \right)\\
\mathrm{integrand}_{12,\sigma_1} &=& \frac{1}{2} \left(
\frac{x_{57}^2 x_{28}^2}{x_{17}^2 x_{27}^2
x_{37}^2 x_{47}^2 x_{78}^2 x_{18}^2 x_{68}^2 x_{58}^2 x_{48}^2} \hspace{0.7mm}+\hspace{0.7mm} (x_7 \longleftrightarrow x_8) \right)\\
\mathrm{integrand}_{13,\sigma_1} &=& \frac{1}{2} \left(
\frac{x_{57}^2 x_{38}^2}{x_{17}^2 x_{27}^2
x_{37}^2 x_{47}^2 x_{78}^2 x_{18}^2 x_{68}^2 x_{58}^2 x_{48}^2} \hspace{0.7mm}+\hspace{0.7mm} (x_7 \longleftrightarrow x_8) \right)\\
\mathrm{integrand}_{14,\sigma_1} &=& 0 \\
\mathrm{integrand}_{15,\sigma_1} &=& 0 \\
\mathrm{integrand}_{16,\sigma_1} &=& \frac{1}{2} \left( \big(
x_{17}^2 x_{37}^2 x_{57}^2 x_{78}^2
x_{18}^2 x_{68}^2 x_{58}^2 \big)^{-1} \hspace{0.7mm}+\hspace{0.7mm} (x_7 \longleftrightarrow x_8) \right)\\
\mathrm{integrand}_{17,\sigma_1} &=& \frac{1}{2} \left( \big(
x_{17}^2 x_{27}^2 x_{37}^2 x_{78}^2
x_{58}^2 x_{48}^2 x_{38}^2 \big)^{-1} \hspace{0.7mm}+\hspace{0.7mm} (x_7 \longleftrightarrow x_8) \right)\\
\mathrm{integrand}_{18,\sigma_1} &=& \frac{1}{2} \left(
\frac{x_{67}^2}{x_{17}^2 x_{37}^2 x_{47}^2
x_{57}^2 x_{78}^2 x_{18}^2 x_{68}^2 x_{58}^2} \hspace{0.7mm}+\hspace{0.7mm} (x_7 \longleftrightarrow x_8) \right)\\
\mathrm{integrand}_{19,\sigma_1} &=& \frac{1}{2} \left( \big(
x_{17}^2 x_{27}^2 x_{47}^2
x_{57}^2 x_{78}^2 x_{18}^2 x_{68}^2 x_{58}^2 \big)^{-1} \hspace{0.7mm}+\hspace{0.7mm} (x_7 \longleftrightarrow x_8) \right)\\
\mathrm{integrand}_{20,\sigma_1} &=& \frac{1}{2} \left( \big(
x_{17}^2 x_{27}^2 x_{37}^2
x_{47}^2 x_{78}^2 x_{18}^2 x_{68}^2 x_{48}^2 \big)^{-1} \hspace{0.7mm}+\hspace{0.7mm} (x_7 \longleftrightarrow x_8) \right)\\
\mathrm{integrand}_{21,\sigma_1} &=& \frac{1}{2} \left( \big(
x_{17}^2 x_{27}^2 x_{37}^2
x_{47}^2 x_{78}^2 x_{18}^2 x_{68}^2 x_{58}^2 \big)^{-1} \hspace{0.7mm}+\hspace{0.7mm} (x_7 \longleftrightarrow x_8) \right)\\
\mathrm{integrand}_{22,\sigma_1} &=& \frac{1}{2} \left( \big(
x_{17}^2 x_{37}^2 x_{47}^2
x_{57}^2 x_{78}^2 x_{18}^2 x_{68}^2 x_{58}^2 \big)^{-1} \hspace{0.7mm}+\hspace{0.7mm} (x_7 \longleftrightarrow x_8) \right)\\
\mathrm{integrand}_{23,\sigma_1} &=& \frac{1}{2} \left( \big(
x_{17}^2 x_{27}^2 x_{37}^2
x_{47}^2 x_{78}^2 x_{18}^2 x_{68}^2 x_{58}^2 x_{48}^2 \big)^{-1} \hspace{0.7mm}+\hspace{0.7mm} (x_7 \longleftrightarrow x_8) \right)\\
\mathrm{integrand}_{24,\sigma_1} &=& \frac{1}{2} \left(
\frac{x_{38}^2}{x_{17}^2 x_{27}^2 x_{37}^2 x_{47}^2 x_{78}^2
x_{18}^2 x_{68}^2 x_{58}^2 x_{48}^2} \hspace{0.7mm}+\hspace{0.7mm}
(x_7 \longleftrightarrow x_8) \right)\: .
\end{array} \label{eq:integrands_of_basis_integrands_permutation_1}
\end{equation}
The integrands for the remaining dihedral permutations $\sigma_j
\in D_6$ can be obtained from eq.
(\ref{eq:integrands_of_basis_integrands_permutation_1}) by
applying
\begin{equation}
\mathrm{integrand}_{i,\sigma_j} (x_1 , x_2, \ldots, x_6 ; x_7,
x_8) \hspace{1mm}=\hspace{1mm} \left\{ \begin{array}{l}
\mathrm{integrand}_{i,\sigma_1} (x_{\sigma_j (1)} , x_{\sigma_j
(2)}, \ldots, x_{\sigma_j (6)} ;
x_7, x_8) \\ \hspace{55mm} \mathrm{for} \hspace{2mm} j = 1, \ldots, 6 \\[2mm]
\mathrm{integrand}_{i,\sigma_1} (x_{\sigma_j (6)} , x_{\sigma_j
(1)}, \ldots, x_{\sigma_j (5)} ; x_7, x_8) \\
\hspace{55mm} \mathrm{for} \hspace{2mm} j = 7, \ldots, 12 \: .
\end{array} \right. \label{eq:dihedral_perm_of_integrand}
\end{equation}

\subsection{Heptacut \#2}\label{sec:heptacut_2}

This heptacut is defined by the on-shell constraints in eqs.
(\ref{eq:on-shell_constraint_1})-(\ref{eq:on-shell_constraint_7})
with the vertex momenta
\begin{equation}
\begin{array}{lll}
K_1 = k_{12} & \hspace{1.2cm} K_2 = k_3 & \hspace{1.2cm} K_3 = 0 \: \phantom{.} \\
K_4 = k_{45} & \hspace{1.2cm} K_5 = k_6 & \hspace{1.2cm} K_6 = 0
\: .
\end{array}
\end{equation}
Applying this heptacut to the right hand side of eq.
(\ref{eq:6gluonampansatz}) leaves the following linear combination
of cut integrals

\begin{figure}[!h]
\begin{center}
\includegraphics[angle=0, width=0.9\textwidth]{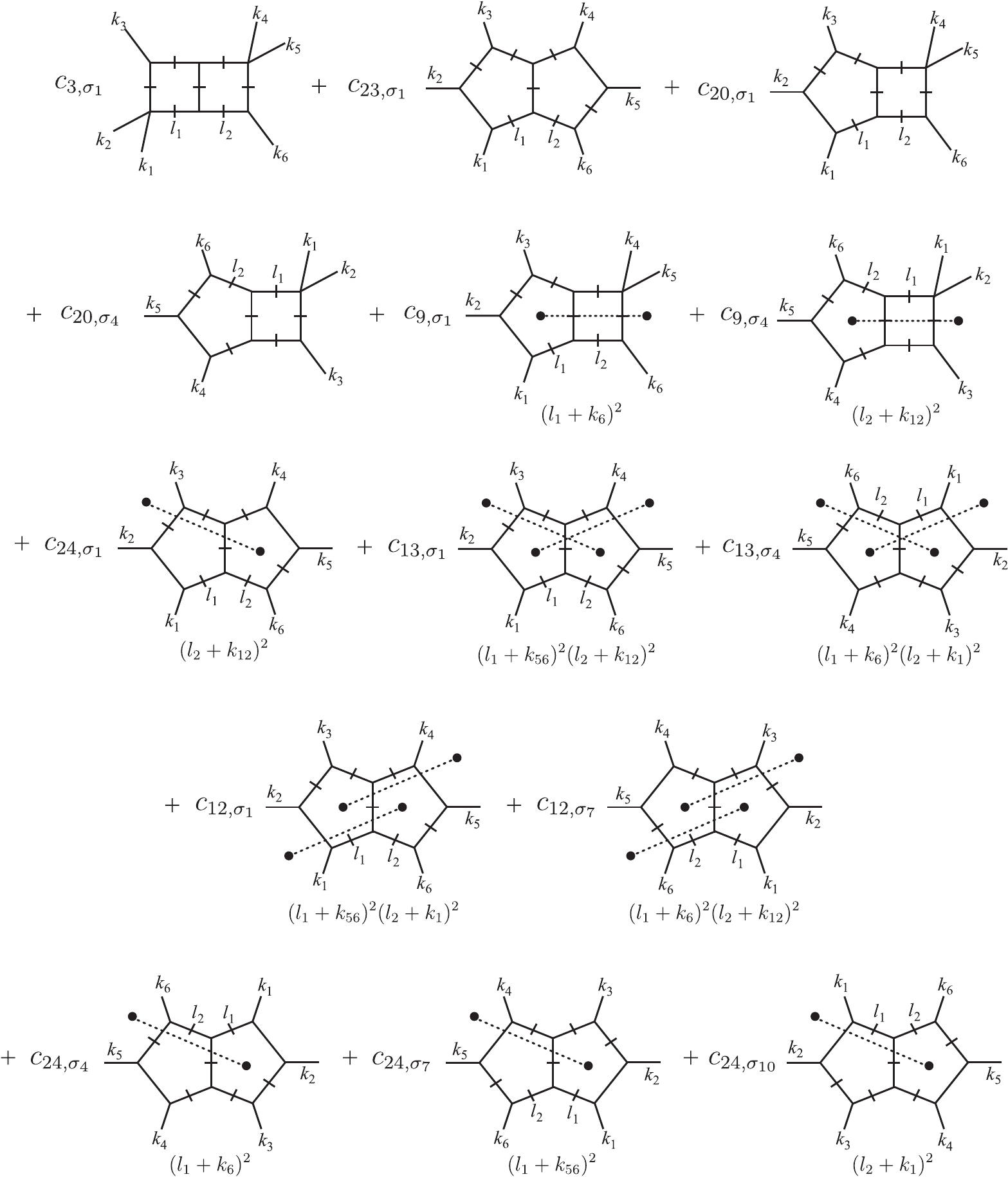}
\end{center}
\end{figure}
\vspace{-0.5cm}

\noindent We define the spinor ratios
\begin{equation}
\begin{array}{lll} P_1 = -\frac{\langle K_1^\flat k_6 \rangle}
{2\langle K_2^\flat k_6 \rangle} \: , \hspace{0.4cm} & P_2 = -
\frac{\langle K_1^\flat k_1 \rangle}{2 \langle K_2^\flat k_1
\rangle} \: , \hspace{0.4cm} & P_3 = -\frac{ K_1^\flat \cdot
k_{56} + \frac{1}{2} s_{56}}{\langle K_2^{\flat -} |
\slashed{k}_{56} | K_1^{\flat -} \rangle} \: , \\
P_4 = - \frac{\langle K_1^\flat K_4^\flat \rangle}
{2 \langle K_2^\flat K_4^\flat \rangle} & & \\
\vspace{-0.2cm} \\ Q_1 = - \frac{\left( 1 + \frac{S_4}{\gamma_2}
\right) [K_5^\flat K_1^\flat]}{2[K_4^\flat K_1^\flat]} \: ,
\hspace{0.4cm} & Q_2 = - \frac{\left( 1 + \frac{S_4}{\gamma_2}
\right) K_5^\flat \cdot k_{56} - \frac{1}{2} s_{56}}{\langle
K_5^{\flat -} | \slashed{k}_{56} | K_4^{\flat -} \rangle} \: ,
\hspace{0.4cm} & Q_3 = - \frac{\left( 1 + \frac{S_4}{\gamma_2}
\right) [K_5^\flat k_1]}{2[K_4^\flat k_1]}
\end{array}
\end{equation}
and their parity conjugates
\begin{equation}
\begin{array}{lll} P_1^\bullet = -\frac{[K_1^\flat k_6]}
{2[K_2^\flat k_6]} \: , \hspace{0.4cm} & P_2^\bullet = -
\frac{[K_1^\flat k_1]}{2[K_2^\flat k_1]} \: , \hspace{0.4cm} &
P_3^\bullet = -\frac{ K_1^\flat \cdot k_{56} + \frac{1}{2}
s_{56}}{\langle K_1^{\flat -} | \slashed{k}_{56} | K_2^{\flat -} \rangle} \: , \\
P_4^\bullet = -\frac{[K_1^\flat K_4^\flat]}{2[K_2^\flat
K_4^\flat]} & & \\
\vspace{-0.2cm} \\ Q_1^\bullet = - \frac{\left( 1 +
\frac{S_4}{\gamma_2} \right) \langle K_5^\flat K_1^\flat
\rangle}{2 \langle K_4^\flat K_1^\flat \rangle} \: ,
\hspace{0.4cm} & Q_2^\bullet = - \frac{\left( 1 +
\frac{S_4}{\gamma_2} \right) K_5^\flat \cdot k_{56} - \frac{1}{2}
s_{56}}{\langle K_4^{\flat -} | \slashed{k}_{56} | K_5^{\flat -}
\rangle} \: , \hspace{0.4cm} & Q_3^\bullet = - \frac{\left( 1 +
\frac{S_4}{\gamma_2} \right) \langle K_5^\flat k_1
\rangle}{2\langle K_4^\flat k_1 \rangle} \: .
\end{array}
\end{equation}
This heptacut belongs to case I treated in Section
\ref{sec:general_double_box}, and there are thus six kinematical
solutions (shown in Fig. \ref{fig:kinematical_solutions_case_I}).
Parametrizing the loop momenta according to eqs.
(\ref{eq:l1_parametrized})-(\ref{eq:l2_parametrized}), the
on-shell constraints in eqs.
(\ref{eq:on-shell_constraint_1})-(\ref{eq:on-shell_constraint_7})
are solved by setting the parameters equal to the values
\begin{equation}
\begin{array}{ll}
\alpha_1 = 1 \: , \hspace{1cm} & \beta_1 = 0\\
\alpha_2 = 0 \: , \hspace{1cm} & \beta_2 = 1 +
\frac{S_4}{\gamma_2}
\end{array}
\end{equation}
and those given in Fig. \ref{fig:kinematical_solutions_case_I}
with
\begin{equation} \beta_3 (z) = - \left( 1 +
\frac{S_4}{\gamma_2}\right) \frac{\langle K_2^\flat K_5^\flat
\rangle (z - P_1)}{2\langle K_2^\flat K_4^\flat \rangle (z - P_4)}
\label{eq:beta_3_heptacut_2}
\end{equation}
for kinematical solution $\mathcal{S}_5$. The heptacut double box
integral $I_{3,\sigma_1}$ is $\sum_{i=1}^6
\oint_{\hspace{0.5mm}\Gamma_i} \hspace{-0.3mm} dz \, J_i(z)$ where
\begin{equation}
J_i(z) \hspace{2mm} = \hspace{2mm} \frac{1}{32 \gamma_1 \gamma_2}
\times \left\{ \begin{array}{ll} \left( \left( 1 +
\frac{S_4}{\gamma_2} \right) \langle K_1^{\flat -} |
\slashed{K}_5^\flat | K_2^{\flat -} \rangle \hspace{0.6mm} z
(z-P_1^\bullet) \right)^{-1} & \hspace{4mm} \mathrm{for} \hspace{3mm} i=2,6 \\
\left( \left( 1 + \frac{S_4}{\gamma_2} \right) \langle K_2^{\flat
-} | \slashed{K}_5^\flat | K_1^{\flat -} \rangle \hspace{0.6mm} z
(z-P_1) \right)^{-1} & \hspace{4mm} \mathrm{for} \hspace{3mm} i=4,5 \\
\left( \langle K_4^{\flat -} | \slashed{K}_1^\flat | K_5^{\flat -}
\rangle \hspace{0.6mm} z
(z-Q_1^\bullet) \right)^{-1} & \hspace{4mm} \mathrm{for} \hspace{3mm} i=1 \\
\left( \langle K_5^{\flat -} | \slashed{K}_1^\flat | K_4^{\flat -}
\rangle \hspace{0.6mm} z (z - Q_1) \right)^{-1} & \hspace{4mm}
\mathrm{for} \hspace{3mm} i=3 \: .
\end{array} \right.
\end{equation}

\subsubsection{Heptacut \#2 of the right hand side of eq. (\ref{eq:6gluonampansatz})}

The result of applying heptacut \#2 to the right hand side of eq.
(\ref{eq:6gluonampansatz}) is
\begin{equation}
\frac{1}{4} \sum_{i=1}^6 \oint_{\Gamma_i} dz \, J_i (z) K_i(z)
\end{equation}
where the kernels evaluated on the six kinematical solutions are

{\small \begin{eqnarray} K_1 (z) \hspace{-2mm}&=&\hspace{-2mm}
c_{3,\sigma_1} + \frac{1}{4} \frac{c_{23,\sigma_1}}{\langle
K_1^{\flat -} | \slashed{k}_1 | K_2^{\flat -} \rangle \langle
K_4^{\flat -} | \slashed{k}_{56} | K_5^{\flat -} \rangle
(P_1^\bullet - P_2^\bullet) (z - Q_2^\bullet)} - \frac{1}{2}
\frac{c_{20,\sigma_1}}{\langle K_1^{\flat -}| \slashed{k}_1 |
K_2^{\flat -} \rangle (P_1^\bullet-P_2^\bullet)} \nonumber \\[2mm]
&\phantom{=}& \hspace{-10mm} + \frac{1}{2} \frac{c_{24,\sigma_1}
\langle K_4^{\flat -} | \slashed{k}_{12} | K_5^{\flat -} \rangle
(z - Q_1^\bullet) \hspace{0.6mm}+\hspace{0.6mm} c_{24,\sigma_7}
\langle K_1^{\flat -} | \slashed{k}_{56} | K_2^{\flat -} \rangle
(P_1^\bullet - P_3^\bullet) \hspace{0.6mm}+\hspace{0.6mm}
c_{24,\sigma_{10}} \langle K_4^{\flat -} | \slashed{k}_1 |
K_5^{\flat -} \rangle (z - Q_3^\bullet)}{\langle K_1^{\flat -} |
\slashed{k}_1 | K_2^{\flat -} \rangle \langle K_4^{\flat -} |
\slashed{k}_{56} | K_5^{\flat -} \rangle
(P_1^\bullet - P_2^\bullet) (z - Q_2^\bullet)} \nonumber \\[2mm]
&\phantom{=}&  \hspace{-10mm} + \frac{c_{13,\sigma_1} \langle
K_1^{\flat -} | \slashed{k}_{56} | K_2^{\flat -} \rangle \langle
K_4^{\flat -} | \slashed{k}_{12} | K_5^{\flat -} \rangle
(P_1^\bullet - P_3^\bullet) (z - Q_1^\bullet)}{\langle K_1^{\flat
-} | \slashed{k}_1 | K_2^{\flat -} \rangle \langle K_4^{\flat -} |
\slashed{k}_{56} | K_5^{\flat -} \rangle (P_1^\bullet -
P_2^\bullet) (z - Q_2^\bullet)} - \frac{1}{2}
\frac{c_{20,\sigma_4}}{\langle K_4^{\flat -}| \slashed{k}_{56} |
K_5^{\flat -} \rangle (z - Q_2^\bullet)} \nonumber \\[2mm]
&\phantom{=}& \hspace{-10mm} + \frac{c_{12,\sigma_1} \langle
K_1^{\flat -} | \slashed{k}_{56} | K_2^{\flat -} \rangle \langle
K_4^{\flat -} | \slashed{k}_1 | K_5^{\flat -} \rangle (P_1^\bullet
- P_3^\bullet) (z - Q_3^\bullet)}{\langle K_1^{\flat -} |
\slashed{k}_1 | K_2^{\flat -} \rangle \langle K_4^{\flat -} |
\slashed{k}_{56} | K_5^{\flat -} \rangle (P_1^\bullet -
P_2^\bullet) (z - Q_2^\bullet)} - \frac{c_{9,\sigma_4} \langle
K_4^{\flat -} | \slashed{k}_{12} | K_5^{\flat -} \rangle (z -
Q_1^\bullet)}{\langle K_4^{\flat -}| \slashed{k}_{56} | K_5^{\flat
-} \rangle (z - Q_2^\bullet)}
\end{eqnarray}}

{\small\begin{eqnarray} K_2(z) \hspace{-2mm}&=&\hspace{-2mm}
c_{3,\sigma_1} + \frac{1}{4} \frac{c_{23,\sigma_1}}{\langle
K_1^{\flat -} | \slashed{k}_1 | K_2^{\flat -} \rangle \langle
K_4^{\flat -} | \slashed{k}_{56} | K_5^{\flat -} \rangle (z -
P_2^\bullet) (Q_1^\bullet - Q_2^\bullet)} - \frac{1}{2}
\frac{c_{20,\sigma_1}}{\langle
K_1^{\flat -}| \slashed{k}_1 | K_2^{\flat -} \rangle (z-P_2^\bullet)} \nonumber \\[2mm]
&\phantom{=}& \hspace{-10mm}  + \frac{1}{2} \frac{c_{24,\sigma_4}
\langle K_1^{\flat -} | \slashed{k}_6 | K_2^{\flat -} \rangle (z -
P_1^\bullet) \hspace{0.6mm}+\hspace{0.6mm} c_{24,\sigma_7} \langle
K_1^{\flat -} | \slashed{k}_{56} | K_2^{\flat -} \rangle (z -
P_3^\bullet) \hspace{0.6mm}+\hspace{0.6mm} c_{24,\sigma_{10}}
\langle K_4^{\flat -} | \slashed{k}_1 | K_5^{\flat -} \rangle
(Q_1^\bullet - Q_3^\bullet)}{\langle K_1^{\flat -} | \slashed{k}_1
| K_2^{\flat -} \rangle \langle K_4^{\flat -} | \slashed{k}_{56} |
K_5^{\flat -} \rangle (z - P_2^\bullet) (Q_1^\bullet - Q_2^\bullet)} \nonumber \\[2mm]
&\phantom{=}& \hspace{-10mm} + \frac{c_{13,\sigma_4} \langle
K_1^{\flat -}| \slashed{k}_6 | K_2^{\flat -} \rangle \langle
K_4^{\flat -} | \slashed{k}_1 | K_5^{\flat -} \rangle (z -
P_1^\bullet) (Q_1^\bullet - Q_3^\bullet)}{\langle K_1^{\flat -} |
\slashed{k}_1 | K_2^{\flat -} \rangle \langle K_4^{\flat -} |
\slashed{k}_{56} | K_5^{\flat -} \rangle (z - P_2^\bullet)
(Q_1^\bullet - Q_2^\bullet)} - \frac{1}{2}
\frac{c_{20,\sigma_4}}{\langle K_4^{\flat -}| \slashed{k}_{56} |
K_5^{\flat -} \rangle (Q_1^\bullet - Q_2^\bullet)} \nonumber \\[2mm]
&\phantom{=}& \hspace{-10mm} + \frac{c_{12,\sigma_1} \langle
K_1^{\flat -} | \slashed{k}_{56} | K_2^{\flat -} \rangle \langle
K_4^{\flat -} | \slashed{k}_1 | K_5^{\flat -} \rangle (z -
P_3^\bullet) (Q_1^\bullet - Q_3^\bullet)}{\langle K_1^{\flat -} |
\slashed{k}_1 | K_2^{\flat -} \rangle \langle K_4^{\flat -} |
\slashed{k}_{56} | K_5^{\flat -} \rangle (z - P_2^\bullet)
(Q_1^\bullet - Q_2^\bullet)} - \frac{c_{9,\sigma_1} \langle
K_1^{\flat -} | \slashed{k}_6 | K_2^{\flat -} \rangle (z -
P_1^\bullet)}{\langle K_1^{\flat -}| \slashed{k}_1 | K_2^{\flat -}
\rangle (z-P_2^\bullet)} \\[4mm]
K_3 (z) \hspace{-2mm}&=&\hspace{-2mm} \mbox{parity conjugate of
$K_1(z)$ \hspace{0.3mm} (obtained by applying eqs.
(\ref{eq:parity_conj_rule_1})-(\ref{eq:parity_conj_rule_6}))} \\[3mm]
K_4 (z) \hspace{-2mm}&=&\hspace{-2mm} \mbox{parity conjugate of
$K_2(z)$ \hspace{0.3mm} (obtained by applying eqs.
(\ref{eq:parity_conj_rule_1})-(\ref{eq:parity_conj_rule_6}))} \\[3mm]
K_5(z) \hspace{-2mm}&=&\hspace{-2mm} c_{3,\sigma_1} + \frac{1}{4}
\frac{c_{23,\sigma_1}}{\langle K_2^{\flat -} | \slashed{k}_1 |
K_1^{\flat -} \rangle \langle K_4^{\flat -} | \slashed{k}_{56} |
K_5^{\flat -} \rangle (z - P_2)
\left(\beta_3(z) -Q_2^\bullet \right)} \nonumber \\[2mm]
&\phantom{=}& \hspace{-13mm} + \frac{1}{2} \frac{c_{24,\sigma_1}
\langle K_4^{\flat -} | \slashed{k}_{12} | K_5^{\flat -} \rangle
\left(\beta_3(z) - Q_1^\bullet \right)}{\langle K_2^{\flat -} |
\slashed{k}_1 | K_1^{\flat -} \rangle \langle K_4^{\flat -} |
\slashed{k}_{56} | K_5^{\flat -} \rangle (z - P_2)
\left(\beta_3(z) - Q_2^\bullet \right)} \nonumber \\[2mm]
&\phantom{=}& \hspace{-13mm} + \frac{1}{2} \frac{c_{24,\sigma_4}
\langle K_2^{\flat -} | \slashed{k}_6 | K_1^{\flat -} \rangle (z -
P_1) + c_{24,\sigma_7} \langle K_2^{\flat -} | \slashed{k}_{56} |
K_1^{\flat -} \rangle (z - P_3) + c_{24,\sigma_{10}} \langle
K_4^{\flat -} | \slashed{k}_1 | K_5^{\flat -} \rangle
\left(\beta_3(z) - Q_3^\bullet \right)}{\langle K_2^{\flat -} |
\slashed{k}_1 | K_1^{\flat -} \rangle \langle K_4^{\flat -} |
\slashed{k}_{56} | K_5^{\flat -}
\rangle (z - P_2) \left(\beta_3(z) -Q_2^\bullet \right)} \nonumber \\[2mm]
&\phantom{=}&  \hspace{-13mm} + \frac{c_{13,\sigma_1} \langle
K_2^{\flat -} | \slashed{k}_{56} | K_1^{\flat -} \rangle \langle
K_4^{\flat -} | \slashed{k}_{12} | K_5^{\flat -} \rangle (z - P_3)
\left(\beta_3(z) -Q_1^\bullet \right)}{\langle K_2^{\flat -} |
\slashed{k}_1 | K_1^{\flat -} \rangle \langle K_4^{\flat -} |
\slashed{k}_{56} | K_5^{\flat -} \rangle (z - P_2)
\left(\beta_3(z) -Q_2^\bullet \right)} - \frac{1}{2}
\frac{c_{20,\sigma_1}}{\langle K_2^{\flat -}| \slashed{k}_1 |
K_1^{\flat -} \rangle (z-P_2)} \nonumber \\[2mm]
&\phantom{=}&  \hspace{-13mm} + \frac{c_{13,\sigma_4} \langle
K_2^{\flat -} | \slashed{k}_6 | K_1^{\flat -} \rangle \langle
K_4^{\flat -} | \slashed{k}_1 | K_5^{\flat -} \rangle (z- P_1)
\left(\beta_3(z) -Q_3^\bullet \right)}{\langle K_2^{\flat -} |
\slashed{k}_1 | K_1^{\flat -} \rangle \langle K_4^{\flat -} |
\slashed{k}_{56} | K_5^{\flat -} \rangle (z - P_2)
\left(\beta_3(z) -Q_2^\bullet \right)} - \frac{1}{2}
\frac{c_{20,\sigma_4}}{\langle K_4^{\flat -}| \slashed{k}_{56} |
K_5^{\flat -} \rangle \left(\beta_3(z) -Q_2^\bullet \right)} \nonumber \\[2mm]
&\phantom{=}&  \hspace{-13mm} + \frac{c_{12,\sigma_1} \langle
K_2^{\flat -} | \slashed{k}_{56} | K_1^{\flat -} \rangle \langle
K_4^{\flat -} | \slashed{k}_1 | K_5^{\flat -}\rangle (z - P_3)
\left(\beta_3(z) -Q_3^\bullet \right)}{\langle K_2^{\flat -} |
\slashed{k}_1 | K_1^{\flat -} \rangle \langle K_4^{\flat -} |
\slashed{k}_{56} | K_5^{\flat -} \rangle (z - P_2)
\left(\beta_3(z) -Q_2^\bullet \right)} - \frac{c_{9,\sigma_1}
\langle K_2^{\flat -} | \slashed{k}_6 | K_1^{\flat -} \rangle (z -
P_1)}{\langle K_2^{\flat -}| \slashed{k}_1 | K_1^{\flat -} \rangle
(z-P_2)} \nonumber \\[2mm]
&\phantom{=}&  \hspace{-13mm} + \frac{c_{12,\sigma_7} \langle
K_2^{\flat -} | \slashed{k}_6 | K_1^{\flat -} \rangle \langle
K_4^{\flat -} | \slashed{k}_{12} | K_5^{\flat -} \rangle (z- P_1)
\left(\beta_3(z) -Q_1^\bullet \right)}{\langle K_2^{\flat -} |
\slashed{k}_1 | K_1^{\flat -} \rangle \langle K_4^{\flat -} |
\slashed{k}_{56} | K_5^{\flat -} \rangle (z - P_2)
\left(\beta_3(z) -Q_2^\bullet \right)} - \frac{c_{9,\sigma_4}
\langle K_4^{\flat -} | \slashed{k}_{12} | K_5^{\flat -} \rangle
\left(\beta_3(z) - Q_1^\bullet \right)}{\langle K_4^{\flat -}|
\slashed{k}_{56} | K_5^{\flat -} \rangle \left(\beta_3(z) -
Q_2^\bullet \right)} \phantom{aaaa} \\[4mm]
K_6 (z) \hspace{-2mm}&=&\hspace{-2mm} \mbox{parity conjugate of
$K_5(z)$ \hspace{0.3mm} (obtained by applying eqs.
(\ref{eq:parity_conj_rule_1})-(\ref{eq:parity_conj_rule_6}))}
\end{eqnarray}}
where $\beta_3 (z)$ is given in eq. (\ref{eq:beta_3_heptacut_2}).

\subsubsection{Heptacut \#2 of the left hand side of eq. (\ref{eq:6gluonampansatz})}

The result of applying heptacut \#2 to the left hand side of eq.
(\ref{eq:6gluonampansatz}) is
\begin{equation}
i \sum_{i=1}^6 \oint_{\Gamma_i} dz \, J_i (z) \left.
\prod_{j=1}^6 A_j^\mathrm{tree}(z) \right|_{\mathcal{S}_i}
\end{equation}
where, assuming without loss of generality the external helicities are $(1^-, 2^-, 3^+, 4^+, 5^+, 6^+)$,
the cut amplitude evaluated on the six different
kinematical solutions yields
\begin{equation}
\left. \prod_{j=1}^6 A_j^\mathrm{tree}(z) \right|_{\mathcal{S}_i}
\hspace{3mm}=\hspace{3mm} \frac{i}{16} A^\mathrm{tree}_{--++++}
\times \left\{ \begin{array}{ll} \frac{1}{J_3(z)} \left(
\frac{1}{z - Q_1} - \frac{1}{z - Q_2} \right) & \hspace{3mm} \mathrm{for} \hspace{3mm} i = 3 \\[2mm]
\frac{1}{J_4(z)} \left(
\frac{1}{z - P_1} - \frac{1}{z - P_2} \right) & \hspace{3mm} \mathrm{for} \hspace{3mm} i = 4 \\[2mm]
0 & \hspace{3mm} \mathrm{for} \hspace{3mm} i = 1,2,5,6 \: .
\end{array} \right.
\end{equation}

\clearpage

\subsection{Heptacut \#3}\label{sec:heptacut_3}

This heptacut is defined by the on-shell constraints in eqs.
(\ref{eq:on-shell_constraint_1})-(\ref{eq:on-shell_constraint_7})
with the vertex momenta
\begin{equation}
\begin{array}{lll}
K_1 = k_{12} & \hspace{1.2cm} K_2 = k_3 & \hspace{1.2cm} K_3 = 0 \: \phantom{.} \\
K_4 = k_4 & \hspace{1.2cm} K_5 = k_{56} & \hspace{1.2cm} K_6 = 0
\: .
\end{array}
\end{equation}
Applying this heptacut to the right hand side of eq.
(\ref{eq:6gluonampansatz}) leaves the following linear combination
of cut integrals

\begin{figure}[!h]
\begin{center}
\includegraphics[angle=0, width=0.9\textwidth]{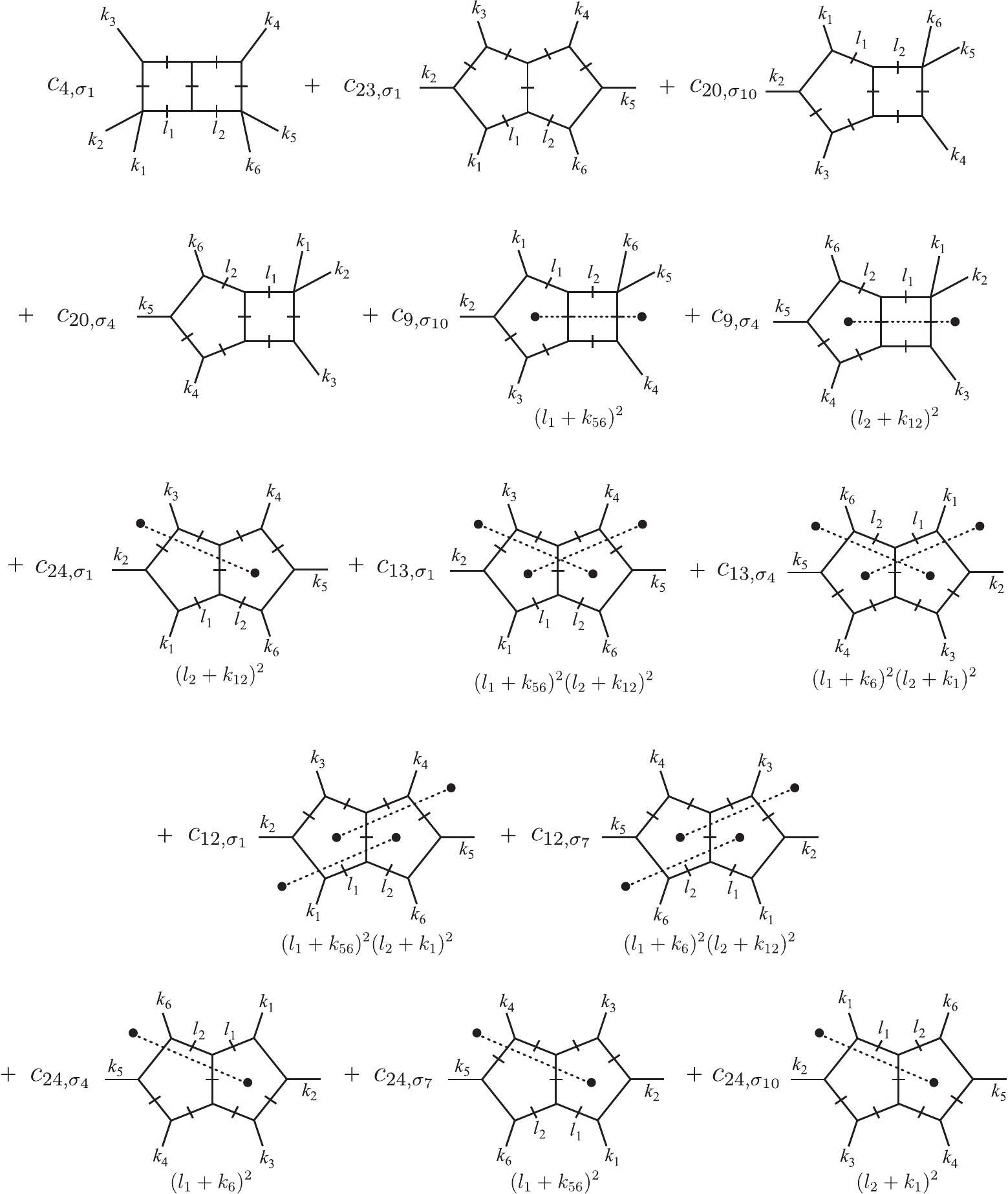}
\end{center}
\end{figure}
\vspace{-0.5cm}

\noindent We define the spinor ratios
\begin{equation}
\begin{array}{lll} P_1 = -\frac{\langle K_1^\flat K_5^\flat \rangle}
{2\langle K_2^\flat K_5^\flat \rangle} \: , \hspace{0.6cm} & P_2 =
- \frac{\langle K_1^\flat k_1 \rangle}{2 \langle K_2^\flat k_1
\rangle} \: , \hspace{0.6cm} & P_3 = - \frac{\langle K_1^\flat
K_4^\flat \rangle} {2 \langle K_2^\flat K_4^\flat \rangle} \: , \\
P_4 = -\frac{\langle K_1^\flat k_6\rangle}{2\langle K_2^\flat k_6\rangle} & & \\
\vspace{-0.2cm} \\ Q_1 = - \frac{[K_1^\flat
K_5^\flat]}{2[K_1^\flat K_4^\flat]} \: , \hspace{0.6cm} & Q_2 =
-\frac{[K_5^\flat k_6]}{2[K_4^\flat k_6]} \: , \hspace{0.6cm} &
Q_3 = -\frac{[K_5^\flat k_1]}{2[K_4^\flat k_1]}
\end{array}
\end{equation}
and their parity conjugates
\begin{equation}
\begin{array}{lll} P_1^\bullet = -\frac{[K_1^\flat K_5^\flat]}
{2[K_2^\flat K_5^\flat]} \: , \hspace{0.6cm} & P_2^\bullet = -
\frac{[K_1^\flat k_1]}{2[K_2^\flat k_1]} \: , \hspace{0.6cm} &
P_3^\bullet = - \frac{[K_1^\flat K_4^\flat]}{2[K_2^\flat
K_4^\flat]} \\
P_4^\bullet = -\frac{[K_1^\flat k_6]}{2[K_2^\flat k_6]} & & \\
\vspace{-0.2cm} \\ Q_1^\bullet = - \frac{\langle K_1^\flat
K_5^\flat \rangle}{2\langle K_1^\flat K_4^\flat \rangle} \: ,
\hspace{0.6cm} & Q_2^\bullet = -\frac{\langle K_5^\flat k_6
\rangle}{2\langle K_4^\flat k_6 \rangle} \: , \hspace{0.6cm} &
Q_3^\bullet = -\frac{\langle K_5^\flat k_1 \rangle}{2\langle
K_4^\flat k_1 \rangle} \: .
\end{array}
\end{equation}
This heptacut belongs to case I treated in Section
\ref{sec:general_double_box}, and there are thus six kinematical
solutions (shown in Fig. \ref{fig:kinematical_solutions_case_I}).
Parametrizing the loop momenta according to eqs.
(\ref{eq:l1_parametrized})-(\ref{eq:l2_parametrized}), the
on-shell constraints in eqs.
(\ref{eq:on-shell_constraint_1})-(\ref{eq:on-shell_constraint_7})
are solved by setting the parameters equal to the values
\begin{equation}
\begin{array}{ll}
\alpha_1 = 1 \: , \hspace{1cm} & \beta_1 = 0\\
\alpha_2 = 0 \: , \hspace{1cm} & \beta_2 = 1
\end{array}
\end{equation}
and those given in Fig. \ref{fig:kinematical_solutions_case_I}
with
\begin{equation}
\beta_3 (z) = -\frac{\langle K_2^\flat K_5^\flat \rangle (z -
P_1)}{2\langle K_2^\flat K_4^\flat \rangle (z - P_3)}
\label{eq:beta_3_heptacut_3}
\end{equation}
for kinematical solution $\mathcal{S}_5$. The heptacut double box
integral $I_{4,\sigma_1}$ is $\sum_{i=1}^6
\oint_{\hspace{0.5mm}\Gamma_i} \hspace{-0.3mm} dz \, J_i(z)$ where
\begin{equation}
J_i(z) \hspace{2mm} = \hspace{2mm} \frac{1}{32 \gamma_1 \gamma_2}
\times \left\{ \begin{array}{ll} \left( \langle K_1^{\flat -}|
\slashed{K}_5^\flat | K_2^{\flat -} \rangle \hspace{0.6mm} z (z -
P_1^\bullet) \right)^{-1} & \hspace{4mm} \mathrm{for} \hspace{3mm} i=2,6 \\
\left( \langle K_2^{\flat -}| \slashed{K}_5^\flat | K_1^{\flat -}
\rangle \hspace{0.6mm} z (z -
P_1) \right)^{-1} & \hspace{4mm} \mathrm{for} \hspace{3mm} i=4,5 \\
\left( \langle K_4^{\flat -}| \slashed{K}_1^\flat | K_5^{\flat -}
\rangle \hspace{0.6mm} z (z -
Q_1^\bullet) \right)^{-1} & \hspace{4mm} \mathrm{for} \hspace{3mm} i=1 \\
\left( \langle K_5^{\flat -}| \slashed{K}_1^\flat | K_4^{\flat -}
\rangle \hspace{0.6mm} z (z - Q_1) \right)^{-1} & \hspace{4mm}
\mathrm{for} \hspace{3mm} i=3 \: .
\end{array} \right.
\end{equation}

\subsubsection{Heptacut \#3 of the right hand side of eq. (\ref{eq:6gluonampansatz})}

The result of applying heptacut \#3 to the right hand side of eq.
(\ref{eq:6gluonampansatz}) is
\begin{equation}
\frac{1}{4} \sum_{i=1}^6 \oint_{\Gamma_i} dz \, J_i (z) K_i(z)
\end{equation}
where the kernels evaluated on the six kinematical solutions are

{\small\begin{eqnarray} K_1(z) \hspace{-2mm}&=&\hspace{-2mm}
c_{4,\sigma_1} + \frac{1}{4} \frac{c_{23, \sigma_1}}{\langle
K_1^{\flat -} | \slashed{k}_1 | K_2^{\flat -} \rangle \langle
K_4^{\flat -} | \slashed{k}_6 | K_5^{\flat -} \rangle (P_1^\bullet
-P_2^\bullet)(z - Q_2^\bullet)} - \frac{c_{9,\sigma_4} \langle
K_4^{\flat -} | \slashed{k}_{12} | K_5^{\flat -} \rangle
(z-Q_1^\bullet)}{\langle K_4^{\flat -}| \slashed{k}_6 |
K_5^{\flat -} \rangle (z-Q_2^\bullet)}\nonumber\\[2mm]
&\phantom{=}& \hspace{-13mm} + \frac{1}{2} \frac{c_{24, \sigma_1}
\langle K_4^{\flat -}|\slashed{k}_{12} | K_5^{\flat -} \rangle (z
- Q_1^\bullet) \hspace{0.6mm}+\hspace{0.6mm} c_{24, \sigma_4}
\langle K_1^{\flat -} | \slashed{k}_6 | K_2^{\flat -} \rangle
(P_1^\bullet - P_4^\bullet) \hspace{0.6mm}+\hspace{0.6mm} c_{24,
\sigma_{10}} \langle K_4^{\flat -} | \slashed{k}_1 | K_5^{\flat -}
\rangle (z - Q_3^\bullet)}{\langle K_1^{\flat -} | \slashed{k}_1 |
K_2^{\flat -} \rangle \langle K_4^{\flat -} | \slashed{k}_6 |
K_5^{\flat -} \rangle (P_1^\bullet-P_2^\bullet)(z - Q_2^\bullet)} \nonumber \\[2mm]
&\phantom{=}& \hspace{-13mm} + \frac{c_{13, \sigma_4} \langle
K_1^{\flat -} | \slashed{k}_6 | K_2^{\flat -} \rangle \langle
K_4^{\flat -} | \slashed{k}_1 | K_5^{\flat -} \rangle (P_1^\bullet
-P_4^\bullet)(z - Q_3^\bullet)}{\langle K_1^{\flat -} |
\slashed{k}_1 | K_2^{\flat -} \rangle \langle K_4^{\flat -} |
\slashed{k}_6 | K_5^{\flat -} \rangle (P_1^\bullet -P_2^\bullet)(z
- Q_2^\bullet)} \nonumber - \frac{1}{2}
\frac{c_{20,\sigma_{10}}}{\langle K_1^{\flat -}| \slashed{k}_1 |
K_2^{\flat -} \rangle (P_1^\bullet-P_2^\bullet)} \\[2mm]
&\phantom{=}& \hspace{-13mm} + \frac{c_{12, \sigma_7} \langle
K_1^{\flat -} | \slashed{k}_6 | K_2^{\flat -} \rangle \langle
K_4^{\flat -} | \slashed{k}_{12} | K_5^{\flat -} \rangle
(P_1^\bullet -P_4^\bullet)(z - Q_1^\bullet)}{\langle K_1^{\flat -}
| \slashed{k}_1 | K_2^{\flat -} \rangle \langle K_4^{\flat -} |
\slashed{k}_6 | K_5^{\flat -} \rangle (P_1^\bullet -P_2^\bullet)(z
- Q_2^\bullet)} - \frac{1}{2} \frac{c_{20,\sigma_4}}{\langle
K_4^{\flat -}| \slashed{k}_6 | K_5^{\flat -} \rangle (z -
Q_2^\bullet)}
\end{eqnarray}}

{\small\begin{eqnarray} K_2(z) \hspace{-2mm}&=&\hspace{-2mm}
c_{4,\sigma_1} + \frac{1}{4} \frac{c_{23, \sigma_1}}{\langle
K_1^{\flat -} | \slashed{k}_1 | K_2^{\flat -} \rangle \langle
K_4^{\flat -} | \slashed{k}_6 | K_5^{\flat -} \rangle
(z-P_2^\bullet)(Q_1^\bullet - Q_2^\bullet)} -
\frac{c_{9,\sigma_{10}} \langle K_1^{\flat -} | \slashed{k}_{56} |
K_2^{\flat -} \rangle (z-P_1^\bullet)}{\langle K_1^{\flat -}|
\slashed{k}_1 | K_2^{\flat -} \rangle (z-P_2^\bullet)} \nonumber \\[2mm]
&\phantom{=}& \hspace{-13mm} + \frac{1}{2} \frac{c_{24, \sigma_4}
\langle K_1^{\flat -} | \slashed{k}_6 | K_2^{\flat -} \rangle (z -
P_4^\bullet) \hspace{0.6mm}+\hspace{0.6mm} c_{24, \sigma_7}
\langle K_1^{\flat -} | \slashed{k}_{56} | K_2^{\flat -} \rangle
(z - P_1^\bullet) \hspace{0.6mm}+\hspace{0.6mm} c_{24,
\sigma_{10}} \langle K_4^{\flat -} | \slashed{k}_1 | K_5^{\flat -}
\rangle (Q_1^\bullet - Q_3^\bullet)}{\langle K_1^{\flat -} |
\slashed{k}_1 | K_2^{\flat -} \rangle \langle K_4^{\flat -} |
\slashed{k}_6 | K_5^{\flat -}
\rangle (z-P_2^\bullet)(Q_1^\bullet - Q_2^\bullet)} \nonumber \\[2mm]
&\phantom{=}& \hspace{-13mm} + \frac{c_{13, \sigma_4} \langle
K_1^{\flat -} | \slashed{k}_6 | K_2^{\flat -} \rangle \langle
K_4^{\flat -} | \slashed{k}_1 | K_5^{\flat -} \rangle (z -
P_4^\bullet) (Q_1^\bullet - Q_3^\bullet)}{\langle K_1^{\flat -} |
\slashed{k}_1 | K_2^{\flat -} \rangle \langle K_4^{\flat -} |
\slashed{k}_6 | K_5^{\flat -} \rangle (z-P_2^\bullet)(Q_1^\bullet
- Q_2^\bullet)} - \frac{1}{2} \frac{c_{20,\sigma_{10}}}{\langle
K_1^{\flat -}| \slashed{k}_1 |
K_2^{\flat -} \rangle (z-P_2^\bullet)} \nonumber \\[2mm]
&\phantom{=}& \hspace{-13mm} + \frac{c_{12, \sigma_1} \langle
K_1^{\flat -} | \slashed{k}_{56} | K_2^{\flat -} \rangle \langle
K_4^{\flat -} | \slashed{k}_1 | K_5^{\flat -} \rangle (z -
P_1^\bullet) (Q_1^\bullet - Q_3^\bullet)}{\langle K_1^{\flat -} |
\slashed{k}_1 | K_2^{\flat -} \rangle \langle K_4^{\flat -} |
\slashed{k}_6 | K_5^{\flat -} \rangle (z-P_2^\bullet) (Q_1^\bullet
- Q_2^\bullet)} - \frac{1}{2} \frac{c_{20,\sigma_4}}{\langle
K_4^{\flat -}| \slashed{k}_6 | K_5^{\flat -} \rangle
(Q_1^\bullet-Q_2^\bullet)} \\[3mm]
K_3(z) \hspace{-2mm}&=&\hspace{-2mm} \mbox{parity conjugate of
$K_1(z)$ \hspace{0.3mm} (obtained by applying eqs.
(\ref{eq:parity_conj_rule_1})-(\ref{eq:parity_conj_rule_6}))} \\[2mm]
K_4(z) \hspace{-2mm}&=&\hspace{-2mm} \mbox{parity conjugate of
$K_2(z)$ \hspace{0.3mm} (obtained by applying eqs.
(\ref{eq:parity_conj_rule_1})-(\ref{eq:parity_conj_rule_6}))} \\[3mm]
K_5(z) \hspace{-2mm}&=&\hspace{-2mm} c_{4,\sigma_1} + \frac{1}{4}
\frac{c_{23, \sigma_1}}{\langle K_2^{\flat -} | \slashed{k}_1 |
K_1^{\flat -} \rangle \langle K_4^{\flat -} | \slashed{k}_6 |
K_5^{\flat -}
\rangle (z-P_2)\left(\beta_3(z) -Q_2^\bullet \right)} \nonumber \\[2mm]
&\phantom{=}& \hspace{-13mm} + \frac{1}{2} \frac{c_{24, \sigma_1}
\langle K_4^{\flat -}|\slashed{k}_{12} | K_5^{\flat -} \rangle
\left(\beta_3(z) - Q_1^\bullet \right)}{\langle K_2^{\flat -} |
\slashed{k}_1 | K_1^{\flat -} \rangle \langle K_4^{\flat -} |
\slashed{k}_6 | K_5^{\flat -} \rangle (z-P_2)
\left(\beta_3(z) - Q_2^\bullet \right)} \nonumber \\[2mm]
&\phantom{=}& \hspace{-13mm} + \frac{1}{2} \frac{c_{24,\sigma_4}
\langle K_2^{\flat -} | \slashed{k}_6 | K_1^{\flat -} \rangle (z -
P_4) \hspace{0.6mm}+\hspace{0.6mm} c_{24, \sigma_7} \langle
K_2^{\flat -} | \slashed{k}_{56} | K_1^{\flat -} \rangle (z - P_1)
\hspace{0.6mm}+\hspace{0.6mm} c_{24, \sigma_{10}} \langle
K_4^{\flat -} | \slashed{k}_1 | K_5^{\flat -} \rangle \left(
\beta_3(z) - Q_3^\bullet \right)}{\langle K_2^{\flat -} |
\slashed{k}_1 | K_1^{\flat -} \rangle \langle K_4^{\flat -} |
\slashed{k}_6 | K_5^{\flat -}
\rangle (z-P_2)\left(\beta_3 (z) - Q_2^\bullet \right)} \nonumber \\[2mm]
&\phantom{=}& \hspace{-13mm} + \frac{c_{13,\sigma_1} \langle
K_2^{\flat -} | \slashed{k}_{56}| K_1^{\flat -} \rangle \langle
K_4^{\flat -} | \slashed{k}_{12} | K_5^{\flat -} \rangle (z-P_1)
\left(\beta_3(z) -Q_1^\bullet \right)}{\langle K_2^{\flat -} |
\slashed{k}_1 | K_1^{\flat -} \rangle \langle K_4^{\flat -} |
\slashed{k}_6 | K_5^{\flat -} \rangle (z-P_2) \left(\beta_3(z)
-Q_2^\bullet \right)} - \frac{1}{2}
\frac{c_{20,\sigma_{10}}}{\langle K_2^{\flat -}| \slashed{k}_1 |
K_1^{\flat -} \rangle (z-P_2)} \nonumber \\[2mm]
&\phantom{=}& \hspace{-13mm} + \frac{c_{13, \sigma_4} \langle
K_2^{\flat -} | \slashed{k}_6 | K_1^{\flat -} \rangle \langle
K_4^{\flat -} | \slashed{k}_1 | K_5^{\flat -} \rangle
(z-P_4)\left(\beta_3(z) -Q_3^\bullet \right)}{\langle K_2^{\flat
-} | \slashed{k}_1 | K_1^{\flat -} \rangle \langle K_4^{\flat -} |
\slashed{k}_6 | K_5^{\flat -} \rangle (z-P_2) \left(\beta_3(z) -
Q_2^\bullet \right)} - \frac{1}{2} \frac{c_{20,\sigma_4}}{\langle
K_4^{\flat -}| \slashed{k}_6 |
K_5^{\flat -} \rangle \left(\beta_3(z) -Q_2^\bullet \right)} \nonumber \\[2mm]
&\phantom{=}& \hspace{-13mm} + \frac{c_{12, \sigma_1} \langle
K_2^{\flat -} | \slashed{k}_{56} | K_1^{\flat -} \rangle \langle
K_4^{\flat -} | \slashed{k}_1 | K_5^{\flat -} \rangle
(z-P_1)\left(\beta_3 (z) -Q_3^\bullet \right)}{\langle K_2^{\flat
-} | \slashed{k}_1 | K_1^{\flat -} \rangle \langle K_4^{\flat -} |
\slashed{k}_6 | K_5^{\flat -} \rangle (z-P_2) \left(\beta_3(z)
-Q_2^\bullet \right)} - \frac{c_{9,\sigma_{10}} \langle K_2^{\flat
-} | \slashed{k}_{56} | K_1^{\flat -} \rangle (z-P_1)}{\langle
K_2^{\flat -}| \slashed{k}_1 | K_1^{\flat -}
\rangle (z-P_2)} \nonumber \\[2mm]
&\phantom{=}& \hspace{-13mm} + \frac{c_{12, \sigma_7} \langle
K_2^{\flat -} | \slashed{k}_6 | K_1^{\flat -} \rangle \langle
K_4^{\flat -} | \slashed{k}_{12} | K_5^{\flat -} \rangle
(z-P_4)\left(\beta_3(z) -Q_1^\bullet \right)}{\langle K_2^{\flat
-} | \slashed{k}_1 | K_1^{\flat -} \rangle \langle K_4^{\flat -} |
\slashed{k}_6 | K_5^{\flat -} \rangle (z-P_2) \left(\beta_3(z)
-Q_2^\bullet \right)} - \frac{c_{9,\sigma_4} \langle K_4^{\flat -}
| \slashed{k}_{12} | K_5^{\flat -} \rangle \left(\beta_3(z)
-Q_1^\bullet \right)}{\langle K_4^{\flat -}| \slashed{k}_6 |
K_5^{\flat -} \rangle \left(\beta_3(z) -Q_2^\bullet \right)} \\[2mm]
K_6(z) \hspace{-2mm}&=&\hspace{-2mm} \mbox{parity conjugate of
$K_5(z)$ \hspace{0.3mm} (obtained by applying eqs.
(\ref{eq:parity_conj_rule_1})-(\ref{eq:parity_conj_rule_6}))}
\end{eqnarray}}

\noindent where $\beta_3 (z)$ is given in eq.
(\ref{eq:beta_3_heptacut_3}).

\subsubsection{Heptacut \#3 of the left hand side of eq. (\ref{eq:6gluonampansatz})}

The result of applying heptacut \#3 to the left hand side of eq.
(\ref{eq:6gluonampansatz}) is
\begin{equation}
i \sum_{i=1}^6 \oint_{\Gamma_i} dz \, J_i (z) \left.
\prod_{j=1}^6 A_j^\mathrm{tree}(z) \right|_{\mathcal{S}_i}
\end{equation}
where, assuming without loss of generality the external helicities are $(1^-, 2^-, 3^+, 4^+, 5^+, 6^+)$,
the cut amplitude evaluated on the six different
kinematical solutions yields
\begin{equation}
\left. \prod_{j=1}^6 A_j^\mathrm{tree}(z) \right|_{\mathcal{S}_i}
\hspace{3mm}=\hspace{3mm} -\frac{i}{16} A^\mathrm{tree}_{--++++}
\times \left\{ \begin{array}{ll} \frac{1}{J_5(z)} \left(
\frac{1}{z - P_2} - \frac{1}{z - P_4} \right) & \hspace{3mm} \mathrm{for} \hspace{3mm} i = 5 \\[2mm]
0 & \hspace{3mm} \mathrm{for} \hspace{3mm} i = 1,2,3,4,6 \: .
\end{array} \right.
\end{equation}

\clearpage

\subsection{Heptacut \#4}\label{sec:heptacut_4}

This heptacut is defined by the on-shell constraints in eqs.
(\ref{eq:on-shell_constraint_1})-(\ref{eq:on-shell_constraint_7})
with the vertex momenta
\begin{equation}
\begin{array}{lll}
K_1 = k_1 & \hspace{1.2cm} K_2 = k_2 & \hspace{1.2cm} K_3 = 0 \: \phantom{.}\\
K_4 = k_{34} & \hspace{1.2cm} K_5 = k_5 & \hspace{1.2cm} K_6 = k_6
\: .
\end{array}
\end{equation}
Applying this heptacut to the right hand side of eq.
(\ref{eq:6gluonampansatz}) leaves the following linear combination
of cut integrals
\begin{figure}[!h]
\begin{center}
\includegraphics[angle=0, width=0.9\textwidth]{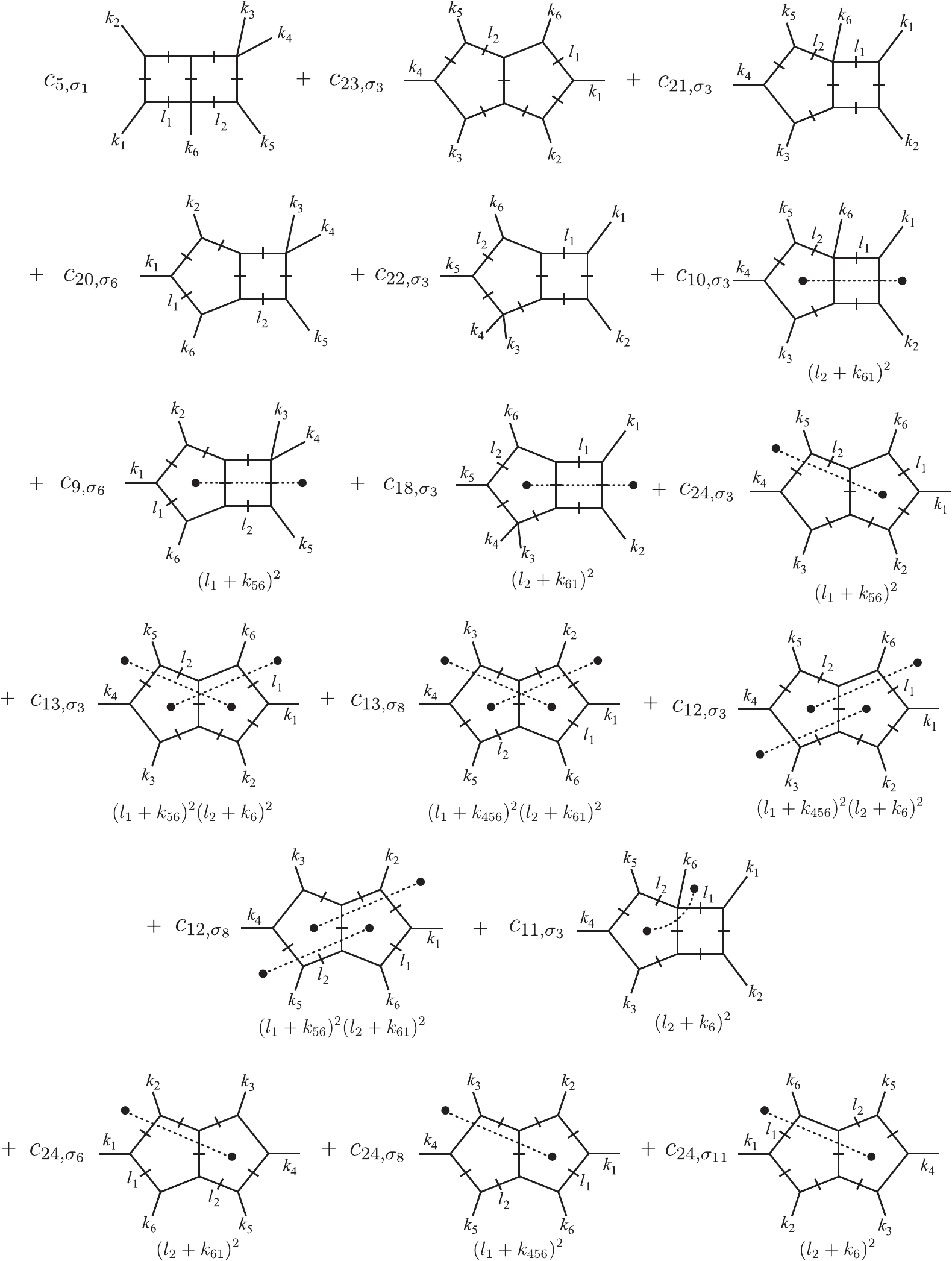}
\end{center}
\end{figure}
\vspace{-0.5cm}

\noindent We define the spinor ratios
\begin{equation}
\begin{array}{lll}
P_1 = - \frac{\left( 1 + \frac{S_4}{\gamma_2}\right) K_5^\flat
\cdot (K_1^\flat + k_6) + K_1^\flat \cdot k_6}{\left\langle
K_2^{\flat -}\left| \left( 1 +
\frac{S_4}{\gamma_2}\right)\slashed{K}_5^\flat + \slashed{k}_6
\right| K_1^{\flat -} \right\rangle} \: , \hspace{0.3cm} & P_2 = -
\frac{K_1^\flat \cdot k_6}{\langle K_2^{\flat -}| \slashed{k}_6 |
K_1^{\flat -} \rangle} \: , \hspace{0.3cm} & P_3 = -
\frac{K_1^\flat \cdot k_{56} + \frac{1}{2}s_{56}}{\langle
K_2^{\flat -} | \slashed{k}_{56} | K_1^{\flat -} \rangle} \\ P_4 =
-\frac{\langle K_4^{\flat -}| \slashed{K}_1^\flat + \slashed{k}_6
| K_5^{\flat -} \rangle}{2 \langle K_2^\flat K_4^\flat \rangle
[K_5^\flat K_1^\flat]} \: , \hspace{0.3cm} & P_5 = -
\frac{K_1^\flat \cdot k_{456} + \frac{1}{2}s_{456}}{\langle K_2^{\flat -} | \slashed{k}_{456} | K_1^{\flat -} \rangle} & \\
\vspace{-0.15cm} \\ Q_1 = -\frac{\left( 1 +
\frac{S_4}{\gamma_2}\right) K_5^\flat \cdot (K_1^\flat + k_6) +
K_1^\flat \cdot k_6}{\langle K_5^{\flat -}| \slashed{K}_1^\flat +
\slashed{k}_6 | K_4^{\flat -} \rangle} \: , \hspace{0.3cm} & Q_2 =
-\frac{\left( 1 + \frac{S_4}{\gamma_2}\right) K_5^\flat \cdot
k_{45} - \frac{1}{2} s_{45}}{\langle K_5^{\flat -}|
\slashed{k}_{45} | K_4^{\flat -} \rangle} \: , \hspace{0.3cm} &
Q_3 = - \frac{\left( 1 + \frac{S_4}{\gamma_2} \right) K_5^\flat
\cdot k_6}{\langle K_5^{\flat-}| \slashed{k}_6 | K_4^{\flat -}
\rangle} \\ \vspace{-0.4cm} \\ Q_4 = - \frac{\left( 1 +
\frac{S_4}{\gamma_2} \right) K_5^\flat \cdot k_1}{\langle
K_5^{\flat -} | \slashed{k}_1 | K_4^{\flat -} \rangle} \: ,
\hspace{0.3cm} & &
\end{array}
\end{equation}
and their parity conjugates
\begin{equation}
\begin{array}{lll} P_1^\bullet = - \frac{\left( 1 +
\frac{S_4}{\gamma_2}\right) K_5^\flat \cdot (K_1^\flat + k_6) +
K_1^\flat \cdot k_6}{\left\langle K_1^{\flat -}\left| \left( 1 +
\frac{S_4}{\gamma_2}\right)\slashed{K}_5^\flat + \slashed{k}_6
\right| K_2^{\flat -} \right\rangle} \: , \hspace{0.3cm} &
P_2^\bullet = - \frac{K_1^\flat \cdot k_6}{\langle K_1^{\flat -}|
\slashed{k}_6 | K_2^{\flat -} \rangle} \: , \hspace{0.3cm} &
P_3^\bullet = - \frac{K_1^\flat \cdot k_{56} +
\frac{1}{2}s_{56}}{\langle K_1^{\flat -} | \slashed{k}_{56} |
K_2^{\flat -} \rangle} \\ P_4^\bullet = -\frac{\langle K_5^{\flat
-}| \slashed{K}_1^\flat + \slashed{k}_6 | K_4^{\flat -} \rangle}{2
[K_2^\flat K_4^\flat] \langle K_5^\flat K_1^\flat \rangle} \: ,
\hspace{0.3cm} & P_5^\bullet = -\frac{K_1^\flat \cdot k_{456}
+ \frac{1}{2}s_{456}}{\langle K_1^{\flat -} | \slashed{k}_{456} | K_2^{\flat -} \rangle} & \\
\vspace{-0.15cm} \\ Q_1^\bullet = -\frac{\left( 1 +
\frac{S_4}{\gamma_2}\right) K_5^\flat \cdot (K_1^\flat + k_6) +
K_1^\flat \cdot k_6}{\langle K_4^{\flat -}| \slashed{K}_1^\flat +
\slashed{k}_6 | K_5^{\flat -} \rangle} \: , \hspace{0.3cm} &
Q_2^\bullet = -\frac{\left( 1 + \frac{S_4}{\gamma_2}\right)
K_5^\flat \cdot k_{45} - \frac{1}{2} s_{45}}{\langle K_4^{\flat
-}| \slashed{k}_{45} | K_5^{\flat -} \rangle} \: , \hspace{0.3cm}
& Q_3^\bullet = - \frac{\left( 1 + \frac{S_4}{\gamma_2} \right)
K_5^\flat \cdot k_6}{\langle K_4^{\flat-}| \slashed{k}_6 |
K_5^{\flat -} \rangle} \\ \vspace{-0.4cm} \\ Q_4^\bullet = -
\frac{\left( 1 + \frac{S_4}{\gamma_2} \right) K_5^\flat \cdot
k_1}{\langle K_4^{\flat -} | \slashed{k}_1 | K_5^{\flat -}
\rangle} \: , \hspace{0.3cm} & &
\end{array}
\end{equation}
This heptacut belongs to case I treated in Section
\ref{sec:general_double_box}, and there are thus six kinematical
solutions (shown in Fig. \ref{fig:kinematical_solutions_case_I}).
Parametrizing the loop momenta according to eqs.
(\ref{eq:l1_parametrized})-(\ref{eq:l2_parametrized}), the
on-shell constraints in eqs.
(\ref{eq:on-shell_constraint_1})-(\ref{eq:on-shell_constraint_7})
are solved by setting the parameters equal to the values
\begin{equation}
\begin{array}{ll}
\alpha_1 = 1 \: , \hspace{1cm} & \beta_1 = 0\\
\alpha_2 = 0 \: , \hspace{1cm} & \beta_2 = 1 +
\frac{S_4}{\gamma_2}
\end{array}
\end{equation}
and those given in Fig. \ref{fig:kinematical_solutions_case_I}
with
\begin{equation}
\beta_3 (z) = \frac{Q_3^\bullet (P_2 - P_4)(z-P_1)}{(P_2 - P_1)
(z-P_4)} \label{eq:beta_3_heptacut_4}
\end{equation}
for kinematical solution $\mathcal{S}_5$. The heptacut double box
integral $I_{5,\sigma_1}$ is $\sum_{i=1}^6
\oint_{\hspace{0.5mm}\Gamma_i} \hspace{-0.3mm} dz \, J_i(z)$ where
\begin{equation}
J_i(z) \hspace{0.5mm}=\hspace{0.5mm} \frac{1}{32 \gamma_1
\gamma_2} \times \left\{ \begin{array}{ll} \left( \left\langle
K_1^{\flat -} \left| \left( 1 + \frac{S_4}{\gamma_2} \right)
\slashed{K}_5^\flat + \slashed{k}_6 \right| K_2^{\flat -}
\right\rangle \hspace{0.6mm}
z (z - P_1^\bullet) \right)^{-1} & \hspace{1mm} \mathrm{for} \hspace{3mm} i=2,6 \\
\left( \left\langle K_2^{\flat -} \left| \left( 1 +
\frac{S_4}{\gamma_2} \right) \slashed{K}_5^\flat + \slashed{k}_6
\right| K_1^{\flat -} \right\rangle \hspace{0.6mm}
z (z - P_1) \right)^{-1} & \hspace{1mm} \mathrm{for} \hspace{3mm} i=4,5 \\
\left( \langle K_4^{\flat -} | \slashed{K}_1^\flat + \slashed{k}_6
| K_5^{\flat -} \rangle
\hspace{0.6mm} z (z - Q_1^\bullet) \right)^{-1} & \hspace{1mm} \mathrm{for} \hspace{3mm} i=1 \\
\left( \langle K_5^{\flat -} | \slashed{K}_1^\flat + \slashed{k}_6
| K_4^{\flat -} \rangle \hspace{0.6mm} z (z - Q_1) \right)^{-1} &
\hspace{1mm} \mathrm{for} \hspace{3mm} i=3 \: .
\end{array} \right.
\end{equation}

\subsubsection{Heptacut \#4 of the right hand side of eq. (\ref{eq:6gluonampansatz})}

The result of applying heptacut \#4 to the right hand side of eq.
(\ref{eq:6gluonampansatz}) is
\begin{equation}
\frac{1}{4} \sum_{i=1}^6 \oint_{\Gamma_i} dz \, J_i (z) K_i(z)
\end{equation}
where the kernels evaluated on the six kinematical solutions are
{\small\begin{eqnarray} K_1(z) \hspace{-2mm}&=&\hspace{-2mm}
c_{5,\sigma_1} - \frac{1}{4} \frac{c_{23, \sigma_3}}{\langle
K_1^{\flat-} | \slashed{k}_6 | K_2^{\flat -} \rangle \langle
K_4^{\flat -} | \slashed{k}_{45} | K_5^{\flat -} \rangle
(P_1^\bullet - P_2^\bullet) (z - Q_2^\bullet)} - \frac{1}{2}
\frac{c_{21,\sigma_3}}{\langle K_4^{\flat -} | \slashed{k}_{45} |
K_5^{\flat -} \rangle (z - Q_2^\bullet)} \nonumber \\[2mm]
&\phantom{=}& \hspace{-13mm}  + \frac{1}{2}
\frac{c_{20,\sigma_6}}{\langle K_1^{\flat -} | \slashed{k}_6 |
K_2^{\flat -} \rangle (P_1^\bullet - P_2^\bullet)}  + \frac{1}{2}
\frac{c_{22,\sigma_3}}{\langle K_4^{\flat -} | \slashed{k}_6 |
K_5^{\flat -} \rangle (z - Q_3^\bullet)} \nonumber \\[2mm]
&\phantom{=}& \hspace{-13mm} - \frac{1}{2} \frac{c_{24, \sigma_3}
\langle K_1^{\flat -} | \slashed{k}_{56} | K_2^{\flat -} \rangle
(P_1^\bullet - P_3^\bullet)}{\langle K_1^{\flat -} | \slashed{k}_6
| K_2^{\flat -} \rangle \langle K_4^{\flat -} | \slashed{k}_{45} |
K_5^{\flat -} \rangle (P_1^\bullet - P_2^\bullet) (z - Q_2^\bullet)} \nonumber \\[2mm]
&\phantom{=}& \hspace{-13mm} - \frac{1}{2} \frac{c_{24, \sigma_6}
\langle K_4^{\flat -} | \slashed{k}_{61} | K_5^{\flat -} \rangle
(z - Q_1^\bullet) \hspace{0.6mm}+\hspace{0.6mm} c_{24, \sigma_8}
\langle K_1^{\flat -} | \slashed{k}_{456} | K_2^{\flat -} \rangle
(P_1^\bullet - P_5^\bullet) \hspace{0.6mm}+\hspace{0.6mm} c_{24,
\sigma_{11}} \langle K_4^{\flat -} | \slashed{k}_6 | K_5^{\flat -}
\rangle (z - Q_3^\bullet)}{\langle K_1^{\flat-} | \slashed{k}_6 |
K_2^{\flat -} \rangle \langle K_4^{\flat -} | \slashed{k}_{45} |
K_5^{\flat -} \rangle (P_1^\bullet - P_2^\bullet) (z - Q_2^\bullet)} \nonumber \\[2mm]
&\phantom{=}& \hspace{-13mm} - \frac{c_{13, \sigma_3} \langle
K_1^{\flat -} | \slashed{k}_{56} | K_2^{\flat -} \rangle \langle
K_4^{\flat -}| \slashed{k}_6 | K_5^{\flat -} \rangle (P_1^\bullet
- P_3^\bullet)(z - Q_3^\bullet)}{\langle K_1^{\flat -} |
\slashed{k}_6 | K_2^{\flat -} \rangle \langle K_4^{\flat -} |
\slashed{k}_{45} | K_5^{\flat -} \rangle (P_1^\bullet -
P_2^\bullet) (z - Q_2^\bullet)} - \frac{c_{10, \sigma_3} \langle
K_4^{\flat -} | \slashed{k}_{61} | K_5^{\flat -} \rangle
(z-Q_1^\bullet)}{\langle K_4^{\flat -} | \slashed{k}_{45} |
K_5^{\flat -} \rangle (z - Q_2^\bullet)} \nonumber \\[2mm]
&\phantom{=}& \hspace{-13mm} - \frac{c_{13, \sigma_8} \langle
K_1^{\flat -} | \slashed{k}_{456} | K_2^{\flat -} \rangle \langle
K_4^{\flat -}| \slashed{k}_{61} | K_5^{\flat -} \rangle
(P_1^\bullet - P_5^\bullet)(z - Q_1^\bullet)}{\langle K_1^{\flat
-} | \slashed{k}_6 | K_2^{\flat -} \rangle \langle K_4^{\flat -} |
\slashed{k}_{45} | K_5^{\flat -} \rangle (P_1^\bullet -
P_2^\bullet) (z - Q_2^\bullet)} + \frac{c_{9, \sigma_6} \langle
K_1^{\flat -} | \slashed{k}_{56} | K_2^{\flat -} \rangle
(P_1^\bullet- P_3^\bullet)}{\langle K_1^{\flat -}| \slashed{k}_6 |
K_2^{\flat -} \rangle (P_1^\bullet - P_2^\bullet)} \nonumber \\[2mm]
&\phantom{=}& \hspace{-13mm} - \frac{c_{12, \sigma_3} \langle
K_1^{\flat -} | \slashed{k}_{456} | K_2^{\flat -} \rangle \langle
K_4^{\flat -}| \slashed{k}_6 | K_5^{\flat -} \rangle (P_1^\bullet
- P_5^\bullet)(z - Q_3^\bullet)}{\langle K_1^{\flat -} |
\slashed{k}_6 | K_2^{\flat -} \rangle \langle K_4^{\flat -} |
\slashed{k}_{45} | K_5^{\flat -} \rangle (P_1^\bullet -
P_2^\bullet) (z - Q_2^\bullet)} + \frac{c_{18,\sigma_3} \langle
K_4^{\flat -} | \slashed{k}_{61} | K_5^{\flat -} \rangle (z -
Q_1^\bullet)}{\langle K_4^{\flat -} |
\slashed{k}_6 | K_5^{\flat -} \rangle (z - Q_3^\bullet)} \nonumber \\[2mm]
&\phantom{=}& \hspace{-13mm} - \frac{c_{12, \sigma_8} \langle
K_1^{\flat -} | \slashed{k}_{56} | K_2^{\flat -} \rangle \langle
K_4^{\flat -}| \slashed{k}_{61} | K_5^{\flat -} \rangle
(P_1^\bullet - P_3^\bullet)(z - Q_1^\bullet)}{\langle K_1^{\flat
-} | \slashed{k}_6 | K_2^{\flat -} \rangle \langle K_4^{\flat -} |
\slashed{k}_{45} | K_5^{\flat -} \rangle (P_1^\bullet -
P_2^\bullet) (z - Q_2^\bullet)} - \frac{c_{11,\sigma_3} \langle
K_4^{\flat -} | \slashed{k}_6 | K_5^{\flat -} \rangle (z -
Q_3^\bullet)}{\langle K_4^{\flat -} | \slashed{k}_{45} |
K_5^{\flat -} \rangle (z - Q_2^\bullet)} \\[8mm]
K_2(z) \hspace{-2mm}&=&\hspace{-2mm} c_{5,\sigma_1} - \frac{1}{4}
\frac{c_{23, \sigma_3}}{\langle K_1^{\flat-} | \slashed{k}_6 |
K_2^{\flat -} \rangle \langle K_4^{\flat -} | \slashed{k}_{45} |
K_5^{\flat -} \rangle (z - P_2^\bullet) (Q_1^\bullet -
Q_2^\bullet)} - \frac{1}{2} \frac{c_{21,\sigma_3}}{\langle
K_4^{\flat -} | \slashed{k}_{45} |
K_5^{\flat -} \rangle (Q_1^\bullet - Q_2^\bullet)}  \nonumber \\[2mm]
&\phantom{=}& \hspace{-10mm} + \frac{1}{2}
\frac{c_{20,\sigma_6}}{\langle K_1^{\flat -} | \slashed{k}_6 |
K_2^{\flat -} \rangle (z - P_2^\bullet)} + \frac{1}{2}
\frac{c_{22,\sigma_3}}{\langle K_4^{\flat -} | \slashed{k}_6 |
K_5^{\flat -} \rangle (Q_1^\bullet - Q_3^\bullet)} + \frac{c_{9,
\sigma_6}\langle K_1^{\flat -} | \slashed{k}_{56} | K_2^{\flat -}
\rangle (z- P_3^\bullet)}{\langle K_1^{\flat -}| \slashed{k}_6 |
K_2^{\flat -} \rangle (z - P_2^\bullet)} \nonumber \\[2mm]
&\phantom{=}& \hspace{-10mm} - \frac{1}{2} \frac{c_{24, \sigma_3}
\langle K_1^{\flat -} | \slashed{k}_{56} | K_2^{\flat -} \rangle
(z-P_3^\bullet) + c_{24, \sigma_8} \langle K_1^{\flat -} |
\slashed{k}_{456} | K_2^{\flat -} \rangle (z - P_5^\bullet) +
c_{24, \sigma_{11}} \langle K_4^{\flat -} | \slashed{k}_6 |
K_5^{\flat -} \rangle (Q_1^\bullet - Q_3^\bullet)}{\langle
K_1^{\flat-} | \slashed{k}_6 | K_2^{\flat -} \rangle \langle
K_4^{\flat -} | \slashed{k}_{45} |
K_5^{\flat -} \rangle (z - P_2^\bullet) (Q_1^\bullet - Q_2^\bullet)} \nonumber \\[2mm]
&\phantom{=}& \hspace{-10mm} - \frac{c_{13, \sigma_3} \langle
K_1^{\flat -} | \slashed{k}_{56} | K_2^{\flat -} \rangle \langle
K_4^{\flat -}| \slashed{k}_6 | K_5^{\flat -} \rangle (z -
P_3^\bullet)(Q_1^\bullet - Q_3^\bullet)}{\langle K_1^{\flat -} |
\slashed{k}_6 | K_2^{\flat -} \rangle \langle K_4^{\flat -} |
\slashed{k}_{45} | K_5^{\flat -} \rangle (z - P_2^\bullet)
(Q_1^\bullet - Q_2^\bullet)} \nonumber - \frac{c_{11,\sigma_3}
\langle K_4^{\flat -} | \slashed{k}_6 | K_5^{\flat -} \rangle
(Q_1^\bullet - Q_3^\bullet)}{\langle K_4^{\flat -} |
\slashed{k}_{45} | K_5^{\flat -} \rangle (Q_1^\bullet - Q_2^\bullet)}\\[2mm]
&\phantom{=}& \hspace{-10mm} - \frac{c_{12, \sigma_3} \langle
K_1^{\flat-} | \slashed{k}_{456} | K_2^{\flat -} \rangle \langle
K_4^{\flat -} | \slashed{k}_6 | K_5^{\flat -} \rangle (z -
P_5^\bullet) (Q_1^\bullet - Q_3^\bullet)}{\langle K_1^{\flat-} |
\slashed{k}_6 | K_2^{\flat -} \rangle \langle K_4^{\flat -} |
\slashed{k}_{45} | K_5^{\flat -} \rangle (z - P_2^\bullet)
(Q_1^\bullet - Q_2^\bullet)} \phantom{aaaaa} \\[6mm]
K_3(z) \hspace{-2mm}&=&\hspace{-2mm} \mbox{parity conjugate of
$K_1(z)$ \hspace{0.3mm} (obtained by applying eqs.
(\ref{eq:parity_conj_rule_1})-(\ref{eq:parity_conj_rule_6}))} \\[3mm]
K_4(z) \hspace{-2mm}&=&\hspace{-2mm} \mbox{parity conjugate of
$K_2(z)$ \hspace{0.3mm} (obtained by applying eqs.
(\ref{eq:parity_conj_rule_1})-(\ref{eq:parity_conj_rule_6}))}
\end{eqnarray}}

{\small\begin{eqnarray} K_5(z) \hspace{-2mm}&=&\hspace{-2mm}
c_{5,\sigma_1} - \frac{1}{2} \frac{c_{21,\sigma_3}}{\langle
K_4^{\flat -} | \slashed{k}_{45} | K_5^{\flat -} \rangle (\beta_3
(z) - Q_2^\bullet)} + \frac{1}{2} \frac{c_{20,\sigma_6}}{\langle
K_2^{\flat -} | \slashed{k}_6 |
K_1^{\flat -} \rangle (z - P_2)} \nonumber \\[2mm]
&\phantom{=}& \hspace{-10mm} + \frac{1}{2}
\frac{c_{22,\sigma_3}}{\langle K_4^{\flat -} | \slashed{k}_6 |
K_5^{\flat -} \rangle (\beta_3 (z) - Q_3^\bullet)} - \frac{1}{4}
\frac{c_{23, \sigma_3}}{\langle K_2^{\flat-} | \slashed{k}_6 |
K_1^{\flat -} \rangle \langle K_4^{\flat -} | \slashed{k}_{45} |
K_5^{\flat -} \rangle (z - P_2) (\beta_3 (z) - Q_2^\bullet)} \nonumber \\[2mm]
&\phantom{=}& \hspace{-10mm} - \frac{c_{10, \sigma_3} \langle
K_4^{\flat -} | \slashed{k}_{61} | K_5^{\flat -} \rangle (\beta_3
(z) - Q_1^\bullet) }{\langle K_4^{\flat -} | \slashed{k}_{45} |
K_5^{\flat -} \rangle (\beta_3 (z) - Q_2^\bullet)} + \frac{c_{9,
\sigma_6}\langle K_2^{\flat -} | \slashed{k}_{56} | K_1^{\flat -}
\rangle (z- P_3)}{\langle K_2^{\flat -}| \slashed{k}_6 | K_1^{\flat -} \rangle (z - P_2)}\nonumber \\[2mm]
&\phantom{=}& \hspace{-10mm} + \frac{c_{18,\sigma_3} \langle
K_4^{\flat -} | \slashed{k}_{61} | K_5^{\flat -} \rangle (\beta_3
(z) - Q_1^\bullet)}{\langle K_4^{\flat -} | \slashed{k}_6 |
K_5^{\flat -} \rangle (\beta_3 (z) - Q_3^\bullet)}  - \frac{1}{2}
\frac{c_{24, \sigma_3} \langle K_2^{\flat -} | \slashed{k}_{56} |
K_1^{\flat -} \rangle (z-P_3)}{\langle K_2^{\flat -} |
\slashed{k}_6 | K_1^{\flat -} \rangle \langle K_4^{\flat -} |
\slashed{k}_{45} |
K_5^{\flat -} \rangle (z - P_2) (\beta_3 (z) - Q_2^\bullet)}\nonumber \\[2mm]
&\phantom{=}& \hspace{-10mm} - \frac{1}{2} \frac{c_{24, \sigma_6}
\langle K_4^{\flat -} | \slashed{k}_{61} | K_5^{\flat -} \rangle
(\beta_3 (z) - Q_1^\bullet)}{\langle K_2^{\flat-} | \slashed{k}_6
| K_1^{\flat -} \rangle \langle K_4^{\flat -} | \slashed{k}_{45} |
K_5^{\flat -} \rangle (z - P_2) (\beta_3 (z) - Q_2^\bullet)} -
\frac{c_{11,\sigma_3} \langle K_4^{\flat -} | \slashed{k}_6 |
K_5^{\flat -} \rangle (\beta_3 (z) - Q_3^\bullet)}{\langle
K_4^{\flat -} | \slashed{k}_{45} |
K_5^{\flat -} \rangle (\beta_3 (z) - Q_2^\bullet)} \nonumber \\[2mm]
&\phantom{=}& \hspace{-10mm} - \frac{1}{2} \frac{c_{24, \sigma_8}
\langle K_2^{\flat -} | \slashed{k}_{456} | K_1^{\flat -} \rangle
(z - P_5)}{\langle K_2^{\flat-} | \slashed{k}_6 | K_1^{\flat -}
\rangle \langle K_4^{\flat -} | \slashed{k}_{45} | K_5^{\flat -}
\rangle (z - P_2) (\beta_3 (z) - Q_2^\bullet)} \nonumber \\[2mm]
&\phantom{=}& \hspace{-10mm} - \frac{1}{2} \frac{c_{24,
\sigma_{11}} \langle K_4^{\flat -} | \slashed{k}_6 | K_5^{\flat -}
\rangle (\beta_3 (z) - Q_3^\bullet)}{\langle K_2^{\flat-} |
\slashed{k}_6 | K_1^{\flat -} \rangle \langle K_4^{\flat -} |
\slashed{k}_{45} | K_5^{\flat -}
\rangle (z - P_2) (\beta_3 (z) - Q_2^\bullet)} \nonumber \\[2mm]
&\phantom{=}& \hspace{-10mm} - \frac{c_{13, \sigma_3} \langle
K_2^{\flat -} | \slashed{k}_{56} | K_1^{\flat -} \rangle \langle
K_4^{\flat -}| \slashed{k}_6 | K_5^{\flat -} \rangle (z -
P_3)(\beta_3 (z) - Q_3^\bullet)}{\langle K_2^{\flat -} |
\slashed{k}_6 | K_1^{\flat -} \rangle \langle K_4^{\flat -} |
\slashed{k}_{45} | K_5^{\flat -} \rangle (z - P_2)
(\beta_3 (z) - Q_2^\bullet)} \nonumber \\[2mm]
&\phantom{=}& \hspace{-10mm} - \frac{c_{13, \sigma_8} \langle
K_2^{\flat -} | \slashed{k}_{456} | K_1^{\flat -} \rangle \langle
K_4^{\flat -}| \slashed{k}_{61} | K_5^{\flat -} \rangle (z -
P_5)(\beta_3 (z) - Q_1^\bullet)}{\langle K_2^{\flat -} |
\slashed{k}_6 | K_1^{\flat -} \rangle \langle K_4^{\flat -} |
\slashed{k}_{45} | K_5^{\flat -} \rangle (z - P_2)
(\beta_3 (z) - Q_2^\bullet)} \nonumber \\[2mm]
&\phantom{=}& \hspace{-10mm} - \frac{c_{12, \sigma_3} \langle
K_2^{\flat -} | \slashed{k}_{456} | K_1^{\flat -} \rangle \langle
K_4^{\flat -}| \slashed{k}_6 | K_5^{\flat -} \rangle (z -
P_5)(\beta_3 (z) - Q_3^\bullet)}{\langle K_2^{\flat -} |
\slashed{k}_6 | K_1^{\flat -} \rangle \langle K_4^{\flat -} |
\slashed{k}_{45} | K_5^{\flat -} \rangle (z - P_2)
(\beta_3 (z) - Q_2^\bullet)} \nonumber \\[2mm]
&\phantom{=}& \hspace{-10mm} - \frac{c_{12, \sigma_8} \langle
K_2^{\flat -} | \slashed{k}_{56} | K_1^{\flat -} \rangle \langle
K_4^{\flat -}| \slashed{k}_{61} | K_5^{\flat -} \rangle (z -
P_3)(\beta_3 (z) - Q_1^\bullet)}{\langle K_2^{\flat -} |
\slashed{k}_6 | K_1^{\flat -} \rangle \langle K_4^{\flat -} |
\slashed{k}_{45} | K_5^{\flat -} \rangle (z - P_2) (\beta_3 (z) -
Q_2^\bullet)}  \\[4mm]
K_6(z) \hspace{-2mm}&=&\hspace{-2mm} \mbox{parity conjugate of
$K_5(z)$ \hspace{0.3mm} (obtained by applying eqs.
(\ref{eq:parity_conj_rule_1})-(\ref{eq:parity_conj_rule_6}))}
\end{eqnarray}}
where $\beta_3 (z)$ is given in eq. (\ref{eq:beta_3_heptacut_4}).

\subsubsection{Heptacut \#4 of the left hand side of eq. (\ref{eq:6gluonampansatz})}

The result of applying heptacut \#4 to the left hand side of eq.
(\ref{eq:6gluonampansatz}) is
\begin{equation}
i \sum_{i=1}^6 \oint_{\Gamma_i} dz \, J_i (z) \left.
\prod_{j=1}^6 A_j^\mathrm{tree}(z) \right|_{\mathcal{S}_i}
\end{equation}
where, assuming without loss of generality the external helicities are $(1^-, 2^-, 3^+, 4^+, 5^+, 6^+)$,
the cut amplitude evaluated on the six different
kinematical solutions yields
\begin{equation}
\left. \prod_{j=1}^6 A_j^\mathrm{tree}(z) \right|_{\mathcal{S}_i}
\hspace{3mm}=\hspace{3mm} -\frac{i}{16} A^\mathrm{tree}_{--++++}
\times \left\{ \begin{array}{ll} \frac{1}{J_4(z)} \left(
\frac{1}{z} - \frac{1}{z - P_2} \right) & \hspace{3mm} \mathrm{for} \hspace{3mm} i = 4 \\[2mm]
\frac{1}{J_6(z)} \left(
\frac{1}{z} - \frac{1}{z - P_5^\bullet} \right) & \hspace{3mm} \mathrm{for} \hspace{3mm} i = 6 \\[2mm]
0 & \hspace{3mm} \mathrm{for} \hspace{3mm} i = 1,2,3,5 \: .
\end{array} \right.
\end{equation}

\clearpage

\subsection{Heptacut \#5}\label{sec:heptacut_5}

This heptacut is defined by the on-shell constraints in eqs.
(\ref{eq:on-shell_constraint_1})-(\ref{eq:on-shell_constraint_7})
with the vertex momenta
\begin{equation}
\begin{array}{lll}
K_1 = k_{12} & \hspace{1.2cm} K_2 = k_3 & \hspace{1.2cm} K_3 = 0 \: \phantom{.} \\
K_4 = k_4 & \hspace{1.2cm} K_5 = k_5 & \hspace{1.2cm} K_6 = k_6 \:
.
\end{array}
\end{equation}
Applying this heptacut to the right hand side of eq.
(\ref{eq:6gluonampansatz}) leaves the following linear combination
of cut integrals

\begin{figure}[!h]
\begin{center}
\includegraphics[angle=0, width=0.9\textwidth]{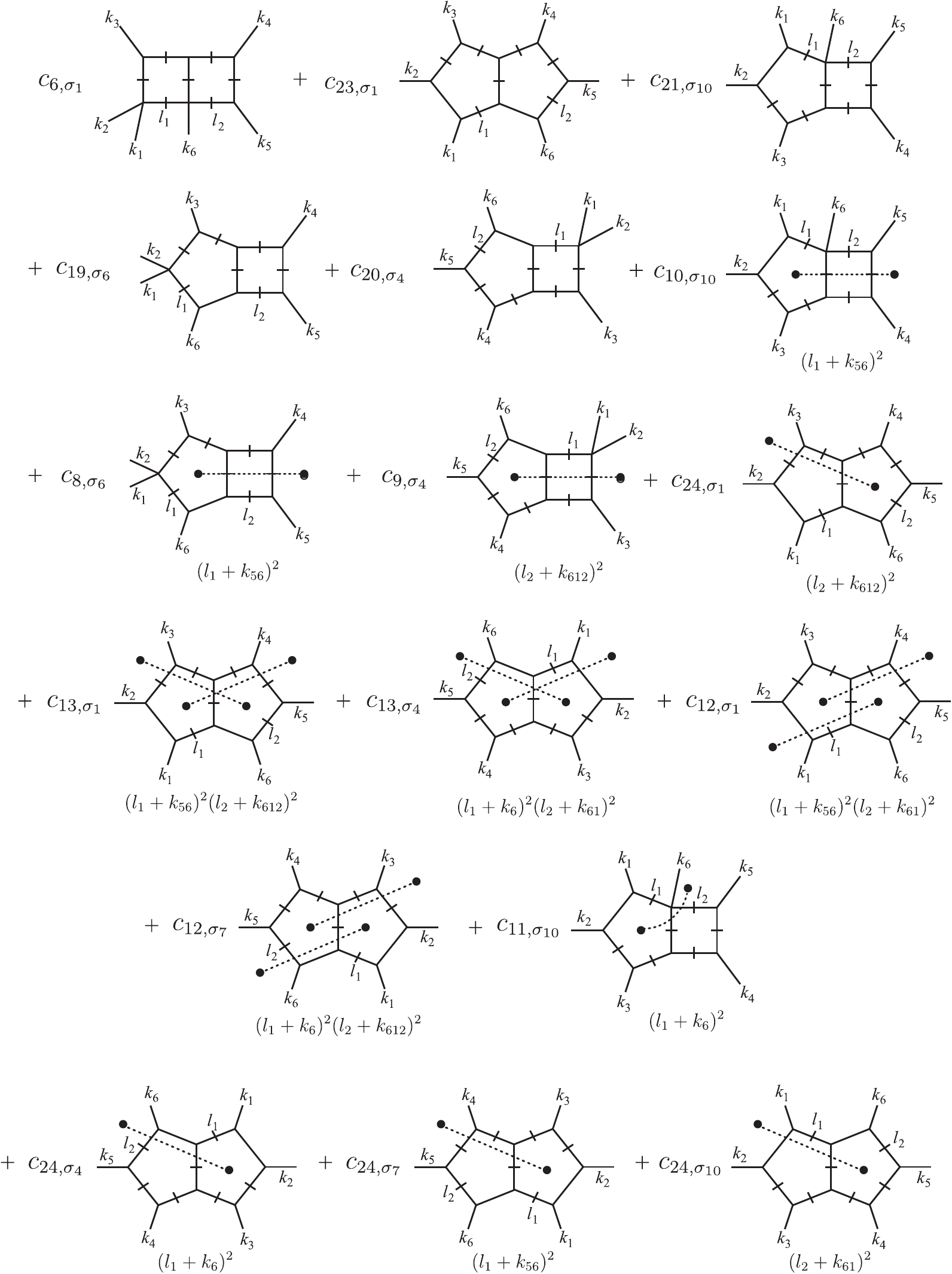}
\end{center}
\end{figure}
\vspace{-0.5cm}

\noindent We define the spinor ratios
\begin{equation}
\begin{array}{lll} P_1 = -\frac{K_5^\flat \cdot k_6 + (K_5^\flat + k_6)
\cdot K_1^\flat}{\langle K_2^{\flat -} | \slashed{K}_5^\flat +
\slashed{k}_6 | K_1^{\flat -} \rangle} \: , \hspace{0.3cm} & P_2 =
- \frac{\langle K_1^\flat k_1 \rangle}{2 \langle K_2^\flat k_1
\rangle} \: , \hspace{0.3cm} & P_3 = -\frac{\langle K_1^\flat k_6
\rangle}{2\langle K_2^\flat k_6 \rangle} \\
P_4 = - \frac{\langle K_1^\flat k_5 \rangle}{2\langle K_2^\flat
k_5 \rangle} \: , \hspace{0.3cm} & P_5 = - \frac{\langle
K_4^{\flat -} | \slashed{K}_1^\flat + \slashed{k}_6 | K_5^{\flat
-} \rangle}{2 \langle K_2^\flat K_4^\flat \rangle
[K_5^\flat K_1^\flat]} &  \\
\vspace{-0.15cm} \\
Q_1 = -\frac{K_1^\flat \cdot k_6 + K_5^\flat \cdot (K_1^\flat +
k_6)}{\langle K_5^{\flat -} | \slashed{K}_1^\flat + \slashed{k}_6
| K_4^{\flat -}\rangle} \: , \hspace{0.3cm} & Q_2 =
-\frac{[K_5^\flat k_6]}{2[K_4^\flat k_6]} \: , \hspace{0.3cm} &
Q_3 = -\frac{K_5^\flat \cdot k_{61} + \frac{1}{2} s_{61}} {\langle
K_5^{\flat -} | \slashed{k}_{61} | K_4^{\flat -} \rangle}
\end{array}
\end{equation}
and their parity conjugates
\begin{equation}
\begin{array}{lll} P_1^\bullet = -\frac{K_5^\flat \cdot k_6 + (K_5^\flat + k_6)
\cdot K_1^\flat}{\langle K_1^{\flat -} | \slashed{K}_5^\flat +
\slashed{k}_6 | K_2^{\flat -} \rangle} \: , \hspace{0.3cm} &
P_2^\bullet = - \frac{[K_1^\flat k_1]}{2[K_2^\flat k_1]} \: ,
\hspace{0.3cm} & P_3^\bullet = -\frac{[K_1^\flat k_6]}
{2[K_2^\flat k_6]} \\
P_4^\bullet = - \frac{[K_1^\flat k_5]}{2[K_2^\flat k_5]} \: ,
\hspace{0.3cm} & P_5^\bullet = - \frac{\langle K_5^{\flat -} |
\slashed{K}_1^\flat + \slashed{k}_6 | K_4^{\flat -} \rangle}{2
[K_2^\flat K_4^\flat] \langle K_5^\flat K_1^\flat \rangle} &  \\
\vspace{-0.15cm} \\
Q_1^\bullet = -\frac{K_1^\flat \cdot k_6 + K_5^\flat \cdot
(K_1^\flat + k_6)}{\langle K_4^{\flat -} | \slashed{K}_1^\flat +
\slashed{k}_6 | K_5^{\flat -}\rangle} \: , \hspace{0.3cm} &
Q_2^\bullet = -\frac{\langle K_5^\flat k_6 \rangle}{2 \langle
K_4^\flat k_6 \rangle} \: , \hspace{0.3cm} & Q_3^\bullet =
-\frac{K_5^\flat \cdot k_{61} + \frac{1}{2} s_{61}} {\langle
K_4^{\flat -} | \slashed{k}_{61} | K_5^{\flat -} \rangle} \: .
\end{array}
\end{equation}
This heptacut belongs to case I treated in Section
\ref{sec:general_double_box}, and there are thus six kinematical
solutions (shown in Fig. \ref{fig:kinematical_solutions_case_I}).
Parametrizing the loop momenta according to eqs.
(\ref{eq:l1_parametrized})-(\ref{eq:l2_parametrized}), the
on-shell constraints in eqs.
(\ref{eq:on-shell_constraint_1})-(\ref{eq:on-shell_constraint_7})
are solved by setting the parameters equal to the values
\begin{equation}
\begin{array}{ll}
\alpha_1 = 1 \: , \hspace{1cm} & \beta_1 = 0\\
\alpha_2 = 0 \: , \hspace{1cm} & \beta_2 = 1
\end{array}
\end{equation}
and those given in Fig. \ref{fig:kinematical_solutions_case_I}
with
\begin{equation}
\beta_3(z) = \frac{Q_2^\bullet (P_3 - P_5)(z-P_1)}{(P_3 -
P_1)(z-P_5)} \label{eq:beta_3_heptacut_5}
\end{equation}
for kinematical solution $\mathcal{S}_5$. The heptacut double box
integral $I_{6,\sigma_1}$ is $\sum_{i=1}^6
\oint_{\hspace{0.5mm}\Gamma_i} \hspace{-0.3mm} dz \, J_i(z)$ where
\begin{equation}
J_i(z) \hspace{2mm} = \hspace{2mm} \frac{1}{32 \gamma_1 \gamma_2}
\times \left\{ \begin{array}{ll} \left( \langle K_1^{\flat -} |
\slashed{K}_5^\flat + \slashed{k}_6 | K_2^{\flat -} \rangle
\hspace{0.6mm} z (z - P_1^\bullet) \right)^{-1} & \hspace{4mm} \mathrm{for} \hspace{3mm} i=2,6 \\
\left( \langle K_2^{\flat -} | \slashed{K}_5^\flat + \slashed{k}_6
| K_1^{\flat -} \rangle
\hspace{0.6mm} z (z - P_1) \right)^{-1} & \hspace{4mm} \mathrm{for} \hspace{3mm} i=4,5 \\
\left( \langle K_4^{\flat -} | \slashed{K}_1^\flat + \slashed{k}_6
| K_5^{\flat -} \rangle
\hspace{0.6mm} z (z - Q_1^\bullet) \right)^{-1} & \hspace{4mm} \mathrm{for} \hspace{3mm} i=1 \\
\left( \langle K_5^{\flat -} | \slashed{K}_1^\flat + \slashed{k}_6
| K_4^{\flat -} \rangle \hspace{0.6mm} z (z - Q_1) \right)^{-1} &
\hspace{4mm} \mathrm{for} \hspace{3mm} i=3 \: .
\end{array} \right.
\end{equation}

\subsubsection{Heptacut \#5 of the right hand side of eq. (\ref{eq:6gluonampansatz})}

The result of applying heptacut \#5 to the right hand side of eq.
(\ref{eq:6gluonampansatz}) is
\begin{equation}
\frac{1}{4} \sum_{i=1}^6 \oint_{\Gamma_i} dz \, J_i (z) K_i(z)
\end{equation}
where the kernels evaluated on the six kinematical solutions are

{\small\begin{eqnarray} K_1(z) \hspace{-2mm}&=&\hspace{-2mm}
c_{6,\sigma_1} - \frac{1}{2} \frac{c_{21, \sigma_{10}}}{\langle
K_1^{\flat -} | \slashed{k}_1 | K_2^{\flat -} \rangle (P_1^\bullet
- P_2^\bullet)} + \frac{1}{2} \frac{c_{19, \sigma_6}}{\langle
K_1^{\flat -} | \slashed{k}_6 | K_2^{\flat -}
\rangle (P_1^\bullet - P_3^\bullet)} \nonumber \\[2mm]
&\phantom{=}&  \hspace{5mm} + \hspace{0.5mm}\frac{1}{2}
\frac{c_{20, \sigma_4}}{\langle K_4^{\flat -} | \slashed{k}_6 |
K_5^{\flat -} \rangle (z - Q_2^\bullet)} - \frac{1}{4}
\frac{c_{23, \sigma_1}}{\langle K_1^{\flat -} | \slashed{k}_1 |
K_2^{\flat -} \rangle \langle K_4^{\flat -} | \slashed{k}_6 |
K_5^{\flat -} \rangle (P_1^\bullet - P_2^\bullet) (z - Q_2^\bullet)} \nonumber \\[2mm]
&\phantom{=}&  \hspace{-13mm} - \frac{1}{2} \frac{c_{24, \sigma_1}
\langle K_4^{\flat -} | \slashed{k}_{612} | K_5^{\flat -} \rangle
(z - Q_1^\bullet) + c_{24, \sigma_4} \langle K_1^{\flat -} |
\slashed{k}_6 | K_2^{\flat -} \rangle (P_1^\bullet - P_3^\bullet)
+ c_{24, \sigma_{10}} \langle K_4^{\flat -} | \slashed{k}_{61} |
K_5^{\flat -} \rangle (z - Q_3^\bullet)}{\langle K_1^{\flat -} |
\slashed{k}_1 | K_2^{\flat -} \rangle \langle K_4^{\flat -} |
\slashed{k}_6 | K_5^{\flat -}
\rangle (P_1^\bullet - P_2^\bullet) (z - Q_2^\bullet)} \nonumber \\[2mm]
&\phantom{=}&  \hspace{-13mm} - \hspace{0.5mm} \frac{c_{13,
\sigma_4} \langle K_1^{\flat -} | \slashed{k}_6 | K_2^{\flat -}
\rangle \langle K_4^{\flat -} | \slashed{k}_{61} | K_5^{\flat -}
\rangle (P_1^\bullet - P_3^\bullet) (z - Q_3^\bullet)}{\langle
K_1^{\flat -} | \slashed{k}_1 | K_2^{\flat -} \rangle \langle
K_4^{\flat -} | \slashed{k}_6 | K_5^{\flat -} \rangle (P_1^\bullet
- P_2^\bullet) (z - Q_2^\bullet)} + \frac{c_{9, \sigma_4} \langle
K_4^{\flat -} | \slashed{k}_{612} | K_5^{\flat -} \rangle (z -
Q_1^\bullet)}{\langle K_4^{\flat -} | \slashed{k}_6 | K_5^{\flat
-} \rangle (z - Q_2^\bullet)} \nonumber \\[2mm]
&\phantom{=}&  \hspace{-13mm} - \hspace{0.5mm} \frac{c_{12,
\sigma_7} \langle K_1^{\flat -} | \slashed{k}_6 | K_2^{\flat -}
\rangle \langle K_4^{\flat -} | \slashed{k}_{612} | K_5^{\flat -}
\rangle (P_1^\bullet - P_3^\bullet) (z - Q_1^\bullet)}{\langle
K_1^{\flat -} | \slashed{k}_1 | K_2^{\flat -} \rangle \langle
K_4^{\flat -} | \slashed{k}_6 | K_5^{\flat -} \rangle (P_1^\bullet
- P_2^\bullet) (z - Q_2^\bullet)} - \frac{c_{11, \sigma_{10}}
\langle K_1^{\flat -} | \slashed{k}_6 | K_2^{\flat -} \rangle
(P_1^\bullet - P_3^\bullet)}{\langle K_1^{\flat -} | \slashed{k}_1
| K_2^{\flat -} \rangle (P_1^\bullet - P_2^\bullet)} \\[20mm]
K_2(z) \hspace{-2mm}&=&\hspace{-2mm} c_{6,\sigma_1} - \frac{1}{2}
\frac{c_{21, \sigma_{10}}}{\langle K_1^{\flat -} | \slashed{k}_1 |
K_2^{\flat -} \rangle (z - P_2^\bullet)} + \frac{1}{2}
\frac{c_{19, \sigma_6}}{\langle K_1^{\flat -} | \slashed{k}_6 |
K_2^{\flat -} \rangle (z - P_3^\bullet)} + \frac{1}{2}
\frac{c_{20, \sigma_4}}{\langle K_4^{\flat -} | \slashed{k}_6 |
K_5^{\flat -} \rangle (Q_1^\bullet - Q_2^\bullet)} \nonumber \\[2mm]
&\phantom{=}&  \hspace{-10mm} - \frac{1}{4} \frac{c_{23,
\sigma_1}}{\langle K_1^{\flat -} | \slashed{k}_1 | K_2^{\flat -}
\rangle \langle K_4^{\flat -} | \slashed{k}_6 | K_5^{\flat -}
\rangle (z - P_2^\bullet) (Q_1^\bullet - Q_2^\bullet)} +
\frac{c_{8, \sigma_6} \langle K_1^{\flat -} | \slashed{k}_{56} |
K_2^{\flat -} \rangle (z - P_1^\bullet)}{\langle K_1^{\flat -} |
\slashed{k}_6 | K_2^{\flat -} \rangle (z - P_3^\bullet)}\nonumber \\[2mm]
&\phantom{=}&  \hspace{-10mm} - \frac{1}{2} \frac{c_{24, \sigma_4}
\langle K_1^{\flat -} | \slashed{k}_6 | K_2^{\flat -} \rangle (z -
P_3^\bullet) + c_{24, \sigma_7} \langle K_1^{\flat -} |
\slashed{k}_{56} | K_2^{\flat -} \rangle (z - P_1^\bullet) +
c_{24, \sigma_{10}} \langle K_4^{\flat -} | \slashed{k}_{61} |
K_5^{\flat -} \rangle (Q_1^\bullet - Q_3^\bullet)}{\langle
K_1^{\flat -} | \slashed{k}_1 | K_2^{\flat -} \rangle \langle
K_4^{\flat -} | \slashed{k}_6 | K_5^{\flat -}
\rangle (z - P_2^\bullet) (Q_1^\bullet - Q_2^\bullet)}\nonumber \\[2mm]
&\phantom{=}&  \hspace{-10mm} - \hspace{0.5mm} \frac{c_{13,
\sigma_4} \langle K_1^{\flat -} | \slashed{k}_6 | K_2^{\flat -}
\rangle \langle K_4^{\flat -} | \slashed{k}_{61} | K_5^{\flat -}
\rangle (z - P_3^\bullet) (Q_1^\bullet - Q_3^\bullet)}{\langle
K_1^{\flat -} | \slashed{k}_1 | K_2^{\flat -} \rangle \langle
K_4^{\flat -} | \slashed{k}_6 | K_5^{\flat -} \rangle (z -
P_2^\bullet) (Q_1^\bullet - Q_2^\bullet)} - \frac{c_{10,
\sigma_{10}} \langle K_1^{\flat -} | \slashed{k}_{56} | K_2^{\flat
-} \rangle (z - P_1^\bullet)}{\langle K_1^{\flat -} |
\slashed{k}_1 | K_2^{\flat -} \rangle (z - P_2^\bullet)}\nonumber \\[2mm]
&\phantom{=}&  \hspace{-10mm} - \hspace{0.5mm} \frac{c_{12,
\sigma_1} \langle K_1^{\flat -} | \slashed{k}_{56} | K_2^{\flat -}
\rangle \langle K_4^{\flat -} | \slashed{k}_{61} | K_5^{\flat -}
\rangle (z - P_1^\bullet) (Q_1^\bullet - Q_3^\bullet)}{\langle
K_1^{\flat -} | \slashed{k}_1 | K_2^{\flat -} \rangle \langle
K_4^{\flat -} | \slashed{k}_6 | K_5^{\flat -} \rangle (z -
P_2^\bullet) (Q_1^\bullet - Q_2^\bullet)} - \frac{c_{11,
\sigma_{10}} \langle K_1^{\flat -} | \slashed{k}_6 | K_2^{\flat -}
\rangle (z - P_3^\bullet)}{\langle K_1^{\flat -} | \slashed{k}_1 |
K_2^{\flat -} \rangle (z - P_2^\bullet)} \phantom{aaaaa} \\[10mm]
K_3(z) \hspace{-2mm}&=&\hspace{-2mm} \mbox{parity conjugate of
$K_1(z)$ \hspace{0.3mm} (obtained by applying eqs.
(\ref{eq:parity_conj_rule_1})-(\ref{eq:parity_conj_rule_6}))} \\[10mm]
K_4(z) \hspace{-2mm}&=&\hspace{-2mm} \mbox{parity conjugate of
$K_2(z)$ \hspace{0.3mm} (obtained by applying eqs.
(\ref{eq:parity_conj_rule_1})-(\ref{eq:parity_conj_rule_6}))}
\end{eqnarray}}

{\small\begin{eqnarray} K_5(z) \hspace{-2mm}&=&\hspace{-2mm}
c_{6,\sigma_1} - \frac{1}{2} \frac{c_{21, \sigma_{10}}}{\langle
K_2^{\flat -} | \slashed{k}_1 | K_1^{\flat -} \rangle (z - P_2)} +
\frac{1}{2} \frac{c_{19, \sigma_6}}{\langle K_2^{\flat -} |
\slashed{k}_6 | K_1^{\flat -} \rangle (z - P_3)} +
\hspace{0.5mm}\frac{1}{2} \frac{c_{20, \sigma_4}}{\langle
K_4^{\flat -} | \slashed{k}_6 |
K_5^{\flat -} \rangle (\beta_3 (z) - Q_2^\bullet)}\nonumber \\[2mm]
&\phantom{=}&  \hspace{-13mm} - \frac{1}{4} \frac{c_{23,
\sigma_1}}{\langle K_2^{\flat -} | \slashed{k}_1 | K_1^{\flat -}
\rangle \langle K_4^{\flat -} | \slashed{k}_6 | K_5^{\flat -}
\rangle (z - P_2) (\beta_3 (z) - Q_2^\bullet)} + \frac{c_{8,
\sigma_6} \langle K_2^{\flat -} | \slashed{k}_{56} | K_1^{\flat -}
\rangle (z - P_1)}{\langle K_2^{\flat -} | \slashed{k}_6 |
K_1^{\flat -} \rangle (z - P_3)} \nonumber \\[2mm]
&\phantom{=}&  \hspace{-13mm}  + \frac{c_{9, \sigma_4} \langle
K_4^{\flat -} | \slashed{k}_{612} | K_5^{\flat -} \rangle (\beta_3
(z) - Q_1^\bullet)}{\langle K_4^{\flat -} | \slashed{k}_6 |
K_5^{\flat -} \rangle (\beta_3 (z) - Q_2^\bullet)} -
\hspace{0.5mm}\frac{1}{2} \frac{c_{24, \sigma_1} \langle
K_4^{\flat -} | \slashed{k}_{612} | K_5^{\flat -} \rangle (\beta_3
(z) - Q_1^\bullet)}{\langle K_2^{\flat -} | \slashed{k}_1 |
K_1^{\flat -} \rangle \langle K_4^{\flat -} | \slashed{k}_6 |
K_5^{\flat -} \rangle (z - P_2) (\beta_3 (z) - Q_2^\bullet)}\nonumber \\[2mm]
&\phantom{=}&  \hspace{-13mm} - \frac{1}{2} \frac{c_{24, \sigma_4}
\langle K_2^{\flat -} | \slashed{k}_6 | K_1^{\flat -} \rangle (z -
P_3) \hspace{0.6mm}+\hspace{0.6mm} c_{24, \sigma_7} \langle
K_2^{\flat -} | \slashed{k}_{56} | K_1^{\flat -} \rangle (z - P_1)
\hspace{0.6mm}+\hspace{0.6mm} c_{24, \sigma_{10}} \langle
K_4^{\flat -} | \slashed{k}_{61} | K_5^{\flat -} \rangle (\beta_3
(z) - Q_3^\bullet)}{\langle K_2^{\flat -} | \slashed{k}_1 |
K_1^{\flat -} \rangle \langle K_4^{\flat -} | \slashed{k}_6 |
K_5^{\flat -} \rangle (z - P_2) (\beta_3 (z) - Q_2^\bullet)} \nonumber \\[2mm]
&\phantom{=}&  \hspace{-13mm} - \hspace{0.5mm} \frac{c_{13,
\sigma_1} \langle K_2^{\flat -} | \slashed{k}_{56} | K_1^{\flat -}
\rangle \langle K_4^{\flat -} | \slashed{k}_{612} | K_5^{\flat -}
\rangle (z - P_1) (\beta_3 (z) - Q_1^\bullet)}{\langle K_2^{\flat
-} | \slashed{k}_1 | K_1^{\flat -} \rangle \langle K_4^{\flat -} |
\slashed{k}_6 | K_5^{\flat -} \rangle (z - P_2) (\beta_3 (z) -
Q_2^\bullet)} - \frac{c_{10, \sigma_{10}} \langle K_2^{\flat -} |
\slashed{k}_{56} | K_1^{\flat -} \rangle (z - P_1)}{\langle
K_2^{\flat -} | \slashed{k}_1 | K_1^{\flat -} \rangle (z - P_2)} \nonumber \\[2mm]
&\phantom{=}&  \hspace{-13mm} - \hspace{0.5mm} \frac{c_{13,
\sigma_4} \langle K_2^{\flat -} | \slashed{k}_6 | K_1^{\flat -}
\rangle \langle K_4^{\flat -} | \slashed{k}_{61} | K_5^{\flat -}
\rangle (z - P_3) (\beta_3 (z) - Q_3^\bullet)}{\langle K_2^{\flat
-} | \slashed{k}_1 | K_1^{\flat -} \rangle \langle K_4^{\flat -} |
\slashed{k}_6 | K_5^{\flat -} \rangle (z - P_2) (\beta_3 (z) -
Q_2^\bullet)} - \frac{c_{11, \sigma_{10}} \langle K_2^{\flat -} |
\slashed{k}_6 | K_1^{\flat -} \rangle (z - P_3)}{\langle
K_2^{\flat -} | \slashed{k}_1 | K_1^{\flat -} \rangle (z - P_2)}\nonumber \\[2mm]
&\phantom{=}&  \hspace{-13mm} - \hspace{0.5mm} \frac{c_{12,
\sigma_1} \langle K_2^{\flat -} | \slashed{k}_{56} | K_1^{\flat -}
\rangle \langle K_4^{\flat -} | \slashed{k}_{61} | K_5^{\flat -}
\rangle (z - P_1) (\beta_3 (z) - Q_3^\bullet)}{\langle K_2^{\flat
-} | \slashed{k}_1 | K_1^{\flat -} \rangle \langle K_4^{\flat -} |
\slashed{k}_6 | K_5^{\flat -} \rangle (z -
P_2) (\beta_3 (z) - Q_2^\bullet)} \nonumber \\[2mm]
&\phantom{=}&  \hspace{-13mm} - \hspace{0.5mm} \frac{c_{12,
\sigma_7} \langle K_2^{\flat -} | \slashed{k}_6 | K_1^{\flat -}
\rangle \langle K_4^{\flat -} | \slashed{k}_{612} | K_5^{\flat -}
\rangle (z - P_3) (\beta_3 (z) - Q_1^\bullet)}{\langle K_2^{\flat
-} | \slashed{k}_1 | K_1^{\flat -} \rangle \langle K_4^{\flat -} |
\slashed{k}_6 | K_5^{\flat -} \rangle (z - P_2) (\beta_3 (z) -
Q_2^\bullet)} \phantom{aaaaa} \\[6mm]
K_6(z) \hspace{-2mm}&=&\hspace{-2mm} \mbox{parity conjugate of
$K_5(z)$ \hspace{0.3mm} (obtained by applying eqs.
(\ref{eq:parity_conj_rule_1})-(\ref{eq:parity_conj_rule_6}))}
\end{eqnarray}}
where $\beta_3 (z)$ is given in eq. (\ref{eq:beta_3_heptacut_5}).

\subsubsection{Heptacut \#5 of the left hand side of eq. (\ref{eq:6gluonampansatz})}

The result of applying heptacut \#5 to the left hand side of eq.
(\ref{eq:6gluonampansatz}) is
\begin{equation}
i \sum_{i=1}^6 \oint_{\Gamma_i} dz \, J_i (z) \left.
\prod_{j=1}^6 A_j^\mathrm{tree}(z) \right|_{\mathcal{S}_i}
\end{equation}
where, assuming without loss of generality the external helicities are $(1^-, 2^-, 3^+, 4^+, 5^+, 6^+)$,
the cut amplitude evaluated on the six different
kinematical solutions yields
\begin{equation}
\left. \prod_{j=1}^6 A_j^\mathrm{tree}(z) \right|_{\mathcal{S}_i}
\hspace{3mm}=\hspace{3mm} -\frac{i}{16} A^\mathrm{tree}_{--++++}
\times \left\{ \begin{array}{ll} \frac{1}{J_4(z)} \left(
\frac{1}{z - P_2} - \frac{1}{z - P_3} \right) & \hspace{3mm} \mathrm{for} \hspace{3mm} i = 4 \\[2mm]
0 & \hspace{3mm} \mathrm{for} \hspace{3mm} i = 1,2,3,5,6 \: .
\end{array} \right.
\end{equation}

\clearpage

\subsection{Heptacut \#6}\label{sec:heptacut_6}

This heptacut is defined by the on-shell constraints in eqs.
(\ref{eq:on-shell_constraint_1})-(\ref{eq:on-shell_constraint_7})
with the vertex momenta
\begin{equation}
\begin{array}{lll}
K_1 = k_1 & \hspace{1.2cm} K_2 = k_2 & \hspace{1.2cm} K_3 = k_3 \: \phantom{.} \\
K_4 = k_4 & \hspace{1.2cm} K_5 = k_5 & \hspace{1.2cm} K_6 = k_6 \:
.
\end{array}
\end{equation}
Applying this heptacut to the right hand side of eq.
(\ref{eq:6gluonampansatz}) leaves the following linear combination
of cut integrals

\begin{figure}[!h]
\begin{center}
\includegraphics[angle=0, width=0.9\textwidth]{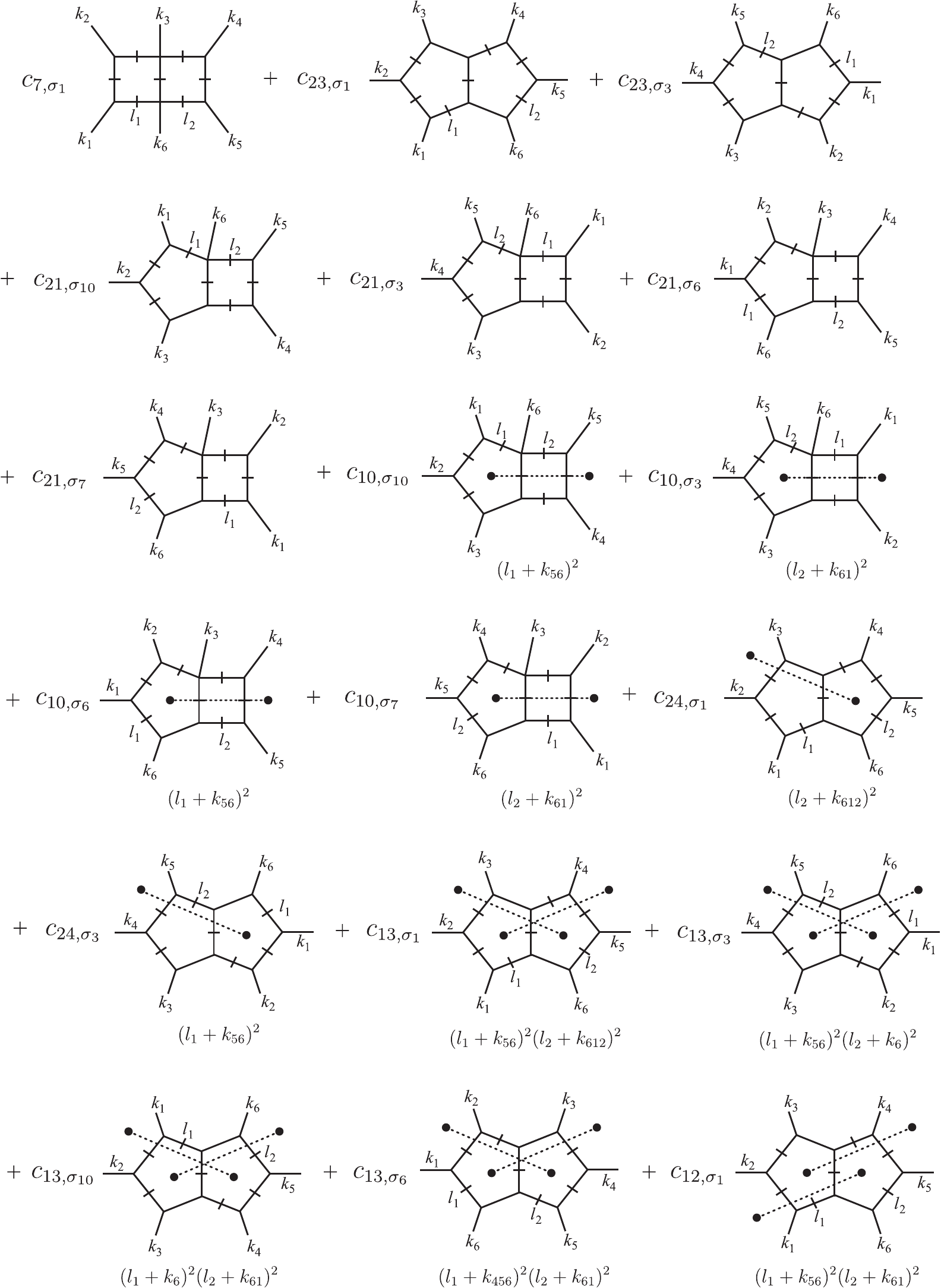}
\end{center}
\end{figure}

\clearpage

\begin{figure}[!h]
\begin{center}
\includegraphics[angle=0, width=0.9\textwidth]{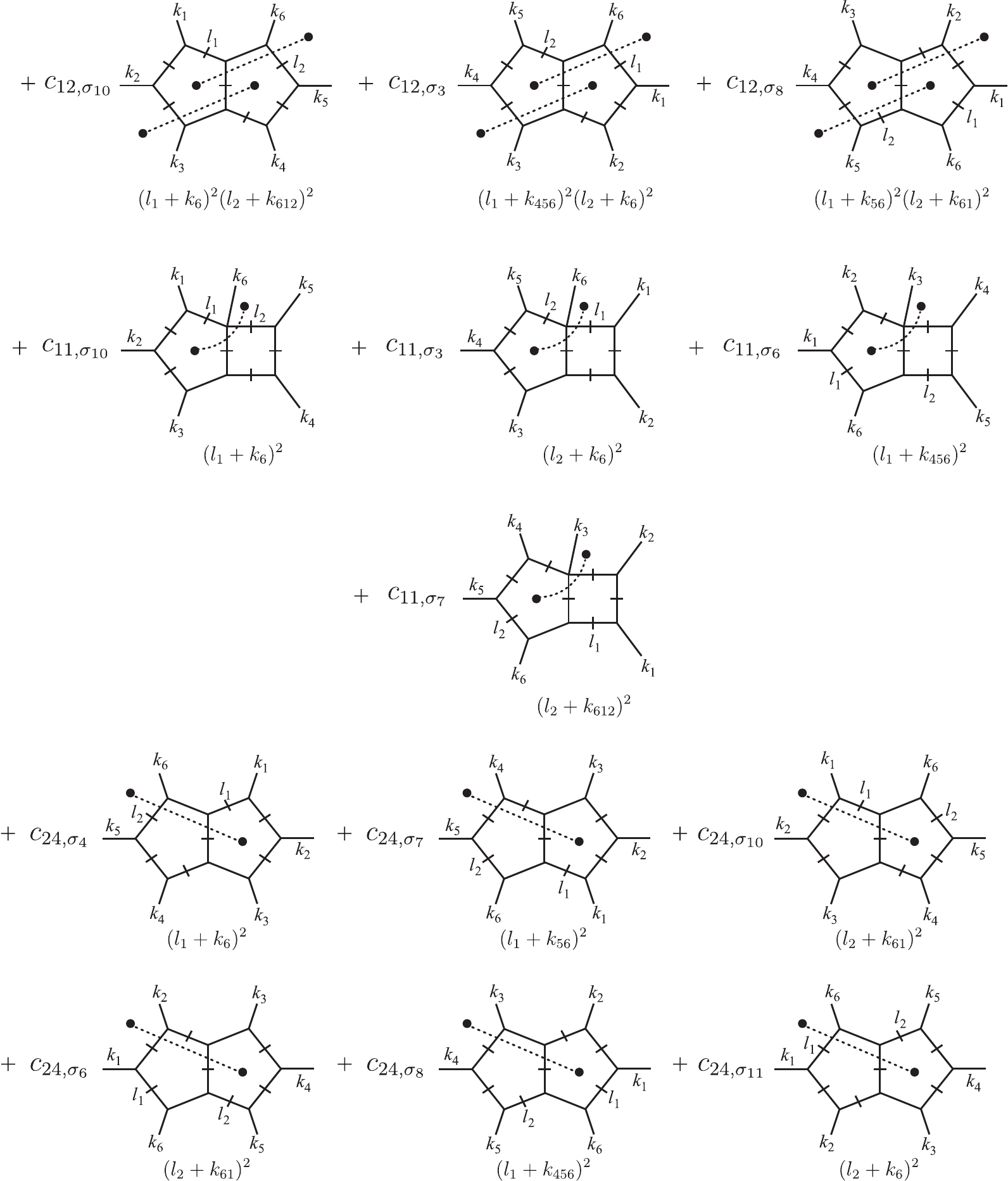}
\end{center}
\end{figure}
\vspace{-0.5cm}

We define the spinor ratios
\begin{equation}
\begin{array}{lll} P_1 = -\frac{K_5^\flat \cdot (K_1^\flat + k_6)
+ K_1^\flat \cdot k_6}{\langle K_2^{\flat -} | \slashed{K}_5^\flat
+ \slashed{k}_6 | K_1^{\flat -} \rangle} \: , \hspace{0.3cm} & P_2
= - \frac{K_1^\flat \cdot k_{123} - \frac{1}{2} s_{123}}{\langle
K_2^{\flat -} | \slashed{k}_{123} | K_1^{\flat -} \rangle} \: ,
\hspace{0.3cm} & P_3 = -\frac{\langle K_1^\flat k_6 \rangle}
{2\langle K_2^\flat k_6 \rangle} \\
P_4 = - \frac{\langle K_1^\flat k_5 \rangle}{2\langle K_2^\flat
k_5 \rangle} \: , \hspace{0.3cm} & P_5 = -\frac{\langle K_5^{\flat
-}| \slashed{K}_1^\flat + \slashed{k}_6 | K_4^{\flat -}
\rangle}{2[K_1^\flat K_4^\flat] \langle K_5^\flat K_2^\flat
\rangle} \: , \hspace{0.3cm} & P_6 = -\frac{\langle K_4^{\flat -}|
\slashed{K}_1^\flat + \slashed{k}_6 | K_5^{\flat -}
\rangle }{2\langle K_2^\flat K_4^\flat \rangle [K_5^\flat K_1^\flat]} \\
\vspace{-0.15cm} \\
Q_1 = -\frac{K_5^\flat \cdot (K_1^\flat + k_6) + K_1^\flat \cdot
k_6}{\langle K_5^{\flat -} | \slashed{K}_1^\flat + \slashed{k}_6 |
K_4^{\flat -}\rangle} \: , \hspace{0.3cm} & Q_2 = -
\frac{K_5^\flat \cdot k_{345} - \frac{1}{2} s_{345}}{\langle
K_5^{\flat -} | \slashed{k}_{345} | K_4^{\flat -} \rangle} \: ,
\hspace{0.3cm} & Q_3 = -\frac{[K_5^\flat k_6]}{2[K_4^\flat k_6]} \\
Q_4 = -\frac{[K_5^\flat k_1]}{2[K_4^\flat k_1]} \: ,
\hspace{0.3cm} & Q_5 = -\frac{\langle K_2^{\flat -}|
\slashed{K}_5^\flat + \slashed{k}_6 | K_1^{\flat -}
\rangle}{2[K_1^\flat K_4^\flat]
\langle K_5^\flat K_2^\flat \rangle} \hspace{0.3cm} & \\
\end{array}
\end{equation}
and their parity conjugates
\begin{equation}
\begin{array}{lll} P_1^\bullet = -\frac{K_5^\flat \cdot (K_1^\flat + k_6)
+ K_1^\flat \cdot k_6}{\langle K_1^{\flat -} | \slashed{K}_5^\flat
+ \slashed{k}_6 | K_2^{\flat -} \rangle} \: , \hspace{0.3cm} &
P_2^\bullet = - \frac{K_1^\flat \cdot k_{123} - \frac{1}{2}
s_{123}}{\langle K_1^{\flat -} | \slashed{k}_{123} | K_2^{\flat -}
\rangle} \: , \hspace{0.3cm} & P_3^\bullet = -\frac{[K_1^\flat
k_6]}{2[K_2^\flat k_6]} \\
P_4^\bullet = - \frac{[K_1^\flat k_5]}{2[K_2^\flat k_5]} \: ,
\hspace{0.3cm} & P_5^\bullet = -\frac{\langle K_4^{\flat -}|
\slashed{K}_1^\flat + \slashed{k}_6 | K_5^{\flat -}
\rangle}{2\langle K_1^\flat K_4^\flat \rangle [K_5^\flat
K_2^\flat]} \: , \hspace{0.3cm} & P_6^\bullet = -\frac{\langle
K_5^{\flat -}| \slashed{K}_1^\flat + \slashed{k}_6 | K_4^{\flat -}
\rangle }{2[K_2^\flat K_4^\flat]\langle K_5^\flat K_1^\flat \rangle} \\
\vspace{-0.15cm} \\
Q_1^\bullet = -\frac{K_5^\flat \cdot (K_1^\flat + k_6) + K_1^\flat
\cdot k_6}{\langle K_4^{\flat -} | \slashed{K}_1^\flat +
\slashed{k}_6 | K_5^{\flat -}\rangle} \: , \hspace{0.3cm} &
Q_2^\bullet = - \frac{K_5^\flat \cdot k_{345} - \frac{1}{2}
s_{345}}{\langle K_4^{\flat -} | \slashed{k}_{345} | K_5^{\flat -}
\rangle} \: , \hspace{0.3cm} & Q_3^\bullet = -\frac{\langle
K_5^\flat k_6 \rangle}
{2\langle K_4^\flat k_6 \rangle} \\
Q_4^\bullet = -\frac{\langle K_5^\flat k_1 \rangle}{2\langle
K_4^\flat k_1 \rangle} \: , \hspace{0.3cm} & Q_5^\bullet =
-\frac{\langle K_1^{\flat -}| \slashed{K}_5^\flat + \slashed{k}_6
| K_2^{\flat -} \rangle}{2\langle K_1^\flat K_4^\flat \rangle
[K_5^\flat K_2^\flat]} \hspace{0.3cm} & \\
\end{array}
\end{equation}
This heptacut belongs to case II treated in Section
\ref{sec:general_double_box}, and there are thus four kinematical
solutions (shown in Fig.
\ref{fig:kinematical_solutions_case_II}). Parametrizing the loop
momenta according to eqs.
(\ref{eq:l1_parametrized})-(\ref{eq:l2_parametrized}), the
on-shell constraints in eqs.
(\ref{eq:on-shell_constraint_1})-(\ref{eq:on-shell_constraint_7})
are solved by setting the parameters equal to the values
\begin{equation}
\begin{array}{ll}
\alpha_1 = 1 \: , \hspace{1cm} & \beta_1 = 0\\
\alpha_2 = 0 \: , \hspace{1cm} & \beta_2 = 1
\end{array}
\end{equation}
and those given in Fig. \ref{fig:kinematical_solutions_case_II}.
The heptacut double box integral $I_{7,\sigma_1}$ is $\sum_{i=1}^4
\oint_{\hspace{0.5mm}\Gamma_i} \hspace{-0.3mm} dz \, J_i(z)$ where
\begin{equation}
J_i(z) \hspace{2mm} = \hspace{2mm} \frac{1}{32 \gamma_1 \gamma_2}
\times \left\{ \begin{array}{ll} \left( \langle K_4^{\flat -} |
\slashed{K}_1^\flat + \slashed{k}_6 | K_5^{\flat -} \rangle
\hspace{0.6mm} z (z - Q_1^\bullet) \right)^{-1} & \hspace{4mm} \mathrm{for} \hspace{3mm} i=1 \\
\left( \langle K_5^{\flat -} | \slashed{K}_1^\flat + \slashed{k}_6
| K_4^{\flat -} \rangle \hspace{0.6mm} z (z - Q_1) \right)^{-1} &
\hspace{4mm} \mathrm{for} \hspace{3mm} i=2 \\
\left( \langle K_2^{\flat -} | \slashed{K}_5^\flat + \slashed{k}_6
| K_1^{\flat -} \rangle
\hspace{0.6mm} z (z - P_1) \right)^{-1} & \hspace{4mm} \mathrm{for} \hspace{3mm} i=3 \\
\left( \langle K_1^{\flat -} | \slashed{K}_5^\flat + \slashed{k}_6
| K_2^{\flat -} \rangle \hspace{0.6mm} z (z - P_1^\bullet)
\right)^{-1} & \hspace{4mm} \mathrm{for} \hspace{3mm} i=4 \: .
\end{array} \right.
\end{equation}

\subsubsection{Heptacut \#6 of the right hand side of eq. (\ref{eq:6gluonampansatz})}

The result of applying heptacut \#6 to the right hand side of eq.
(\ref{eq:6gluonampansatz}) is
\begin{equation}
\frac{1}{4} \sum_{i=1}^4 \oint_{\Gamma_i} dz \, J_i (z) K_i(z)
\end{equation}
where the kernels evaluated on the four kinematical solutions are

{\small \begin{eqnarray} K_1(z) \hspace{-2mm}&=&\hspace{-2mm}
c_{7,\sigma_1} - \frac{1}{4} \frac{c_{23,\sigma_1}}{\langle
K_1^{\flat -} | \slashed{k}_{123} | K_2^{\flat -} \rangle \langle
K_4^{\flat -} | \slashed{k}_6 | K_5^{\flat -} \rangle (\alpha_3
(z) - P_2^\bullet) (z - Q_3^\bullet)} - \frac{1}{2} \frac{c_{21,
\sigma_{10}}}{\langle K_1^{\flat -} | \slashed{k}_{123} |
K_2^{\flat -} \rangle (\alpha_3 (z) - P_2^\bullet)}  \nonumber \\[2mm]
&\phantom{=}&  \hspace{-14mm} - \frac{1}{4}
\frac{c_{23,\sigma_3}}{\langle K_1^{\flat -} | \slashed{k}_6 |
K_2^{\flat -} \rangle \langle K_4^{\flat -} | \slashed{k}_{345} |
K_5^{\flat -} \rangle (\alpha_3 (z) - P_3^\bullet) (z -
Q_2^\bullet)} - \frac{1}{2} \frac{c_{21, \sigma_3}}{\langle
K_4^{\flat -} | \slashed{k}_{345} | K_5^{\flat -} \rangle (z -
Q_2^\bullet)} \nonumber \\[2mm]
&\phantom{=}&  \hspace{-14mm} + \frac{1}{2} \frac{c_{21,
\sigma_6}}{\langle K_1^{\flat -} | \slashed{k}_6 | K_2^{\flat -}
\rangle (\alpha_3 (z) - P_3^\bullet)} + \frac{1}{2} \frac{c_{21,
\sigma_7}}{\langle K_4^{\flat -} | \slashed{k}_6 | K_5^{\flat -}
\rangle (z - Q_3^\bullet)} - \frac{c_{10, \sigma_{10}} \langle
K_1^{\flat -} | \slashed{k}_{56} | K_2^{\flat -} \rangle (\alpha_3
(z) - P_1^\bullet)}{\langle K_1^{\flat -} | \slashed{k}_{123} |
K_2^{\flat -} \rangle (\alpha_3 (z) - P_2^\bullet)} \nonumber \\[2mm]
&\phantom{=}& \hspace{-14mm}  - \frac{1}{2} \frac{c_{24, \sigma_1}
\langle K_4^{\flat -} | \slashed{k}_{612} | K_5^{\flat -} \rangle
(z - Q_2^\bullet) + c_{24,\sigma_4}\langle K_1^{\flat -} |
\slashed{k}_6 | K_2^{\flat -} \rangle (\alpha_3(z) -
P_3^\bullet)}{\langle K_1^{\flat -} | \slashed{k}_{123} |
K_2^{\flat -} \rangle \langle K_4^{\flat -} | \slashed{k}_6 |
K_5^{\flat -} \rangle (\alpha_3 (z) - P_2^\bullet) (z - Q_3^\bullet)}  \nonumber \\[2mm]
&\phantom{=}& \hspace{-14mm} - \frac{1}{2} \frac{c_{24,\sigma_7}
\langle K_1^{\flat -} | \slashed{k}_{56} | K_2^{\flat -} \rangle
(\alpha_3(z) - P_1^\bullet) + c_{24,\sigma_{10}} \langle
K_4^{\flat -} | \slashed{k}_{61} | K_5^{\flat -} \rangle (z -
Q_1^\bullet)}{\langle K_1^{\flat -} | \slashed{k}_{123} |
K_2^{\flat -} \rangle \langle K_4^{\flat -} | \slashed{k}_6 |
K_5^{\flat -} \rangle (\alpha_3 (z) - P_2^\bullet) (z - Q_3^\bullet)} \nonumber \\[2mm]
&\phantom{=}&  \hspace{-14mm} - \frac{c_{10, \sigma_3} \langle
K_4^{\flat -} | \slashed{k}_{61} | K_5^{\flat -} \rangle (z -
Q_1^\bullet)}{\langle K_4^{\flat -} | \slashed{k}_{345} |
K_5^{\flat -} \rangle (z - Q_2^\bullet)} - \frac{1}{2}
\frac{c_{24, \sigma_3} \langle K_1^{\flat -} | \slashed{k}_{56} |
K_2^{\flat -} \rangle (\alpha_3 (z) - P_1^\bullet) +
c_{24,\sigma_6} \langle K_4^{\flat -} | \slashed{k}_{61} |
K_5^{\flat -} \rangle (z - Q_1^\bullet)}{\langle K_1^{\flat -} |
\slashed{k}_6 | K_2^{\flat -} \rangle \langle K_4^{\flat -} |
\slashed{k}_{345} | K_5^{\flat -} \rangle
(\alpha_3 (z) - P_3^\bullet) (z - Q_2^\bullet)} \nonumber \\[2mm]
&\phantom{=}&  \hspace{-14mm} - \frac{1}{2} \frac{c_{24,\sigma_8}
\langle K_1^{\flat -} | \slashed{k}_{456} | K_2^{\flat -} \rangle
(\alpha_3(z) - P_2^\bullet) + c_{24,\sigma_{11}} \langle
K_4^{\flat -} | \slashed{k}_6 | K_5^{\flat -} \rangle (z -
Q_3^\bullet)}{\langle K_1^{\flat -} | \slashed{k}_6 | K_2^{\flat
-} \rangle \langle K_4^{\flat -} | \slashed{k}_{345} | K_5^{\flat
-} \rangle (\alpha_3 (z) - P_3^\bullet) (z - Q_2^\bullet)} \nonumber
\end{eqnarray}}
{\small \begin{eqnarray}
&\phantom{=}&  \hspace{-14mm} + \frac{c_{10, \sigma_6} \langle
K_1^{\flat -} | \slashed{k}_{56} | K_2^{\flat -} \rangle (\alpha_3
(z) - P_1^\bullet)}{\langle K_1^{\flat -} | \slashed{k}_6 |
K_2^{\flat -} \rangle (\alpha_3 (z) - P_3^\bullet)} - \frac{c_{13,
\sigma_1} \langle K_1^{\flat -} | \slashed{k}_{56} | K_2^{\flat -}
\rangle \langle K_4^{\flat -} | \slashed{k}_{612} | K_5^{\flat -}
\rangle (\alpha_3 (z) - P_1^\bullet)(z - Q_2^\bullet)}{\langle
K_1^{\flat -} | \slashed{k}_{123} | K_2^{\flat -} \rangle \langle
K_4^{\flat -} | \slashed{k}_6 | K_5^{\flat -} \rangle (\alpha_3
(z) - P_2^\bullet) (z - Q_3^\bullet)} \nonumber \\[2mm]
&\phantom{=}&  \hspace{-14mm} + \frac{c_{10, \sigma_7} \langle
K_4^{\flat -} | \slashed{k}_{61} | K_5^{\flat -} \rangle (z -
Q_1^\bullet)} {\langle K_4^{\flat -} | \slashed{k}_6 | K_5^{\flat
-} \rangle (z - Q_3^\bullet)} - \frac{c_{13, \sigma_3} \langle
K_1^{\flat -} | \slashed{k}_{56} | K_2^{\flat -} \rangle \langle
K_4^{\flat -} | \slashed{k}_6 | K_5^{\flat -} \rangle (\alpha_3
(z) - P_1^\bullet)(z - Q_3^\bullet)}{\langle K_1^{\flat -} |
\slashed{k}_6 | K_2^{\flat -} \rangle \langle K_4^{\flat -} |
\slashed{k}_{345} | K_5^{\flat -} \rangle (\alpha_3 (z) -
P_3^\bullet) (z - Q_2^\bullet)} \nonumber \\[2mm]
&\phantom{=}& \hspace{-14mm} - \frac{c_{13, \sigma_{10}} \langle
K_1^{\flat -} | \slashed{k}_6 | K_2^{\flat -} \rangle \langle
K_4^{\flat -} | \slashed{k}_{61} | K_5^{\flat -} \rangle (\alpha_3
(z) - P_3^\bullet)(z - Q_1^\bullet)}{\langle K_1^{\flat -} |
\slashed{k}_{123} | K_2^{\flat -} \rangle \langle K_4^{\flat -} |
\slashed{k}_6 | K_5^{\flat -} \rangle (\alpha_3 (z) -
P_2^\bullet) (z - Q_3^\bullet)} \nonumber \\[2mm]
&\phantom{=}&  \hspace{-14mm} - \frac{c_{13, \sigma_6} \langle
K_1^{\flat -} | \slashed{k}_{456} | K_2^{\flat -} \rangle \langle
K_4^{\flat -} | \slashed{k}_{61} | K_5^{\flat -} \rangle (\alpha_3
(z) - P_2^\bullet)(z - Q_1^\bullet)}{\langle K_1^{\flat -} |
\slashed{k}_6 | K_2^{\flat -} \rangle \langle K_4^{\flat -} |
\slashed{k}_{345} | K_5^{\flat -} \rangle (\alpha_3 (z) -
P_3^\bullet) (z - Q_2^\bullet)} \nonumber \\[2mm]
&\phantom{=}&  \hspace{-14mm} - \frac{c_{11, \sigma_{10}} \langle
K_1^{\flat -} | \slashed{k}_6 | K_2^{\flat -} \rangle (\alpha_3
(z) - P_3^\bullet)}{\langle K_1^{\flat -} | \slashed{k}_{123} |
K_2^{\flat -} \rangle (\alpha_3 (z) - P_2^\bullet)} - \frac{c_{12,
\sigma_1} \langle K_1^{\flat -} | \slashed{k}_{56} | K_2^{\flat -}
\rangle \langle K_4^{\flat -} | \slashed{k}_{61} | K_5^{\flat -}
\rangle (\alpha_3 (z) - P_1^\bullet)(z - Q_1^\bullet)}{\langle
K_1^{\flat -} | \slashed{k}_{123} | K_2^{\flat -} \rangle \langle
K_4^{\flat -} | \slashed{k}_6 | K_5^{\flat -} \rangle (\alpha_3
(z) - P_2^\bullet) (z - Q_3^\bullet)} \nonumber \\[2mm]
&\phantom{=}&  \hspace{-14mm} - \frac{c_{11, \sigma_3} \langle
K_4^{\flat -} | \slashed{k}_6 | K_5^{\flat -} \rangle (z -
Q_3^\bullet)}{\langle K_4^{\flat -} | \slashed{k}_{345} |
K_5^{\flat -} \rangle (z - Q_2^\bullet)} - \frac{c_{12,
\sigma_{10}} \langle K_1^{\flat -} | \slashed{k}_6 | K_2^{\flat -}
\rangle \langle K_4^{\flat -} | \slashed{k}_{612} | K_5^{\flat -}
\rangle (\alpha_3 (z) - P_3^\bullet)(z - Q_2^\bullet)}{\langle
K_1^{\flat -} | \slashed{k}_{123} | K_2^{\flat -} \rangle \langle
K_4^{\flat -} | \slashed{k}_6 | K_5^{\flat -} \rangle (\alpha_3
(z) - P_2^\bullet) (z - Q_3^\bullet)} \nonumber \\[2mm]
&\phantom{=}&  \hspace{-14mm} + \frac{c_{11, \sigma_6} \langle
K_1^{\flat -} | \slashed{k}_{456} | K_2^{\flat -} \rangle
(\alpha_3 (z) - P_2^\bullet)}{\langle K_1^{\flat -} |
\slashed{k}_6 | K_2^{\flat -} \rangle (\alpha_3 (z) -
P_3^\bullet)} - \frac{c_{12, \sigma_3} \langle K_1^{\flat -} |
\slashed{k}_{456} | K_2^{\flat -} \rangle \langle K_4^{\flat -} |
\slashed{k}_6 | K_5^{\flat -} \rangle (\alpha_3 (z) -
P_2^\bullet)(z - Q_3^\bullet)}{\langle K_1^{\flat -} |
\slashed{k}_6 | K_2^{\flat -} \rangle \langle K_4^{\flat -} |
\slashed{k}_{345} | K_5^{\flat -} \rangle (\alpha_3 (z) -
P_3^\bullet) (z - Q_2^\bullet)} \nonumber \\[2mm]
&\phantom{=}&  \hspace{-14mm} + \frac{c_{11, \sigma_7} \langle
K_4^{\flat -} | \slashed{k}_{612} | K_5^{\flat -} \rangle (z -
Q_2^\bullet)}{\langle K_4^{\flat -} | \slashed{k}_6 | K_5^{\flat
-} \rangle (z - Q_3^\bullet)} - \frac{c_{12, \sigma_8} \langle
K_1^{\flat -} | \slashed{k}_{56} | K_2^{\flat -} \rangle \langle
K_4^{\flat -} | \slashed{k}_{61} | K_5^{\flat -} \rangle (\alpha_3
(z) - P_1^\bullet)(z - Q_1^\bullet)}{\langle K_1^{\flat -} |
\slashed{k}_6 | K_2^{\flat -} \rangle \langle K_4^{\flat -} |
\slashed{k}_{345} | K_5^{\flat -} \rangle (\alpha_3 (z) -
P_3^\bullet) (z - Q_2^\bullet)} \\[6mm]
K_2(z) \hspace{-2mm}&=&\hspace{-2mm} \mbox{parity conjugate of
$K_1(z)$ \hspace{0.3mm} (obtained by applying eqs.
(\ref{eq:parity_conj_rule_1})-(\ref{eq:parity_conj_rule_6}))}
\end{eqnarray}}

\noindent where $\alpha_3(z)$ can be read off from the on-shell
values quoted below solution $\mathcal{S}_1$ in Fig.
\ref{fig:kinematical_solutions_case_II}.

{\small \begin{eqnarray} \vspace{-10mm} K_3(z)
\hspace{-2mm}&=&\hspace{-2mm} c_{7,\sigma_1} - \frac{1}{4}
\frac{c_{23,\sigma_1}}{\langle K_2^{\flat -} | \slashed{k}_{123} |
K_1^{\flat -} \rangle \langle K_4^{\flat -} | \slashed{k}_6 |
K_5^{\flat -} \rangle (z - P_2) (\beta_3 (z) - Q_3^\bullet)} -
\frac{1}{2} \frac{c_{21, \sigma_{10}}}{\langle K_2^{\flat -} |
\slashed{k}_{123} | K_1^{\flat -} \rangle (z - P_2)} \nonumber \hspace{7mm} \\
&\phantom{=}& \hspace{-14mm} -
\frac{1}{4} \frac{c_{23,\sigma_3}}{\langle K_2^{\flat -} |
\slashed{k}_6 | K_1^{\flat -} \rangle \langle K_4^{\flat -} |
\slashed{k}_{345} | K_5^{\flat -} \rangle (z - P_3) (\beta_3 (z) -
Q_2^\bullet)} - \frac{1}{2} \frac{c_{21, \sigma_3}}{\langle
K_4^{\flat -} | \slashed{k}_{345} | K_5^{\flat -} \rangle (\beta_3
(z) - Q_2^\bullet)} \nonumber \\[2mm]
&\phantom{=}&  \hspace{-14mm} + \frac{1}{2} \frac{c_{21,
\sigma_6}}{\langle K_2^{\flat -} | \slashed{k}_6 | K_1^{\flat -}
\rangle (z - P_3)} - \frac{1}{2} \frac{c_{24, \sigma_1} \langle
K_4^{\flat -} | \slashed{k}_{612} | K_5^{\flat -} \rangle (\beta_3
(z) - Q_2^\bullet) + c_{24,\sigma_4} \langle K_2^{\flat -} |
\slashed{k}_6 | K_1^{\flat -} \rangle (z - P_3)}{\langle
K_2^{\flat -} | \slashed{k}_{123} | K_1^{\flat -} \rangle \langle
K_4^{\flat -} | \slashed{k}_6 |
K_5^{\flat -} \rangle (z - P_2) (\beta_3 (z) - Q_3^\bullet)}  \nonumber \\[2mm]
&\phantom{=}&  \hspace{-14mm} + \frac{1}{2} \frac{c_{21,
\sigma_7}}{\langle K_4^{\flat -} | \slashed{k}_6 | K_5^{\flat -}
\rangle (\beta_3 (z) - Q_3^\bullet)} - \frac{1}{2}
\frac{c_{24,\sigma_7} \langle K_2^{\flat -} | \slashed{k}_{56} |
K_1^{\flat -} \rangle (z - P_1) + c_{24,\sigma_{10}} \langle
K_4^{\flat -} | \slashed{k}_{61} | K_5^{\flat -} \rangle
(\beta_3(z) - Q_1^\bullet)}{\langle K_2^{\flat -} |
\slashed{k}_{123} | K_1^{\flat -} \rangle \langle K_4^{\flat -} |
\slashed{k}_6 |
K_5^{\flat -} \rangle (z - P_2) (\beta_3 (z) - Q_3^\bullet)} \nonumber \\[2mm]
&\phantom{=}&  \hspace{-14mm}  - \frac{1}{2} \frac{c_{24,
\sigma_3} \langle K_2^{\flat -} | \slashed{k}_{56} | K_1^{\flat -}
\rangle (z - P_1) + c_{24,\sigma_6} \langle K_4^{\flat -} |
\slashed{k}_{61} | K_5^{\flat -} \rangle (\beta_3(z) -
Q_1^\bullet)}{\langle K_2^{\flat -} | \slashed{k}_6 | K_1^{\flat
-} \rangle \langle K_4^{\flat -} |
\slashed{k}_{345} | K_5^{\flat -} \rangle (z - P_3) (\beta_3 (z) - Q_2^\bullet)} \nonumber \\[2mm]
&\phantom{=}&  \hspace{-14mm}  - \frac{1}{2} \frac{c_{24,\sigma_8}
\langle K_2^{\flat -} | \slashed{k}_{456} | K_1^{\flat-} \rangle
(z - P_2) + c_{24,\sigma_{11}} \langle K_4^{\flat -} |
\slashed{k}_6 | K_5^{\flat -} \rangle (\beta_3 (z) -
Q_3^\bullet)}{\langle K_2^{\flat -} | \slashed{k}_6 | K_1^{\flat
-} \rangle \langle K_4^{\flat -} | \slashed{k}_{345} |
K_5^{\flat -} \rangle (z - P_3) (\beta_3 (z) - Q_2^\bullet)} \nonumber \\[2mm]
&\phantom{=}&  \hspace{-14mm} - \frac{c_{10, \sigma_{10}} \langle
K_2^{\flat -} | \slashed{k}_{56} | K_1^{\flat -} \rangle (z -
P_1)}{\langle K_2^{\flat -} | \slashed{k}_{123} | K_1^{\flat -}
\rangle (z - P_2)} - \frac{c_{13, \sigma_1} \langle K_2^{\flat -}
| \slashed{k}_{56} | K_1^{\flat -} \rangle \langle K_4^{\flat -} |
\slashed{k}_{612} | K_5^{\flat -} \rangle (z - P_1)(\beta_3 (z) -
Q_2^\bullet)}{\langle K_2^{\flat -} | \slashed{k}_{123} |
K_1^{\flat -} \rangle \langle K_4^{\flat -} | \slashed{k}_6 |
K_5^{\flat -} \rangle (z - P_2) (\beta_3 (z) - Q_3^\bullet)} \nonumber \\[2mm]
&\phantom{=}&  \hspace{-14mm} - \frac{c_{10, \sigma_3} \langle
K_4^{\flat -} | \slashed{k}_{61} | K_5^{\flat -} \rangle (\beta_3
(z) - Q_1^\bullet)}{\langle K_4^{\flat -} | \slashed{k}_{345} |
K_5^{\flat -} \rangle (\beta_3 (z) - Q_2^\bullet)} - \frac{c_{13,
\sigma_3} \langle K_2^{\flat -} | \slashed{k}_{56} | K_1^{\flat -}
\rangle \langle K_4^{\flat -} | \slashed{k}_6 | K_5^{\flat -}
\rangle (z - P_1)(\beta_3 (z) - Q_3^\bullet)}{\langle K_2^{\flat
-} | \slashed{k}_6 | K_1^{\flat -} \rangle \langle K_4^{\flat -} |
\slashed{k}_{345} | K_5^{\flat -} \rangle (z - P_3) (\beta_3 (z) - Q_2^\bullet)}  \nonumber
\end{eqnarray}}

{\small \begin{eqnarray}
&\phantom{=}&  \hspace{-14mm} + \frac{c_{10, \sigma_6} \langle
K_2^{\flat -} | \slashed{k}_{56} | K_1^{\flat -} \rangle (z -
P_1)}{\langle K_2^{\flat -} | \slashed{k}_6 | K_1^{\flat -}
\rangle (z - P_3)} - \frac{c_{13, \sigma_{10}} \langle K_2^{\flat
-} | \slashed{k}_6 | K_1^{\flat -} \rangle \langle K_4^{\flat -} |
\slashed{k}_{61} | K_5^{\flat -} \rangle (z - P_3)(\beta_3 (z) -
Q_1^\bullet)}{\langle K_2^{\flat -} | \slashed{k}_{123} |
K_1^{\flat -} \rangle \langle K_4^{\flat -} | \slashed{k}_6 |
K_5^{\flat -} \rangle (z -
P_2) (\beta_3 (z) - Q_3^\bullet)} \nonumber \\[2mm]
&\phantom{=}&  \hspace{-14mm} + \frac{c_{10, \sigma_7} \langle
K_4^{\flat -} | \slashed{k}_{61} | K_5^{\flat -} \rangle (\beta_3
(z) - Q_1^\bullet)}{\langle K_4^{\flat -} | \slashed{k}_6 |
K_5^{\flat -} \rangle (\beta_3 (z) - Q_3^\bullet)} - \frac{c_{13,
\sigma_6} \langle K_2^{\flat -} | \slashed{k}_{456} | K_1^{\flat
-} \rangle \langle K_4^{\flat -} | \slashed{k}_{61} | K_5^{\flat
-} \rangle (z - P_2)(\beta_3 (z) - Q_1^\bullet)}{\langle
K_2^{\flat -} | \slashed{k}_6 | K_1^{\flat -} \rangle \langle
K_4^{\flat -} | \slashed{k}_{345} | K_5^{\flat -} \rangle (z -
P_3) (\beta_3 (z) - Q_2^\bullet)} \nonumber \\[2mm]
&\phantom{=}&  \hspace{-14mm} - \frac{c_{11, \sigma_{10}} \langle
K_2^{\flat -} | \slashed{k}_6 | K_1^{\flat -} \rangle (z -
P_3)}{\langle K_2^{\flat -} | \slashed{k}_{123} | K_1^{\flat -}
\rangle (z - P_2)} - \frac{c_{12, \sigma_1} \langle K_2^{\flat -}
| \slashed{k}_{56} | K_1^{\flat -} \rangle \langle K_4^{\flat -} |
\slashed{k}_{61} | K_5^{\flat -} \rangle (z - P_1)(\beta_3 (z) -
Q_1^\bullet)}{\langle K_2^{\flat -} | \slashed{k}_{123} |
K_1^{\flat -} \rangle \langle K_4^{\flat -} |
\slashed{k}_6 | K_5^{\flat -} \rangle (z - P_2) (\beta_3 (z) - Q_3^\bullet)} \nonumber \\[2mm]
&\phantom{=}&  \hspace{-14mm} - \frac{c_{11, \sigma_3} \langle
K_4^{\flat -} | \slashed{k}_6 | K_5^{\flat -} \rangle (\beta_3 (z)
- Q_3^\bullet)}{\langle K_4^{\flat -} | \slashed{k}_{345} |
K_5^{\flat -} \rangle (\beta_3 (z) - Q_2^\bullet)} - \frac{c_{12,
\sigma_{10}} \langle K_2^{\flat -} | \slashed{k}_6 | K_1^{\flat -}
\rangle \langle K_4^{\flat -} | \slashed{k}_{612} | K_5^{\flat -}
\rangle (z - P_3)(\beta_3 (z) - Q_2^\bullet)}{\langle K_2^{\flat
-} | \slashed{k}_{123} | K_1^{\flat -} \rangle \langle K_4^{\flat
-} | \slashed{k}_6 | K_5^{\flat -} \rangle (z - P_2) (\beta_3 (z) - Q_3^\bullet)} \nonumber \\[2mm]
&\phantom{=}&  \hspace{-14mm} + \frac{c_{11, \sigma_6} \langle
K_2^{\flat -} | \slashed{k}_{456} | K_1^{\flat -} \rangle (z -
P_2)}{\langle K_2^{\flat -} | \slashed{k}_6 | K_1^{\flat -}
\rangle (z - P_3)} - \frac{c_{12, \sigma_3} \langle K_2^{\flat -}
| \slashed{k}_{456} | K_1^{\flat -} \rangle \langle K_4^{\flat -}
| \slashed{k}_6 | K_5^{\flat -} \rangle (z - P_2)(\beta_3 (z) -
Q_3^\bullet)}{\langle K_2^{\flat -} | \slashed{k}_6 | K_1^{\flat
-} \rangle \langle K_4^{\flat -} |
\slashed{k}_{345} | K_5^{\flat -} \rangle (z - P_3) (\beta_3 (z) - Q_2^\bullet)} \nonumber \\[2mm]
&\phantom{=}&  \hspace{-14mm} + \frac{c_{11, \sigma_7} \langle
K_4^{\flat -} | \slashed{k}_{612} | K_5^{\flat -} \rangle (\beta_3
(z) - Q_2^\bullet)}{\langle K_4^{\flat -} | \slashed{k}_6 |
K_5^{\flat -} \rangle (\beta_3 (z) - Q_3^\bullet)} - \frac{c_{12,
\sigma_8} \langle K_2^{\flat -} | \slashed{k}_{56} | K_1^{\flat -}
\rangle \langle K_4^{\flat -} | \slashed{k}_{61} | K_5^{\flat -}
\rangle (z - P_1)(\beta_3 (z) - Q_1^\bullet)}{\langle K_2^{\flat
-} | \slashed{k}_6 | K_1^{\flat -} \rangle \langle K_4^{\flat -} |
\slashed{k}_{345} | K_5^{\flat -} \rangle (z - P_3) (\beta_3 (z) -
Q_2^\bullet)} \phantom{aaaaaa} \\[6mm]
K_4(z) \hspace{-2mm}&=&\hspace{-2mm} \mbox{parity conjugate of
$K_3(z)$ \hspace{0.3mm} (obtained by applying eqs.
(\ref{eq:parity_conj_rule_1})-(\ref{eq:parity_conj_rule_6}))}
\end{eqnarray}}
\vspace{-5mm}

\noindent where $\beta_3(z)$ can be read off from the on-shell
values quoted below solution $\mathcal{S}_3$ in Fig.
\ref{fig:kinematical_solutions_case_II}.

\subsubsection{Heptacut \#6 of the left hand side of eq. (\ref{eq:6gluonampansatz})}

The result of applying heptacut \#6 to the left hand side of eq.
(\ref{eq:6gluonampansatz}) is
\begin{equation}
i \sum_{i=1}^4 \oint_{\Gamma_i} dz \, J_i (z) \left.
\prod_{j=1}^6 A_j^\mathrm{tree}(z) \right|_{\mathcal{S}_i}
\end{equation}
where, assuming without loss of generality the external helicities are $(1^-, 2^-, 3^+, 4^+, 5^+, 6^+)$,
the cut amplitude evaluated on the four different
kinematical solutions yields
\begin{equation}
\left. \prod_{j=1}^6 A_j^\mathrm{tree}(z) \right|_{\mathcal{S}_i}
\hspace{3mm}=\hspace{3mm} 0 \hspace{9mm} \mathrm{for} \hspace{4mm}
i = 1,\ldots, 4 \: .
\end{equation}

\clearpage

\subsection{Heptacut \#7}\label{sec:heptacut_7}

Note that in this section we will leave the $\pm$ in
$\gamma_1^\pm, P_i^\pm, Q_i^\pm$ etc. implicit and simply write
$\gamma_1, P_i, Q_i$ etc. for notational simplicity. The heptacut
considered here is defined by the on-shell constraints in eqs.
(\ref{eq:on-shell_constraint_1})-(\ref{eq:on-shell_constraint_7})
with the vertex momenta
\begin{equation}
\begin{array}{lll}
K_1 = k_{12} & \hspace{1.2cm} K_2 = k_{34} & \hspace{1.2cm} K_3 = 0 \\
K_4 = k_5 & \hspace{1.2cm} K_5 = k_6 & \hspace{1.2cm} K_6 = 0 \: .
\end{array}
\end{equation}
Applying this heptacut to the right hand side of eq.
(\ref{eq:6gluonampansatz}) leaves the following linear combination
of cut integrals

\begin{figure}[!h]
\begin{center}
\includegraphics[angle=0, width=0.9\textwidth]{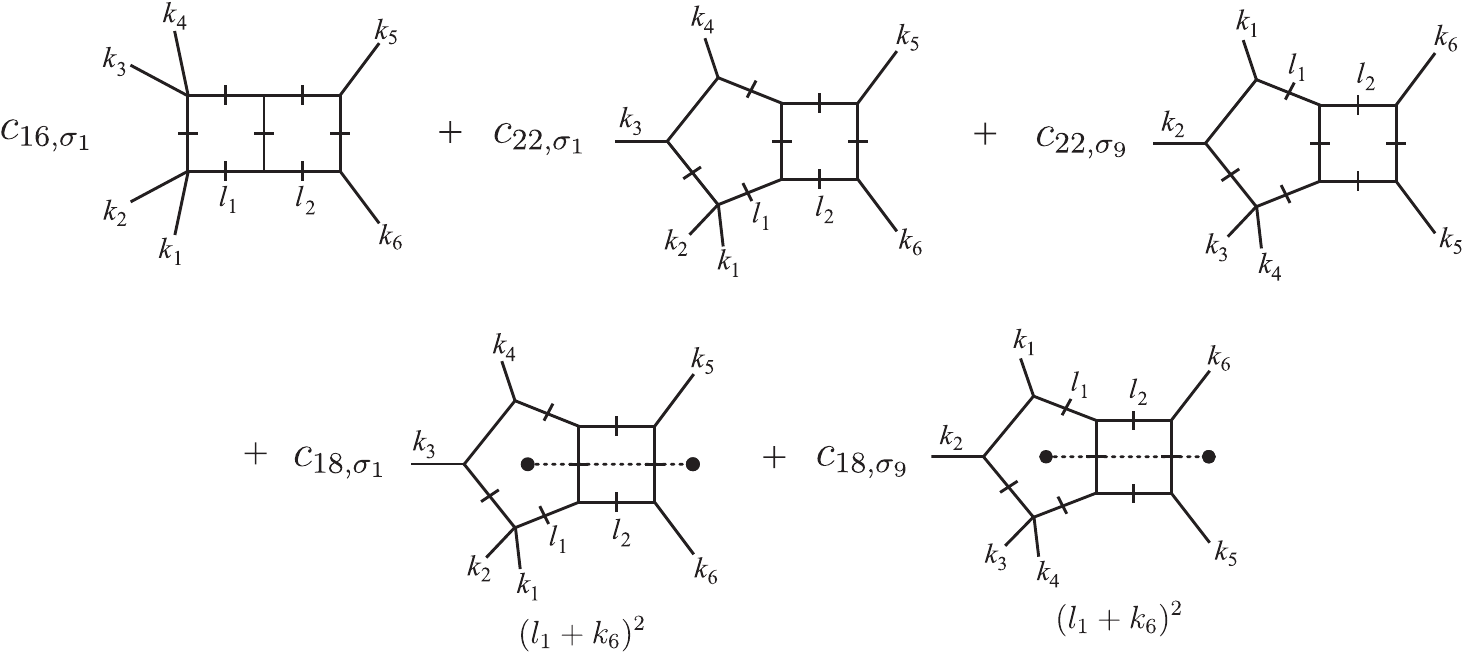}
\end{center}
\end{figure}
\vspace{-0.5cm}

\noindent We define the spinor ratios
\begin{equation}
\begin{array}{ll}
P_1 = - \frac{\langle K_1^\flat K_5^\flat \rangle}{2\langle
K_2^\flat K_5^\flat \rangle} \frac{\gamma_1 (S_2 +
\gamma_1)}{\gamma_1^2 - S_1 S_2} \: , & \hspace{2.5mm} P_2 =
\frac{[K_2^\flat K_5^\flat]}{2[K_1^\flat K_5^\flat]} \frac{S_1 S_2
(1 + S_1/\gamma_1)}{\gamma_1^2 - S_1 S_2} \: ,  \hspace{8mm}
P_3 = \frac{[K_2^\flat K_4^\flat]}{2[K_1^\flat K_4^\flat]}
\frac{S_1 S_2 (1 + S_1/\gamma_1)}{\gamma_1^2 - S_1 S_2} \\[3mm]
P_4 = - \frac{\langle K_1^\flat k_1 \rangle}{2\langle K_2^\flat
k_1 \rangle} \frac{\gamma_1 (S_2 + \gamma_1)}{\gamma_1^2 - S_1
S_2} \: , & \hspace{2.5mm} P_6 = -\frac{\frac{\gamma_1 (S_2 + \gamma_1)}{\gamma_1^2 - S_1
S_2} K_1^\flat \cdot k_3 - \frac{S_1 S_2 (S_1 +
\gamma_1)}{\gamma_1 (\gamma_1^2 - S_1 S_2)} K_2^\flat \cdot k_3 -
\frac{1}{2} (s_{13} + s_{23}) - \sqrt{\Delta}}{2\langle K_2^{\flat
-} | \slashed{k}_3| K_1^{\flat -} \rangle} \: , \\[3mm]
P_5 = \frac{[K_2^\flat k_1]}{2[K_1^\flat k_1]} \frac{S_1 S_2 (1 +
S_1/\gamma_1)}{\gamma_1^2 - S_1 S_2} \: , & \hspace{2.5mm} P_7 = -\frac{\frac{\gamma_1 (S_2 + \gamma_1)}{\gamma_1^2 - S_1
S_2} K_1^\flat \cdot k_3 - \frac{S_1 S_2 (S_1 +
\gamma_1)}{\gamma_1 (\gamma_1^2 - S_1 S_2)} K_2^\flat \cdot k_3 -
\frac{1}{2} (s_{13} + s_{23}) + \sqrt{\Delta}}{2\langle K_2^{\flat
-} | \slashed{k}_3| K_1^{\flat -} \rangle} \: , \\[3mm]
& \hspace{-47.6mm} Q_1 = -\frac{\gamma_1 (S_2 + \gamma_1) \langle K_5^\flat K_1^\flat
\rangle [K_1^\flat K_5^\flat] \hspace{0.6mm}+\hspace{0.6mm} S_1
S_2 (1 + S_1/\gamma_1) \langle K_5^\flat K_2^\flat \rangle
[K_2^\flat K_5^\flat]}{2\left( \gamma_1 (S_2 + \gamma_1) \langle
K_5^\flat K_1^\flat \rangle [K_1^\flat K_4^\flat]
\hspace{0.6mm}+\hspace{0.6mm} S_1 S_2 (1 + S_1/\gamma_1) \langle
K_5^\flat K_2^\flat \rangle [K_2^\flat K_4^\flat]\right)} \: ,
\hspace{4mm} Q_2 = -\frac{[K_1^\flat K_5^\flat]}{2[K_1^\flat
K_4^\flat]}
\end{array}
\end{equation}
and their parity conjugates
\begin{equation}
\begin{array}{ll}
P_1^\bullet = - \frac{[K_1^\flat K_5^\flat]}{2[K_2^\flat
K_5^\flat]} \frac{\gamma_1 (S_2 + \gamma_1)}{\gamma_1^2 - S_1 S_2}
\: , & \hspace{2.5mm} P_2^\bullet = \frac{\langle K_2^\flat
K_5^\flat \rangle}{2\langle K_1^\flat K_5^\flat \rangle} \frac{S_1
S_2 (1 + S_1/\gamma_1)}{\gamma_1^2 - S_1 S_2} \: ,
\hspace{8mm} P_3^\bullet = \frac{\langle K_2^\flat K_4^\flat
\rangle}{2\langle K_1^\flat K_4^\flat \rangle}
\frac{S_1 S_2 (1 + S_1/\gamma_1)}{\gamma_1^2 - S_1 S_2} \\[3mm]
P_4^\bullet = - \frac{[K_1^\flat k_1]}{2[K_2^\flat k_1]}
\frac{\gamma_1 (S_2 + \gamma_1)}{\gamma_1^2 - S_1 S_2} \: , &
\hspace{2.5mm} P_6^\bullet = -\frac{\frac{\gamma_1 (S_2 + \gamma_1)}{\gamma_1^2 -
S_1 S_2} K_1^\flat \cdot k_3 - \frac{S_1 S_2 (S_1 +
\gamma_1)}{\gamma_1 (\gamma_1^2 - S_1 S_2)} K_2^\flat \cdot k_3 -
\frac{1}{2} (s_{13} + s_{23}) - \sqrt{\Delta}}{2\langle K_1^{\flat
-} | \slashed{k}_3| K_2^{\flat -} \rangle} \: , \\[3mm]
P_5^\bullet = \frac{\langle K_2^\flat k_1
\rangle}{2\langle K_1^\flat k_1 \rangle} \frac{S_1 S_2 (1 +
S_1/\gamma_1)}{\gamma_1^2 - S_1 S_2} \: , & \hspace{2.5mm}
P_7^\bullet = -\frac{\frac{\gamma_1 (S_2 + \gamma_1)}{\gamma_1^2 -
S_1 S_2} K_1^\flat \cdot k_3 - \frac{S_1 S_2 (S_1 +
\gamma_1)}{\gamma_1 (\gamma_1^2 - S_1 S_2)} K_2^\flat \cdot k_3 -
\frac{1}{2} (s_{13} + s_{23}) + \sqrt{\Delta}}{2\langle K_1^{\flat
-} | \slashed{k}_3| K_2^{\flat -} \rangle} \: , \\[3mm]
& \hspace{-48.3mm} Q_1^\bullet = -\frac{\gamma_1 (S_2 + \gamma_1) \langle K_5^\flat
K_1^\flat \rangle [K_1^\flat K_5^\flat]
\hspace{0.6mm}+\hspace{0.6mm} S_1 S_2 (1 + S_1/\gamma_1) \langle
K_5^\flat K_2^\flat \rangle [K_2^\flat K_5^\flat]}{2\left(
\gamma_1 (S_2 + \gamma_1) \langle K_4^\flat K_1^\flat \rangle
[K_1^\flat K_5^\flat] \hspace{0.6mm}+\hspace{0.6mm} S_1 S_2 (1 +
S_1/\gamma_1) \langle K_4^\flat K_2^\flat \rangle [K_2^\flat
K_5^\flat]\right)} \: , \hspace{4mm} Q_2^\bullet = -\frac{\langle
K_1^\flat K_5^\flat \rangle}{2 \langle K_1^\flat K_4^\flat
\rangle}
\end{array}
\end{equation}
where the discriminant appearing in $P_6, P_7, P_6^\bullet,
P_7^\bullet$ is given by
\begin{equation}
\begin{array}{c}
\Delta \hspace{1mm}=\hspace{1mm} \left(\frac{\gamma_1 (S_2 +
\gamma_1)}{\gamma_1^2 - S_1 S_2} K_1^\flat \cdot k_3 - \frac{S_1
S_2 (1 + S_1/\gamma_1)}{\gamma_1^2 - S_1 S_2} K_2^\flat \cdot k_3
-
\frac{1}{2}(s_{13} + s_{23}) \right)^2 \\[2mm]
\hspace{-6mm} + \frac{4 S_1 S_2 (S_1 + \gamma_1)(S_2 +
\gamma_1)}{(\gamma_1^2 - S_1 S_2)^2} (K_1^\flat \cdot k_3)
(K_2^\flat \cdot k_3) \: .
\end{array}
\end{equation}
This heptacut belongs to case III treated in Section
\ref{sec:general_double_box}, and there are thus four kinematical
solutions (shown in Fig. \ref{fig:kinematical_solutions_case_III}).
Parametrizing the loop momenta according to eqs.
(\ref{eq:l1_parametrized})-(\ref{eq:l2_parametrized}), the
on-shell constraints in eqs.
(\ref{eq:on-shell_constraint_1})-(\ref{eq:on-shell_constraint_7})
are solved by setting the parameters equal to the values
\begin{equation}
\begin{array}{ll}
\alpha_1 = \frac{\gamma_1 (S_2 + \gamma_1)}{\gamma_1^2 - S_1 S_2} \: , \hspace{1cm} & \beta_1 =
0\\[2.5mm]
\alpha_2 = \frac{S_1 S_2 (S_1 + \gamma_1)}{\gamma_1 (S_1 S_2 -
\gamma_1^2)} \: , \hspace{1cm} & \beta_2 = 1
\end{array}
\end{equation}
and those given in Fig. \ref{fig:kinematical_solutions_case_III}.
The heptacut double box integral $I_{16,\sigma_1}$ is $\frac{1}{2}
\sum_{\pm} \sum_{i=1}^4 \oint_{\hspace{0.5mm}\Gamma_i}
\hspace{-0.3mm} dz \, J_i^\pm(z)$ where
\begin{equation}
J_i^\pm(z) \hspace{0.5mm}=\hspace{0.5mm} \left\{ \begin{array}{l}
\hspace{-1mm} \left( 32\gamma_2 \left( (S_2 + \gamma_1) \langle
K_4^{\flat -} | \slashed{K}_1^\flat | K_5^{\flat -} \rangle +
\frac{S_1 S_2}{\gamma_1} \left(1 + \frac{S_1}{\gamma_1} \right)
\langle K_4^{\flat -} | \slashed{K}_2^\flat | K_5^{\flat -}
\rangle \right) z(z - Q_1^\bullet) \right)^{-1} \\ \hspace{118mm} \mathrm{for} \hspace{2mm} i=1,2 \\[5mm]
\hspace{-1mm} \left( 32\gamma_2 \left( (S_2 + \gamma_1) \langle
K_5^{\flat -} | \slashed{K}_1^\flat | K_4^{\flat -} \rangle +
\frac{S_1 S_2}{\gamma_1} \left(1 + \frac{S_1}{\gamma_1} \right)
\langle K_5^{\flat -} | \slashed{K}_2^\flat | K_4^{\flat -}
\rangle \right) z(z - Q_1) \right)^{-1} \\ \hspace{118mm}
\mathrm{for} \hspace{2mm} i=3,4
\end{array} \right.
\end{equation}
and where we recall that the $\pm$ in $\gamma_1^\pm, K_{1
\pm}^\flat, Q_1^\pm$ etc. have here been left implicit.

\subsubsection{Heptacut \#7 of the right hand side of eq. (\ref{eq:6gluonampansatz})}

The result of applying heptacut \#7 to the right hand side of eq.
(\ref{eq:6gluonampansatz}) is
\begin{equation}
\frac{1}{8} \sum_{\pm} \sum_{i=1}^4 \oint_{\Gamma_i} dz \, J_i^\pm
(z) K_i^\pm(z)
\end{equation}
where the kernels evaluated on the four kinematical solutions are
{\small \begin{eqnarray} K_1^\pm (z) \hspace{-2.7mm}&=&\hspace{-2.7mm}
c_{16, \sigma_1} - \frac{1}{2} \frac{c_{22,\sigma_1}
P_1^\bullet}{\langle K_1^{\flat -} | \slashed{k}_3 | K_2^{\flat -}
\rangle (P_1^\bullet - P_6^\bullet)(P_1^\bullet - P_7^\bullet)} -
\frac{1}{2} \frac{c_{22,\sigma_9} P_1^\bullet}{\langle K_1^{\flat
-} | \slashed{k}_1 | K_2^{\flat -} \rangle
(P_1^\bullet - P_4^\bullet)(P_1^\bullet - P_5^\bullet)} \\[1mm]
K_2^\pm (z) \hspace{-2.7mm}&=&\hspace{-2.7mm} c_{16, \sigma_1} -
\frac{1}{2} \frac{c_{22,\sigma_1} \alpha_3(z)}{\langle K_1^{\flat
-} | \slashed{k}_3 | K_2^{\flat -} \rangle (\alpha_3(z) -
P_6^\bullet)(\alpha_3(z) - P_7^\bullet)} - \frac{1}{2}
\frac{c_{22,\sigma_9} \alpha_3(z)}{\langle K_1^{\flat -} |
\slashed{k}_1 | K_2^{\flat -} \rangle
(\alpha_3(z) - P_4^\bullet)(\alpha_3(z) - P_5^\bullet)} \nonumber \\[1mm]
&\phantom{=}& \hspace{-15mm}- \frac{c_{18,\sigma_1} \langle
K_1^{\flat -} | \slashed{k}_6 | K_2^{\flat -} \rangle (\alpha_3(z)
- P_1^\bullet) (\alpha_3(z) - P_2^\bullet)}{\langle K_1^{\flat -}
| \slashed{k}_3 | K_2^{\flat -} \rangle (\alpha_3(z) -
P_6^\bullet)(\alpha_3(z) - P_7^\bullet)}
\hspace{-0.3mm}-\hspace{-0.3mm} \frac{c_{18,\sigma_9} \langle
K_1^{\flat -} | \slashed{k}_6 | K_2^{\flat -} \rangle (\alpha_3(z)
- P_1^\bullet) (\alpha_3(z) - P_2^\bullet)}{\langle K_1^{\flat -}
| \slashed{k}_1 | K_2^{\flat
-} \rangle (\alpha_3(z) - P_4^\bullet)(\alpha_3(z) - P_5^\bullet)} \phantom{aaaaaa}\\[4mm]
K_3^\pm(z) \hspace{-2.7mm}&=&\hspace{-2.7mm} \mbox{parity conjugate of
$K_1^\pm(z)$ \hspace{0.3mm} (obtained by applying eqs.
(\ref{eq:parity_conj_rule_1})-(\ref{eq:parity_conj_rule_6}))} \\[2mm]
K_4^\pm(z) \hspace{-2.7mm}&=&\hspace{-2.7mm} \mbox{parity conjugate of
$K_2^\pm(z)$ \hspace{0.3mm} (obtained by applying eqs.
(\ref{eq:parity_conj_rule_1})-(\ref{eq:parity_conj_rule_6}))}
\end{eqnarray}}
{\vskip -4mm}

\noindent where $\alpha_3(z)$ is given in Fig.
\ref{fig:kinematical_solutions_case_III}.

\subsubsection{Heptacut \#7 of the left hand side of eq. (\ref{eq:6gluonampansatz})}

The result of applying heptacut \#7 to the left hand side of eq.
(\ref{eq:6gluonampansatz}) is
\begin{equation}
\frac{i}{2} \sum_{\pm} \sum_{i=1}^4 \oint_{\Gamma_i} dz \,
J_i^\pm (z) \left. \prod_{j=1}^6 A_j^\mathrm{tree}(z)
\right|_{\mathcal{S}_i}
\end{equation}
where, assuming without loss of generality the external helicities are $(1^-, 2^-, 3^+, 4^+, 5^+, 6^+)$,
the cut amplitude evaluated on the four different
kinematical solutions yields
\begin{equation}
\left. \prod_{j=1}^6 A_j^\mathrm{tree}(z) \right|_{\mathcal{S}_i}
\hspace{3mm}=\hspace{3mm} 0 \hspace{9mm} \mathrm{for} \hspace{4mm}
i = 1,\ldots, 4 \: .
\end{equation}

\clearpage

\subsection{Heptacut \#8}\label{sec:heptacut_8}

This heptacut is defined by the on-shell constraints in eqs.
(\ref{eq:on-shell_constraint_1})-(\ref{eq:on-shell_constraint_7})
with the vertex momenta
\begin{equation}
\begin{array}{lll}
K_1 = k_1 & \hspace{1.2cm} K_2 = k_2 & \hspace{1.2cm} K_3 = 0 \: \phantom{.} \\
K_4 = k_3 & \hspace{1.2cm} K_5 = k_4 & \hspace{1.2cm} K_6 = k_{56}
\: .
\end{array}
\end{equation}
Applying this heptacut to the right hand side of eq.
(\ref{eq:6gluonampansatz}) leaves the following linear combination
of cut integrals

\begin{figure}[!h]
\begin{center}
\includegraphics[angle=0, width=0.9\textwidth]{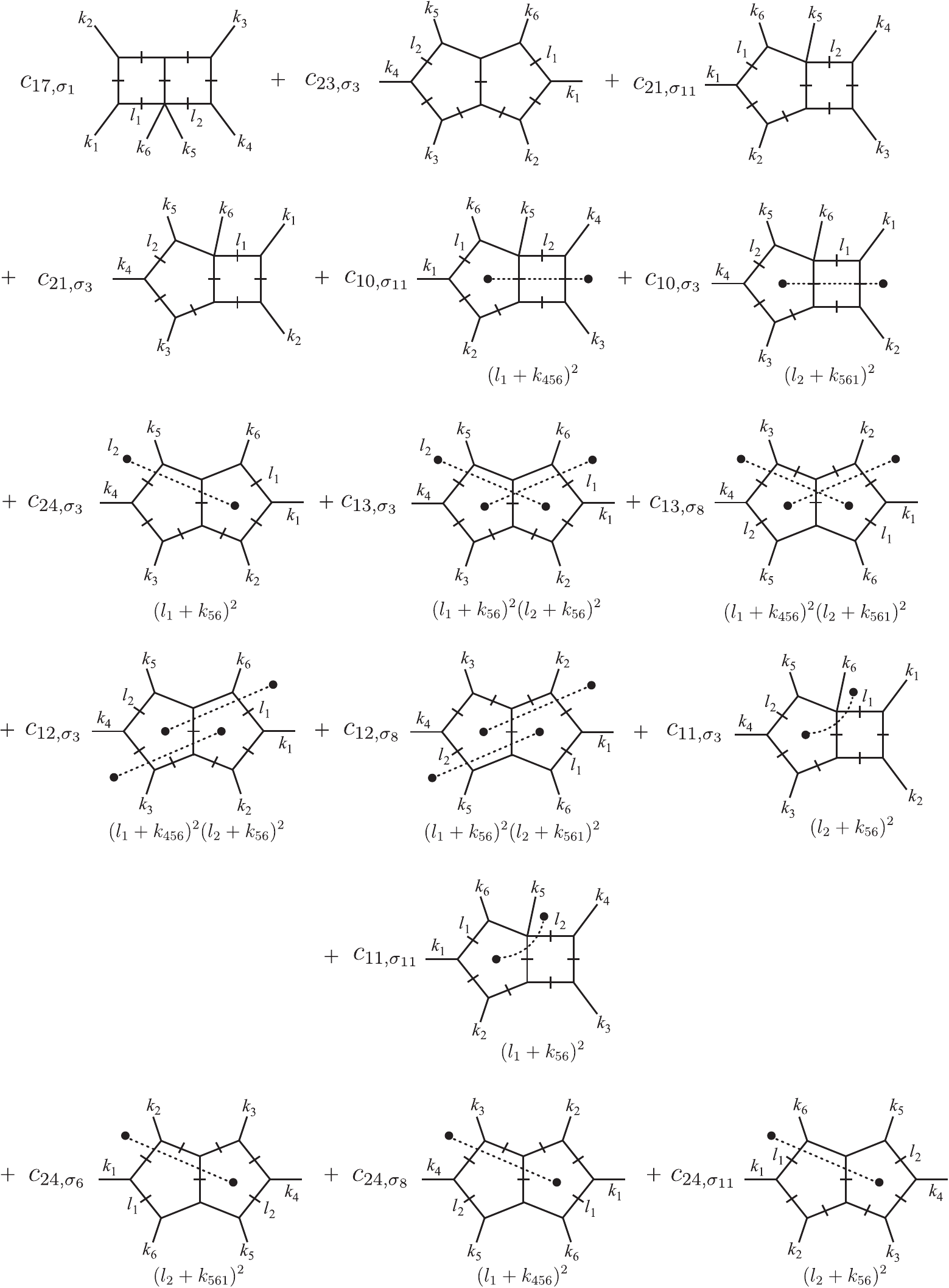}
\end{center}
\end{figure}
\vspace{-0.5cm}

\noindent We define the spinor ratios
\begin{equation}
\begin{array}{lll} P_1 = -\frac{K_5^\flat \cdot K_6 +
\frac{1}{2} S_6 + (K_5^\flat + K_6) \cdot K_1^\flat}{\langle
K_2^{\flat -} | \slashed{K}_5^\flat + \slashed{K}_6 | K_1^{\flat
-} \rangle} \: , \hspace{0.2cm} & P_2 = - \frac{K_1^\flat \cdot
k_6}{\langle K_2^{\flat -}| \slashed{k}_6 | K_1^{\flat -} \rangle}
\: , \hspace{0.2cm} & P_3 = -\frac{K_1^\flat \cdot k_{56} +
\frac{1}{2} s_{56}}{\langle
K_2^{\flat -} | \slashed{k}_{56}| K_1^{\flat -} \rangle} \\
P_4 = -\frac{\langle K_4^{\flat -} | \slashed{K}_1^\flat + \slashed{K}_6 |
K_5^{\flat -} \rangle}{2 \langle K_2^\flat K_4^\flat \rangle [K_5^\flat K_1^\flat]} & & \\
\vspace{-0.2cm} \\ Q_1 = -\frac{K_1^\flat \cdot K_6 + \frac{1}{2}
S_6 + (K_1^\flat + K_6) \cdot K_5^\flat}{\langle K_5^{\flat
-}|\slashed{K}_1^\flat + \slashed{K}_6 | K_4^{\flat -} \rangle} \:
, \hspace{0.2cm} & Q_2 = -\frac{K_5^\flat \cdot k_5}{\langle
K_5^{\flat -} | \slashed{k}_5 | K_4^{\flat -} \rangle} \: ,
\hspace{0.2cm} & Q_3 = - \frac{K_5^\flat \cdot k_{56} +
\frac{1}{2} s_{56}}{\langle
K_5^{\flat -} | \slashed{k}_{56} | K_4^{\flat -} \rangle}  \\
\end{array}
\end{equation}
and their parity conjugates
\begin{equation}
\begin{array}{lll} P_1^\bullet = -\frac{K_5^\flat \cdot K_6 +
\frac{1}{2} S_6 + (K_5^\flat + K_6) \cdot K_1^\flat}{\langle
K_1^{\flat -} | \slashed{K}_5^\flat + \slashed{K}_6 | K_2^{\flat
-} \rangle} \: , \hspace{0.2cm} & P_2^\bullet = - \frac{K_1^\flat
\cdot k_6}{\langle K_1^{\flat -}| \slashed{k}_6 | K_2^{\flat -}
\rangle} \: , \hspace{0.2cm} & P_3^\bullet = -\frac{K_1^\flat
\cdot k_{56} + \frac{1}{2} s_{56}}
{\langle K_1^{\flat -} | \slashed{k}_{56} | K_2^{\flat -} \rangle} \\
P_4^\bullet = -\frac{\langle K_5^{\flat -} | \slashed{K}_1^\flat +
\slashed{K}_6 |
K_4^{\flat -} \rangle}{2 [K_2^\flat K_4^\flat] \langle K_5^\flat K_1^\flat \rangle} & &  \\
\vspace{-0.2cm} \\ Q_1^\bullet = -\frac{K_1^\flat \cdot K_6 +
\frac{1}{2} S_6 + (K_1^\flat + K_6) \cdot K_5^\flat}{\langle
K_4^{\flat -}|\slashed{K}_1^\flat + \slashed{K}_6 | K_5^{\flat -}
\rangle} \: , \hspace{0.2cm} & Q_2^\bullet = -\frac{K_5^\flat
\cdot k_5}{\langle K_4^{\flat -} | \slashed{k}_5 | K_5^{\flat -}
\rangle} \: , \hspace{0.2cm} & Q_3^\bullet = - \frac{K_5^\flat
\cdot k_{56} + \frac{1}{2} s_{56}}{\langle K_4^{\flat -} |
\slashed{k}_{56} | K_5^{\flat -} \rangle} \: . \hspace{0.2cm}
\end{array}
\end{equation}
This heptacut belongs to case I treated in Section
\ref{sec:general_double_box}, and there are thus six kinematical
solutions (shown in Fig. \ref{fig:kinematical_solutions_case_I}).
Parametrizing the loop momenta according to eqs.
(\ref{eq:l1_parametrized})-(\ref{eq:l2_parametrized}), the
on-shell constraints in eqs.
(\ref{eq:on-shell_constraint_1})-(\ref{eq:on-shell_constraint_7})
are solved by setting the parameters equal to the values
\begin{equation}
\begin{array}{ll}
\alpha_1 = 1 \: , \hspace{1cm} & \beta_1 = 0\\
\alpha_2 = 0 \: , \hspace{1cm} & \beta_2 = 1
\end{array}
\end{equation}
and those given in Fig. \ref{fig:kinematical_solutions_case_I}
with
\begin{equation}
\beta_3(z) = -\frac{\langle K_2^{\flat -} | \slashed{K}_5^\flat +
\slashed{K}_6 | K_1^{\flat -} \rangle (z - P_1)}{2 \langle
K_2^\flat K_4^\flat \rangle [K_5^\flat K_1^\flat] (z - P_4)}
\label{eq:beta_3_heptacut_8}
\end{equation}
for kinematical solution $\mathcal{S}_5$. The heptacut double box
integral $I_{17,\sigma_1}$ is $\sum_{i=1}^6
\oint_{\hspace{0.5mm}\Gamma_i} \hspace{-0.3mm} dz \, J_i(z)$ where
\begin{equation}
J_i(z) \hspace{2mm} = \hspace{2mm} \frac{1}{32 \gamma_1 \gamma_2}
\times \left\{ \begin{array}{ll} \left( \langle K_1^{\flat -} |
\slashed{K}_5^\flat + \slashed{K}_6 | K_2^{\flat -} \rangle
\hspace{0.6mm} z (z-P_1^\bullet) \right)^{-1} & \hspace{4mm} \mathrm{for} \hspace{3mm} i=2,6 \\
\left( \langle K_2^{\flat -} | \slashed{K}_5^\flat + \slashed{K}_6
| K_1^{\flat -} \rangle
\hspace{0.6mm} z (z-P_1) \right)^{-1} & \hspace{4mm} \mathrm{for} \hspace{3mm} i=4,5 \\
\left( \langle K_4^{\flat -} | \slashed{K}_1^\flat + \slashed{K}_6
| K_5^{\flat -} \rangle
\hspace{0.6mm} z (z-Q_1^\bullet) \right)^{-1} & \hspace{4mm} \mathrm{for} \hspace{3mm} i=1 \\
\left( \langle K_5^{\flat -} | \slashed{K}_1^\flat + \slashed{K}_6
| K_4^{\flat -} \rangle \hspace{0.6mm} z (z-Q_1) \right)^{-1} &
\hspace{4mm} \mathrm{for} \hspace{3mm} i=3 \: .
\end{array} \right.
\end{equation}

\subsubsection{Heptacut \#8 of the right hand side of eq. (\ref{eq:6gluonampansatz})}

The result of applying heptacut \#8 to the right hand side of eq.
(\ref{eq:6gluonampansatz}) is
\begin{equation}
\frac{1}{4} \sum_{i=1}^6 \oint_{\Gamma_i} dz \, J_i (z) K_i(z)
\end{equation}
where the kernels evaluated on the six kinematical solutions are

{\small\begin{eqnarray} K_1(z) \hspace{-2mm}&=&\hspace{-2mm}
c_{17,\sigma_1} + \frac{1}{2} \frac{c_{21,\sigma_{11}}}{\langle
K_1^{\flat -} | \slashed{k}_6 | K_2^{\flat -} \rangle (P_1^\bullet
- P_2^\bullet)} + \frac{1}{2} \frac{c_{21,\sigma_3}}{\langle
K_4^{\flat -} | \slashed{k}_5 | K_5^{\flat -} \rangle (z - Q_2^\bullet)} \nonumber \\[2mm]
&\phantom{=}& \hspace{-13mm} + \frac{1}{4}
\frac{c_{23,\sigma_3}}{\langle K_1^{\flat -} | \slashed{k}_6 |
K_2^{\flat -} \rangle \langle K_4^{\flat -} | \slashed{k}_5 |
K_5^{\flat -} \rangle (P_1^\bullet - P_2^\bullet) (z -
Q_2^\bullet)} + \frac{c_{10,\sigma_3} \langle K_4^{\flat -} |
\slashed{k}_{561} | K_5^{\flat -} \rangle (z -
Q_1^\bullet)}{\langle K_4^{\flat -}|
\slashed{k}_5 | K_5^{\flat -} \rangle (z - Q_2^\bullet)}\nonumber \\[2mm]
&\phantom{=}& \hspace{-13mm} + \frac{1}{2} \frac{c_{24,\sigma_3}
\langle K_1^{\flat -} | \slashed{k}_{56} | K_2^{\flat -} \rangle
(P_1^\bullet - P_3^\bullet) + c_{24,\sigma_6} \langle K_4^{\flat
-} | \slashed{k}_{561} | K_5^{\flat -} \rangle (z - Q_1^\bullet) +
c_{24,\sigma_{11}} \langle K_4^{\flat -} | \slashed{k}_{56} |
K_5^{\flat -} \rangle (z - Q_3^\bullet)}{\langle K_1^{\flat -} |
\slashed{k}_6 | K_2^{\flat -} \rangle \langle K_4^{\flat -} |
\slashed{k}_5 | K_5^{\flat -} \rangle (P_1^\bullet - P_2^\bullet) (z - Q_2^\bullet)} \nonumber \\[2mm]
&\phantom{=}& \hspace{-13mm} + \frac{c_{13,\sigma_3} \langle
K_1^{\flat -} | \slashed{k}_{56} | K_2^{\flat -}\rangle \langle
K_4^{\flat -} | \slashed{k}_{56} | K_5^{\flat -} \rangle
(P_1^\bullet - P_3^\bullet) (z - Q_3^\bullet)}{\langle K_1^{\flat
-} | \slashed{k}_6 | K_2^{\flat -} \rangle \langle K_4^{\flat -} |
\slashed{k}_5 | K_5^{\flat -} \rangle (P_1^\bullet-P_2^\bullet) (z
- Q_2^\bullet)} + \frac{c_{11,\sigma_3} \langle K_4^{\flat -} |
\slashed{k}_{56} | K_5^{\flat -} \rangle (z -
Q_3^\bullet)}{\langle K_4^{\flat -} | \slashed{k}_5 | K_5^{\flat
-} \rangle (z - Q_2^\bullet)} \nonumber \\[2mm]
&\phantom{=}& \hspace{-13mm} + \frac{c_{12,\sigma_8} \langle
K_1^{\flat -} | \slashed{k}_{56} | K_2^{\flat -}\rangle \langle
K_4^{\flat -} | \slashed{k}_{561} | K_5^{\flat -} \rangle
(P_1^\bullet - P_3^\bullet) (z - Q_1^\bullet)}{\langle K_1^{\flat
-} | \slashed{k}_6 | K_2^{\flat -} \rangle \langle K_4^{\flat -} |
\slashed{k}_5 | K_5^{\flat -} \rangle (P_1^\bullet-P_2^\bullet) (z
- Q_2^\bullet)} + \frac{c_{11,\sigma_{11}} \langle K_1^{\flat -} |
\slashed{k}_{56} | K_2^{\flat -} \rangle (P_1^\bullet -
P_3^\bullet)}{\langle K_1^{\flat -} | \slashed{k}_6 | K_2^{\flat
-} \rangle (P_1^\bullet - P_2^\bullet)} \phantom{aaaaaa} \\[20mm]
K_2(z) \hspace{-2mm}&=&\hspace{-2mm} c_{17,\sigma_1} + \frac{1}{2}
\frac{c_{21,\sigma_{11}}}{\langle K_1^{\flat -} | \slashed{k}_6 |
K_2^{\flat -} \rangle (z - P_2^\bullet)} + \frac{1}{2}
\frac{c_{21,\sigma_3}}{\langle
K_4^{\flat -} | \slashed{k}_5 | K_5^{\flat -} \rangle (Q_1^\bullet - Q_2^\bullet)} \nonumber \\[2mm]
&\phantom{=}& \hspace{-11mm} + \frac{1}{4}
\frac{c_{23,\sigma_3}}{\langle K_1^{\flat -} | \slashed{k}_6 |
K_2^{\flat -} \rangle \langle K_4^{\flat -} | \slashed{k}_5 |
K_5^{\flat -} \rangle (z - P_2^\bullet) (Q_1^\bullet -
Q_2^\bullet)} + \frac{c_{10,\sigma_{11}} \langle K_1^{\flat -} |
\slashed{k}_{456} | K_2^{\flat -} \rangle (z -
P_1^\bullet)}{\langle K_1^{\flat -} | \slashed{k}_6 | K_2^{\flat
-} \rangle (z-P_2^\bullet)}\nonumber \\[2mm]
&\phantom{=}& \hspace{-11mm} + \frac{1}{2} \frac{c_{24,\sigma_3}
\langle K_1^{\flat -} | \slashed{k}_{56} | K_2^{\flat -} \rangle
(z - P_3^\bullet) \hspace{0.6mm}+\hspace{0.6mm} c_{24,\sigma_8}
\langle K_1^{\flat -} | \slashed{k}_{456} | K_2^{\flat -} \rangle
(z - P_1^\bullet) \hspace{0.6mm}+\hspace{0.6mm} c_{24,\sigma_{11}}
\langle K_4^{\flat -} | \slashed{k}_{56} | K_5^{\flat -} \rangle
(Q_1^\bullet - Q_3^\bullet)}{\langle K_1^{\flat -} | \slashed{k}_6
| K_2^{\flat -} \rangle \langle K_4^{\flat -} | \slashed{k}_5 |
K_5^{\flat -}
\rangle (z - P_2^\bullet) (Q_1^\bullet - Q_2^\bullet)} \nonumber \\[2mm]
&\phantom{=}& \hspace{-11mm} + \frac{c_{13,\sigma_3} \langle
K_1^{\flat -} | \slashed{k}_{56} | K_2^{\flat -}\rangle \langle
K_4^{\flat -} | \slashed{k}_{56} | K_5^{\flat -} \rangle (z -
P_3^\bullet) (Q_1^\bullet - Q_3^\bullet)}{\langle K_1^{\flat -} |
\slashed{k}_6 | K_2^{\flat -} \rangle \langle K_4^{\flat -} |
\slashed{k}_5 | K_5^{\flat -} \rangle (z-P_2^\bullet) (Q_1^\bullet
- Q_2^\bullet)} + \frac{c_{11,\sigma_3} \langle K_4^{\flat -} |
\slashed{k}_{56} | K_5^{\flat -} \rangle (Q_1^\bullet -
Q_3^\bullet)}{\langle K_4^{\flat -} | \slashed{k}_5
| K_5^{\flat -} \rangle (Q_1^\bullet - Q_2^\bullet)} \nonumber \\[2mm]
&\phantom{=}& \hspace{-11mm} + \frac{c_{12,\sigma_3} \langle
K_1^{\flat -} | \slashed{k}_{456} | K_2^{\flat -}\rangle \langle
K_4^{\flat -} | \slashed{k}_{56} | K_5^{\flat -} \rangle (z -
P_1^\bullet) (Q_1^\bullet - Q_3^\bullet)}{\langle K_1^{\flat -} |
\slashed{k}_6 | K_2^{\flat -} \rangle \langle K_4^{\flat -} |
\slashed{k}_5 | K_5^{\flat -} \rangle (z-P_2^\bullet) (Q_1^\bullet
- Q_2^\bullet)} + \frac{c_{11,\sigma_{11}} \langle K_1^{\flat -} |
\slashed{k}_{56} | K_2^{\flat -} \rangle (z -
P_3^\bullet)}{\langle K_1^{\flat -} | \slashed{k}_6 | K_2^{\flat
-} \rangle (z - P_2^\bullet)} \phantom{aaaaaa} \\[10mm]
K_3(z) \hspace{-2mm}&=&\hspace{-2mm} \mbox{parity conjugate of
$K_1(z)$ \hspace{0.3mm} (obtained by applying eqs.
(\ref{eq:parity_conj_rule_1})-(\ref{eq:parity_conj_rule_6}))} \\[10mm]
K_4(z) \hspace{-2mm}&=&\hspace{-2mm} \mbox{parity conjugate of
$K_2(z)$ \hspace{0.3mm} (obtained by applying eqs.
(\ref{eq:parity_conj_rule_1})-(\ref{eq:parity_conj_rule_6}))}
\end{eqnarray}}

{\small \begin{eqnarray} K_5(z) \hspace{-2mm}&=&\hspace{-2mm}
c_{17,\sigma_1} + \frac{1}{4} \frac{c_{23,\sigma_3}}{\langle
K_2^{\flat -} | \slashed{k}_6 | K_1^{\flat -} \rangle \langle
K_4^{\flat -} | \slashed{k}_5 | K_5^{\flat -} \rangle (z - P_2)
(\beta_3 (z) - Q_2^\bullet)}  \nonumber \\[2mm]
&\phantom{=}& \hspace{-14mm} + \frac{1}{2} \frac{c_{24,\sigma_3}
\langle K_2^{\flat -} | \slashed{k}_{56} | K_1^{\flat -} \rangle
(z - P_3) + c_{24,\sigma_6} \langle K_4^{\flat-} |
\slashed{k}_{561} | K_5^{\flat -} \rangle (\beta_3(z) -
Q_1^\bullet)}{\langle K_2^{\flat -} | \slashed{k}_6 | K_1^{\flat
-} \rangle \langle K_4^{\flat -} | \slashed{k}_5 | K_5^{\flat -}
\rangle (z-P_2) (\beta_3 (z) - Q_2^\bullet)} + \frac{1}{2}
\frac{c_{21,\sigma_{11}}}{\langle K_2^{\flat -} | \slashed{k}_6 |
K_1^{\flat -} \rangle (z - P_2)} \nonumber \\[2mm]
&\phantom{=}& \hspace{-14mm} + \frac{1}{2} \frac{c_{24,\sigma_8}
\langle K_2^{\flat -} | \slashed{k}_{456} | K_1^{\flat -} \rangle
(z - P_1) + c_{24,\sigma_{11}} \langle K_4^{\flat -} |
\slashed{k}_{56} | K_5^{\flat -} \rangle (\beta_3 (z) -
Q_3^\bullet)}{\langle K_2^{\flat -} | \slashed{k}_6 | K_1^{\flat
-} \rangle \langle K_4^{\flat -} | \slashed{k}_5 | K_5^{\flat -}
\rangle (z - P_2) (\beta_3 (z) - Q_2^\bullet)} + \frac{1}{2}
\frac{c_{21,\sigma_3}}{\langle K_4^{\flat -} |
\slashed{k}_5 | K_5^{\flat -} \rangle (\beta_3 (z) - Q_2^\bullet)} \nonumber \\[2mm]
&\phantom{=}& \hspace{-14mm} + \frac{c_{13,\sigma_3} \langle
K_2^{\flat -} | \slashed{k}_{56} | K_1^{\flat -}\rangle \langle
K_4^{\flat -} | \slashed{k}_{56} | K_5^{\flat -} \rangle (z - P_3)
(\beta_3 (z) - Q_3^\bullet)}{\langle K_2^{\flat -} | \slashed{k}_6
| K_1^{\flat -} \rangle \langle K_4^{\flat -} | \slashed{k}_5 |
K_5^{\flat -} \rangle (z-P_2) (\beta_3 (z) - Q_2^\bullet)} +
\frac{c_{10,\sigma_{11}} \langle K_2^{\flat -} | \slashed{k}_{456}
| K_1^{\flat -} \rangle (z - P_1)}{\langle K_2^{\flat -} |
\slashed{k}_6 | K_1^{\flat -} \rangle (z-P_2)}\nonumber \\[2mm]
&\phantom{=}& \hspace{-14mm} + \frac{c_{13,\sigma_8} \langle
K_2^{\flat -} | \slashed{k}_{456} | K_1^{\flat -}\rangle \langle
K_4^{\flat -} | \slashed{k}_{561} | K_5^{\flat -} \rangle (z -
P_1) (\beta_3 (z) - Q_1^\bullet)}{\langle K_2^{\flat -} |
\slashed{k}_6 | K_1^{\flat -} \rangle \langle K_4^{\flat -} |
\slashed{k}_5 | K_5^{\flat -} \rangle (z-P_2) (\beta_3 (z) -
Q_2^\bullet)} + \frac{c_{10,\sigma_3} \langle K_4^{\flat -} |
\slashed{k}_{561} | K_5^{\flat -} \rangle (\beta_3 (z) -
Q_1^\bullet)}{\langle K_4^{\flat -}| \slashed{k}_5 | K_5^{\flat -}
\rangle (\beta_3 (z)
- Q_2^\bullet)} \nonumber \\[2mm]
&\phantom{=}& \hspace{-14mm} + \frac{c_{12,\sigma_3} \langle
K_2^{\flat -} | \slashed{k}_{456} | K_1^{\flat -}\rangle \langle
K_4^{\flat -} | \slashed{k}_{56} | K_5^{\flat -} \rangle (z - P_1)
(\beta_3 (z) - Q_3^\bullet)}{\langle K_2^{\flat -} | \slashed{k}_6
| K_1^{\flat -} \rangle \langle K_4^{\flat -} | \slashed{k}_5 |
K_5^{\flat -} \rangle (z-P_2) (\beta_3 (z) - Q_2^\bullet)} +
\frac{c_{11,\sigma_3} \langle K_4^{\flat -} | \slashed{k}_{56} |
K_5^{\flat -} \rangle (\beta_3 (z) - Q_3^\bullet)}{\langle
K_4^{\flat -} | \slashed{k}_5 | K_5^{\flat -} \rangle (\beta_3 (z) - Q_2^\bullet)} \nonumber \\[2mm]
&\phantom{=}& \hspace{-14mm} + \frac{c_{12,\sigma_8} \langle
K_2^{\flat -} | \slashed{k}_{56} | K_1^{\flat -}\rangle \langle
K_4^{\flat -} | \slashed{k}_{561} | K_5^{\flat -} \rangle (z -
P_3) (\beta_3 (z) - Q_1^\bullet)}{\langle K_2^{\flat -} |
\slashed{k}_6 | K_1^{\flat -} \rangle \langle K_4^{\flat -} |
\slashed{k}_5 | K_5^{\flat -} \rangle (z-P_2) (\beta_3 (z) -
Q_2^\bullet)} + \frac{c_{11,\sigma_{11}} \langle K_2^{\flat -} |
\slashed{k}_{56} | K_1^{\flat -} \rangle (z - P_3)}{\langle
K_2^{\flat -} | \slashed{k}_6 | K_1^{\flat -} \rangle (z - P_2)}
\phantom{aaaaaa} \\[6mm]
K_6(z) \hspace{-2mm}&=&\hspace{-2mm} \mbox{parity conjugate of
$K_5(z)$ \hspace{0.3mm} (obtained by applying eqs.
(\ref{eq:parity_conj_rule_1})-(\ref{eq:parity_conj_rule_6}))}
\end{eqnarray}}

\noindent where $\beta_3 (z)$ is given in eq.
(\ref{eq:beta_3_heptacut_8}).

\subsubsection{Heptacut \#8 of the left hand side of eq. (\ref{eq:6gluonampansatz})}

The result of applying heptacut \#8 to the left hand side of eq.
(\ref{eq:6gluonampansatz}) is
\begin{equation}
i \sum_{i=1}^6 \oint_{\Gamma_i} dz \, J_i (z) \left.
\prod_{j=1}^6 A_j^\mathrm{tree}(z) \right|_{\mathcal{S}_i}
\end{equation}
where, assuming without loss of generality the external helicities are $(1^-, 2^-, 3^+, 4^+, 5^+, 6^+)$,
the cut amplitude evaluated on the six different
kinematical solutions yields
\begin{equation}
\left. \prod_{j=1}^6 A_j^\mathrm{tree}(z) \right|_{\mathcal{S}_i}
\hspace{3mm}=\hspace{3mm} -\frac{i}{16} A^\mathrm{tree}_{--++++}
\times \left\{ \begin{array}{ll} \frac{1}{J_1(z)} \left(
\frac{1}{z} - \frac{1}{z - Q_2^\bullet} \right) & \hspace{3mm} \mathrm{for} \hspace{3mm} i = 1 \\[2mm]
\frac{1}{J_4(z)} \left(
\frac{1}{z} - \frac{1}{z - P_2} \right) & \hspace{3mm} \mathrm{for} \hspace{3mm} i = 4 \\[2mm]
\frac{1}{J_6(z)} \left(
\frac{1}{z} - \frac{1}{z - P_1^\bullet} \right) & \hspace{3mm} \mathrm{for} \hspace{3mm} i = 6 \\[2mm]
0 & \hspace{3mm} \mathrm{for} \hspace{3mm} i = 2,3,5 \: .
\end{array} \right.
\end{equation}

\clearpage

\subsection{Heptacut \#9}\label{sec:heptacut_9}

This heptacut is defined by the on-shell constraints in eqs.
(\ref{eq:on-shell_constraint_factorized_1})-(\ref{eq:on-shell_constraint_factorized_7}).
Applying it to the right hand side of eq.
(\ref{eq:6gluonampansatz}) leaves the following linear combination
of cut integrals

\begin{figure}[!h]
\begin{center}
\includegraphics[angle=0, width=0.9\textwidth]{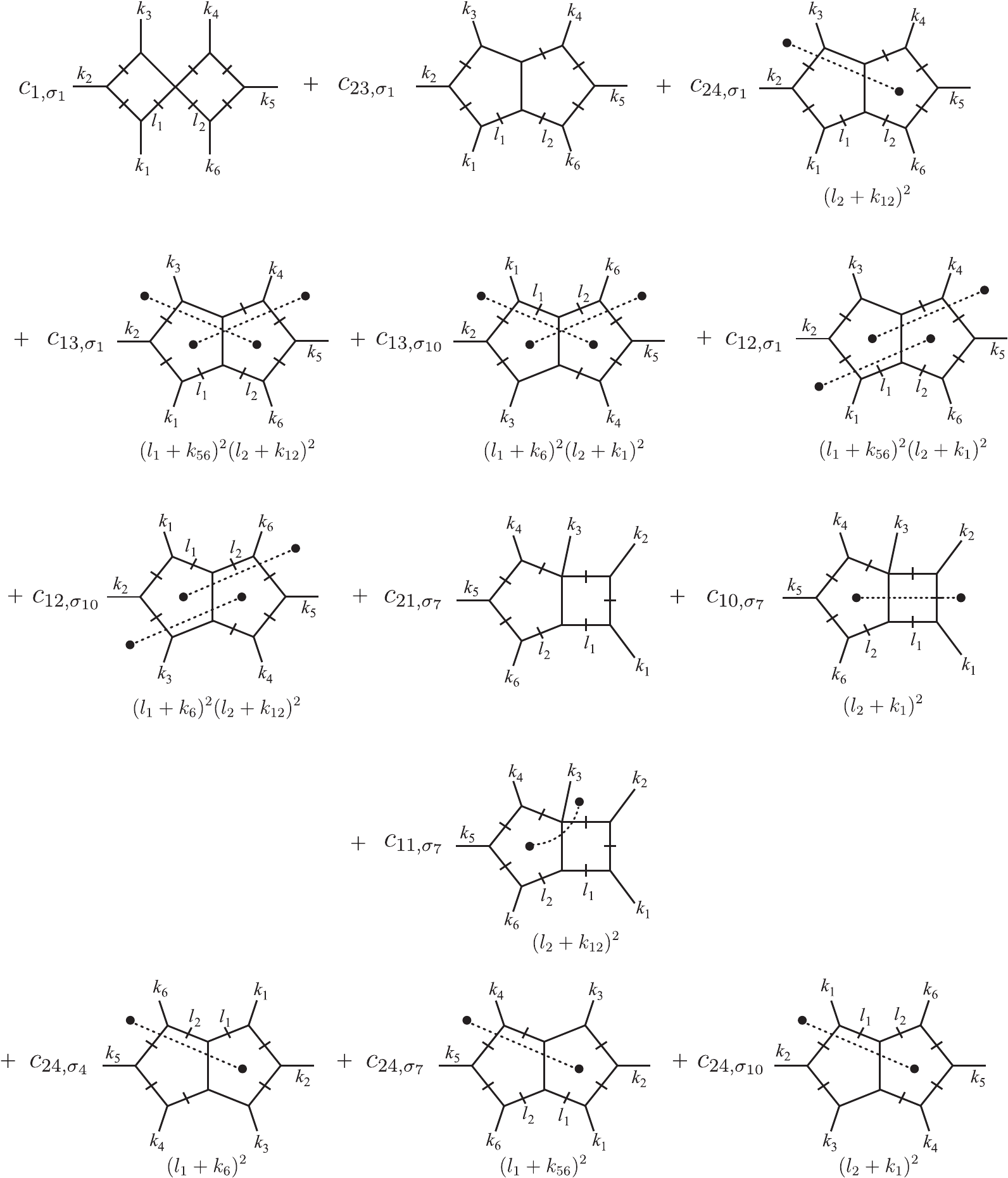}
\end{center}
\end{figure}
\vspace{-0.5cm}

\noindent We define the spinor ratios
\begin{equation}
\begin{array}{lll} P_1 = \frac{[k_3 k_2]}{2[k_3 k_1]} \: ,
\hspace{10mm} & P_2 = - \frac{\langle k_1 k_6 \rangle}{2\langle
k_2 k_6 \rangle} \: , \hspace{10mm} & P_3 = - \frac{k_1 \cdot
k_{56} + \frac{1}{2} s_{56}}{\langle
k_2^- | \slashed{k}_{56} | k_1^- \rangle} \: , \\
P_4 = -\frac{\langle k_1 k_6 \rangle + \frac{[k_5 k_4]}{[k_6 k_4]}
\langle k_1 k_5 \rangle}{2\left(\langle k_2 k_6 \rangle +
\frac{[k_5 k_4]}{[k_6 k_4]} \langle k_2 k_5 \rangle \right)} \hspace{10mm} & & \\ \vspace{-0.2cm} \\
Q_1 = \frac{\langle k_4 k_5 \rangle}{2 \langle k_4 k_6 \rangle} \:
, & Q_2 = -\frac{k_6 \cdot k_{12} + \frac{1}{2} s_{12}}{\langle
k_6^- | \slashed{k}_{12} | k_5^- \rangle} \: , \hspace{10mm} & Q_3
=
-\frac{[k_1 k_6]}{2[k_1 k_5]} \\
\end{array}
\end{equation}
and their parity conjugates
\begin{equation}
\begin{array}{lll} P_1^\bullet = \frac{\langle k_3 k_2 \rangle}
{2\langle k_3 k_1 \rangle} \: , \hspace{10mm} & P_2^\bullet = -
\frac{[k_1 k_6]}{2[k_2 k_6]} \: , \hspace{10mm} & P_3^\bullet = -
\frac{k_1 \cdot k_{56} + \frac{1}{2} s_{56}}{\langle
k_1^- | \slashed{k}_{56} | k_2^- \rangle} \: , \\
P_4^\bullet = -\frac{[k_1 k_6] + \frac{\langle k_5 k_4
\rangle}{\langle k_6 k_4 \rangle} [k_1 k_5]}{2\left([k_2 k_6] +
\frac{\langle k_5 k_4 \rangle}{\langle k_6 k_4 \rangle} [k_2
k_5]\right)} \hspace{10mm} & & \\ \vspace{-0.2cm} \\
Q_1^\bullet = \frac{[k_4 k_5]}{2[k_4 k_6]} \: , & Q_2^\bullet =
-\frac{k_6 \cdot k_{12} + \frac{1}{2} s_{12}}{\langle k_5^- |
\slashed{k}_{12} | k_6^- \rangle} \: , \hspace{10mm} & Q_3^\bullet
= -\frac{\langle k_1
k_6 \rangle}{2 \langle k_1 k_5 \rangle} \: . \\
\end{array}
\end{equation}
This heptacut was treated in Section
\ref{sec:factorized_double_box}, and there are four kinematical
solutions (shown in Fig.
\ref{fig:solutions_to_heptacut_constaints_for_factorized_DB}).
Parametrizing the loop momenta according to eqs.
(\ref{eq:loop_param_factDB1})-(\ref{eq:loop_param_factDB2}), the
on-shell constraints in eqs.
(\ref{eq:on-shell_constraint_factorized_1})-(\ref{eq:on-shell_constraint_factorized_7})
are solved by setting the parameters equal to the values
\begin{equation}
\begin{array}{ll}
\alpha_1 = 1 \: , \hspace{1cm} & \beta_1 = 0\\
\alpha_2 = 0 \: , \hspace{1cm} & \beta_2 = 1
\end{array}
\end{equation}
and those given in Fig.
\ref{fig:solutions_to_heptacut_constaints_for_factorized_DB}. The
Jacobian associated with the heptacut
(\ref{eq:on-shell_constraint_factorized_1})-(\ref{eq:on-shell_constraint_factorized_7})
is
\begin{equation}
J_i(z) \hspace{1mm}=\hspace{1mm} - \frac{1}{16 s_{12} s_{45}
s_{56}} \frac{1}{z} \hspace{10mm} \mathrm{for} \hspace{4mm}
i=1,\ldots,4 \: .
\end{equation}

\subsubsection{Heptacut \#9 of the right hand side of eq. (\ref{eq:6gluonampansatz})}

The result of applying heptacut \#9 to the right hand side of eq.
(\ref{eq:6gluonampansatz}) is
\begin{equation}
\frac{1}{4} \sum_{i=1}^4 \oint_{\Gamma_i} dz \, J_i (z) K_i(z)
\end{equation}
where the kernels evaluated on the four kinematical solutions are

{\small \begin{eqnarray} K_1(z) \hspace{-2mm}&=&\hspace{-2mm}
-\frac{1}{2\langle k_1^- | \slashed{k}_3 | k_2^- \rangle (z -
P_1^\bullet)} \Bigg[ \hspace{0.5mm} c_{1,\sigma_1} + \frac{1}{2}
\left( \langle k_1^- | \slashed{k}_6 | k_2^- \rangle + \langle k_1
k_5 \rangle [k_6 k_2] \frac{[k_5
k_4]}{[k_6 k_4]} \right)^{-1} \frac{1}{z - P_2^\bullet}\nonumber \\[2mm]
&\phantom{=}& \hspace{-13mm} \times \left( c_{23,\sigma_1} +
4c_{13,\sigma_{10}} \langle k_1^- | \slashed{k}_6 | k_2^- \rangle
\langle k_5^- | \slashed{k}_1 | k_6^- \rangle (z - P_2^\bullet)
\left( Q_1^\bullet - Q_3^\bullet \right) \right. \nonumber \\[2mm]
&\phantom{=}& \hspace{2.1cm} + \hspace{0.7mm} 4c_{12,\sigma_1}
\langle k_1^- | \slashed{k}_{56} | k_2^- \rangle \langle k_5^- |
\slashed{k}_1 | k_6^- \rangle (z - P_3^\bullet) \left( Q_1^\bullet - Q_3^\bullet \right) \nonumber \\[2mm]
&\phantom{=}& \hspace{2.1cm} + \hspace{0.7mm} 4c_{12,\sigma_{10}}
\langle k_1^- | \slashed{k}_6 | k_2^- \rangle \langle k_5^- |
\slashed{k}_{12} | k_6^- \rangle (z - P_2^\bullet) \left(
Q_1^\bullet - Q_2^\bullet \right) \nonumber \\[2mm]
&\phantom{=}& \hspace{2.1cm} - 2\langle k_1^- | \slashed{k}_3 |
k_2^- \rangle (z - P_1^\bullet)\left( c_{21,\sigma_7} +
\hspace{0.7mm} 2c_{10,\sigma_7} \langle k_5^- | \slashed{k}_1 |
k_6^-\rangle \left( Q_1^\bullet - Q_3^\bullet \right) \nonumber \right.\\[2mm]
&\phantom{=}& \left. \hspace{6.75cm} + \hspace{1.3mm}
2c_{11,\sigma_7} \langle k_5^- | \slashed{k}_{12} | k_6^-\rangle
\left( Q_1^\bullet - Q_2^\bullet \right) \right) \nonumber \\[2mm]
&\phantom{=}& \hspace{2.1cm} + \hspace{0.7mm} 2 c_{24,\sigma_1}
\langle k_5^- | \slashed{k}_{12} | k_6^- \rangle \left( Q_1^\bullet - Q_2^\bullet \right) \nonumber \\[2mm]
&\phantom{=}& \hspace{2.1cm} + \hspace{0.7mm} 2 c_{24,\sigma_4}
\langle k_1^- | \slashed{k}_6 | k_2^- \rangle (z - P_2^\bullet) +
2 c_{24,\sigma_7} \langle k_1^- | \slashed{k}_{56} | k_2^- \rangle
(z - P_3^\bullet) \nonumber \\[2mm]
&\phantom{=}& \hspace{2.1cm} + \hspace{0.7mm} 2 c_{24,\sigma_{10}}
\langle k_5^- | \slashed{k}_1 | k_6^- \rangle \left( Q_1^\bullet - Q_3^\bullet \right) \nonumber \\
&\phantom{=}& \left. \hspace{2.1cm} + \hspace{0.7mm} 4
c_{13,\sigma_1} \langle k_1^- | \slashed{k}_{56} | k_2^- \rangle
\langle k_5^- | \slashed{k}_{12} | k_6^- \rangle (z-P_3^\bullet)
\left( Q_1^\bullet - Q_2^\bullet \right) \right) \Bigg]
\end{eqnarray}}

{\small \begin{eqnarray} K_2 (z) \hspace{-2mm}&=&\hspace{-2mm}
-\frac{1}{2\langle k_1^- | \slashed{k}_3 | k_2^- \rangle (z -
P_1^\bullet)} \Bigg[ \hspace{0.5mm} c_{1,\sigma_1} + \frac{1}{2}
\left( \langle k_1^- | \slashed{k}_6 | k_2^- \rangle + \langle k_1
k_6 \rangle [k_5 k_2] \frac{\langle
k_5 k_4 \rangle}{\langle k_6 k_4\rangle} \right)^{-1} \frac{1}{z - P_4^\bullet}\nonumber \\[2mm]
&\phantom{=}& \hspace{0.6cm} \times \left( c_{23,\sigma_1} +
4c_{13,\sigma_{10}} \langle k_1^- | \slashed{k}_6 | k_2^- \rangle
\langle k_6^- | \slashed{k}_1 | k_5^- \rangle (z - P_2^\bullet)
\left( Q_1 - Q_3 \right) \right. \nonumber \\[2mm]
&\phantom{=}& \hspace{2.1cm} + \hspace{0.7mm} 4c_{12,\sigma_1}
\langle k_1^- | \slashed{k}_{56} | k_2^- \rangle \langle k_6^- |
\slashed{k}_1 | k_5^- \rangle (z - P_3^\bullet) \left( Q_1 - Q_3 \right) \nonumber \\[2mm]
&\phantom{=}& \hspace{2.1cm} + \hspace{0.7mm} 4c_{12,\sigma_{10}}
\langle k_1^- | \slashed{k}_6 | k_2^- \rangle \langle k_6^- |
\slashed{k}_{12} | k_5^- \rangle (z - P_2^\bullet) \left( Q_1 - Q_2 \right) \nonumber \\[2mm]
&\phantom{=}& \hspace{2.1cm} - 2\langle k_1^- | \slashed{k}_3 |
k_2^- \rangle (z - P_1^\bullet)\left( c_{21,\sigma_7} +
\hspace{0.7mm} 2c_{10,\sigma_7} \langle k_6^- | \slashed{k}_1 |
k_5^-\rangle \left( Q_1 - Q_3 \right) \nonumber \right.\\[2mm]
&\phantom{=}& \left. \hspace{6.75cm} + \hspace{1.3mm}
2c_{11,\sigma_7} \langle k_6^- | \slashed{k}_{12} | k_5^-\rangle
\left( Q_1 - Q_2 \right) \right) \nonumber \\[2mm]
&\phantom{=}& \hspace{2.1cm} + \hspace{0.7mm} 2 c_{24,\sigma_1}
\langle k_6^- | \slashed{k}_{12} | k_5^- \rangle \left( Q_1 - Q_2 \right) \nonumber \\[2mm]
&\phantom{=}& \hspace{2.1cm} + \hspace{0.7mm} 2 c_{24,\sigma_4}
\langle k_1^- | \slashed{k}_6 | k_2^- \rangle (z - P_2^\bullet) +
2 c_{24,\sigma_7} \langle k_1^- | \slashed{k}_{56} | k_2^- \rangle
(z - P_3^\bullet) \nonumber \\[2mm]
&\phantom{=}& \hspace{2.1cm} + \hspace{0.7mm} 2 c_{24,\sigma_{10}}
\langle k_6^- | \slashed{k}_1 | k_5^- \rangle \left( Q_1 - Q_3 \right) \nonumber \\
&\phantom{=}& \left. \hspace{2.1cm} + \hspace{0.7mm} 4
c_{13,\sigma_1} \langle k_1^- | \slashed{k}_{56} | k_2^- \rangle
\langle k_6^- | \slashed{k}_{12} | k_5^- \rangle (z-P_3^\bullet)
\left( Q_1 - Q_2 \right) \right) \Bigg] \\[4mm]
K_3(z) \hspace{-2mm}&=&\hspace{-2mm} \mbox{parity conjugate of
$K_2(z)$ \hspace{0.3mm} (obtained by applying eqs.
(\ref{eq:parity_conj_rule_1})-(\ref{eq:parity_conj_rule_6}))} \\[4mm]
K_4(z) \hspace{-2mm}&=&\hspace{-2mm} \mbox{parity conjugate of
$K_1(z)$ \hspace{0.3mm} (obtained by applying eqs.
(\ref{eq:parity_conj_rule_1})-(\ref{eq:parity_conj_rule_6}))} \: .
\end{eqnarray}}

\subsubsection{Heptacut \#9 of the left hand side of eq. (\ref{eq:6gluonampansatz})}

The result of applying heptacut \#9 to the left hand side of eq.
(\ref{eq:6gluonampansatz}) is
\begin{equation}
i \sum_{i=1}^4 \oint_{\Gamma_i} dz \, J_i (z) \left.
\prod_{j=1}^6 A_j^\mathrm{tree}(z) \right|_{\mathcal{S}_i}
\end{equation}
where, assuming without loss of generality the external helicities are $(1^-, 2^-, 3^+, 4^+, 5^+, 6^+)$,
the cut amplitude evaluated on the four different
kinematical solutions yields
\begin{equation}
\left. \prod_{j=1}^6 A_j^\mathrm{tree}(z) \right|_{\mathcal{S}_i}
\hspace{3mm}=\hspace{3mm} \frac{i}{16} A^\mathrm{tree}_{--++++}
\times \left\{ \begin{array}{ll} \frac{1}{J_2(z)} \left(
\frac{1}{z } - \frac{1}{z - P_1^\bullet} \right) & \hspace{3mm} \mathrm{for} \hspace{3mm} i = 2 \\[2mm]
\frac{1}{J_4(z)} \left(
\frac{1}{z} - \frac{1}{z - P_2} \right) & \hspace{3mm} \mathrm{for} \hspace{3mm} i = 4 \\[2mm]
0 & \hspace{3mm} \mathrm{for} \hspace{3mm} i = 1,3 \: .
\end{array} \right.
\end{equation}

\end{document}